\pdfoutput=1
\documentclass[double]{usthesis}

\usepackage[ascii]{inputenc}
\usepackage[T1]{fontenc}
\usepackage[english]{babel}
\usepackage{amsmath,amssymb,amsfonts,textcomp}
\usepackage{calc}
\usepackage[table]{xcolor}
\usepackage{setspace}
\usepackage[numbers,sort&compress]{natbib}
\usepackage{acronym}
\usepackage{units}
\usepackage[pdftex]{graphicx}
\usepackage{pdflscape}
\usepackage{epigraph}
\usepackage[boxed,linesnumbered]{algorithm2e}
\usepackage{booktabs}
\usepackage{tikz}
\usepackage{pgfplots}
\usepackage{subfig}
\usepackage{fancyref}
\usepackage{listings}
\usepackage{longtable}
\usepackage{colortbl}
\usepackage{array}
\usepackage{dcolumn}
\usepackage{fancyhdr}

\usepackage[colorlinks,linkcolor=black, citecolor=gray, pdfusetitle,plainpages=false, a4paper]{hyperref}

\newcolumntype{L}[1]{>{\raggedright\arraybackslash}p{#1}}
\newcolumntype{R}[1]{>{\raggedleft\arraybackslash}p{#1}}
\definecolor{dkgreen}{rgb}{0,0.6,0}
\usetikzlibrary{shapes,backgrounds,matrix}
\newcommand\textstyleHTMLCode[1]{\texttt{#1}}
\newcommand\Maxima[1]{\texttt{#1}}
\renewcommand\theenumi{\arabic{enumi}}

\renewcommand\labelenumi{\theenumi.}

\newcommand{\degree}{\ensuremath{^\circ}\xspace}
\newcommand\textstyleYear[1]{\textbf{#1.}\,\,\,}
\renewcommand{\vec}[1]{\boldsymbol{#1}}

\newcolumntype{A}{>{\columncolor[rgb]{0.7,0.9,1}\arraybackslash}c}
\newcolumntype{B}{>{\columncolor[rgb]{1,0.85,1}\arraybackslash}c}
\newcolumntype{C}{>{\columncolor[rgb]{0.8,1,0.8}\arraybackslash}c}

\usepackage[avantgarde]{quotchap}
\usepackage[calcwidth]{titlesec}

\fancypagestyle{plain}{
\fancyhf{}
\fancyhead[R]{\color{gray}\sffamily\bfseries \rightmark}
\fancyfoot[C]{\bfseries \thepage}

}

\title{Direct-Write Digital Holography}
\subtitle{Development and research of a hologram printer}
\author{John Peter Tapsell}
\qualification{Doctor of Philosophy}
\submitdate{April 2008}

\begin{document}

\Opening
\pagestyle{fancy}
\renewcommand\chapterheadstartvskip{\vspace*{0\baselineskip}}
\newpage
\pagenumbering{arabic}

\chapter{Introduction}
\label{chap:intro}
\section{Holography}

To watch a person's first interaction with a hologram is a truly fascinating experience. They cannot help but try to reach out and touch the object that they can see projected out in front of them, but know is not really there.  A large full colour hologram can be a beautiful, albeit expensive, piece of artwork, projecting a \ac{3D} holographic image producing all of the visual impressions of depth and realism that is found in real scenes.  Holography has traditionally required a real object from which to make the hologram, but recent advances in digital holography have allowed the production of realistic holographic images of scenes and objects that exist only in the mind of the artist.  Digital holography is already being used in numerous applications - from artwork, to product advertising, to data visualization.

Holography is the technique of recording \ac{3D} images by recording and replaying optical wavefronts.  \citet{gabor1965rpo} developed the mathematical toolkit for holography just under 50 years ago.  It took a further 20 years for technology to advance sufficiently to allow researchers to reliably produce \ac{3D} holograms using a laser\citep{gabor1972h}.  Optical holographic imaging is the traditional technique of first illuminating a firm rigid object with a coherent laser source.  The scattered light from the object falls upon a coated plate, interfering with a second mutually-coherent beam.  The resultant microscopic interference pattern is recorded by the high resolution light sensitive emulsion on the plate\citep{gabor1972h, gabor1968tdh}.  After chemical development, the plate can be viewed to reveal a \ac{3D} image of the original object, on a 1:1 scale.

\bigskip

With the goal of producing holograms that were not 1:1 scale images of real objects, researchers looked at using \ac{LCD} displays as a replacement of the real object.  Although these produce an inherently \ac{2D} image, by printing millions of different \ac{2D} images in a dot-matrix style with a suitable image transformation, a \ac{3D} or \ac{2.5D} hologram image can be composed.

\bigskip

This thesis investigates possible improvements to an existing digital monochrome holographic printer.  The architecture of a digitally-based holographic printer is directed by three main goals: (1) To produce holograms of models created on a computer by an artist with no special knowledge of holography, (2) to have as high a resolution, contrast and fidelity as possible, and (3) to produce said holograms at a commercially viable rate and cost.
\clearpage
\section{Overview of thesis}
The following original research is documented in this thesis dissertation:

\begin{itemize}
	\item A description of a digital hologram printer, intended to aid with the production of a practical hologram printer.
	\item An analytical examination of the two main components of the lens system, allowing for the final pixel size and shape to be predicted accurately, and thus predicting the energy density on the holographic film.  The problem of speckle is also addressed and qualitatively assessed.  This knowledge becomes crucial for increasing the resolution, and thus decreasing the pixel size.
	\item Mechanical and software improvements to a holographic printer design based upon a sensitivity analysis, qualitative experience, and quantitative research.
	\item The architecture and implementation of a temperature-energy feedback system designed to improve stability of the pulsed laser, a key component in the holographic printer.
	\item A case study analysis of the unwanted side effects of the angular intensity distribution of a hologram pixel on its apparent intensity.
	\item Demonstration of improvements to produce a high resolution hologram, recorded at \unit[532]{nm}.
	\item Demonstration of the feasibility of using a high contrast reflective \ac{LCOS} display system over the older lower contrast transmissive \ac{LCD} display system.
\end{itemize}

The following \textbf{Background} chapter provides a brief historical overview of holography, focussing on the path that led to digital holography, as well as a more detailed look at recent research in the field of digital holographic printers.

\chapter{Background}
\label{chapter:background}

\section{Definition}
What is a hologram? The term hologram has been used (often incorrectly) to mean many
things {--} encompassing everything from a projected \ac{3D} image floating
in the air, to lenticular posters, to sparkly Christmas wrapping paper.

The Merriam{}-Webster dictionary defines it as:
\begin{quote}
`A three-dimensional image
reproduced from a pattern of interference produced by a split coherent
beam of radiation (as a laser); '
\end{quote}

For the purposes of this thesis, a hologram is defined to mean that the viewer can
see an apparently \acl{3D} virtual image with at least horizontal parallax when looking
at a hologram device with the image being produced from a pattern of
interference.

For a better understanding of how a hologram works, a brief look at the history and theory
of traditional photography is required.

\section{A brief introduction to photography}
A traditional photograph is created when ordinary white light is
captured by a light sensitive emulsion attached to some substrate. A
camera captures light that is emitted or reflected from the target
objects and is focused by a series of lenses to create a real image on
the emulsion. The emulsion is made of a light sensitive mixture
(typically involving silver or chalk) that undergoes a chemical
reaction whose reaction rate has some approximately linear correlation to the intensity of the light
incident upon it. The film can then undergo wet chemical processing to make a
permanent image.

In this way, the film captures the intensity of the light. Phase
information about the light is discarded, requiring that a particular
point in the scene is set permanently as the focus point. 

\bigskip

\textstyleYear{1724}J.H. Schultz discovered that
some silver salts, for example silver chloride and silver nitrate,
darken when exposed to light.

\begin{figure}[htp]
\centering
\includegraphics[width=10cm]{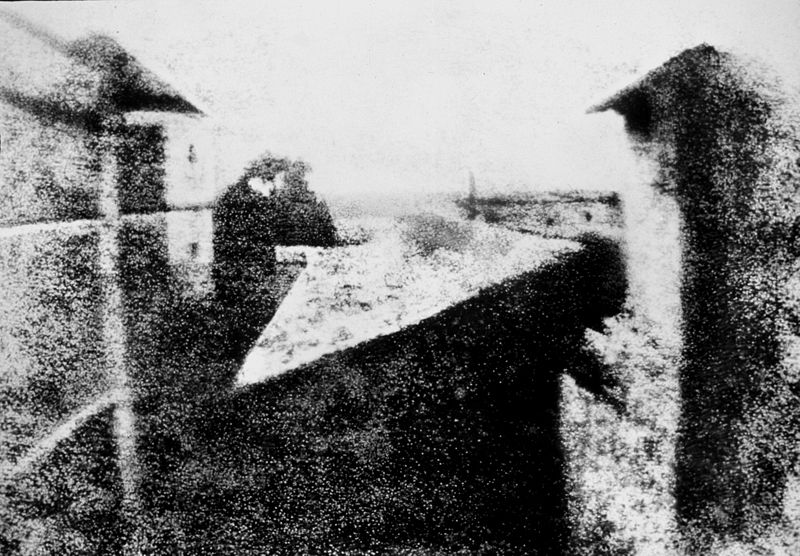}
\caption[First permanent photograph, by Ni\'epce in 1826]{First permanent photograph, by Ni\'epce in 1826 (Public domain) }
\label{fig:firstphotograph}
\end{figure}

\clearpage

\textstyleYear{1826}It took over a hundred years before
Joseph Nic\'ephore Ni\'epce used the silver salts to make a permanent photograph
based on these principles (See \Fref{fig:firstphotograph}). It took 8 hours to expose
this photograph.

\bigskip

\textstyleYear{1840}\citet{talbot1841aca} developed a process
to create a negative image first, creating the today{}-common
nomenclature `photograph', `negative' and `positive'.

\bigskip

\textstyleYear{1851}\citet{archer1852cp} discovered\footnote{For historical accuracy, I feel obliged to note that the Collodion process was first suggested by Mr. Le Gray\citep{archer1852cp} but published by Archer first.} by using a process he coined the Wet Plate Collodion process the exposure time could be drastically reduced.

\bigskip

\textstyleYear{1884}\citet{1884pf} invented the photographic
film, replacing the photographic plate.

\bigskip

\textstyleYear{1975}Analogue photography does not change
fundamentally until modern digital photography is invented. \citet{dillon1976icf} uses solid{}-state \ac{CCD} image sensor chips to capture an image.

\bigskip

\textstyleYear{1990}The first true commercial digital
camera is released {--} the Dycam Model 1; using an improved \ac{CCD} image
sensor. 

\section{History of holography}
This section looks at development of the field of holography.
Holograms in many different forms have been developed in the second
half of the 20\textsuperscript{th} century. The basic idea idea has
always been the same {--} to create a diffraction grating on some
material in a way that ultimately ends up with a viewable picture.

Arguably a hologram is synonymous with a diffraction grating {--} at the
technical level they are the same thing. But the distinction is
analogous to the difference between a painting and a piece of paper
with paint on it. The latter is technically the same as the painting,
but may contain no picture or image that can be interpreted by a human
eye.

Because of the technical similarity between a hologram and a diffraction
grating, the advances of each tend to go hand-in-hand. So to give
an explanation of holography, a brief description of a diffraction
grating is in order.

\bigskip

A diffraction grating is a surface covered by a pattern of parallel
lines, with the distance between the lines ideally of the order of the
wavelength of visible light. Light incident on the diffraction
grating is bent or reflected due to diffraction, or absorbed.  The
light acts as a wave, to produce a image or a pattern. Most
literature considers diffraction gratings with a series of regularly
space lines, producing just a change in angle of the light. However
throughout this thesis the term diffraction grating will refer
to a more general grating with an arbitrary{}-spaced series of lines.

\bigskip

A brief history of the diffraction grating:

\bigskip

\textstyleYear{Approx 1660}James Gregory noticed that bird
feathers produced diffraction patterns, creating iridescent
colors. Reflection diffraction gratings appear quite commonly in
nature, from butterfly wings, to peacock feathers and even many
beetles. Birds, for example, grow their feathers in such a way as to
form a diffraction grating, creating a beautiful visual effect in order
to attract mates.

\bigskip

\textstyleYear{1785}The first man-made diffraction
grating was made around 1785 by David Rittenhouse, who used 50
hairs strung between two finely threaded screws.

\textstyleYear{1803}Thomas Young used two thin slits to
demonstrate that light behaves as a wave, diffracting and
constructively and destructively interfering {--} principles fundamental to the field of holography.

\begin{figure}[htp]
\centering
\includegraphics[width=10cm]{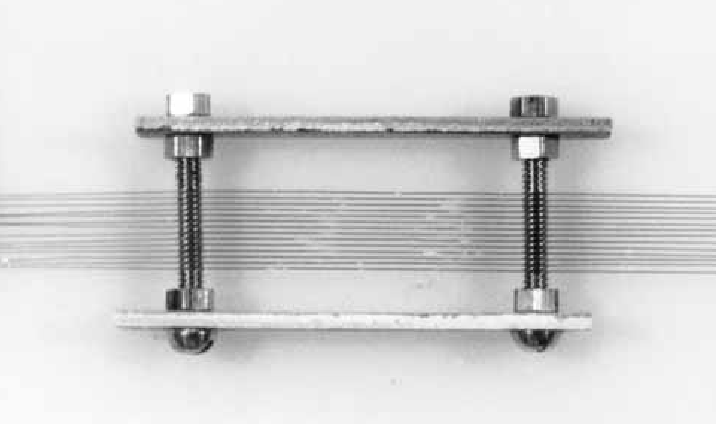}
\caption{Wire diffraction grating, similar to that made by Fraunhofer}
\label{fig:wiregrating}
\end{figure}

\textstyleYear{1821}Joseph von Fraunhofer produced a
diffraction grating using a similar method to that which Rittenhouse used, but using wire (\Fref{fig:wiregrating}). Fraunhofer used his diffraction grating to discover the absorption `Fraunhofer lines' in the solar spectrum \citep{greenslade:76}. 

\bigskip

\begin{figure}[htp]
\centering
\includegraphics[width=5cm]{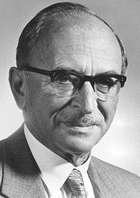}
\caption{Dennis Gabor}
\label{fig:gabor}
\end{figure}

With the wave-like nature of light starting to be understood, the stage is set for invention of holography. This starts with Dennis Gabor (\Fref{fig:gabor})
in 1947.

Dennis Gabor was a brilliant British/Hungarian scientist with an
interest in the way that light behaves. Even as a child he was
fascinated by Abbe's theory of the microscope and by
Gabriel Lippmann's method of color photography. 

Before he had entered university, he had already repeated many modern
(at that time) experiments on wireless X{}-rays and radioactivity, with
his brother George Gabor.

\bigskip

He moved to Germany for his education, and during his time at
university, he invented one of the first high speed cathode ray
oscillographs, and made the first iron{}-shrouded magnetic electron
lens. He worked for a while in this field, but left Germany when
Hitler came into power. He ultimately ended up in England, and
despite the depression managed to find work at a research company,
British Thomson{}-Houston Co., where he would work for many years.

\bigskip

\textstyleYear{1946}During his time at the company, \citet{gabor1946tc}
wrote the first papers on communication theory, as well as many other
subjects. Although on the surface it seems that communication theory
has nothing to do with creating pretty \ac{3D} images, it actually the underlying
theory of how holography works.  \citet{gabor1944spa} went on to first develop stereoscopic
cinematography, and then on to creating basic flat holograms -- although
at the time his was goal was to improve the resolution of the electron
microscope \citep{gabor1965ite,gabor1949mrw,gabor1948nmp}.

\bigskip

The first hologram made by Gabor was an in{}-line plane transmission
hologram \citep{gabor1969ph}. In-line means that the reference beam and object beam
come from the same direction. This was a requirement for Gabor
because his light source was a mercury arc lamp. The light was
filtered (he used the \unit[546]{nm} mercury green line) and squeezed through a
pinhole to make it quasi-coherent. The resulting light had a very
small coherence length -- just enough to make an in{}-line hologram.

\bigskip

A plane hologram, as opposed to a white{}-light hologram, means that the
hologram has to be reconstructed (viewed) in the same monochromatic
light.

\bigskip

A transmission hologram is one which is viewed with a light being
transmitted through the hologram. The replay light (the light to view
it again) has to illuminate the hologram at the same angle, but in the opposite direction, that the reference beam was at when exposing the hologram. The virtual image of the object
appears at the original object position. Since for in{}-line
holograms the object and reference beams come from the same direction,
this meant that to view Gabor's in-line holograms,
the viewing light had to be shining straight into the viewer's eyes, or else projected onto a surface \citep{gabor1969tdp}. And for
Gabor, this viewing light had to be his dim filtered mercury{}-arc
lamp.

\bigskip

Despite the problems with his holograms, \citet{gabor1965rpo} had proved that the
interference pattern carries all the information about the original
object, and that from the interference pattern you can reconstruct the entire image of the original object. It was for these concepts that in 1971 Gabor was
awarded the Nobel Prize for Physics.

\bigskip

Although the first holograms were very interesting, and generated some
talk in the scientific world, Gabor was much too early. Lasers still
had not been invented, and the reliance on mercury arc lamps meant that it
took several hours to expose even a small hologram.

\bigskip

\begin{figure}[htp]
\centering
\includegraphics[width=10cm]{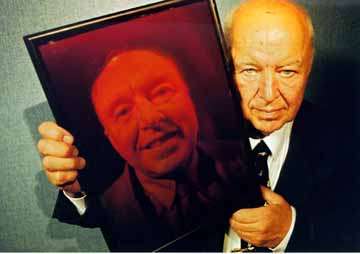}
\caption{Yuri Denisyuk}
\label{fig:denisyuk}
\end{figure}

\textstyleYear{1958}Another key player in the development of the field of
holography was Yuri Denisyuk (\Fref{fig:denisyuk}). Denisyuk was a Russian scientist that
started the work on `interference
photography' (\ac{2D} holograms) in 1958. After several
years he published his work in the Soviet Union, however it was largely
ignored.

An American professor, Emmett Leith, who happened to the same age as
Yuri, independently also created a hologram (\citet{leith1961ntw}). This sparked off an
interest in holography in the US, bringing attention on to
Denisyuk's work.

\bigskip

But Leith's original work was still restricted to using Mercury arc lamps,
holding back development of the field in general. Fortunately this situation
only lasted a couple of years.

\bigskip

\begin{figure}[tp]
\centering
\includegraphics[width=5cm]{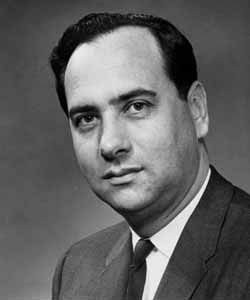}
\caption{Theodore H. Maiman}
\label{fig:maiman}
\end{figure}

\textstyleYear{1960}The laser was invented. Maiman (\Fref{fig:maiman})
created the first laser (\textit{light amplification by stimulated
emission of radiation}) providing a powerful source of the coherent,
monochromatic light\citep{maiman1960sor, maiman1961soe}. This reduced the required hologram exposure
times from many hours to a few seconds, not to mention the huge impact
it had on almost every aspect of science.

Maiman's laser was based on a synthetic ruby crystal,
and built in Hughes Research Laboratory. Its importance was
immediately recognized, and the legal rights were fought over.

\bigskip

Maiman left the company to form his own company, Korad, to build lasers
and to further develop the technology.

\bigskip

\textstyleYear{1962}Leith and Upatnieks experimented with the
lasers to also create transmission holograms. Originally these were
also in{}-line transmission, bringing along all the mentioned problems
with viewing such a hologram. However the laser had a much longer
coherence length, allowing them to independently discover off{}-axis
holography (Gabor had already proposed this some 12-14 years
before). Off-axis holography is a technique where the laser beam is split into an
object and reference beam. The object beam illuminates the object,
and is scattered onto the film. The mutually coherent reference beam is shone onto the
film, from the same side but at a different angle. The two beams
interfere and the interference pattern recorded. This means that on
replay, the reconstructing beam is no longer coming from the same place
as the virtual object.

\citet{Leith:62, leith1964wrd} applied communication theory\citep{gabor1946tc} and the reconstruction process of \citet{gabor1965rpo} to produce a mathematical analysis of wavefront reconstruction in three dimensions.

\bigskip

\textstyleYear{1968}Two-step white light transmission holography, also known as rainbow
holography, is invented by Stephen Benton \citep{benton1977wlt}.  This technique allowed transmission holograms to be viewable in ordinary light.  \citet{benton1980osw} went on to discover one-step white light transmission holography. 

At this stage, holography begins to branch out and to really take
off. The problem with creating the holograms optically is that it
requires an actual physical object, and that the object must be of the
same size as the hologram. The larger the object, the more difficult it is to keep
it stationary (To within an order of the wavelength of light) during an
exposure, and the more powerful the laser required. Objects up to
the size of a car could be just about be created, taking many hours to
expose, but holograms of people and natural soft objects was
impossible. The invention of the pulsed laser, and thus pulsed laser
holography, happened rather quickly, allowing for moving objects to
captured (For example portraits of people by \citet{bjelkhagen1992hpm}).

\bigskip

\citet{debitetto1970b} invents a new method of creating a hologram using slits.
The motivation is that the eye only requires horizontal parallax in
order to see something as `\ac{3D}'.
\citet{1968ApPhL..12..176D} showed that by using a thin vertical slit held against the
film, with a photograph or other image behind it, a realistic hologram
could be made. This technique has an additional benefit that there is
no chromatic aberration due to the vertical parallax.

\bigskip

A basic holographic printer utilizing an \ac{LCD} screen \citep{patent:3832027}
used an optical vibration isolation table with a split{}-beam \ac{CW} laser.
The object beam transmissively illuminates an \ac{LCD} screen and projects the image onto a diffusive
screen. This screen acts as the object in traditional holography 
illuminating the hologram plate. However only a small area of the
hologram plate is illuminated {--} the rest is blocked off with a large slit aperture. The image
is changed on the \ac{LCD} screen, and another area of the hologram plate is
illuminated. In this fashion the hologram plate is exposed to
multiple images {--} each image offering a slightly different
perspective. The images can be either taken by a camera or be computer
generated.

\bigskip

The mutually-coherent reference beam is arranged to simultaneously illuminate the hologram plate, generating the required wave-interference spatial pattern on the light
sensitive emulsion. The area exposed for illumination can be one of 2D array of
rectangles/squares, producing full{}-parallax holograms at the cost of longer
print times, or one of a 1D array of slits, producing single{}-parallax holograms.

This technique however relies on a \ac{CW} laser and a diffusive screen, making
it very sensitive to vibrations.

\bigskip

\citet{patent:4206965} side steps the need for a diffusion screen by
directly exposing the laser light onto the holographic plate. But this still
suffers from vibration problems.

\bigskip

\textstyleYear{1969} Benton develops the rainbow hologram, requiring two recording steps (termed
two-step holography).  A master hologram is recorded from a real object with conventional off-axis holographic techniques.  The rainbow hologram is then recorded from the image of the master \citep{patent:3832027}.

The technique of recording a master hologram offers a significant advantage; it allows for the mass production of the rainbow hologram without the physical object, making commercial holography a real possibility.  Commercial machines are still available today for the purpose of mass production from a master hologram\citep{grichine:203}.  Two-step holography also has various beneficial side{}-effects. Because the recording of the hologram is done in two steps, the copy can be a slightly blurred version of the master, hiding any pixillation effects, hiding any seams, etc.  It also allows objects in holograms to appear to lie both inside and outside the hologram plane, since the copy hologram can be placed in
the master hologram's real object plane.

\bigskip

The draw back, however, is that two-step holography requires a high powered laser. The
larger the hologram, the more powerful the laser required (or the
longer the exposure time), as ideally the whole hologram needs to be
copied at the same time.  Two-step holography also introduces significant complexity into the design, making it impractical for a small studio.

\citet{chen1978osr} used imaging by a lens to produce a one-step process for creating rainbow holograms in a single step. This method had the advantages that:

\begin{itemize}
\item Conceptually and mechanically fairly simple -- a hologram can be directly printed.
\item The laser does not need to be as powerful -- a small part of the image can be exposed at a time , particularly when creating the hologram in slits.
\item Tiling holograms is possible, to make particularly large holograms.
\item Non-standard viewing windows are possibly.
\item In the future, one-step holography could use dry-processing (e.g.
photosensitive polymers \citep{patent:EP0697631})
\end{itemize}

\bigskip

\textstyleYear{1970}Salvador Dali used a ruby pulsed laser to
produce holographic works of art. The quality of holograms increased
rapidly over the years, as new emulsions were made available.

\bigskip

By this stage, people wanted more flexibility in what they could
create holograms of. To create a hologram of a large building, say,
or a fictional monster, a small miniature model had to be created.

\bigskip

One way to get around this was to make holograms of photos. \citet{King:70} showed that by using a different photograph for each angle, the photographs could be optically multiplexed to compose a hologram.  With this method a building or an outside
scene could be effectively holographed\footnote{Holographed {--} like photographed,
but with holograms.}.

\bigskip

By this time, computers were starting to become useful and available.
For \citet{King:70} and \citet{patent:3843225}, the next logical step was to use a computer to
generate the images of a scene, and multiplex those images onto a
hologram.

\bigskip

\textstyleYear{Late 1970's}An alternative
technique to produce holograms of fictional objects, was to be more
direct. Gabor had already shown, some 25 years earlier, that the
interference pattern could be calculated. Using computers, or
otherwise, the interference pattern for an arbitrary scene could be
calculated, and then mechanically or chemically\citep{gale1976gpp} etched in some way onto a material.

\bigskip

This now allows for the creation of holograms of arbitrary objects. A
hologram of any object, real or fictional, can now be created, as long
as you can work out the interference patten for it.

\bigskip

There are many drawbacks to commercially producing holograms by
mechanically etching on interference lines however. There is
mechanical abrasion, resulting in requiring frequent tip changes and
mechanical upkeep. It is also very slow to produce even small
holograms, and the groove spacing is unlikely to be anywhere near the
wavelength of visible light, resulting in a low fidelity hologram.

\bigskip

Despite the commercial drawbacks, it was a success in demonstrating
truly that the interference pattern for an object could be calculated
and a true hologram produced.

\bigskip

Briefly looking at the next twenty years of development in holography
field:

\bigskip

\textstyleYear{1979}\citet{mcgrew1983fch}, working with the Diffraction
Company, develops an embossing mass production technique for surface
relief holograms. McGrew went on to form his own company, Light
Impressions Inc, which was the first company to bring the embossed
hologram to the commercial market with a set of embossed images of \ac{3D}
subjects.

\bigskip

\textstyleYear{Early 1980's}Benton proposes the use of
a movie camera mounted on a linear rail to obtain images of on object at
different angle, as opposed to rotating object on turn table\citep{pizzanelli:ddw}.

\bigskip

\textstyleYear{1988}F. Iwata and K. Ohuma, of Toppan
Printing Co Ltd, made a novel process for making animated diffractive
patterns by making lots of small tiny dots of gratings. Although
crude, it is effective. It allows for very fine diffraction gratings,
giving a high efficiency, while maintaining a crude control of the
image (by deciding whether to put a dot in a particular place, etc.)

\bigskip

\citet{patent:5822092} and \citet{patent:5291317} independently improved on this by modulating the object, allowing crude \ac{3D} images to be built up of `pixels'.

\bigskip

This thesis follows a similar method as pioneered by Davis and Newswanger {--} building up a hologram by
printing in a `dot matrix' style of
`pixels' in a grid. Throughout this thesis, the printed `pixels' on the hologram will be referred to as 
as holopixels, to emphasize their iridescent nature.

\bigskip

\textstyleYear{1992}By the early 1990's, computers
and \ac{LCD} screens were a lot more ubiquitous. \citet{spierings:52} 
replaced film in their Holoprinter with an \ac{LCD} screen. 
\citet{nuland:9} went one step further over
the course of a year, and make color \ac{3D} stereograms using an \ac{LCD} in a
single step 

\bigskip

\textstyleYear{1994}\citet{Yamaguchi:94, yamaguchi:50} describes a monochrome one-step holographic printer based on a \ac{CW} laser. This can print small full parallax white-light reflection
holograms, but takes 2 seconds per pixel {--} e.g. 36 hours to do a
reflection hologram of 320x224 holopixels.

\bigskip

\textstyleYear{1998}Some 20 years before, Fujio Iwata et al (\citep{kodera1992hj,ohnuma1989fcr})
developed a mechanical machine to etch on diffraction gratings on to a
substrate. They could calculate the diffraction gratings required by
using Gabor's work done 25 years before them.
\citet{patent:4701006} and \citet{hamano:2} modernize the etching technique by using a high powered electron beam to etch
the grating on. This reduces the mechanical wear and tear, and speeds
up the process. However the electron beam is itself very expensive
to produce and still very slow. It is also computationally difficult, requiring
a significant amount of computing power.

\section{Digital versus analogue}

It is interesting to see the parallels between photography and
holography. Both have become increasing digitalized, with photography
leading the way. 

\bigskip

Photography, as we have seen, is traditionally the recording of the
intensity of light on a light sensitive film. By using color filters
we can also record the color of the light, allowing for a crude image
suitable for the human eye.

\bigskip

The film can then be chemically processed, with the end result being an
analogue photograph.

\bigskip

The analogue camera had various important drawbacks:

\begin{itemize}
\item The photographer is forced to wait a long time before the results of a shot can be seen. There were a few solutions {--} most notably the Polaroid camera which automatically developed the film straight after a photograph was taken.
\item Film had to be bought continually {--} a re-occurring cost
\item Negatives also had to be stored along with the photographs in case reprints were needed at a later time.
\end{itemize}

\bigskip

There are many other factors to consider, and the debate between
analogue and digital photography is still ongoing today. 

\bigskip

The 1990s saw the advent of digital photography. Digital photography allowed for
photographs to be captured with \ac{CCD} based image sensors. The photographs could be
modified and then printed to paper if needed.

\chapter{Design of a digital hologram printer}\label{chap:design}
This chapter considers the technical design of a digital hologram printer, beginning with a description of the optical and mechanical components for both the object and reference beam paths.  The lens systems in the object beam are examined quantitatively, concluding with a straightforward set of instructions for adjusting the lens system.  The chapter finishes with a qualitative examination of the lens system based on testing three different lens arrays.

\section{Overview}
Detailed is a digital hologram printer capable of recording transmission or
reflection holograms in an off-axis geometry for subsequent developing and bleaching in order to produce
white-light viewable holograms. The printer consists of: a pulsed
laser source arranged to produce a visible-light laser beam; a lens
system to direct each beam pulse to a photosensitive medium; a display
system to modulate the object beam; a two dimensional positioning track
to position the photosensitive medium relative to said lens system; and
a computer control system.

The use of a pulsed laser offers the advantage of printing without
sensitivity to vibrations or slight temperature changes. The printer
utilizes a long-cavity frequency-doubled \ac{NDYAG} pulsed laser which
can produce a stable second harmonic TEM$_{00}$ coherent \unit[532]{nm} (visible green) beam.

The lens system splits the beam with a Brewster-angle polarising beam
splitter into the mutually-coherent object and reference beams. 
The display system is a reflective \ac{LCOS} display or a transmissive \ac{LCD}, placed
downstream of the object beam, and upstream of the photosensitive
medium.

The use of a reflective \ac{LCOS} as the display system for a hologram
printer is particularly advantageous. The high efficiency and high
contrast, compared to the less efficient transmissive \ac{LCD}, allows 
ultimately for higher contrast upon hologram replay and increased
energy economy during writing, allowing for a less powerful laser source to be
used. The high resolution on the \ac{LCOS} allows for the hologram to have a large depth of
view.

Typically a silver halide green{}-light photosensitive emulsion on a glass
substrate is placed upsteam in the Fourier plane of the display
system, recording the spatial standing diffraction pattern between the
reference and object beams. This records a small `pixel' of the order
of \unit[1]{mm} in diameter. The motorized two-dimensional positioning tracks
translate the plate into position for the next pixel to be printed.
The plate can then be developed and bleached chemically to produce a
white{}-light viewable reflection hologram.

\bigskip

The hologram printer design outlined in this chapter is based upon the
digital hologram printer detailed by \citet{ratcliffepatent}.
The patent consists of the general schematic for a monochromatic hologram
printer, but lacks key information required for implementation and modification. As is
typical for any complicated machine, intricate knowledge is required
for correct fabrication; alignment mistakes will result in a bad
hologram (e.g. containing dim, missing or bad pixels).

Because of the design's sensitivity to the layout, a methodical and
detailed setup that incorporates previous experience is required. The
patent also misses vital theoretical information required for modifying
the machine. The lens system used around the microlens array and display
system is sensitive to position. This chapter details the setup
required to produce high{}-fidelity images while wasting the minimum amount of
energy and controlling the energy density and pixel size of the beam exposed to the
holographic plate. This is accomplished with a mix of quantitative and
qualitative analysis of the lens system.

\bigskip

\begin{figure}[htp]
\centering
\includegraphics[angle=90,width=\textwidth]{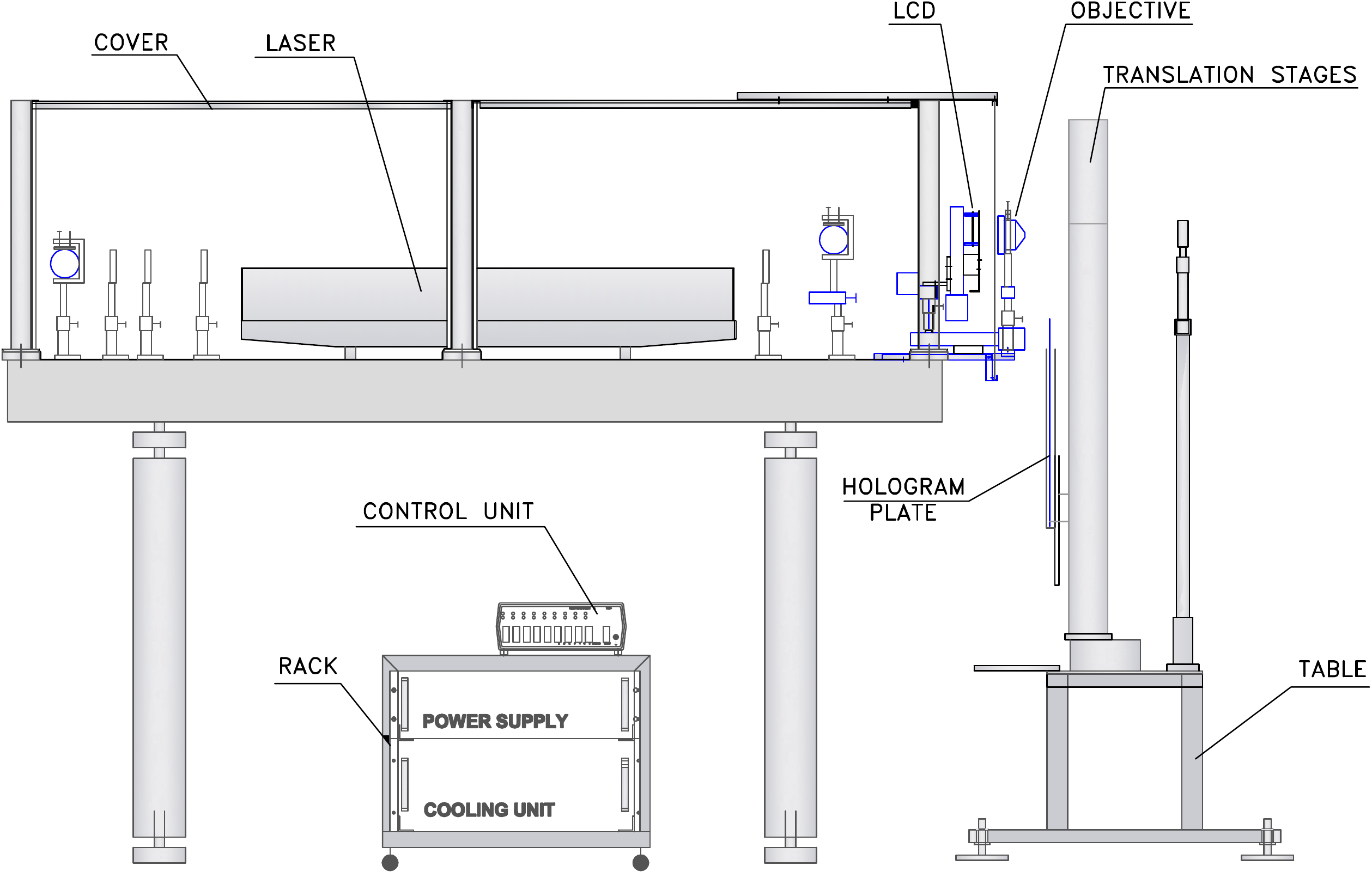}
\caption[Hologram printer by Ratcliffe et al.]{Hologram printer with \ac{LCD} display system by \citet{ratcliffepatent}.  For full-page schematics, see also \Fref{fig:hrip_lcos_big} and \Fref{fig:hrip_lcd_big}. }
\label{fig:hrip_lcd}
\end{figure}

The basis of this work is the green pulsed-laser holographic printer detailed
by \citet{ratcliffepatent} and illustrated in \Fref{fig:hrip_lcd}. The said hologram printer can print \unit[1.0]{mm}$\times$\unit[1.0]{mm} sized pixels onto a photosensitive glass plate at
a maximum rate of 4 pixels per second. The majority of the energy
from the laser is lost at various points in the design. The
mechanical setup suffered from mechanical vibrations. The laser was
also unstable with between 10\% to 30\% pulse{}-to{}-pulse variation in energy.
This produces noticeable changes in
the hologram. The goal of the work reported was to overcome these
shortcomings, increase the printing speed and decrease the pixel size.

\clearpage

\section{Detail of system components}
This section provides a detailed description for the building of digital hologram printer for
writing composite 1-step holograms.

\begin{figure}[htp]
\centering
\includegraphics[angle=90,width=\textwidth]{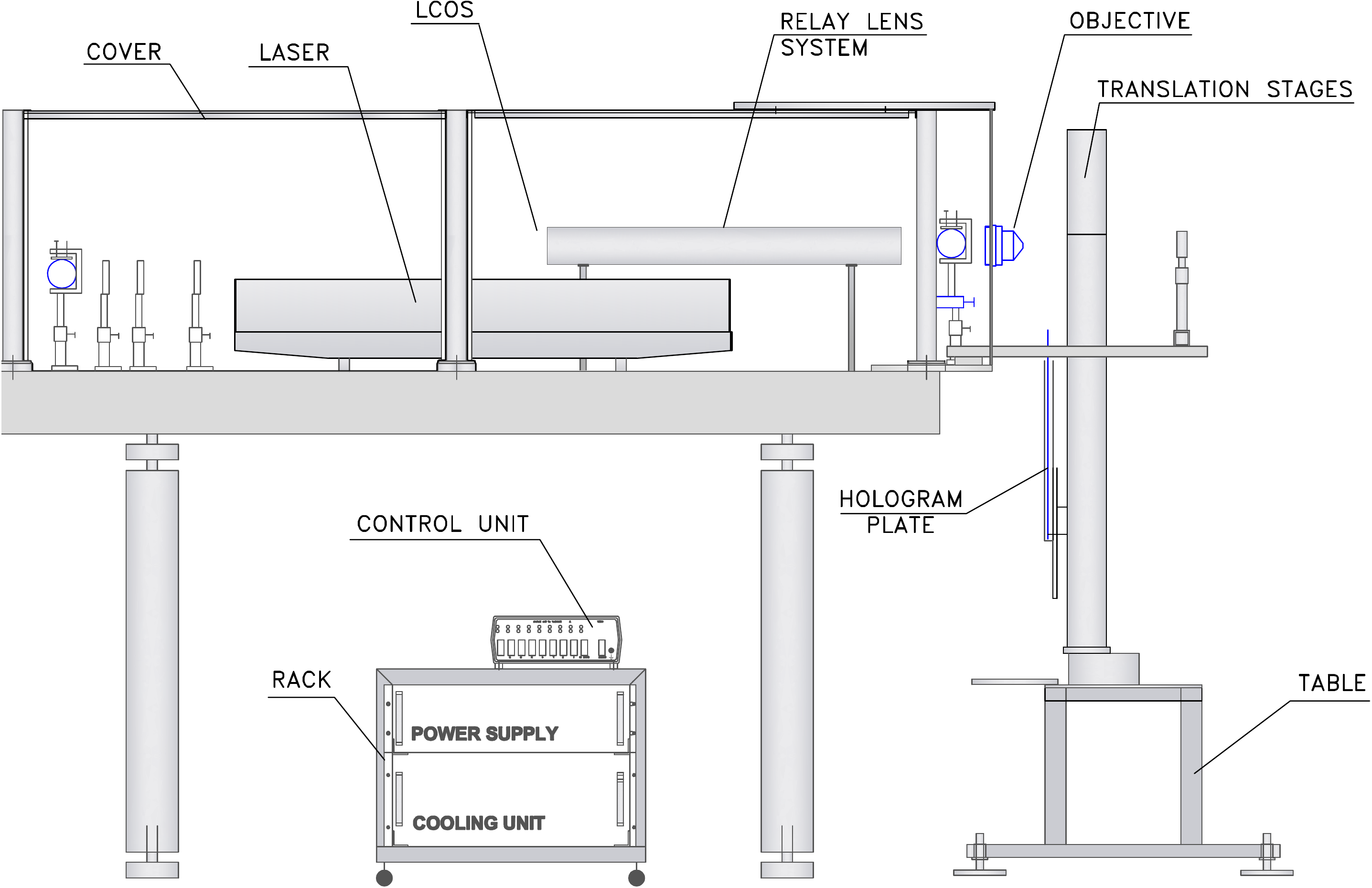}
\caption[Hologram Printer with LCOS]{Side-on orthographic projection of final hologram printer design with \ac{LCOS} display system.  For full-page schematics, see also \Fref{fig:hrip_lcos_big} and \Fref{fig:hrip_lcd_big}. }
\label{fig:hrip_lcos}
\end{figure}

\Fref{fig:hrip_lcos} shows how the final hologram printer looked, after the various modifications, shown here for reference.  The laser and various lens systems sit on an optical table.  The second harmonic laser beam is split into two beams, and transported to the hologram plate.  The object beam goes via the display system and the objective, while the reference beam is reflected off of the rear mirror and strikes the plate from the opposite direction. The hologram plate can be spatially translated by the pair of translation stages, to print each holopixel.

\begin{figure}[htp]
\centering
\includegraphics[angle=90,width=\textwidth]{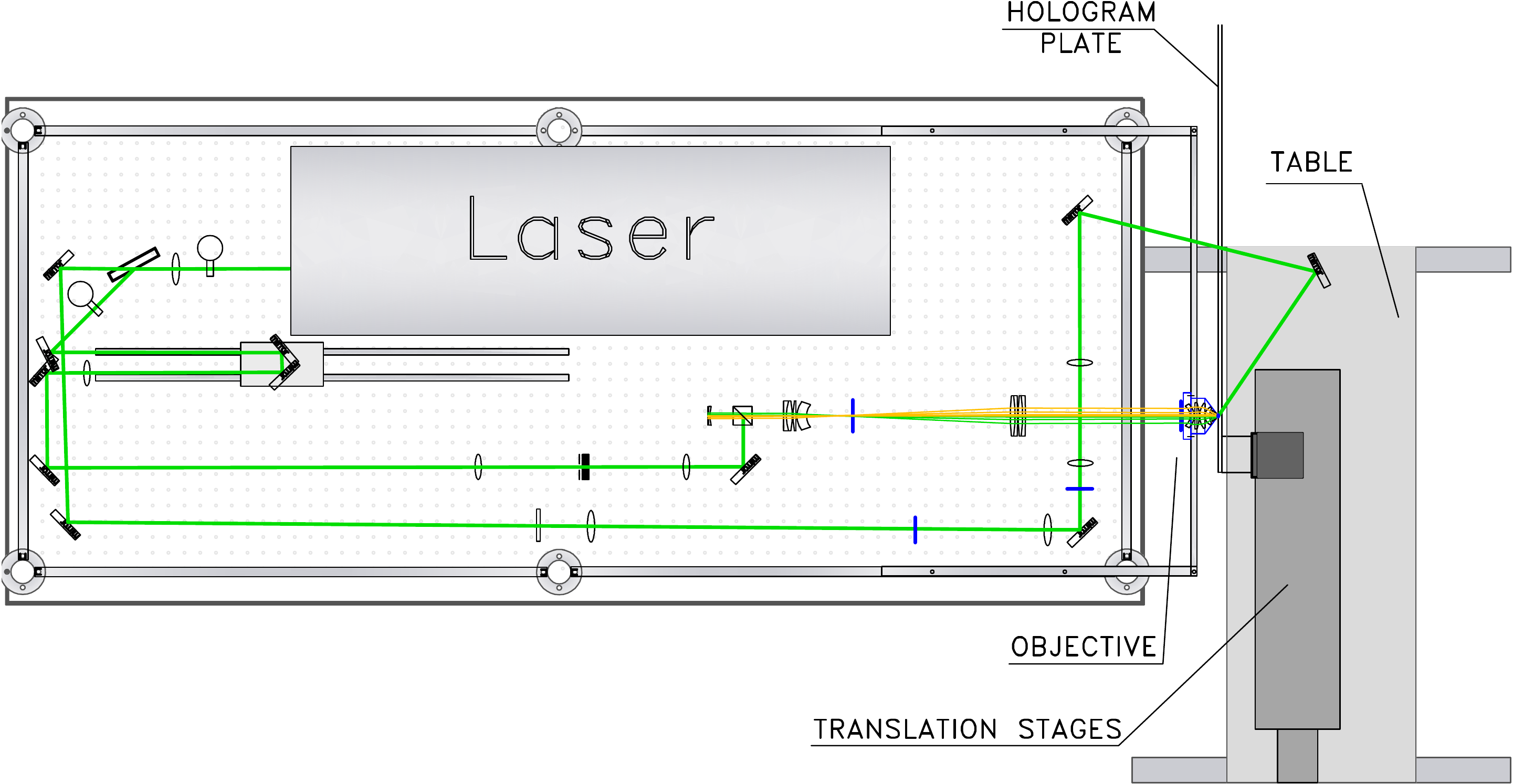}
\caption[Top-down orthographic projection of Hologram Printer with LCOS]{Top-down orthographic projection of final hologram printer design with \ac{LCOS} display system.  For full-page labelled schematics see also \Fref{fig:hrip_lcos_top_big} and \Fref{fig:hrip_lcos_top_dimensions_big}. }
\label{fig:hrip_lcos_top}
\end{figure}

\subsection{Laser system}

Most lasers will rest on their own bread board which in turn stands on
legs. The height of the laser can be adjusted; however it is a lot
more flexible to adjust the height of the beam by using a system of
mirrors instead.

The object beam is raised to the correct height using a triangle of
three mirrors, each higher than the last. The beam then passes
through a motorized half wave plate which rotates the polarization.
The beam then passes through a Brewster-angle split beam polariser.
The p-component of the laser light is transmitted through the
split-beam polariser, while the s-component is reflected.

It is very important to get the angle of the Brewster polariser correct,
otherwise circular polarization of the beam will result. To align
correctly, the angle of the \textonehalf-wave plate is set to
maximize the amount of reflected light (so that it polarises the light
into the p{}-direction). This is best done by using an energy meter
to measure the intensity of the light reflected from the split{}-beam
polariser. By eye, check the profile of the transmitted beam is then
checked. If the angle of the split beam polariser is wrong, the beam
profile will look poor. The angle of the split{}-beam polariser is
adjusted to minimize the intensity of the transmitted beam and produce
a good Gaussian beam profile. In this setup the Brewster split beam
polariser was coated with an anti-reflection surface, increasing the
efficiency. For a non-coated split-beam polariser it is possible
that the maximum transmitted energy is not at the Brewster angle \citep{Hecht1990}.
In this case, the best results
would probably be obtained by trying to maximise the quality of the
beam, rather than the intensity.

\subsection{Optics for transport of beams}

\subsubsection{Adjustable object beam path length}

\begin{figure}[htp]
\centering
\includegraphics[angle=90, width=\textwidth]{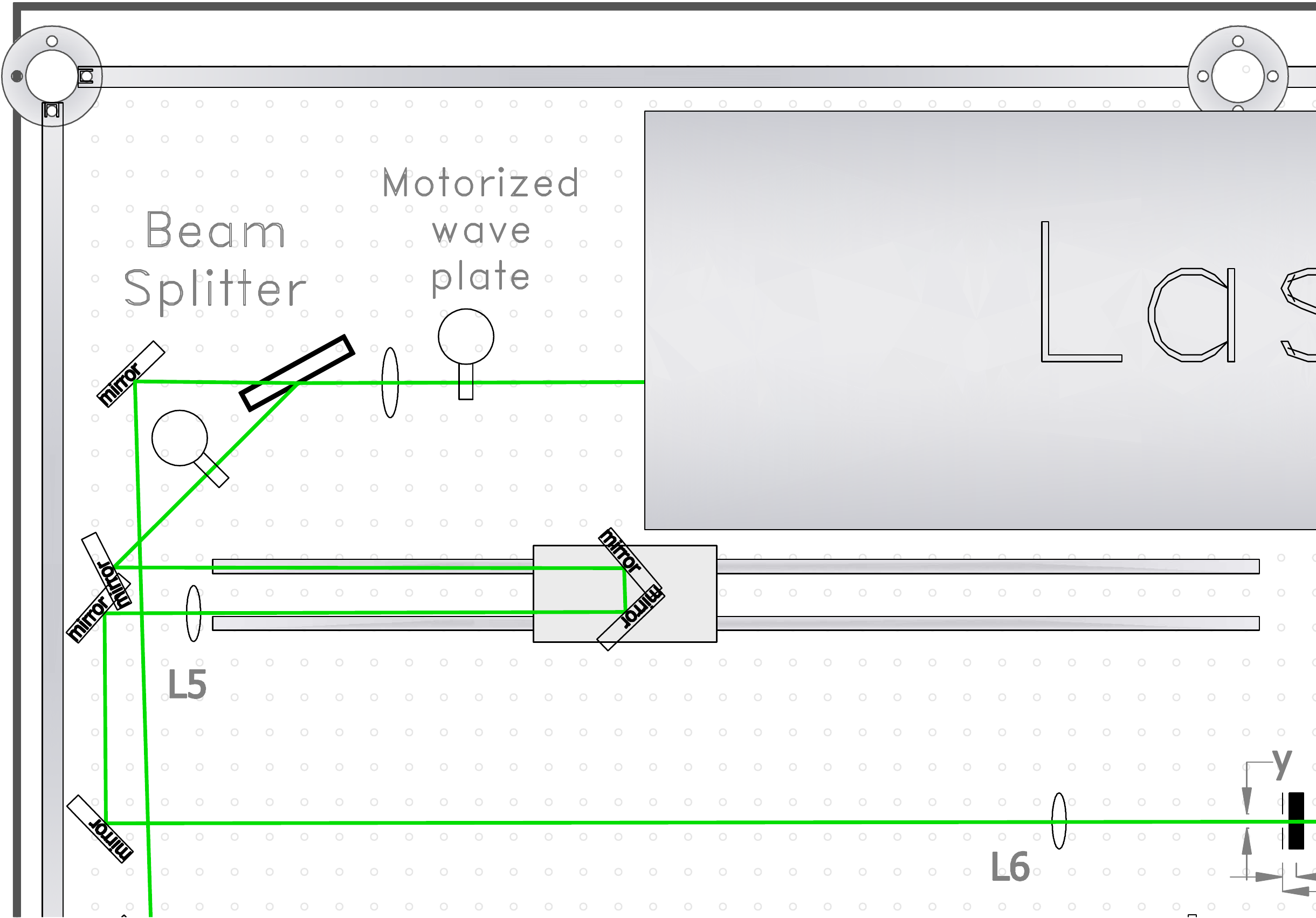}
\caption[Length adjustment system]{Top-down orthographic projection of system to adjust optical-path distance of object beam.  See also \Fref{fig:hrip_rails_big} for a full-page schematic.}
\label{fig:hrip_rails}
\end{figure}

\Fref{fig:hrip_rails} shows that the object beam is reflected between two sets of mirrors, with one set of
mirrors mounted on a sliding base. This allows the distance between
the two sets of mirrors (and hence the overall optical path length of the object
beam) to be easily adjusted, while keeping the correct beam alignment.

It is important to be able to modify the object beam path length such that
the difference in distance travelled by the reference beam and object beam is kept as
small as possible.  This maximises the temporal overlap of the object and reference
beam pulses at the hologram plate. If one beam arrives at the plate before the other, it will
expose the plate without any signal, reducing contrast and diffraction efficiency upon replay.

To setup this system of mirrors, all four mirrors were mounted as indicated, with two of the mirrors
on a sliding base. The base was moved back and forth while watching for beam movement.
The angles of the four mirrors were adjusted until there was no perceptible lateral movement of the beam 
when the base is moved.

\subsubsection{Cleaning up the object beam}

To produce a clean Gaussian spatial profile of the object (and reference) beam if required, a pair of positive lenses (Marked L5 and L6 in \Fref{fig:hrip_rails}) placed at a distance equal to the sum of their focal lengths can be used, with a pinhole aperture (not shown) placed at the their mutual focusing plane to remove any high order defects.  This afocal lens system can have the dual purpose of magnifying the beam as required, where there magnification of the laser beam, $M_\text{laser}$ is dependant on the focal length of L5, $f_\text{L5}$ and the focal length of L6, $f_\text{L6}$ as:
\begin{equation}
M_\text{laser} = \dfrac{f_\text{L5}}{f_\text{L6}}
\end{equation}
\nomenclature{$M_\text{laser}$}{Magnification of laser beam due to lenses L5 and L6}
\nomenclature{$f_\text{L5}$}{Focal length of lens L5}
\nomenclature{$f_\text{L6}$}{Focal length of lens L6}
The beam can thus be magnified such that its diameter closely matches the downstream width of object beam aperture, to maximise overall beam efficiency.

\bigskip

\subsection{Digital display system}

In direct write analogue holography, the object beam illuminates a
physical object. The illuminated beam is reflected off of the object,
modulated by the object, and may be considered to scatter as a series
of spherical waves. A photosensitive plate is positioned as to
capture this light. A reference beam, also incident upon the plate,
interferes with the modulated object beam creating a microscopic interference
pattern (fringe pattern) that encodes the required information about the object. This fringe
pattern is recorded on to the plate as an absorption hologram, to be
developed and bleached to create a white light hologram. On replay of
the hologram, the object image is reconstructed in its original
position.

\bigskip

For digital holography the physical object is replaced with a computer
controlled display system. This ultimately allows the production of a
hologram using a computer generated model or a series of photographs or
some other method that generates a series of suitable images. There are 
various methods for using a computer controlled display.

The display system used was a \ac{LCOS} based \ac{SLM}; a
small, reflective, high contrast, high fill-rate and high resolution panel.  This has the advantages of a small form factor allowing for future miniaturization of the printer; high resolution allowing for holograms with a large depth of field (both in and out of the hologram plane); high efficiency rate increasing overall beam efficiency and allowing for a weaker laser to be used, thus decreasing the cost and increasing the lifetime of the printer; high contrast which in turn allows for high contrast holograms; and a high fill rate of up to \unit[60]{Hz}, allowing for holograms to be printed at up to this rate of holopixels per second.

\begin{figure}[htp]
\centering
\begin{tikzpicture}[join=round]
\draw[draw=gray!40](1.369,.793)--(2.1,-.54);
\filldraw[fill=white,fill opacity=0.6,line width=0,color=white](1.388,.751)--(2.081,-.515)--(-1.388,-.769)--(-2.081,.498)--cycle;
\draw[draw=gray!40](1.187,.78)--(1.917,-.553);
\draw[draw=gray!40](1.004,.767)--(1.734,-.567);
\draw[draw=gray!40](-2.173,.5)--(1.479,.767);
\draw[draw=gray!40](.822,.753)--(1.205,.053);
\draw[draw=gray!40](.639,.74)--(1.022,.04);
\draw[draw=gray!40](.456,.727)--(.84,.027);
\draw[draw=gray!40](.274,.713)--(.621,.08);
\draw[draw=gray!40](.091,.7)--(.438,.067);
\draw[draw=gray!40](-2.136,.433)--(1.515,.7);
\draw[draw=gray!40](-.091,.687)--(.256,.053);
\draw[draw=gray!40](-.274,.673)--(.073,.04);
\draw[draw=gray!40](-.456,.66)--(-.11,.027);
\draw[draw=gray!40](-.639,.647)--(-.256,-.053);
\draw[draw=gray!40](-.822,.633)--(-.091,-.7);
\draw[draw=gray!40](-2.1,.367)--(1.552,.633);
\draw[draw=gray!40](-1.004,.62)--(-.274,-.713);
\draw[draw=gray!40](-1.187,.607)--(-.456,-.727);
\draw[draw=gray!40](-1.369,.593)--(-.639,-.74);
\draw[draw=gray!40](-1.552,.58)--(-.822,-.753);
\draw[draw=gray!40](-1.734,.567)--(-1.004,-.767);
\draw[draw=gray!40](-.146,.44)--(1.588,.567);
\draw[draw=gray!40](-1.917,.553)--(-1.187,-.78);
\draw[draw=gray!40](-2.1,.54)--(-1.369,-.793);
\draw[draw=gray!40](-.11,.373)--(1.625,.5);
\draw[draw=gray!40](-2.063,.3)--(-.146,.44);
\draw[draw=gray!40](-.073,.307)--(1.661,.433);
\draw[draw=gray!40](-2.027,.233)--(-.11,.373);
\draw[draw=gray!40](-.037,.24)--(1.698,.367);
\draw[draw=gray!40](0,.173)--(1.734,.3);
\draw[draw=gray!40](.037,.107)--(1.771,.233);
\draw[draw=gray!40](.073,.04)--(1.807,.167);
\filldraw[fill=blue!20,fill opacity=0.5](.073,.04)--(1.095,2.333)--(1.643,1.333)--(.11,-.027)--cycle;
\draw[draw=gray!40](-1.99,.167)--(-.073,.307);
\draw[draw=gray!40](-.219,.227)--(-.037,.24);
\draw[draw=gray!40](-1.954,.1)--(-.219,.227);
\draw[draw=gray!40](-.183,.16)--(0,.173);
\draw[draw=gray!40](-1.917,.033)--(-.183,.16);
\draw[draw=gray!40](-.146,.093)--(.037,.107);
\draw[draw=gray!40](-1.881,-.033)--(-.146,.093);
\filldraw[fill=blue!20,fill opacity=0.5](-.11,.027)--(-1.643,2.133)--(1.095,2.333)--(.073,.04)--cycle;
\draw[draw=gray!40](-1.844,-.1)--(-.11,.027);
\draw[arrows=->,line width=.4pt](-.091,-.007)--(-2.1,-.153);
\filldraw[fill=blue!20,fill opacity=0.5](-.073,-.04)--(-1.095,1.133)--(-1.643,2.133)--(-.11,.027)--cycle;
\draw[draw=gray!40](-1.807,-.167)--(1.844,.1);
\draw[draw=gray!40](.621,.08)--(1.004,-.62);
\draw[draw=gray!40](.438,.067)--(.822,-.633);
\draw[draw=gray!40](.256,.053)--(.639,-.647);
\draw[draw=gray!40](1.205,.053)--(1.552,-.58);
\draw[draw=gray!40](.073,.04)--(.11,-.027);
\draw[draw=gray!40](1.022,.04)--(1.369,-.593);
\draw[draw=gray!40](-.11,.027)--(.073,.04);
\draw[draw=gray!40](-1.771,-.233)--(1.881,.033);
\draw[draw=gray!40](-.11,.027)--(-.073,-.04);
\draw[draw=gray!40](.84,.027)--(1.187,-.607);
\draw[arrows=-,line width=.4pt](0,0)--(-.091,-.007);
\draw[arrows=->,line width=.4pt](.018,-.033)--(0,0)--(0,2.6);
\filldraw[fill=blue!20,fill opacity=0.5](.11,-.027)--(1.643,1.333)--(-1.095,1.133)--(-.073,-.04)--cycle;
\draw[draw=gray!40](.11,-.027)--(.456,-.66);
\draw[arrows=<-,line width=.4pt](.42,-.767)--(.018,-.033);
\draw[draw=gray!40](-1.734,-.3)--(1.917,-.033);
\draw[draw=gray!40](-.073,-.04)--(.274,-.673);
\draw[draw=gray!40](-.256,-.053)--(.091,-.687);
\draw[draw=gray!40](-1.698,-.367)--(1.954,-.1);
\draw[draw=gray!40](-1.661,-.433)--(1.99,-.167);
\draw[draw=gray!40](-1.625,-.5)--(2.027,-.233);
\draw[draw=gray!40](-1.588,-.567)--(2.063,-.3);
\draw[dashed,line width=2pt,color=green!50](1.03,2.763)--(1.399,3.751);
\draw[draw=gray!40](-1.552,-.633)--(2.1,-.367);
\draw[draw=gray!40](-1.515,-.7)--(2.136,-.433);
\draw[draw=gray!40](-1.479,-.767)--(2.173,-.5);
\filldraw[line width=0,color=black!70](1.425,-5.575)--(-1.425,-5.575)--(-1.425,-8.425)--(1.425,-8.425)--cycle;
\draw[draw=gray,densely dotted](-1.425,-5.575)--(.1,0);
\draw[draw=gray,densely dotted](1.425,-5.575)--(-.1,0);
\filldraw[fill=green,line width=0,color=green](1.125,-5.875)--(-1.125,-5.875)--(-1.125,-8.125)--(1.125,-8.125)--cycle;
\draw[draw=gray!60,line width=.1pt](-1.425,-8.425)--(1.425,-8.425);
\draw[draw=gray!60,line width=.1pt](-1.425,-8.275)--(1.425,-8.275);
\draw[draw=gray!60,line width=.1pt](-1.425,-8.125)--(1.425,-8.125);
\draw[draw=gray!60,line width=.1pt](-1.425,-7.975)--(1.425,-7.975);
\draw[draw=gray!60,line width=.1pt](-1.425,-7.825)--(1.425,-7.825);
\draw[draw=gray!60,line width=.1pt](-1.425,-7.675)--(1.425,-7.675);
\draw[draw=gray!60,line width=.1pt](-1.425,-7.525)--(1.425,-7.525);
\draw[draw=gray!60,line width=.1pt](-1.425,-7.375)--(1.425,-7.375);
\draw[draw=gray!60,line width=.1pt](-1.425,-7.225)--(1.425,-7.225);
\draw[draw=gray!60,line width=.1pt](-1.425,-7.075)--(1.425,-7.075);
\draw[draw=gray!60,line width=.1pt](-1.425,-6.925)--(1.425,-6.925);
\draw[draw=gray!60,line width=.1pt](-1.425,-6.775)--(1.425,-6.775);
\draw[draw=gray!60,line width=.1pt](-1.425,-6.625)--(1.425,-6.625);
\draw[draw=gray!60,line width=.1pt](-1.425,-6.475)--(1.425,-6.475);
\draw[draw=gray!60,line width=.1pt](-1.425,-6.325)--(1.425,-6.325);
\draw[draw=gray!60,line width=.1pt](-1.425,-6.175)--(1.425,-6.175);
\draw[draw=gray!60,line width=.1pt](-1.425,-6.025)--(1.425,-6.025);
\draw[draw=gray!60,line width=.1pt](-1.425,-5.875)--(1.425,-5.875);
\draw[draw=gray!60,line width=.1pt](-1.425,-5.725)--(1.425,-5.725);
\draw[draw=gray!60,line width=.1pt](-1.425,-5.575)--(1.425,-5.575);
\draw[draw=gray!60,line width=.1pt](1.425,-8.425)--(1.425,-5.575);
\draw[draw=gray!60,line width=.1pt](1.275,-8.425)--(1.275,-5.575);
\draw[draw=gray!60,line width=.1pt](1.125,-8.425)--(1.125,-5.575);
\draw[draw=gray!60,line width=.1pt](.975,-8.425)--(.975,-5.575);
\draw[draw=gray!60,line width=.1pt](.825,-8.425)--(.825,-5.575);
\draw[draw=gray!60,line width=.1pt](.675,-8.425)--(.675,-5.575);
\draw[draw=gray!60,line width=.1pt](.525,-8.425)--(.525,-5.575);
\draw[draw=gray!60,line width=.1pt](.375,-8.425)--(.375,-5.575);
\draw[draw=gray!60,line width=.1pt](.225,-8.425)--(.225,-5.575);
\draw[draw=gray!60,line width=.1pt](.075,-8.425)--(.075,-5.575);
\draw[draw=gray!60,line width=.1pt](-.075,-8.425)--(-.075,-5.575);
\draw[draw=gray!60,line width=.1pt](-.225,-8.425)--(-.225,-5.575);
\draw[draw=gray!60,line width=.1pt](-.375,-8.425)--(-.375,-5.575);
\draw[draw=gray!60,line width=.1pt](-.525,-8.425)--(-.525,-5.575);
\draw[draw=gray!60,line width=.1pt](-.675,-8.425)--(-.675,-5.575);
\draw[draw=gray!60,line width=.1pt](-.825,-8.425)--(-.825,-5.575);
\draw[draw=gray!60,line width=.1pt](-.975,-8.425)--(-.975,-5.575);
\draw[draw=gray!60,line width=.1pt](-1.125,-8.425)--(-1.125,-5.575);
\draw[draw=gray!60,line width=.1pt](-1.275,-8.425)--(-1.275,-5.575);
\draw[draw=gray!60,line width=.1pt](-1.425,-8.425)--(-1.425,-5.575);
\draw[arrows=<->,line width=.4pt](1.725,-8.425)--(-1.425,-8.425)--(-1.425,-5.275);

\begin{scope}
\clip (0,-3) circle (2cm and 0.7cm);
\filldraw[fill=gray!4,line width=0.6pt] (0,-4) circle (2cm and 0.7cm);
\end{scope}
\begin{scope}
\clip (0,-4) circle (2cm and 0.7cm);
\draw[line width=0.6pt] (0,-3) circle (2cm and 0.7cm);
\end{scope}
\path (1,-3.5) node[above right] {\begin{footnotesize}Optical Fourier transform\end{footnotesize}};

\path (1.725,-8.425) node[below] {$x_\text{lcos}$}
                   (-1.425,-5.275) node[left] {$y_\text{lcos}$}
                   (-1.425,-8.425) -- (1.725,-8.425) node[pos=0.5, below] {\begin{footnotesize}LCOS\end{footnotesize}};
\path (-2.1,-.153) node[left] {$x_\text{hologram}$}
                   (.42,-.767) node[below] {$y_\text{hologram}$}
                   (0,2.6) node[above] {$z$};

\begin{scope}
\clip (1.655,4.437) -- (1.683,3.6) -- (0,0) -- (1.196,3.917);
\filldraw[fill=gray!7, line width=0.6pt] (1.655,4.437) circle (0.6cm and 0.7cm);
\clip (1.655,4.437) circle (0.6cm and 0.7cm);
\filldraw[fill=black] (1.399,3.751) circle (0.15cm and 0.15cm);
\end{scope}
\draw[line width=0.6pt ] (1.683,3.6) -- (1.655,4.437) -- (1.196,3.917);
\draw[line width=0.6pt ] (1.262,3.991) .. controls (1.196,3.917) and (1.12,3.891) ..(1.124,4.028);
\draw[line width=0.6pt ] (1.196,3.917) .. controls (1.13,3.842) and (1.055,3.817) ..(1.059,3.954);

\path (1.683,3.6) node[right] {\begin{footnotesize}eye\end{footnotesize}};

\path (1.095,2.333) node[right] {\begin{footnotesize}viewing window\end{footnotesize}}; 
\end{tikzpicture}
\caption[Lens system]{The Fourier transform of the image on the LCOS is printed on to the hologram as a single 'holopixel'.  Upon replay of the hologram, this recorded Fourier image is reconstructed as an angular intensity distribution.}
\label{fig:scene2}
\end{figure}
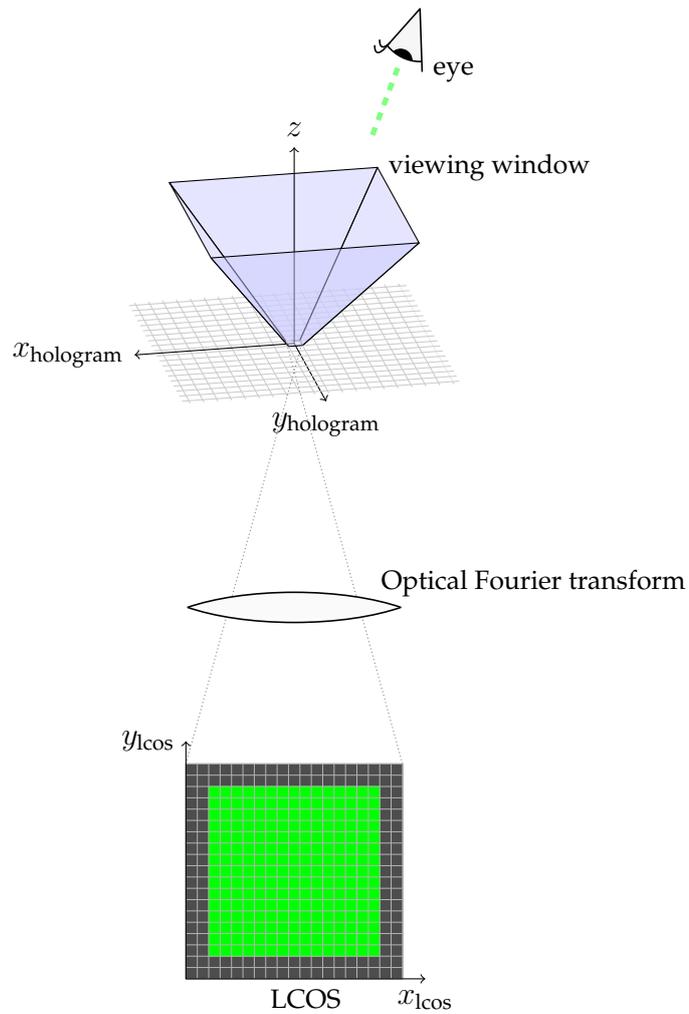

\begin{figure}[htp]
\centering
\includegraphics[width=\textwidth]{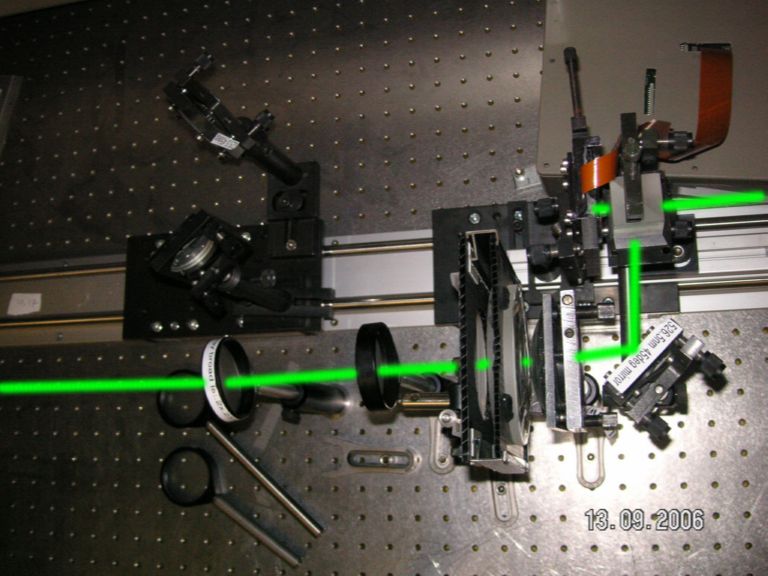}
\caption[Photograph of LCOS being illuminated]{Birds-eye-view photograph of \ac{LCOS} being illuminated, with the beam path indicated.}
\label{fig:photolcos}
\end{figure}

\Fref{fig:scene2} illustrates the logical layout required and \Fref{fig:photolcos} is a photograph of the microlens array system downstream of the \ac{LCOS}.

\clearpage

\section{Analysis of lens system}\label{sec:analysis_of_lens_system}

The object beam is shaped according to several restrictions. The requirements are that:

\begin{itemize}
\item The beam exposes the \ac{LCOS} uniformly and smoothly.
\item Energy losses are minimized {--} the beam should illuminate the whole of the \ac{LCOS} display while minimizing losses.  Thus the beam profile should be rectangular and of the same aspect ratio and scale.
\item The spatial profile of the beam at the plane of the hologram plate emulsion should be of a controllable size and shape.  Typically either circular or square with a width of around \unit[1]{mm} to \unit[0.3]{mm}.
\end{itemize}

\begin{figure}[htp]
\centering
\includegraphics[angle=90, width=\textwidth]{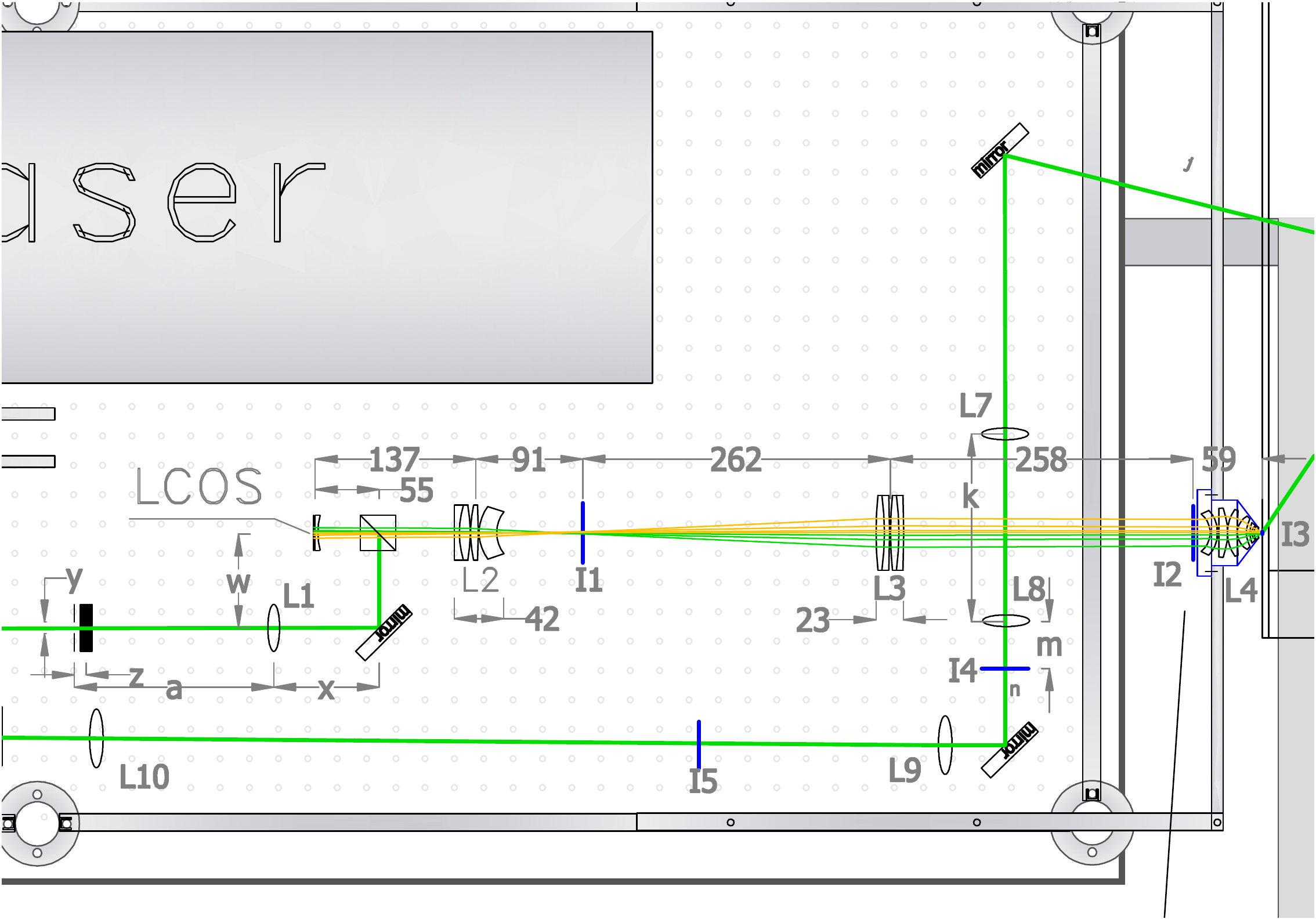}
\caption[Object lens system on printer]{Top-down orthographic projection of object lens system in hologram printer.  See also \Fref{fig:hrip_lens_system_big} for a full-page schematic.}
\label{fig:hrip_lens_system}
\end{figure}

To achieve these aims, an afocal reversing telescopic lens system with an objective compound lens was designed and optimized using the lens software ZEMAX \citep{zemax}.  The lenses were designed and built by Marcin Lesniewski.  The resulting setup is shown in \Fref{fig:hrip_lens_system}.

\bigskip

\bigskip

The object beam is spatially filtered by the rectangular aperture $P$ which has an aperture width $y$, as labelled in \Fref{fig:hrip_lens_system}. The shape of the aperture will ultimately determine the downstream beam's spatial geometry (scaled in size by a factor $1/M_3$) of the pixel in the real image plane I1 and the real image labelled I3 where the photosensitive plate is placed.

\bigskip

The beam travels the distance labelled $z$ to the microlens array
$A$. This is a rectangularly-packed microlens array with spherically curved lenslets. The plane geometry of the lenslets in the microlens array determines the spatial shape of the beam on the
\ac{LCOS} display. Since the \ac{LCOS} display system is rectangular, the lenslets are also chosen
to be rectangular, with the same aspect ratio.

The expanded beam enters the positive lens L1 downstream of the microlens array, is reflected off the right-angled coated mirror, and continues to the polarising beam splitter.

\bigskip

The beam's polarization is already orientated in the direction such that all of the beam's energy is directed by the polarizing beam splitter to the \ac{LCOS} display system. The lenses between the microlens array and \ac{LCOS} are arranged such that the beam arrives at the display system with the correct spatial
geometry and scaling to match the display region on the \ac{LCOS}.  Preferably the beam should have an even spatial beam intensity profile with minimal defects and speckle in order to produce a replay viewing window with an even intensity.  It is conceivable to desire a non-even beam distribution (for example Gaussian) for the purpose of having a brighter optimal replay angle at the sacrifice of dimmer non-optimal replay angles.  Such possibilities are not considered further here, and an even spatial beam intensity profile is assumed to be desired.

\bigskip

The \ac{LCOS} twists the polarization of the light at each pixel, to build up
a spatial image. For a full intensity pixel, the polarization is rotated 90
degrees. For a black pixel, the polarization is not rotated. For
greyscale, it is only partially rotated.

A thin weak lens in front of the \ac{LCOS} display acts as a field lens to correct for
the final image curvature so that the final image from the objective lens L4 is focused on a flat
plane.  A schematic for the physical mount for the \ac{LCOS} display, field lens and beam splitter is given in Appendix~\ref{chap:lcos_mount}, \Fref{fig:lcos_mount}.  The light beam, now containing the image that was on the \ac{LCOS}, is now reflected back through the split beam polarising cube, splitting
the image into an intensity encoded image and its negative. The
negative bounces back along the original path, harmlessly being
ignored. The positive image is then sent through a custom \ac{LCOS}
telecentric afocal reversing lens system to be projected on to the
hologram plate.

\bigskip

The purpose of the \ac{LCOS} lens system is to project the image of the \ac{LCOS}
onto a virtual hemisphere, mapping the two dimensional pixel spatial
coordinates to longitude and latitude spherical coordinates, projected
on to the holographic film. 

Upon replay of the final developed hologram, each holographic pixel is iridescent {--} 
the brightness of the pixel changes with the angle at which it is viewed from, with the
angular intensity profile matching the original spatial pattern that was on the \ac{LCOS}.

This projection from the flat spatial plane (I1) at the \ac{LCOS} to a hemispherical plane at 
the hologram plate (I3) is achieved with the telescopic afocal reversing lens system and objective lens.
This lens setup is shown in Figure~\ref{fig:raytraced} and \Fref{fig:raytraced2}, with sample ray{}-traced light paths overlaid on the figure.

\begin{landscape}
\begin{figure}
\centering
\includegraphics[angle=-90, clip=true, trim=11.4cm 4.9cm 7.7cm 3.1cm, width=\hsize]{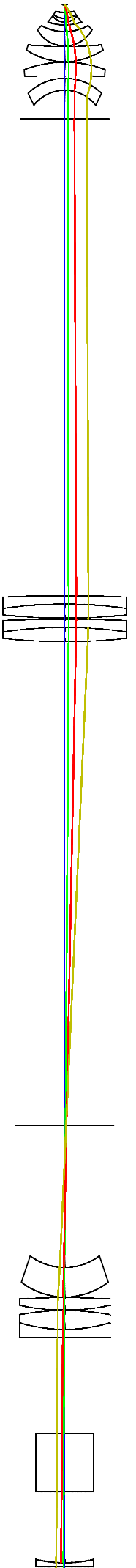}
\caption[Ray-traced lens setup 1]{Ray-traced lens setup for object beam afocal telescopic reversing lens system.  Rays drawn leaving parallel and spatially separated from the \ac{LCOS} on the left and meeting at a point on the hologram plane on the right - the Fourier transform.}
\label{fig:raytraced}
\end{figure}

\begin{figure}
\centering
\includegraphics[angle=-90, clip=true, trim=11.4cm 4.9cm 7.7cm 3.1cm, width=\hsize]{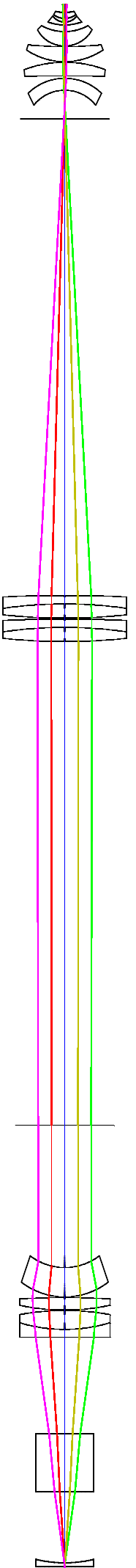}
\caption[Ray-traced lens setup 2]{Ray-traced lens setup for object beam afocal telescopic reversing lens system.  Rays drawn leaving from the same spatial point but at different angles from the \ac{LCOS} on the left and striking parallel and spatially separated on the hologram plane on the right - the Fourier transform.}{See Appendix~\ref{sec:opticalfourier} for more detailed ray-traced diagrams of this lens system.}
\label{fig:raytraced2}
\end{figure}

\end{landscape}

This complex system of lenses provides near diffraction-limited optical performance, requiring
various different optical elements and several different glass materials to correct for various aberrations.

It is, however, desirable to derive a simple set of formulas that describes this lens system.
This allows us to determine the effect of the set up of the optical system downstream, on the
holopixel that is projected upstream. The size and shape of the final holopixel is 
determined by both the downstream lens system as well as the telescopic afocal lens system.
The lens system was previously set up by a method of trial and error.  This can produce adequate
results for printing at large pixel sizes, however it is prone to error and makes it difficult 
to know what the ultimate holopixel size will be for a given lens setup.

For a large holopixel size (around \unit[1]{mm}), setting up the lens system by
eye is feasible and provides results of sufficient quality. However for smaller
holopixel sizes (less than \unit[0.5]{mm}), the system becomes increasingly
sensitive to the exact placement of the lenses.  Large uncertainties in the 
beam size translate to large uncertainties in the energy density on the photographic
plate.  Finding a 'sweet spot' for the correct pixel size and pixel fidelity frequently results 
in the beam size exceeding that of the optics, resulting in undesirable energy losses.

\bigskip

The purpose of this section is to provide an analytical approach to the
problem. To do this, the telecentric afocal reversing lens system is
treated as two overlapping relay telescopes. The compound
lenses are ray-traced to determine their effective focal lengths.  The small weak lens on the
\ac{LCOS} is a field lens for correcting the curvature on the final image,
and is thus assumed to be negligible for these purposes.

Since the vertically polarised (the negative) part of the image from the
\ac{LCOS} is discarded by the polarised beam splitter, the phrase `\ac{LCOS}
image' will be used to mean the reflected horizontally-polarised 
image (the positive), in order simplify the language required.

Consider the image from the \ac{LCOS} passing through the beam splitter and
then continuing through the lens L2. It forms a real image at I1.
By considering that L2 and L1 approximately form a relay lens system,
it can be seen that the image at L1 will be that of
the aperture P, scaled according to the lens relay formulas (considered
later).  The beam continues through L3, forming an image at I2. By considering
that the lenses L3 and L2 form a relay lens, it can be seen that this image
at I2 is the image of the \ac{LCOS}, again scaled according to the lens
relay formulas.

The system of lenses at L4 take the image at I2 and focus it at
I3 which is thus the Fourier transform plane for the lens system
L4. By considering that L4 and L3 form a relay lens, it can be seen
that this image at I3 is the image that was at I1, and thus is an image
of the aperture P, scaled twice by the two sets of relay lenses. The
details of the lens are discussed in more detail later in \Fref{sec:optical_system_for_lens_array}.

The image at I3 has a geometrically similar shape (i.e. the same shape
but scaled) to that of that of P. It is also the Fourier transform of
the image I2.

\bigskip

The thin lens formula in air is:
\begin{equation}
\frac{1}{S_1}+\frac{1}{S_2}=\frac{1}{f}
\end{equation}

Where  $S_1$  is the distance between an object plane and a thin
lens with focal length $f$ and
$S_2$ is the distance between the thin
lens and the image plane.

For two thin lenses with focal lengths $f_1$ and $f_2$ respectively, separated
by an optical distance of $f_1+f_2$ , the final convergence
of the beam is not altered (making it afocal), but the width
of the beam is magnified by a factor of:
\begin{equation}
M=-\frac{f_1}{f_2}
\end{equation}

Thus the object image plane at distance $f_1$ from the first lens will be
relayed to the image plane at distance $f_2$ from the second lens and inverted.

\bigskip

Compound lenses have a front focal length and a back focal length.  The front focal length
is the distance from the front surface to the principle upstream focal point.  The back focal length 
is the distance from the back surface to the principle downstream focal point.  While this is important for the placement and design of the compound lens, for the purposes of determining the magnification, the front and back focal lengths are not of particular interest.  

The effective focal length for a compound lens is the distance from a principle plane to its corresponding principle focal point\cite{BasicOptics}. 


So if lens 1 has an effective focal length $f_1$ and is downstream of lens 2 which likewise has an effective focal length $f_2$, and if their inner principle planes are separated by a distance of $f_1$ +
$f_2$ then the image plane at a distance $f_1$ from back principle plane of lens 1 is magnified at 
the image plane at a distance $f_{2,f}$ from the front principle lane of lens 2 by a factor of:

\begin{equation}
M=-\frac{f_1}{f_2}
\end{equation}

\bigskip

\begin{landscape}
\begin{figure}
\centering
\includegraphics[width=\hsize]{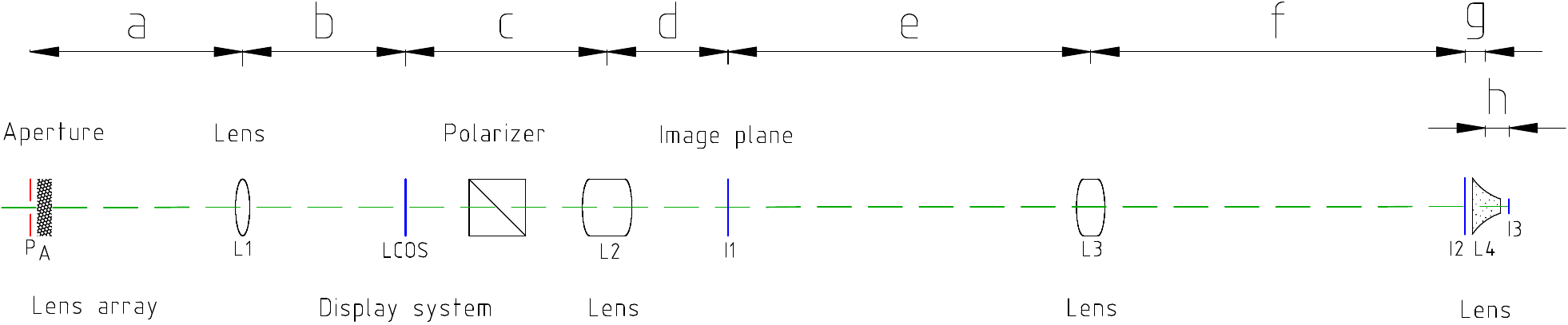}
\caption{Logical diagram of optical beam path layout}
\label{fig:logicallayout}
\end{figure}
\end{landscape}

The object beam lens system is shown schematically in \Fref{fig:logicallayout}. Note
that the actual path of the laser beam from lens L1 to the \ac{LCOS} is
via a mirror and the beam splitter. For diagram simplicity it is
drawn as if the \ac{LCOS} is transmissive. The distance labelled
$b$ should be interpreted as the optical distance from L1 to the
\ac{LCOS}.

Note that L2, L3 and L4 are compound lenses with different front and
back focal distances. As is the convention in optics, the front is
defined as in the direction of the beam, and back as the direction in
which the beam came from.

This complex system of lenses is particularly sensitive to distances between L1 and L3, requiring
an optical collimator for precise alignment.  This can be done by mounting the \ac{LCOS} and its weak correctional mirror together onto the moving platform at one end.


\bigskip

\subsection{Lenses L1 and L2}

The object image plane at aperture P is at a distance of $a$ from the center of
lens L1, which has a effective focal length of $f_1$.  The distance from the aperture P to the closest
principle plane of lens L1 should be approximately equal to the focal length, not exactly, such that the final pixel shape is a slightly defocused image of the aperture, producing a blurred outline on the hologram film.

The light then travels a total distance of $b+c$ to lens L2, via
a mirror, beam splitter, a reflection off of the \ac{LCOS}, twice passing through a small correction field lens on the \ac{LCOS}, and back through the beam splitter.

The lenses L1 and L2 act together as an afocal relay lens, imaging the aperture P at the image plane I1.

It turns out that it is sufficient to use a thin lens at L1.  Thus the principle plane is at the center of L1.  The magnification due to L1 and L2, $M_{P,1}$, can be determined, given that the effective focal length of lens L2 is $f_2$.
\begin{align}
f_1&\approx a\label{eq:f1_approx_a}\\
f_1&=b\label{eq:f1_isequal_b}\\
M_{P,1}&=-\frac{f_1}{f_2}
\end{align}
\nomenclature{$M_{P,1}$}{Magnification of aperture image plane P at image plane I1}

\subsection{Lenses L3 and L4}

The lenses L3 and L4, with effective focal lengths $f_3$ and $f_4$ respectively, likewise act together as an afocal relay lens, imaging the image of the aperture P at I1 to the image plane I3.  The image plane I1 is magnified by a factor of $M_{1,3} = -\frac{f_3}{f_4}$ at image plane I3.
\nomenclature{$M_{1,3}$}{Magnification of image plane I1 at image plane I3}

The image plane P is magnified by a factor of $-\frac{f_1}{f_2}$ at
the real image plane I1, and then magnified again by a factor of $-\frac{f_3}{f_4}$ at
the real image plane I3. So the total magnification, $M_{P,3}$ of the aperture P at the image plane I3 is:
\begin{equation}
M_{P,3}=\frac{f_1 f_3}{f_2 f_4}
\end{equation}
\nomenclature{$M_{P,3}$}{Magnification of aperture image plane P at image plane I3}

So the shape and size of P determines the shape and size of the final
holopixel on the holographic film plate, magnified by a factor $M_3$.  The Zemax software determined the effective local length of the system from the \ac{LCOS} display to the lens L4 to be $f_\text{eff}=\unit[-7.669]{mm}$. Thus:
\begin{align}
M_{P,3}&=-\frac{f_1}{f_\text{eff}}\\
M_{P,3}&=-\frac{f_1}{\unit[-7.669]{mm}}\\
&=\frac{f_1}{\unit[7.669]{mm}}\label{eq:mp3}
\end{align}
\nomenclature{$f_\text{eff}$}{Total effective focal length of system from LCOS display to the Lens L4 as determined by the Zemax software}

\subsection{Lenses L2 and L3}

For completeness, consider the lens system formed by L2 and L3. The
image plane of the \ac{LCOS} is relayed to the image plane I2 by lenses L2 and L3 with effective
focal lengths $f_2$ and $f_3$ respectively.  The image plane I2 has a magnification $M_{lcos,2}$ of:
\begin{align}
M_{lcos,2}&=-\frac{f_2}{f_3}
\end{align}
\nomenclature{$M_{lcos,2}$}{Magnification of LCOS image plane at image plane I2}

\subsection{Focal length approximations}

The surface properties of the compound lenses L2, L3 and L4 were calculated and optimized in the Zemax optics software\footnote{The Zemax software is owned by the \citet{zemax} company and is used to aid in lens design.}.  \Fref{tab:opticalcomponents} in Appendix~\ref{chap:lenscomponents} details the lens materials, separation and clear diameter properties.  

\begin{figure}[htp]
\centering
\includegraphics[angle=-90,width=\textwidth]{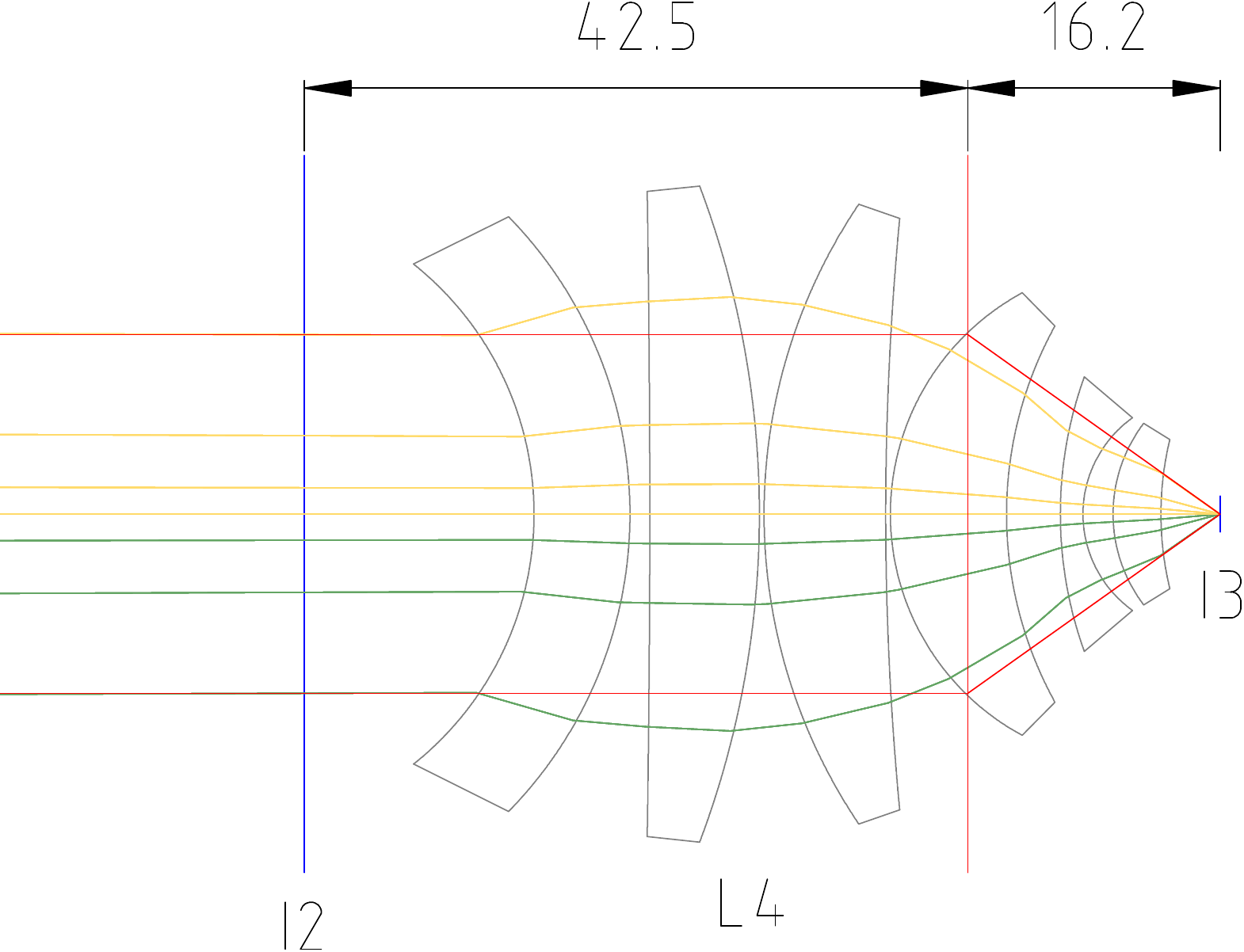}
\caption[Effective focal length of lens L4]{Effective focal length of compound objective lens L4 determined from construction lines (drawn in red).  The focal length is determined to be $f_4=\unit[16.2]{mm}$.  Focal lines and ray traced light paths determined with Zemax software.}
\label{fig:focal_length_L4}
\end{figure}

For lenses L2 and L3, the effective focal length of each of the lenses can be determined from the ray transfer matrices method (See \citet[pages 120-130]{IntroOptics} and the Chapter~Nomenclature on page \pageref{sec:matrixmethod}).  Compound lens L4 is significantly more complex and has a much shorter focal length.  As a result, the paraxial approximation is not valid for this lens, resulting in inaccuracies in the matrix method.  The matrix method calculates the effective focal length of lens L4 to be \unit[19.7]{mm}, whereas a geometrical calculation based on a raytrace produced by Zemax gives the effective focal length as \unit[16.2]{mm}.  This is shown in  \Fref{fig:focal_length_L4}.  To avoid such inaccuracies, the values calculated by Zemax were used exclusively.

\subsection{Optical system for microlens array}\label{sec:optical_system_for_lens_array}

There are additional factors to be taken into account because of the microlens
array close to, and downstream of, the object beam aperture. Without the microlens array, the \ac{LCOS} display would be illuminated with the `raw' Gaussian beam. This means that any dust
etc that was on any of the mirrors would appear as dark spots on the
\ac{LCOS} along with diffraction rings.  The use of a microlens array acts a series of point
sources which expand and overlap downstream, spatially averaging out the noise in the beam.
The microlens array is also used to apodize\footnote{apodize -- to purposely change the input intensity profile} the beam's Gaussian spatial intensity profile into an even spatial intensity profile at the \ac{LCOS} display.

Unfortunately the use of a microlens array introduces an additional problem {--} the overlapping beams are
coherent with each other and so interfere, giving a real speckled image.
This speckle degrades the image of the \ac{LCOS} and thus reduces the effective
resolution. This will reduce the overall depth of field of the
hologram when viewed.

The microlens arrays required in this design are rectangularly packed with an aspect ratio equal or close to that of the \ac{LCOS} such that the downstream beam is similar is shape and size to the \ac{LCOS} display.  This rotationally asymmetric geometry results in rotationally asymmetric speckle.  Rectangular speckle structures are observed in the downstream spatial profile.

\bigskip

In order to analyze the microlens array system, consider the
meridional plane only (the vertical plane that crosses through the
optical axis). The lenslets have the same aspect ratio as the \ac{LCOS}
and the size of the beam (as we will see later) will be a linear
function of the lenslet size, so the beam will have the correct width
if the height is correct.

\bigskip

The Gauss conditions will also be assumed and applied {--} that the beam is parallel
to the optics. The assumption that the lenses are thin lens and that
the angles are small enough that $\sin(x)\approx x$ will also be used.

\bigskip

First consider light rays travelling near the optical axis such that the 
rays pass through a lenslet centered on the optical axis.
The ray passes through this central lenslet which has a focal length of 
$f_\text{lenslet}$.  The ray then travels a distance of
$a$ (see \Fref{fig:logicallayout}), passing through the lens L1 which has a
focal length of $f_1$.  The ray then travels a distance of
$b$ to illuminates the \ac{LCOS} display.
Using the thin lens approximation and the paraxial approximation, the matrix method can be used to examine this system (See \citet[pages 120-130]{IntroOptics} and the Chapter~Nomenclature on page \pageref{sec:matrixmethod}).  Thus using a combination of lens and transfer matrices the path of the rays travelling through the central lenslet onto the \ac{LCOS} can be described.

The matrix method requires an input ray, $\vec{v}_\text{lenslet}$, which can be described as 
$\vec{v}_\text{lenslet}=\left(r, \theta\right)^T$ where $2r$ is the height and $\theta$ is the angle that the ray is travelling from the horizontal. The resulting ray vector on the \ac{LCOS} is thus $\vec{M}_\text{lenslet}\cdot \vec{v}_\text{lenslet}$, where the matrix $M$ for this system is thus:
\nomenclature{$f_n$}{Focal length of lens L$n$}
\nomenclature{$f_\text{lenslet}$}{Focal length of lenslets in microlens array}
\nomenclature{$\vec{v}_\text{lenslet}$}{Input light ray vector at lensarray for central lenslet, for lens analysis}
\nomenclature{$\vec{M}_\text{lenslet}$}{Matrix for ray path from central lenslet to lensarray, for lens analysis}
\bigskip

\begin{equation}
\vec{M}_\text{lenslet}=\left(\begin{matrix}1&b\\0&1\end{matrix}\right)
   \left(\begin{matrix}1&0\\-\dfrac{1}{f_1}&1\end{matrix}\right)
   \left(\begin{matrix}1&a\\0&1\end{matrix}\right)
   \left(\begin{matrix}1&0\\-{\dfrac{1}{f_\text{lenslet}}}&1\end{matrix}\right)
\end{equation}

\bigskip

This can be extended to work for an arbitrary lenslet on the lensarray, by simply extending the ray vector as:

\begin{equation}
\vec{v}_\text{lensarray}=\left(\begin{matrix}r& \theta & s \end{matrix}\right)^T
\end{equation}
\nomenclature{$\vec{v}_\text{lensarray}$}{Input light ray vector at lensarray for an arbitrary lenslet, for lens analysis}
\nomenclature{$\vec{M}_\text{lensarray}$}{Matrix for ray path from lensarray to LCOS for an arbitrary lenslet, for lens analysis}
Where $r$ is the height, $s$ is the height from the center
of the lenslet that it is passing through, and $\theta$ is
the angle that the ray is travelling from the horizontal.

Only the lenslet matrix is concerned with the $d$ parameter so
this parameter is dropped by the lensarray matrix.

The final matrix for the ray path from the lensarray to the \ac{LCOS}, $\vec{M}_\text{lensarray}$, becomes:

\bigskip

\begin{equation}
\vec{M}_\text{lensarray}=\left(\begin{matrix}1&b\\0&1\end{matrix}\right)
   \left(\begin{matrix}1&0\\-{\dfrac{1}{f_1}}&1\end{matrix}\right)
   \left(\begin{matrix}1&a\\0&1\end{matrix}\right)
   \left(\begin{matrix}1&0&0\\0&1&-\dfrac{1}{f_\text{lenslet}}\end{matrix}\right)
\end{equation}

From the matrix method\citep{IntroOptics}, $\vec{M}_\text{lensarray} \cdot \vec{v}_\text{lensarray} = \vec{v}_\text{lcos}$ where $\vec{v}_\text{lcos}$ is the ray vector at the \ac{LCOS}.  Solving:
\nomenclature{$\vec{v}_\text{lcos}$}{Output light ray vector at the LCOS for an arbitrary lenslet}

\begin{equation}
\left(\begin{matrix}
  r-\dfrac{\left(a+b\right)\cdot s}{f_\text{lenslet}}-\dfrac{b\cdot r}{f_1}+\dfrac{a\cdot b\cdot s}{f_\text{lenslet}\cdot f_1} \\
\dfrac{a\cdot s}{f_\text{lenslet}\cdot f_1}-\dfrac{r}{f_1}-\dfrac{s}{f_\text{lenslet}}
\end{matrix}\right)
= \left(\begin{matrix} r_\text{LCOS} \\ \theta_\text{LCOS} \end{matrix}\right)
\end{equation}

\bigskip

For the image of the aperture to be approximately focused, \Fref{eq:f1_approx_a} established the condition that $f_1\approx a\approx b$.

\bigskip

To proceed, it is worth considering some practical implementation
aspects. If  $f_1=b$  then the practical significance is that without the microlens
array, a collimated beam would be focused by lens L1 to a tiny point on
the sensitive \ac{LCOS}. This means that if the microlens array is ever accidentally removed
or knocked over while the laser is on, the \ac{LCOS} will be
instantly destroyed. Given the high cost of the \ac{LCOS} this is clearly
not desirable in machine used for prototyping. From this practical point of view,
it is worth making sure that only $f_1\approx b$.  There are no similar concerns for making 
$f_1=a$.

\bigskip

Examining the matrix $\vec{M}_\text{lensarray}$, it can be noted that if $f_1=a$ then the matrix simplifies considerably to:

\begin{equation}
\left(\begin{matrix}
\dfrac{f_1\cdot s}{-f_\text{lenslet}} \\
\dfrac{r}{f_1}
\end{matrix}\right)
+ r
\left(\begin{matrix}
1-\dfrac{b}{f_1} \\
0
\end{matrix}\right)
=
\left(\begin{matrix}
r_\text{LCOS} \\
\theta_\text{LCOS}
\end{matrix}\right)
\end{equation}

Where the matrix has been split into the main unperturbed component and
its perturbation for $f_1\neq b$, for ease of reading. We can
confirm that the perturbation approaches zero as $f_1$ approaches $b$.

\bigskip

Since $f_1$ is a positive quantity and $f_{lenslet}$ is a negative quantity, the
magnification of the beam is positive so the image is right side up.

\bigskip

Each lenslet has an identical clear height {--} the height of the lenslet.
Consider the top-most and bottom-most ray for a given lenslet of
height $h_\text{lenslet}$. The size of final illumination on the \ac{LCOS} will be
defined by these two rays and thus has a height of:
\nomenclature{$h_\text{lenslet}$}{Height of each lenslet in microlens array}
\begin{equation}
\begin{split}
\left(\frac{f_1\cdot
h_\text{lenslet}/2}{-f_\text{lenslet}}+\left(r+h_\text{lenslet}/2\right)\left(1-\frac{b}{f_1}\right)\right)-\\
\left(\frac{f_1\cdot\left(-h_\text{lenslet}/2\right)}{-f_\text{lenslet}}+\left(r-h_\text{lenslet}/2\right)\left(1-\frac{b}{f_1}\right)\right)\\
=\frac{f_1\cdot
h_\text{lenslet}}{-f_\text{lenslet}}+h_\text{lenslet}\left(1-\frac{b}{f_1}\right)
\end{split}
\end{equation}

Since the lenslets are rectangular with the same aspect ratio as the
\ac{LCOS}, then the width of the illumination beam on the \ac{LCOS} will have the
correct width if the height is correct.

\bigskip

Considering the perturbation part of the matrix, it can be seen that the
rectangular image produced by each lenslet will not exactly overlap.
The size of the beam produced by each lenslet will be identical, but
the lateral position of each illumination rectangle will be different for each lenslet. The difference between the lowest top-most ray and the
highest bottom-most ray can be used to find the height of the area that will be illuminated
by all of the lenslets.

If the beam illuminating the microlens array has a height $y$ (equal to the height of the aperture), then the illumination height $H_\text{lcos}$ is:
\nomenclature{$y$}{Diameter of beam illuminating microlens array}
\begin{equation}
H_\text{lcos}=\frac{f_1\cdot h_\text{lenslet}}{-f_\text{lenslet}}-y\left(1-\frac{b}{f_1}\right)
\label{eq:effectiveIlluminationHeight}
\end{equation}

\bigskip

The previously mentioned matrix $\vec{M}_\text{lensarray}$ describes the required lens properties for a chosen microlens array can be calculated (i.e. calculate the focal length such
that the beam correctly illuminates the \ac{LCOS}).  The required lens is placed at the 
optical distance approximately equal to the lens' focal length.  The \ac{LCOS} should be protected with a covering (or removed) when first testing the inserted lens, to avoid destroying the \ac{LCOS}.

\bigskip

The initial beam is of the order of a centimeter and the focal length 
$f_1$ of the lens L1 is of the order of \unit[30]{cm}, so 
$r \ll f_1$  so 
$\mathit{\theta}_\text{LCOS}\approx 0$ . This is important
because the beam clear height is determined by the clear height of the intermediate mirror and cube beam-splitter.  This puts a restriction on initial beam size, since the larger the initial beam size, the larger the beam divergence.

\bigskip

The magnification of the aperture at the holopixel plane was determined in \Fref{eq:mp3}.  This was based on a total effective focal length, $f_\text{eff}$ of \unit[7.669]{mm}. For a holopixel size of between \unit[1.0]{mm} to \unit[0.3]{mm} and utilizing at least \unit[10.0]{mm} of the microlens array up to a maximum of \unit[30.0]{mm}, the magnification factor should be between 10 to 100.  Given that the distance $b$, and thus the focal length $f_1$, is restricted to approximately \unit[150]{mm}, and using the said magnification formula, we obtain:
\begin{align}
\unit[150]{mm} &\le f_1 < 100\times f_\text{eff}\\
\unit[150]{mm} &\le f_1 < 30\times \unit[7.669]{mm}&\text{For holopixel of size \unit[1.0]{mm}}\\
\unit[150]{mm} &\le f_1 < \unit[230]{mm}&\text{For holopixel of size \unit[1.0]{mm}}\\
33\times\unit[7.669]{mm} &\le f_1 < 100\times \unit[7.669]{mm}&\text{For holopixel of size \unit[0.3]{mm}}\\
\unit[256]{mm} &\le f_1 < \unit[767]{mm}&\text{For holopixel of size \unit[0.3]{mm}}
\end{align}

Rearranging the unperturbed part of \Fref{eq:effectiveIlluminationHeight}:
\begin{align}
f_1 &= -H_\text{lcos} \times \frac{f_\text{lenslet}}{h_\text{lenslet}}
\end{align}

\clearpage

\subsection{Summary}

\definecolor{MyGray}{rgb}{0.96,0.97,0.98}
\makeatletter\newenvironment{summarybox}{%
   \begin{lrbox}{\@tempboxa}\begin{minipage}{\columnwidth}}{\end{minipage}\end{lrbox}%
   \colorbox{MyGray}{\usebox{\@tempboxa}}
}\makeatother

\begin{summarybox}
Choose the required holopixel radius, $r$.  Typically $\unit[1.0]{mm} \ge 2r \ge \unit[0.3]{mm}$.

For a given microlens array with lenslets with a height of $h_\text{lenslet}$ and a focal length of $f_\text{lenslet}$, determine the required focal length of lens L1, $f_1$.
\begin{align}
f_1 &= -H_\text{lcos} \times \frac{f_\text{lenslet}}{h_\text{lenslet}}\label{eq:requiredf1}
\end{align}

Ensure that the chosen microlens array sets $f_1$ within the range:
\begin{align}
\unit[150]{mm} &\le f_1 < \unit[230]{mm}&\text{For holopixel of size \unit[1.0]{mm}}\\
\unit[256]{mm} &\le f_1 < \unit[767]{mm}&\text{For holopixel of size \unit[0.3]{mm}}
\end{align}

Place the lens L1 at a distance $b\approx f_1$, such that $b$ is a few millimeters different from $f_1$, to slightly defocus the holopixel image.  Place the microlens array at a distance $a$ from the lens L1 such the $a=f_1$.

Use a square or circular aperture with a height $y$, placed against the microlens array, where $y$ is:
\begin{align}
y &= 2r\cdot \frac{f_1}{f_\text{eff}} \label{eq:sizeofy}
\end{align}
\end{summarybox}
\clearpage

\subsection{Example microlens array}

The necessary formulas for determining the properties of the microlens array and lenses were established in the previous section. The distance $b$ (From L1 to the \ac{LCOS}, illustrated in \Fref{fig:scene2}) is restricted by the requirement that the clear height of the diverging beam is set by the clear height of the intervening beam splitter and mirror.  The distance is further restricted by the physical minimum spacing between the \ac{LCOS}, beam splitter and mirror.  This restricts the minimum distance of $b$ to approximately \unit[150]{mm}\label{page:minimumb} (\Fref{fig:hrip_lcos_top}).

\begin{figure}[p]
\centering
\includegraphics[width=10cm]{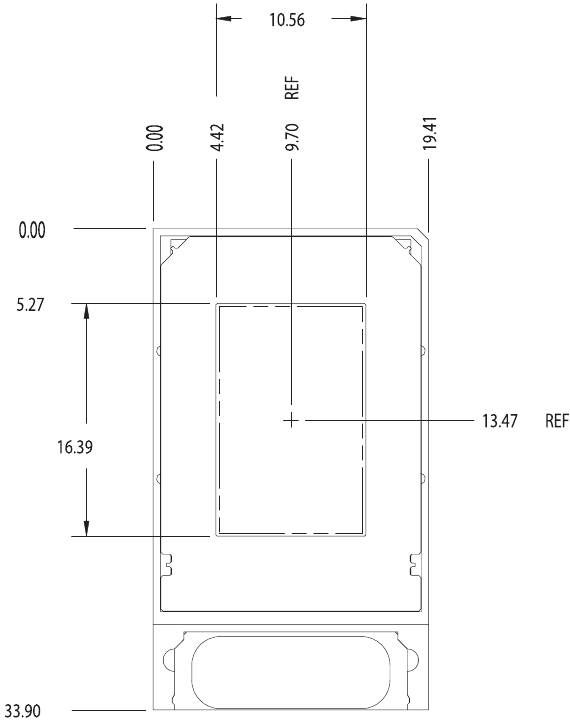}
\caption[Brillian LCOS Display BR1080HC]{Dimensional schematic of the Brillian \acs{LCOS} Display BR1080HC\\ \small{\url{http://www.brilliancorp.com/PDF/projection/BR1080HC_BR_D4.pdf}}}
\label{fig:brillian}
\end{figure}

The display used in this printer is a reflective \ac{LCOS} display made by
the company Brillian. The model is the BR1080HC, and the dimensions
are shown in \Fref{fig:brillian}. The \ac{LCOS} is used in landscape format.

This Brillian \ac{LCOS} has a display screen height of \unit[10.56]{mm} (See \Fref{fig:brillian}), requiring that the illuminating beam has a height equal to this to ensure full illumination without unnecessary loss of energy. With this requirement, $H_\text{lcos}$ = \unit[10.56]{mm}.

\bigskip

Applying the determined values to the unperturbed term in \Fref{eq:effectiveIlluminationHeight}, the desired ratio of lenslet height $h_\text{lenslet}$ to lenslet focal length $f_\text{lenslet}$ is obtained:
\begin{align}
\frac{h_\text{lenslet}}{-f_\text{lenslet}} &\le \frac{\unit[10.56]{mm}}{\unit[150]{mm}}\\
\implies \frac{h_\text{lenslet}}{-f_\text{lenslet}} &\le 0.07 \label{eq:condition}
\end{align}

\begin{table}[p]
\centering
\begin{tabular}{rlll}
\toprule
& \multicolumn{1}{A}{\textbf{First}} & \multicolumn{1}{B}{\textbf{Second}} & \multicolumn{1}{C}{\textbf{Third}}\\
\midrule
Lenslet dimensions: &0.40$\times$\unit[0.46]{mm}&0.20$\times$\unit[0.23]{mm}&0.040$\times$\unit[0.046]{mm}\\
Spherical radius:&\unit[1.3]{mm}&\unit[0.65]{mm}&\unit[0.13]{mm}\\
Required magnification:&26.4&52.8&264.0\\
Focal length $f_\text{lenslet}$:&\unit[-0.4]{mm}&\unit[-0.8]{mm}&\unit[-4.0]{mm}\\
Required $f_1$:&\unit[10.56]{mm}&\unit[42.24]{mm}&\unit[324.92]{mm}\\
Aperture size ($r$=\unit[1.0]{mm}):&\unit[1.38]{mm}&\unit[5.51]{mm}&\unit[42.37]{mm}\\
Aperture size ($r$=\unit[0.3]{mm}):&\unit[0.41]{mm}&\unit[1.65]{mm}&\unit[12.71]{mm}\\
\bottomrule
\end{tabular}
\caption{Microlens array properties}
\label{tab:Microlensarrayproperties}
\end{table}

Consider an \ac{LCOS} with lenslets with a clear height of $h_\text{lenslet}=\unit[0.2]{mm}$ and a focal length of $f_\text{lenslet}=-\unit[3.0]{mm}$.  This satisfies the condition since $\unit[0.2]{mm} \div \unit[3.0]{mm} = 0.67$.

From \Fref{eq:requiredf1}:
\begin{align}
f_1 &= -H_\text{lcos} \times \frac{f_\text{lenslet}}{h_\text{lenslet}}\\
&= \unit[10.56]{mm} \times \frac{\unit[3.0]{mm}}{\unit[0.2]{mm}}\\
&= \unit[158]{mm}
\end{align}

For a \unit[1.0]{mm} pixel the aperture has a height $y$, from \Fref{eq:sizeofy}, of:
\begin{align}
y &= 2r\cdot \frac{f_1}{f_\text{eff}}\\
  &= \unit[1.0]{mm} \cdot \frac{\unit[158]{mm}}{\unit[7.669]{mm}}\\
  &= \unit[20.7]{mm}
\end{align}

To obtain a circular holopixel that is \unit[1.0]{mm} in diameter, the aperture needs to also be circular with a diameter $y$ of \unit[20.7]{mm}.  This would utilize approximately $\frac{20.7^2}{0.2^2} = \unit[10712]{lenslets}$ - sufficient to provide smooth illumination.

\bigskip

Three different microlens arrays that satisfy the condition in \Fref{eq:condition} are experimentally tested in the next section for a qualitative analysis. The three microlens arrays examined are all composed of an array of rectangular lenslets with a dimensional ratio equal to that of the \ac{LCOS}.  The relevant details along with the required focal length of lens L1 and required aperture height calculated by the methods given here, are given in \Fref{tab:Microlensarrayproperties}.

\clearpage
\section{Experimental evaluation of lens array system}\label{sec:experimental_evaluation_lens_array}

In analogue holography a hologram is made by illuminated a physical object with coherent light.  The microscopic roughness of the surface of the object creates speckle which can present a serious visual problem.  The final hologram can appear to 'twinkle' or 'sparkle' and the effective resolution of the hologram is several times lower than the diffraction limit \citep{oliver1963ssa}.

In digital holography using a computer display, the problem of speckle is almost removed.  A dot matrix of 'holopixels' is printed, each with their own independent speckle.  On replay with a white-light source, the human eye does not observe any sparkling effects.  Speckle does pose other problems however.  The speckle affects the angular distribution of intensity for each holopixel, as described below.

There is an additional possible problem -- speckle on image at the Fourier plane may also affect the diffraction efficiency as the light intensity is not evenly distributed across emulsion, thus not fully utilizing all of the grains.

\bigskip

\Fref{sec:optical_system_for_lens_array} analytically considered the microlens array system to determine the optimal distance based upon the lenslet size, aperture size, pixel size, and distances.  However this approach does not take into account the effect of speckle.  The microlens array acts as an array of mutually coherent point light-sources.  These point sources expand and overlap, each illuminating the microlens array, creating an averaged illumination that hides any defects in the beam quality (due to dust particles etc).  An unwanted side-effect however is that the mutually coherent sources interfere with each other creating speckle.

The speckle pattern created on the \ac{LCOS} results in an effectively reduced resolution of the \ac{LCOS} display system, reducing the possible depth of field of the hologram (both into the hologram plane, and out of the hologram plane).  A severe degree of speckle can result in holograms with low fidelity and contrast.

\bigskip

To reduce the effect of speckle, a simple method was required that introduced minimal complexity and cost into the design.  A brief summary of existing methods is presented below, with its applicability to digital holography noted.  As the speckle is affected by the properties of the microlens array and the optics between the microlens array and the \ac{LCOS}, the simplest method was to try different microlens arrays and different positions to qualitatively minimize the effect of speckle.

The structure of the speckle is examined, followed by a subjective assessment of the overall image quality.  The results are summarized, and the microlens array with the larger lenslets (rectangular lenslets of dimensions \unit[0.40]{mm}$\times$\unit[0.46]{mm} with a curvature of radius of \unit[1.3]{mm}) was chosen for future digital holography research.

\bigskip

The raw recorded data is presented in Appendix~\ref{sec:lensarrayappendix}.

\subsection{Previous work}

The problem of speckle due to a coherent source on a microlens array has been tackled many times.  \citet{goodman1976sfp} showed over 30 years ago that speckle can be regarded as arising from a classical random walk in the complex plane.  Reducing the contrast or smoothing the speckle requires diversity in the polarization, space, frequency or time.  In the proceeding years, each of these avenues were investigated with varying success for use in many different fields in science.  \citet{iwai1996src} summarize the 30 years of research into each of these different avenues. The applicability of these approaches to digital holography is very briefly given here.

\bigskip

For pulsed holography, time varying the polarization in a pulse would destroy the diffraction gratings on the hologram.  Compositing both horizontal and vertical polarizations on to a single pixel may possibly work, although it would present practical difficulties.

\bigskip

Increasing spatial diversity has been used with moderate success in analogue holography \citep{yu1973srh}.  This approach could be taken with digital holography by using a spatial mask in front of the microlens array.  This would however result in blocking off a significant portion of the beam, reducing the energy efficiency of the holographic printer.

An alternative to introducing spatial diversity without masking the beam is to introduce random or pseudo-random phase shifts.  For holography, a common approach is to use a spatial phase mask with a pseudorandom sequence of phase shifts.  By keeping some degree of regularity in the phase shifts, speckle can be significantly reduced, while the slight pseudo-randomness removes the noise as intended \citep{kato1973srd,  kato1975srh}.

\bigskip

Varying the frequency or compounding multiple frequencies has been applied successfully in many other fields of science (\citet{trahey1986qas} for example).  It is not clear however how such an approach could be employed by holography.  Recording a hologram with a compound of two (or more) very similar frequencies would result in low diffraction efficiency as the diffracting gratings interfere with one another.  Time varying the frequency would also destroy the diffraction gratings required, aside from the practical complications.

\bigskip

Time varying the beam is a common approach to speckle.  Often the image is time averaged to visually remove speckle, such as employing the use of a rotating diffuser in the imaging plane of the microlens array\citep{Shin:02}, or vibrating an optical fibre\citep{ambar1985msr,ambar1986rss,ambar1986fci} to produce a time-varying phase-shift\citep{wang1998srl}.  This type of approach is however not particularly suitable for pulsed holography where the beam pulse is on the order of tens of nanoseconds.  More complicated approaches have been used such as splitting the speckled beam into several parts and using a delay line to temporally delay each of the beams differently.  Recombining the beam averages the speckle, effectively removing it without loss of resolution \citep{curtis2001lis}.  This introduces significant complexity into the design, and it is not clear if this type of approach would work for holography.

\clearpage
\subsection{Experimental setup}

The following microlens arrays were considered\footnote{A style note: For those with a colour version of this report, the data for the first microlens array is always plotted and referenced in a blue color, the second microlens array uses purple/red, and the third microlens array uses green.  This is purely a visual aid to the reader.}:
\begin{enumerate}
  \renewcommand{\labelenumi}{\color{blue}\textbf{\theenumi.}}
 \item \Fref{tab:lensarray1} examines the first microlens array with each lenslet being rectangular with physical dimensions of \unit[0.40]{mm}$\times$\unit[0.46]{mm}.  These lenslets are spherical lenses with a radius of \unit[1.3]{mm}. 
\renewcommand{\labelenumi}{\color{purple!90!white}\textbf{\theenumi.}}
\item \Fref{tab:lensarray2} examines the second microlens array with each lenslet being rectangular with physical dimensions of \unit[0.20]{mm}$\times$\unit[0.23]{mm}.  These lenslets are spherical lenses with a radius of \unit[0.65]{mm}.
\renewcommand{\labelenumi}{\color{green!50!black}\textbf{\theenumi.}}
\item \Fref{tab:lensarray3} examines the third microlens array with each lenslet being rectangular with physical dimensions of \unit[0.040]{mm}$\times$\unit[0.046]{mm}.  These lenslets are spherical lenses with a radius of \unit[0.13]{mm}.
\end{enumerate}

To test the three microlens array arrays, the lens system was setup as previously described in \Fref{sec:analysis_of_lens_system}.  Each of the three different microlens array arrays were tried at different positions.  The pulsed laser was switched producing a projected image from the final objective lens.  When printing, the photosensitive hologram plate would be placed a few millimeters away from the objective lens, in the Fourier plane of the image of the \ac{LCOS} display system.  This experiment instead removed any hologram plate, and projected the image of the \ac{LCOS} on to a white board \unit[1.2]{m} from the lens objective.

The contrast and intensity of the speckle relative to the displayed image was determined by the properties of microlens array array as well as its position.  The overall 'noise' (intensity strength relative to the displayed image) of the speckle was recorded, along with the size and shape of the beam profile where it illuminates the \ac{LCOS} display system.  

The speckle in the projected image also containted a repeating pattern.  The size of these repeating patterns was measured by measuring the size of a multiple number of these repeating structures with a ruler.

Appendix~\ref{sec:lensarrayappendix} gives the raw recorded data for the three different microlens array arrays at different positions.

\bigskip

As an important experimental note: If the microlens array array focal point is imaged directly onto the \ac{LCOS}, the energy density on the \ac{LCOS} can be sufficiently high to risk breaking the \ac{LCOS}.  This was avoided and noted in the tables were applicable.

\subsection{Image quality}

\begin{figure}[htp]
\centering
\includegraphics[angle=-90,width=\textwidth]{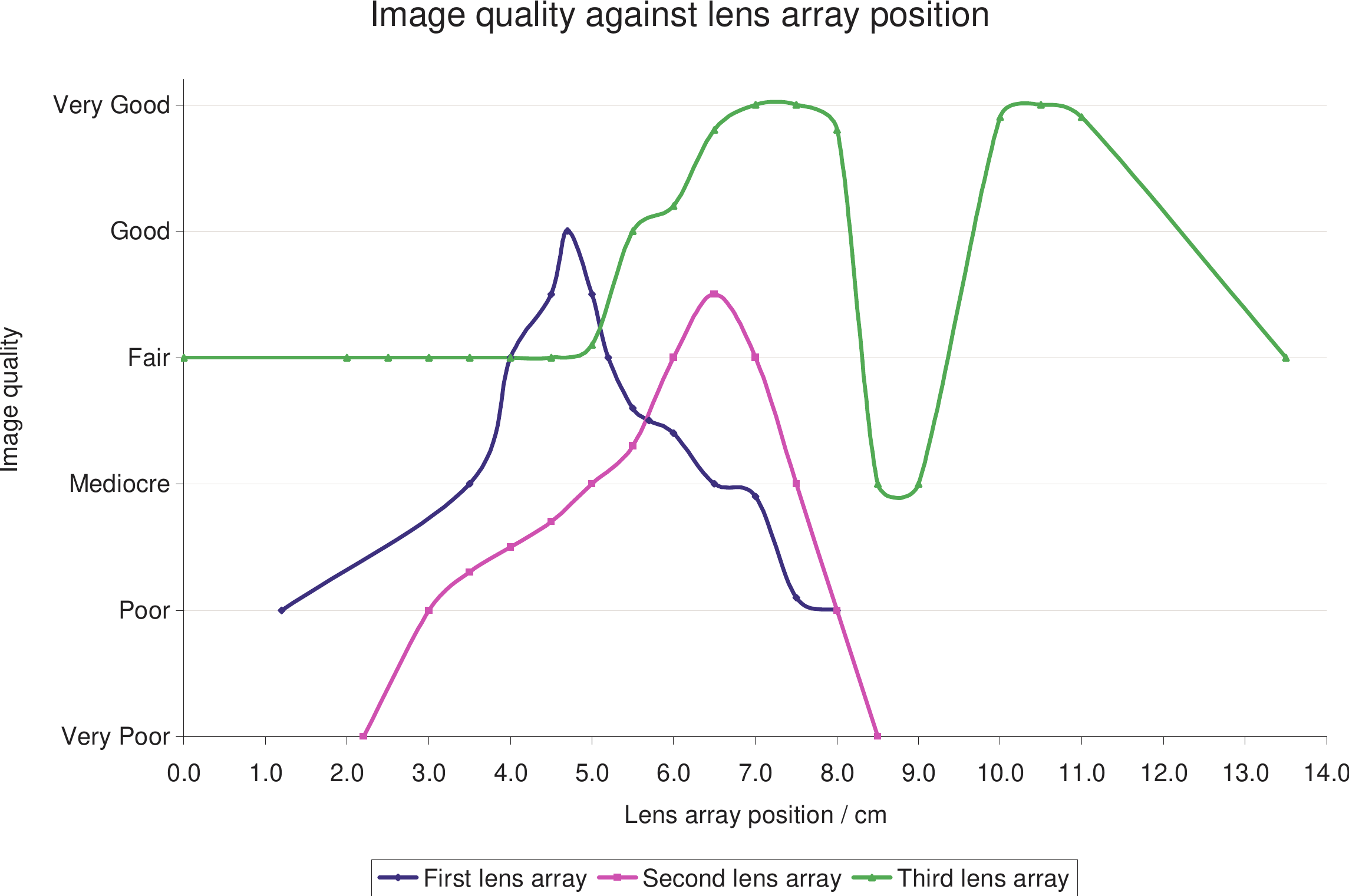}
\caption[Quality of the projected image due to speckle]{The subjectively judged quality of the projected image of the \ac{LCOS} on a white board \unit[1.2]{m} away from the objective lens, due to speckle noise.  An image judged as Very Good would have a nearly unnoticeable amount of visual speckle.  The data is drawn from the recorded data in Appendix~\ref{sec:lensarrayappendix}.}
\label{fig:image_quality_all}
\end{figure}

\begin{figure}[htp]
\centering
\includegraphics[width=\textwidth]{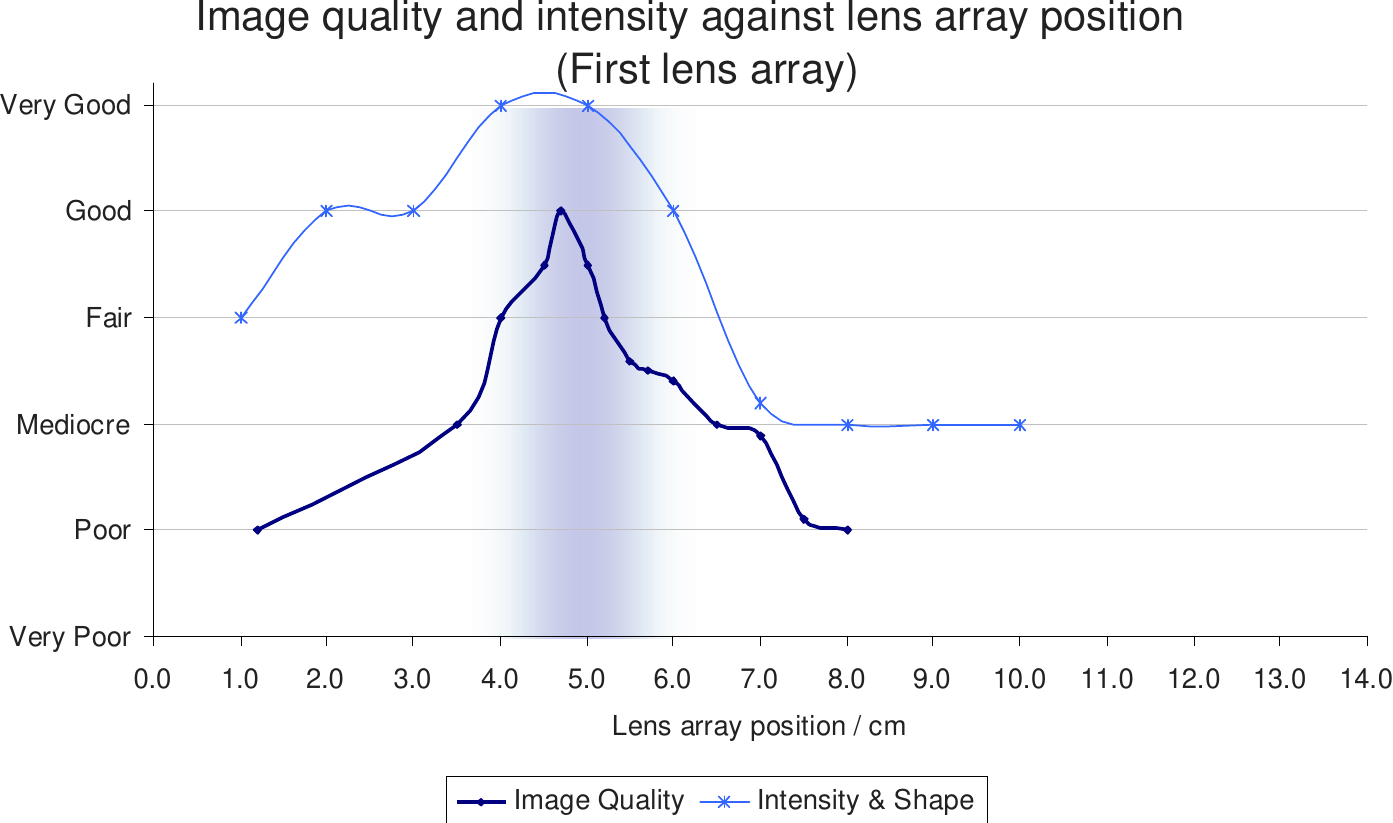}
\caption[Suitable positions for first microlens array]{The size and shape of the beam must be sufficiently close to the size and shape of the \ac{LCOS}.  Additionally the illumination intensity across the \ac{LCOS} must be spatially smooth and even.  These two factors were considered together, using the recorded data in Appendix~\ref{sec:lensarrayappendix}, and drawn here as the `Intensity \& Shape' curve.  This is drawn with the Image Quality curve shown in Figure~\ref{fig:image_quality_all} for comparison.  This graph is shown for the first microlens array.  A suitable region for the distance between the microlens array and closest lens such that the image quality, beam intensity and beam shape are all at least acceptable is between approximately \unit[4.0]{cm} to \unit[6.0]{cm}.}
\label{fig:image_quality_first}
\end{figure}

\begin{figure}[htp]
\centering
\includegraphics[width=\textwidth]{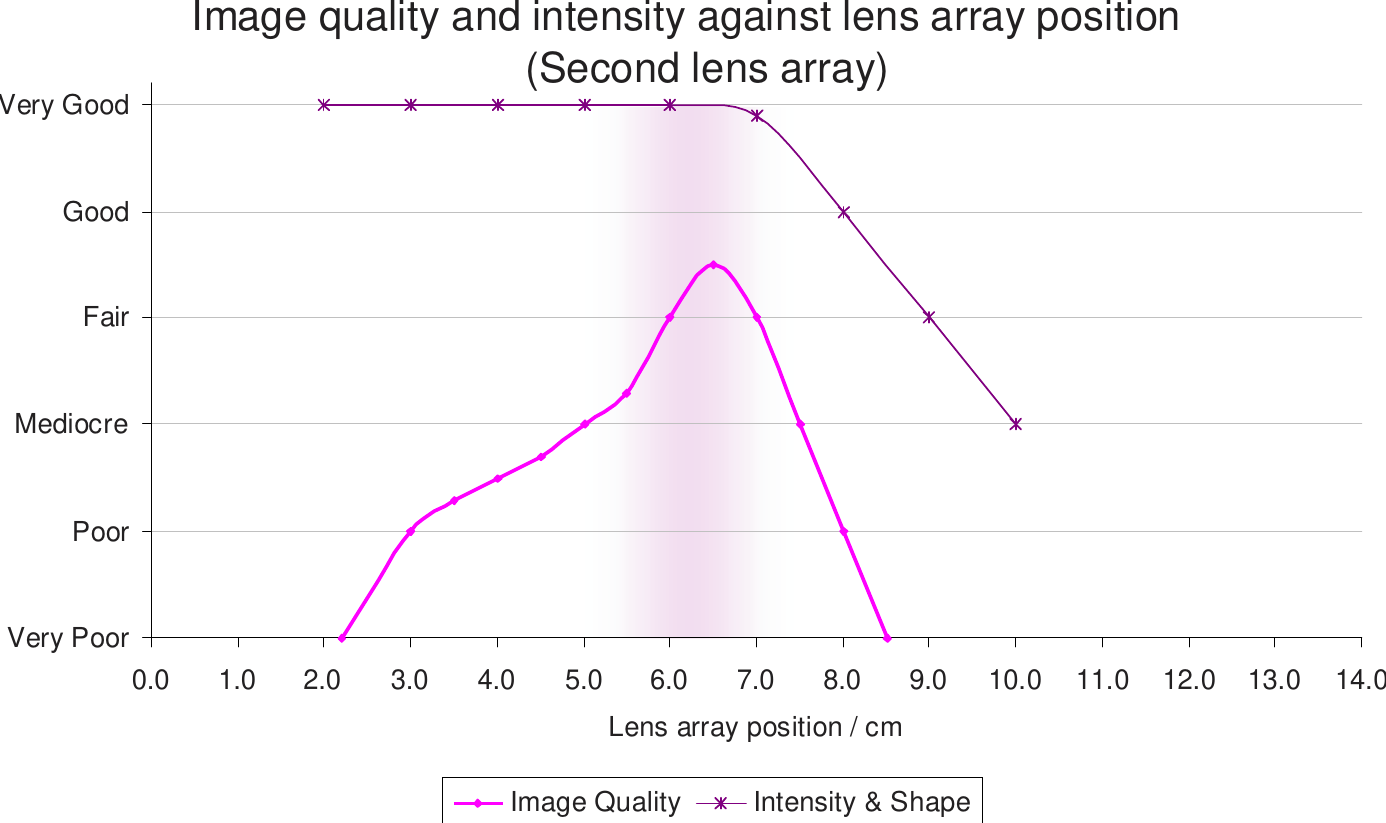}
\caption[Suitable positions for second microlens array]{This graph is similar to \Fref{fig:image_quality_first} and shown for the second microlens array.  A suitable region for the distance between the microlens array and closest lens such that the image quality, beam intensity and beam shape are all at least acceptable is between approximately \unit[5.5]{cm} to \unit[7.5]{cm}.}
\label{fig:image_quality_second}
\end{figure}

\begin{figure}[htp]
\centering
\includegraphics[width=\textwidth]{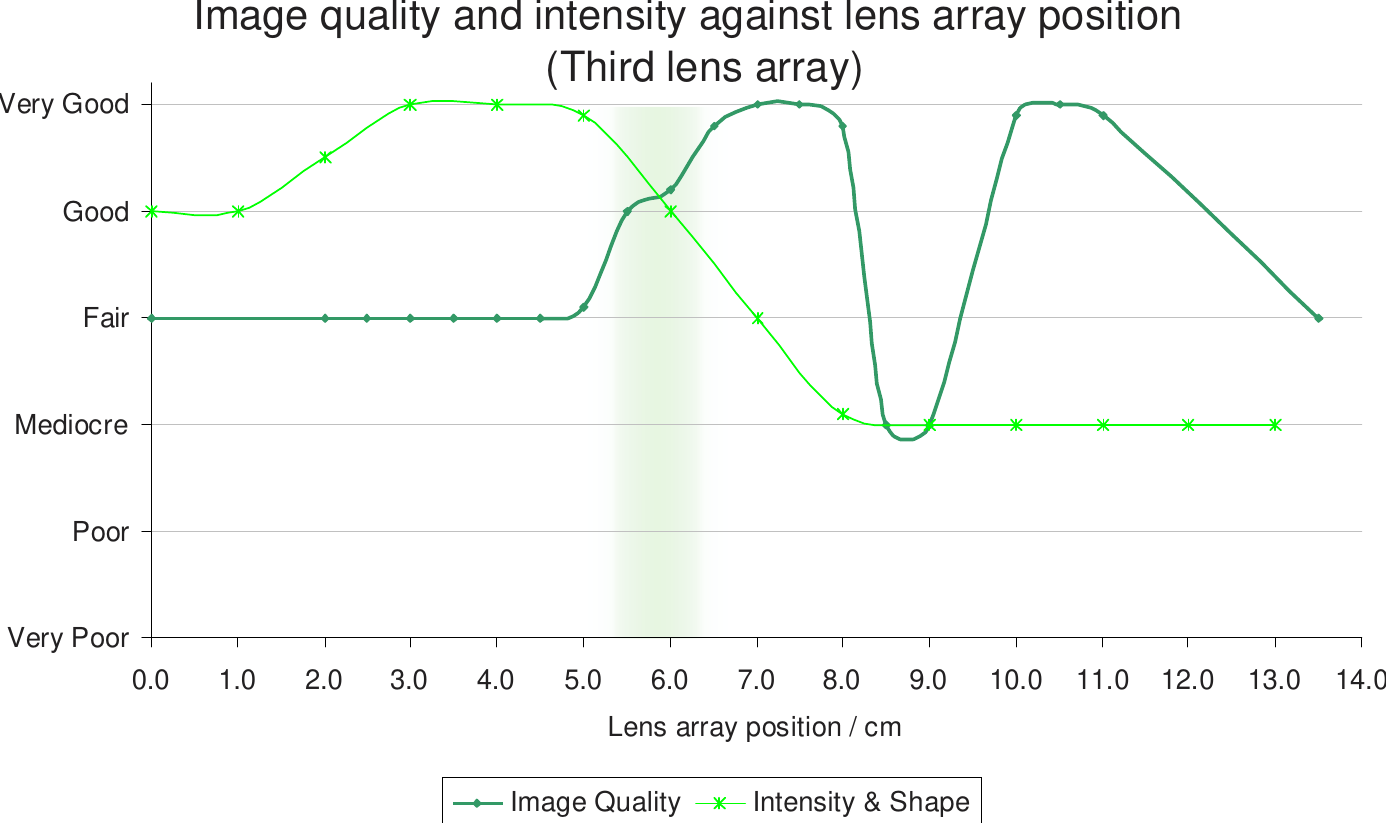}
\caption[Suitable positions for third microlens array]{This graph is similar to \Fref{fig:image_quality_first} and shown for the third microlens array.  A suitable region for the distance between the microlens array and closest lens such that the image quality, beam intensity and beam shape are all at least acceptable is between approximately \unit[5.5]{cm} to \unit[6.5]{cm}.}
\label{fig:image_quality_third}
\end{figure}

\Fref{fig:image_quality_all} graphs the quality of the projected image of the \ac{LCOS} display based upon a subjective judgment of the quality of the image.  A large range of positions was tested, however the range is limited by the \Fref{eq:f1_approx_a}; the microlens array is required to be positioned at approximately the distance of the focal length of the lens that images it onto the \ac{LCOS} display.  \Fref{fig:image_quality_first}, \Fref{fig:image_quality_second} and \Fref{fig:image_quality_third} judge a suitable range for the \ac{LCOS} based upon intensity distribution across the \ac{LCOS} (A smooth, even distribution is required, as opposed to a Gaussian distribution if the lens array is too close) and the shape of the beam (The whole of the \ac{LCOS} should be illuminated with little energy wasted).

\clearpage
\subsection{Speckle repeating structure size}

If the lenslets' focal points are imaged directly onto the \ac{LCOS}, the energy density on the \ac{LCOS} can be sufficiently high to risk breaking the \ac{LCOS}.  This was avoided and noted in the tables were applicable.

\Fref{tab:lensarray_repeating_structure} summarizes the speckle data that is presented in Appendix~\ref{sec:lensarrayappendix}; noting that the uncertainty error was $\pm$\unit[0.5]{cm} for the measurement of the total size of the sample of repeating structures.

\begin{table}[htp]
\centering
\begin{tabular}{R{0.86cm}*{3}{r@{\;}R{1.00cm}r@{}R{1.1cm}}}
\toprule
& \multicolumn{4}{A}{\textbf{First}} & \multicolumn{4}{B}{\textbf{Second}} & \multicolumn{4}{C}{\textbf{Third}}\\ 
Dist. A & 
No. & 
Total Dist 
& 
\multicolumn{2}{c}{Distance} & 
No. & Total Dist & \multicolumn{2}{c}{Distance} & 
No. & Total Dist & \multicolumn{2}{c}{Distance} \\
\midrule
0.0 &
&  &  &  &
&  &  &  &
6.0 & 21.5 & 3.58 & $\pm$0.08 \\
1.2 &
1.0 & 1.5 & 1.5 & $\pm$0.5 &
 &  &  &  &
 &  &  &  \\
2.0 & 
 &  &  &  &
 &  &  &  &
6.0 & 19.7 & 3.28 & $\pm$0.08 \\ 
2.2 &
 &  &  & &
7.0 & 25.5 & 3.64 & $\pm$0.07 &
 &  & & \\
2.5 & 
 &  &  & &
 &  &  & &
7.0 & 23.6 & 3.37 & $\pm$0.07 \\
3.0 & 
 &  &  & &
4.0 & 11.5 & 2.88 & $\pm$0.13 &
8.0 & 25.2 & 3.15 & $\pm$0.06 \\
3.5 & 
10.0 & 14.8 & 1.48 & $\pm$0.05 &
4.0 & 10.0 & 2.50 & $\pm$0.13 &
5.0 & 11.4 & 2.28 & $\pm$0.10 \\ 
4.0 & 
10.0 & 12.5 & 1.25 & $\pm$0.05 &
5.0 & 10.0 & 2.00 & $\pm$0.10 &
8.0 & 18.7 & 2.34 & $\pm$0.06 \\
4.5 & 
10.0 & 12.4 & 1.24 & $\pm$0.05 &
10.0 & 22.0 & 2.20 & $\pm$0.05 &
8.0 & 17.9 & 2.24 & $\pm$0.06 \\
4.7 & 
10.0 & 11.1 & 1.11 & $\pm$0.05 &
 &  &  & &
 &  & & \\
5.0 & 
10.0 & 11.5 & 1.15 & $\pm$0.05 &
10.0 & 22.0 & 2.20 & $\pm$0.05 &
10.0 & 22.0 & 2.20 & $\pm$0.05 \\
5.2 & 
10.0 & 10.5 & 1.05 & $\pm$0.05 &
10.0 & 21.0 & 2.10 & $\pm$0.05 &
 &  & & \\
5.5 & 
10.0 & 10.0 & 1.00 & $\pm$0.05 &
10.0 & 19.0 & 1.90 & $\pm$0.05 &
8.0 & 15.4 & 1.93 & $\pm$0.06 \\
5.7 & 
10.0 & 10.0 & 1.00 & $\pm$0.05 &
 &  &  & &
 &  & & \\
6.0 & 
10.0 & 9.0 & 0.90 & $\pm$0.05 &
10.0 & 17.0 & 1.70 & $\pm$0.05 &
5.0 & 7.8 & 1.56 & $\pm$0.10 \\
6.5 & 
10.0 & 7.8 & 0.78 & $\pm$0.05 &
10.0 & 15.5 & 1.55 & $\pm$0.05 &
6.0 & 9.3 & 1.55 & $\pm$0.08 \\
7.0 & 
10.0 & 4.3 & 0.43 & $\pm$0.05 &
10.0 & 11.2 & 1.12 & $\pm$0.05 &
10.0 & 12.3 & 1.23 & $\pm$0.05 \\
7.5 & 
30.0 & 14.5 & 0.48 & $\pm$0.02 &
10.0 & 9.2 & 0.92 & $\pm$0.05 &
8.0 & 8.7 & 1.09 & $\pm$0.06 \\ 
8.0 & 
10.0 & 3.0 & 0.30 & $\pm$0.05 & 
30.0 & 19.0 & 0.63 & $\pm$0.02 & 
10.0 & 7.1 & 0.71 & $\pm$0.05 \\
8.5 & 
 &  &  & &
 &  &  & &
10.0 & 5.0 & 0.50 & $\pm$0.05 \\
9.0 &
 &  &  & &
 &  &  & &
10.0 & 2.9 & 0.29 & $\pm$0.05 \\
\bottomrule
\end{tabular}
\caption[Size of repeating speckle structure]{Size of repeating speckle structure. Total distance uncertainty is $\pm$\unit[0.5]{cm}.  All values given in cm.  Dist. A is the distance between the microlens array array and the closest lens with an uncertainty of \unit[0.1]{cm}.  'No.' is the number of repeating structures measured for the purpose of reducing uncertainty.}
\label{tab:lensarray_repeating_structure}
\end{table}

\begin{figure}[htp]
\centering
\includegraphics[angle=-90,width=\textwidth]{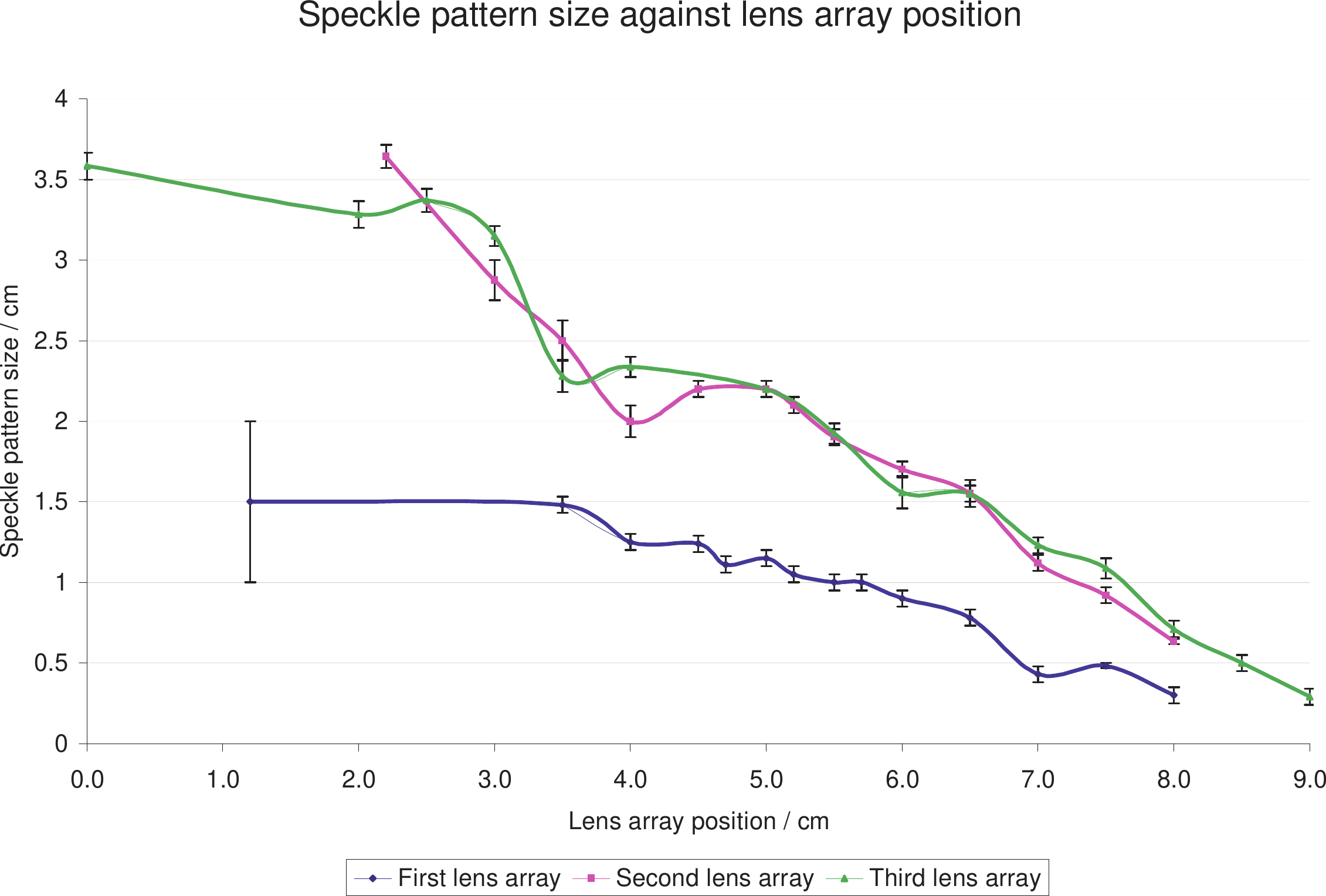}
\caption[Speckle structure against microlens array array position]{Repeating speckle structure size plotting against position of microlens array for each of the three different microlens arrays.}
\label{fig:speckle_against_lens}
\end{figure}

This data is plotted in \Fref{fig:speckle_against_lens}.
\clearpage

The size of the speckle patterns is smallest for the first microlens array, which is the microlens array with the largest lenslets and with the largest radius of curvature.  It is not entirely clear how much effect this structure will have on the final hologram.

\clearpage
\subsection{Microlens array focal plane}

There is another concern regarding the position of the microlens array.  The plane of the aperture that is imaged on to the photosensitive emulsion is very close to the microlens array.  As the lenslets have a small focal distance, the aperture is also very close to the virtual image of the focal plane of the microlens array.  It is my concern that if the imaged focal plane of the microlens array coincides with the plane of the photosensitive emulsion, this will result in an uneven illumination of the film, significantly reducing diffraction efficiency.

To monitor this effect, the data in Appendix~\ref{sec:lensarrayappendix} records the relative distance of the lenslet focal plane from the hologram plate.  This distance could not be measured directly, so to measure this a piece of white card was placed inside the afocal telescopic lens system and moved into position such that the focal point of the microlens array was focused onto the card.  This could be seen clearly as an array of points of light.   The distance between this position and the image plane of the aperture was recorded.  All the distances given below are in terms of the distance measured between the imaged focal plane of the microlens array and the corresponding imaged plane of the aperture in telescopic lens system.  The position of the aperture remained unmoved and the only variable changed was the position of the microlens array.  It is worth noting that moving the microlens array will in turn slightly move the plane that the aperture is focused on near the hologram plane.

Extracting this distance data from the tables for the first, second and third microlens arrays we get \Fref{tab:lensarraydistances}.
\begin{table}[htp]
\centering
\begin{tabular}{R{2cm}R{1.5cm}R{1.5cm}R{1.5cm}}
\toprule
Microlens array & 
\multicolumn{1}{A}{\textbf{First}} & \multicolumn{1}{B}{\textbf{Second}} & \multicolumn{1}{C}{\textbf{Third}}\\ 
dist. (cm) &(cm) & 
(cm) & 
(cm) \\
\midrule
0.0 & & & 12.5 \\
1.2 & 8.0 &  &  \\
2.0 &  &  &10.0 \\
2.2 & & 10.5 &  \\
2.5 &  &  &9.5 \\
3.0 &  & 8.0 & 8.5 \\
3.5 & 5.5 & 8.0 & 8.0 \\ 
4.0 & 4.2 & 7.4 & 7.5 \\ 
4.5 & 4.0 & 6.4 & 5.5 \\ 
4.7 & 2.1 &  &  \\ 
5.0 & 1.5 & 5.0 & 2.1 \\
5.2 & 0.0 &  &  \\
5.5 & -0.8 & 2.0 & 0.2 \\
5.7 & -1.7 &  &  \\
6.0 & -4.0 & 0.0 & -1.8 \\
6.5 & -6.8 & -3.5 & -5.0 \\ 
7.0 & -19.0 & -10.0 & -9.8 \\
7.5 &  & -17.0 & -15.0 \\ 
8.0 &  &  & -17.0 \\
\bottomrule
\end{tabular}
\caption[Distance between microlens array focal plane and aperture]{Distance measured between the imaged focal plane of the microlens array and the corresponding imaged plane of the aperture in telescopic lens system.  All values given with uncertainty of $\pm\unit[0.5]{cm}$.}
\label{tab:lensarraydistances}
\end{table}

\clearpage

\begin{figure}[htp]
\centering
\includegraphics[angle=-90,width=\textwidth]{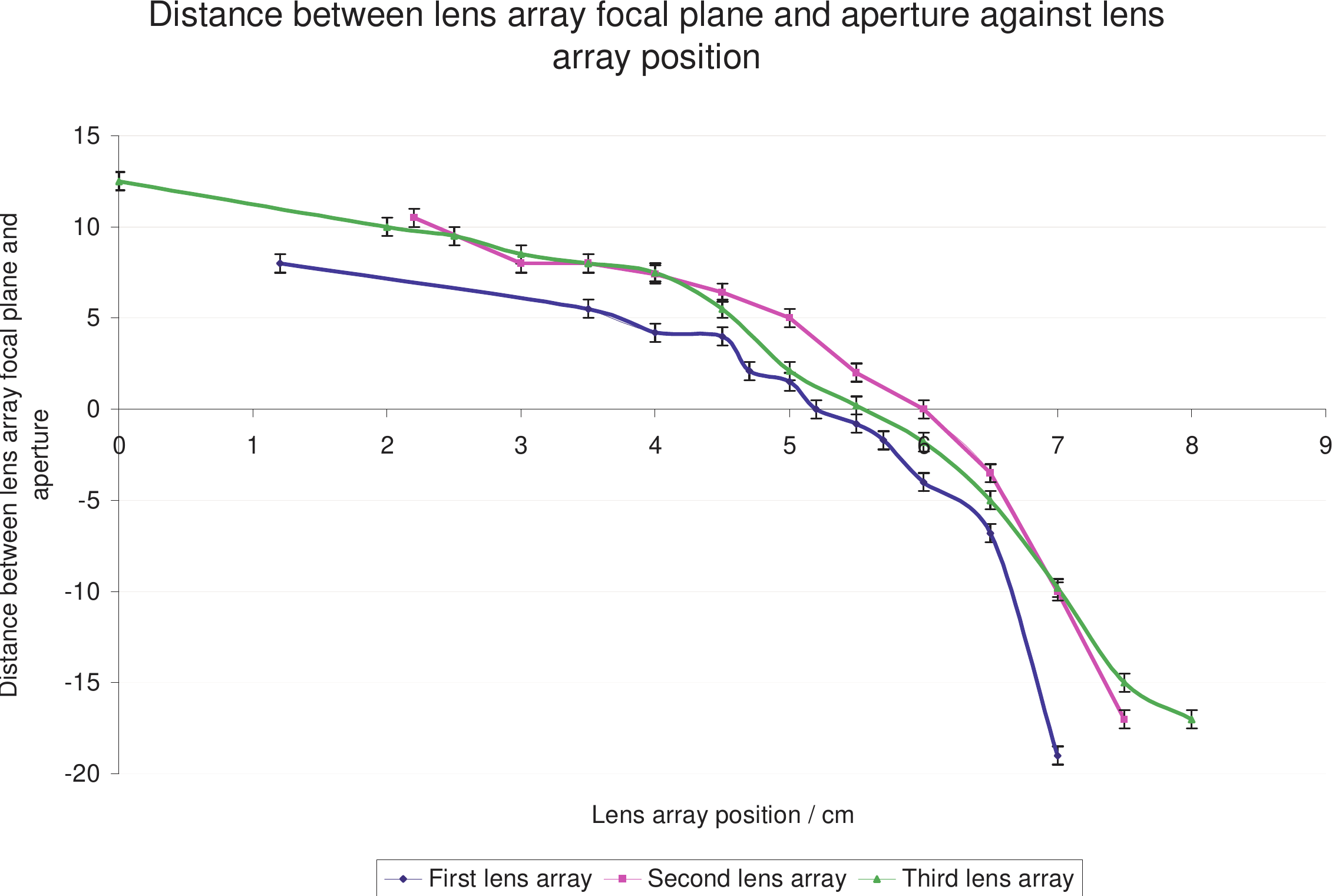}
\caption[Distance between microlens array focal plane and aperture images]{Graph of distance measured between the imaged focal plane of the microlens array and the corresponding imaged plane of the aperture in telescopic lens system.  All values shown with error bar of $\pm\unit[0.5]{cm}$.}
\label{fig:focal_plane_hologram}
\end{figure}

\Fref{fig:focal_plane_hologram} graphs the results given in \Fref{tab:lensarraydistances} with a smoothed connecting line for visual clarity.

\bigskip

\subsection{Summary}

The first microlens array had rectangular lenslets with physical dimensions of \unit[0.40]{mm}$\times$\unit[0.46]{mm} and spherical curvature of radius of \unit[1.3]{mm}. The best position based upon image quality, beam intensity and shape was at approximately \unit[4.8]{cm} from the lens.  This microlens array exhibited the smallest repeating structure size of \unit[1.2]{cm} and in this position the focal plane of the lenslets is at small distance away from the plane of the aperture.

\bigskip

The second microlens array had rectangular lenslets with physical dimensions of \unit[0.20]{mm}$\times$\unit[0.23]{mm} and spherical curvature of radius of \unit[0.23]{mm}. The best position based upon image quality, beam intensity and shape was at approximately \unit[6.7]{cm} from the lens.  At this position, the repeating structure size is approximately \unit[1.4]{cm}, and the focal plane of the lenslets is at a distance away from the plane of the aperture. 

\bigskip

The third microlens array had rectangular lenslets with physical dimensions of  \unit[0.040]{mm}$\times$\unit[0.046]{mm} and spherical curvature of radius of \unit[0.13]{mm}. The best position based upon image quality, beam intensity and shape was at approximately \unit[6.0]{cm} from the lens.  At this position, the repeating structure size is the largest of the three lens arrays at approximately \unit[1.6]{cm}.  The focal plane of the lenslets is close to the plane of the aperture. 

\bigskip

There does not appear to be any significant differences between using the three microlens arrays.  They each exhibited quite different behavior, but when all of the factors are taken into account, there is no clear superior choice.

For future work, the first microlens array was chosen due to the smaller repeating speckle structure.



\clearpage

\section{Minimizing beam energy loss in lens system}

The \ac{RGB} digital hologram printer design by \citet{ratcliffepatent} utilizes a long cavity \ac{NDYAG} laser to provide sufficient energy to illuminate the silver halide emulsion. The energy required from the laser, per color channel, is dependant upon the size of the printed
holopixel, the efficiency of the hologram printer (lens system, display
system) and the sensitivity of the photo-sensitive emulsion to the illumination
wavelength and pulse length used.

\bigskip

For the printing of monochromatic green holograms, a green{}-sensitive monochromatic fine-grain emulsion such as VRP-M (\citet[see][]{Zacharovas2001}) from \citet{Slavich} is sufficient. This particular
emulsion is sensitive to pulse lengths from nanosecond pulses up to \ac{CW} laser emission.
For pulsed radiation with a length of around tens of nanoseconds, the VRP-M emulsion requires an energy density of approximately $\unit[60]{J/cm^2}$  for maximum
diffraction efficiency\footnote{http://www.geola.com/45.asp}.

\bigskip

For the production of true color volume holograms, a panchromatic or dichromatic gelatine emulsion is required,
such as the ultra-fine grain emulsion known as PFG-03C \citep{Zacharovas2001} from \citet{Slavich}
\footnote{A photographic company outside Moscow. See References.}.
The fine size of the grains makes the emulsion less sensitive to radiation, requiring an
energy density approximately $\unit[3.0]{mJ/cm^2}$ of
green-light pulsed radiation\footnote{http://www.geola.com/48.asp}
for maximum diffraction efficiency.  Some characteristics of the Slavich color emulsion have
been presented by \citet{markov:33}, with a further more theoretical analysis on selective characteristics of single layer color holograms by \citet{markov:304}.  An overview of the current state of colour
reflection holography is given by \citet{bjelkhagen:104}.

\bigskip

\citet{ratcliffepatent} proposes the use of a long cavity \ac{NDYAG} pulsed laser, produced by Geola Technologies, which has a repetition rate of up to ten pulses per second and can provide ample energy for such a system (Typically around \unit[15]{mJ} for the second harmonic). In this context, the energy efficiency of the system is not of high importance.

\bigskip

There are however many advantages to minimizing energy loss.  Increasing the laser repetition rate is technologically simpler if the energy in each shot is decreased.  Likewise if a short cavity laser is utilized instead of a long cavity laser, a more stable output is achieved at the sacrifice of shot energy.  The use of a short cavity laser also offers a simpler design, making the system cheaper due to less components required and a shorter build and maintenance period of time.

The biggest loss of energy in the system by \citet{ratcliffepatent} is in the \ac{LCD} display system.  The system detailed in the section utilizes an \ac{LCOS} system to increase the efficiency by approximately 30 percentile points, and a series of magnifying relay lenses to minimize energy loss at the apertures.

\bigskip

An extra system of magnifying relay lenses can be used to reduce the
energy loss on the apertures on both the reference and object beams by
reducing the beam size to the size of the apertures. Such a system of
magnifying relay lenses is also useful to control the size of the object
beam at the plane of the microlens array. The spatial profile of the beam
at the microlens array is geometrically similar to the spatial profile of
the beam on the hologram plate.  To clean the reference beam, a small aperture
was placed at the mutual focal point of the magnifying relay lenses and adjusted
with the aid of a laser beam spatial profiler placed upstream.

\clearpage
\section{Summary}

A digital hologram printer was developed and built based upon the design by \citet{ratcliffepatent}.  The steps required reproduce the creation of a hologram printer are given, mixed in with lessons learned from the accumulated experience of building and maintaining three such printers.  A new display system (\ac{LCOS} display) was investigated and found to have a significant advantage over the \ac{LCD} display system used by \citet{ratcliffepatent}.  The lens systems were analytically examined by various approximations and methods to determine a simple set of formulas to allow the holopixel spatial size to be easily adjusted by the hologram printer operator.

Three microlens arrays were qualitatively analyzed by a series of experiments and they were found to perform almost equally well, demonstrating that the properties of the microlens array do not have a significant effect on the hologram quality, within the parameter space tested.

A brief mention on increasing the energy efficiency of the design was made, with the goal of replacing the long cavity laser with a short cavity laser.  A short cavity laser would have the advantage of increasing the laser energy point-to-point stability while also lowering the overall cost.  Minimizing energy loss is discussed further in Chapter~\ref{chap:speed}.

\begin{figure}[ht]
\centering
\includegraphics[height=\hsize]{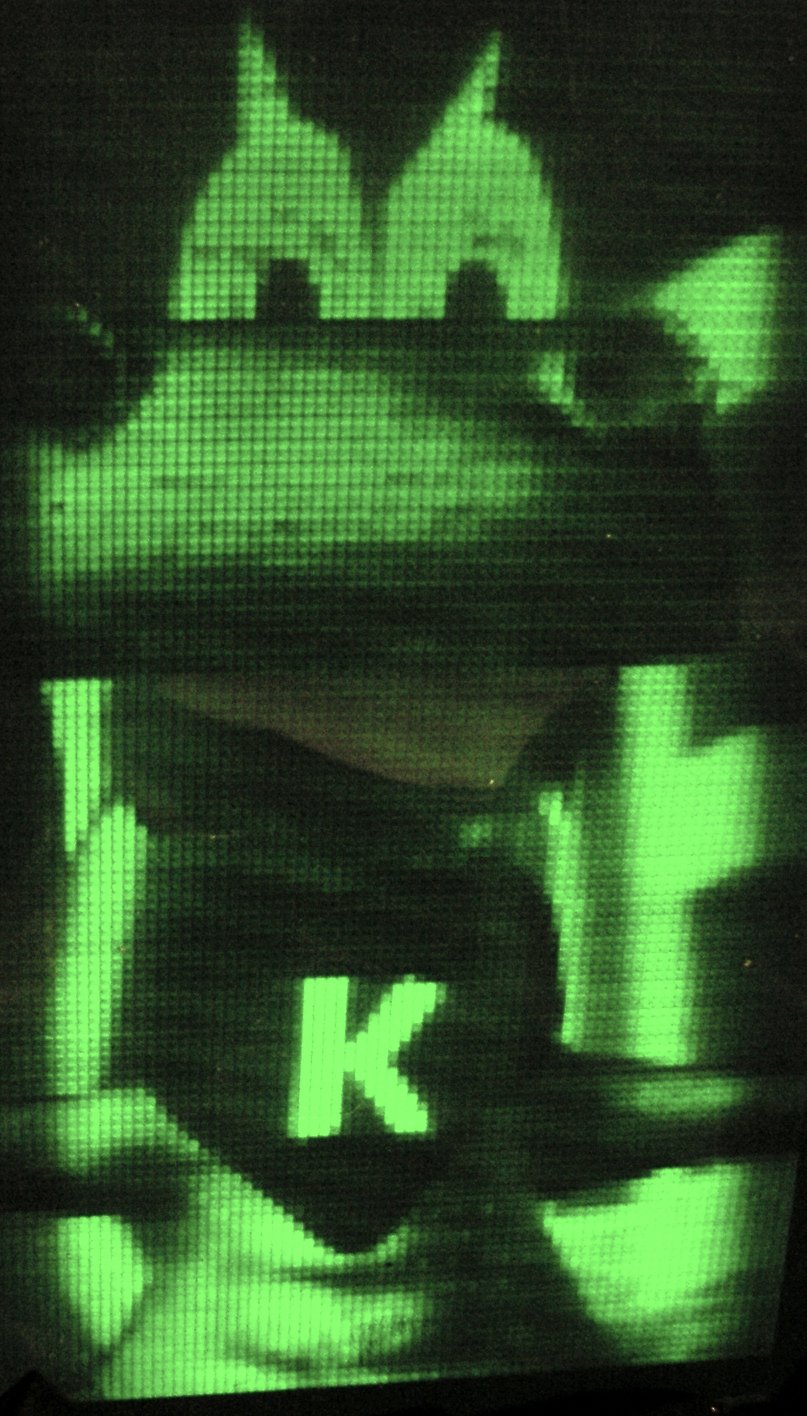}
\caption[Photograph of dragon hologram]{Photograph of green dragon hologram with \unit[1]{mm}$\times$\unit[1]{mm} pixels printed on the described hologram printer.  Note that there are occasional incorrectly-dim pixels due to the laser energy instabilities.  This is discussed and fixed in \Fref{chap:temperature_energy_feedback}.}
\label{fig:greendragon}
\end{figure}

\Fref{fig:greendragon} demonstrates the working hologram printer.

\chapter{White logo}\label{chap:whitelogo}

This chapter looks at a common problem with printing digital holograms that have a large contrast range.  A specific hologram is studied to study this problematic 'ghosting' effect and the cause is identified.  A set of Matlab programs are written to test possible solutions and their effectiveness is determined.

\section{Overview}
\Fref{fig:photo_logo} is a photograph of a white-light-viewable full-color hologram that was printed on a \ac{RGB} hologram printer that is similar in design to that described in this thesis.  In particular, the same Brillian \ac{LCOS} display system was utilized, along with the same lens systems.  The photograph was taken at an angle normal to the hologram.  The intended result was to produce a hologram that looks similar to the rendered image for that angle, as shown in \Fref{fig:white_0360}.

\begin{figure}[htp]
\input{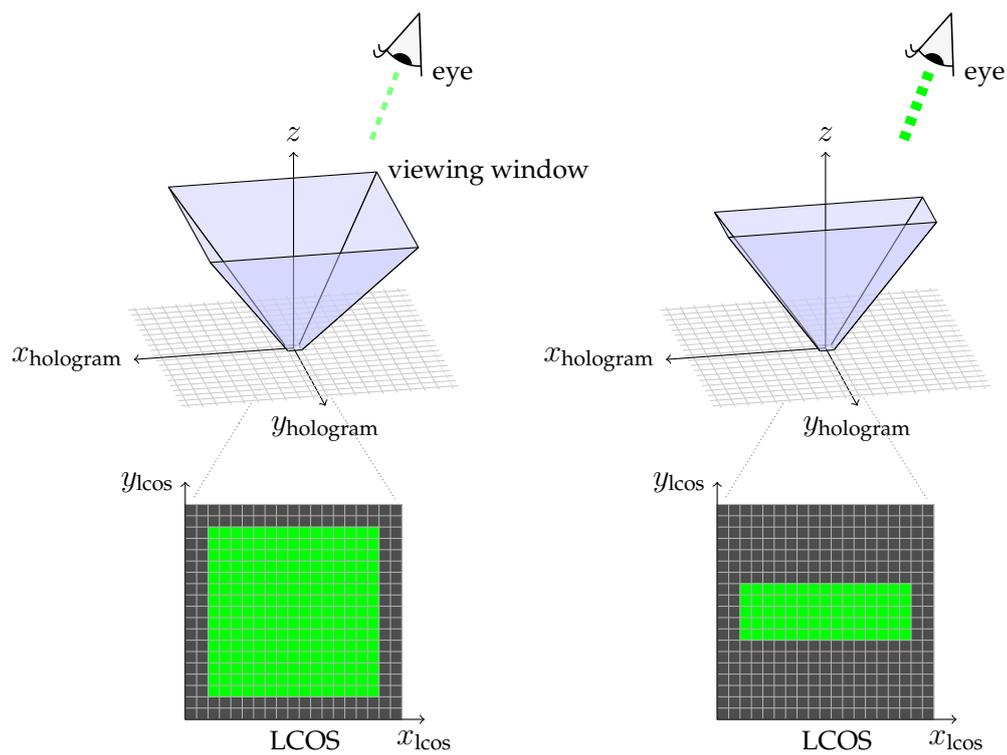}
\caption[Viewing-window comparison]{The Fourier transform of the image on the LCOS is printed on to the hologram as a single 'holopixel'.  Upon replay of the hologram, this recorded Fourier image is reconstructed as an angular intensity distribution.  If the viewing window is smaller, as shown on the right, then the overall intensity as seen by the eye will be larger. }
\label{fig:scene}
\end{figure}

As can be seen in the photograph, the hologram has the highest intensity between the letters.  This visually appears as a `ghosting' or a `shadow' effect.  This effect due to an unintended side effect that is due to the decrease in the effective viewing angle of that holopixel, and thus a decrease in the viewing window.  \Fref{fig:scene} illustrates the problem; the \ac{LCOS} display system spatially modulates the object beam with the angular intensity distribution required for a particular holopixel.  This spatially-modulated beam is optically Fourier transformed and arranged to interfere with the reference beam.  The resulting signal is recorded onto the hologram plate.  After chemical development and bleaching, upon illumination of the hologram, the angular intensity distribution for that holopixel is replayed.  

If the original image on the display system looks similar to that as shown on the left-hand side of Figure~\ref{fig:scene} then the range of viewing angles in which the pixel appears illuminated is large.   If the original image on the display system is significantly smaller than the display, as shown in the right-hand side of Figure~\ref{fig:scene}, then upon replay a smaller range of viewing angles in which the pixel appears illuminated is obtained.

The problem, however, is that the large viewing window results in the total energy being spread over a large area.  Thus the intensity of the light in any particular direction is less than it would be for a small viewing window. So if the two holopixels indicated Figure~\ref{fig:scene} were printed, the viewer would observe the holopixel indicated on the left-hand side as being brighter than the holopixel indicated on the right-hand side, despite that the corresponding pixel on the display system in both cases has the same angle.

\bigskip

To correct for this problem, the intensity of each pixel on the \ac{LCOS} must be adjusted as a function of the entire image on the \ac{LCOS}.  Artists also need to be aware of the limitations - pixels that will illuminate in all directions appear as bright as pixels that only illuminate in very specific directions.

\bigskip

A mathematical analysis of this problem would be presumably be possible based on considering the continuous wavelet Fourier transform and subsequent interference of the \ac{LCOS} image to the recorded holopixel (Such as the wavelet Fresnel techniques by \citet{liebling2004fif}).  The non-linear properties of the photosensitive film would ideally be then taken into account. The determined white-light reflection holograms can then be analyzed with the same wavelet techniques to reconstruct the final image (For example, the work by \citet{sandoz1997wtp, recknagel1998awl} and the statistical approach by \citet{sotthivirat2004pli}).  However such an approach would be both tricky and computationally demanding.  Correct sampling, for an example of the trickiness involved,  is a key issue in such reconstruction algorithms, with extensive research (For example see \citep{allebach1976aed, evans2003rla, jacquot2001hrd}). To avoid such complications, an experimental approach was taken instead.

\bigskip

To research and solve the problem experimentally, three distinct steps were required.  

The first step was to compare the photograph of the White Logo hologram against the rendered images.  This is done in \Fref{sec:photograph_analysis}.

The second step was to precisely determine the relative diffraction efficiency of the pixels of the hologram in a consistent and accurate way.  This required the building of a framework to hold the hologram, a spectrometer, and a halogen lamp, and is explained further in \Fref{sec:spectrometer}.

Thirdly, a mathematical approach to the physical system is required to determine an algorithm to correct the intensity of the pixels shown on the \ac{LCOS}.  This is discussed in \Fref{sec:whitelogofuture}.
\clearpage
\section{Photograph analysis}\label{sec:photograph_analysis}
\begin{figure}[ht]
\centering
\subfloat[Photograph of White Logo hologram]{
\includegraphics[width=\textwidth/4*3]{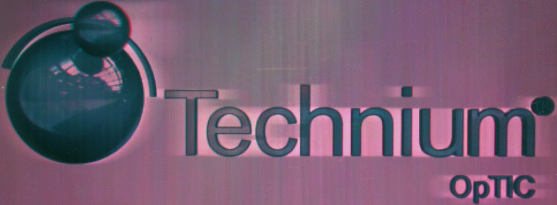}
\label{fig:photo_logo}
}
\qquad
\subfloat[Rendered image \#0663 taken with the virtual camera on the far-right hand side of the track]{
\includegraphics[width=\textwidth/4*3]{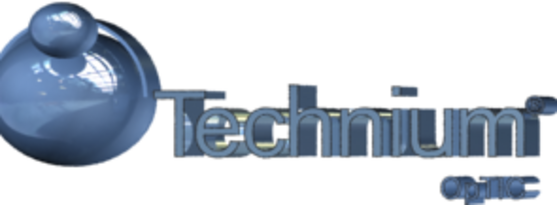}
\label{fig:white_0663}
}
\qquad
\subfloat[Rendered image \#0360 taken with the virtual camera approximately in the center of the virtual track at a similar angle to that used by the real camera that took the photograph]{
\includegraphics[width=\textwidth/4*3]{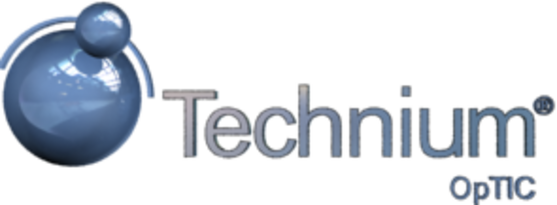}
\label{fig:white_0360}
}
\qquad
\subfloat[Rendered image \#0000 taken with the virtual camera on the far-left hand side of the track]{
\includegraphics[width=\textwidth/4*3]{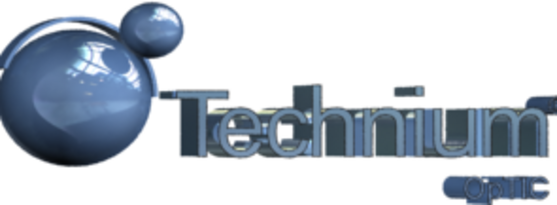}
\label{fig:white_0000}
}
\caption{White Logo photograph and rendered images}
\label{fig:white_logo_figures}
\end{figure}

To better understand the problem, the photograph of the White Logo hologram (\Fref{fig:photo_logo}) was analyzed against the series of rendered images that were used to create the said hologram.  

To compare the photograph and the rendered images programmatically, the photograph and images were carefully cropped and scaled such that the photograph could be matched pixel-wise with the image.  This was done by the following Linux command that utilizes the ImageMagick \textit{convert} program:

\lstset{language=bash, basicstyle=\ttfamily,
    keywordstyle=\color{blue},commentstyle=\color{red},
    stringstyle=\color{dkgreen},
    numbers=none,
    numberstyle=\tiny\color{gray},
    stepnumber=1,
    numbersep=10pt,
    backgroundcolor=\color{white},
    tabsize=4,
    showspaces=false,
    showstringspaces=false }
\begin{lstlisting}
for f in white_0*.png; do 
  convert  $f -crop "295x108+97+115" \
              -geometry 557x205! ../WhiteLogoPngMine/$f
done
\end{lstlisting}

Where the parameters were found through a trial-and-error effort.

\begin{figure}[htp]
\centering
\includegraphics[width=\textwidth]{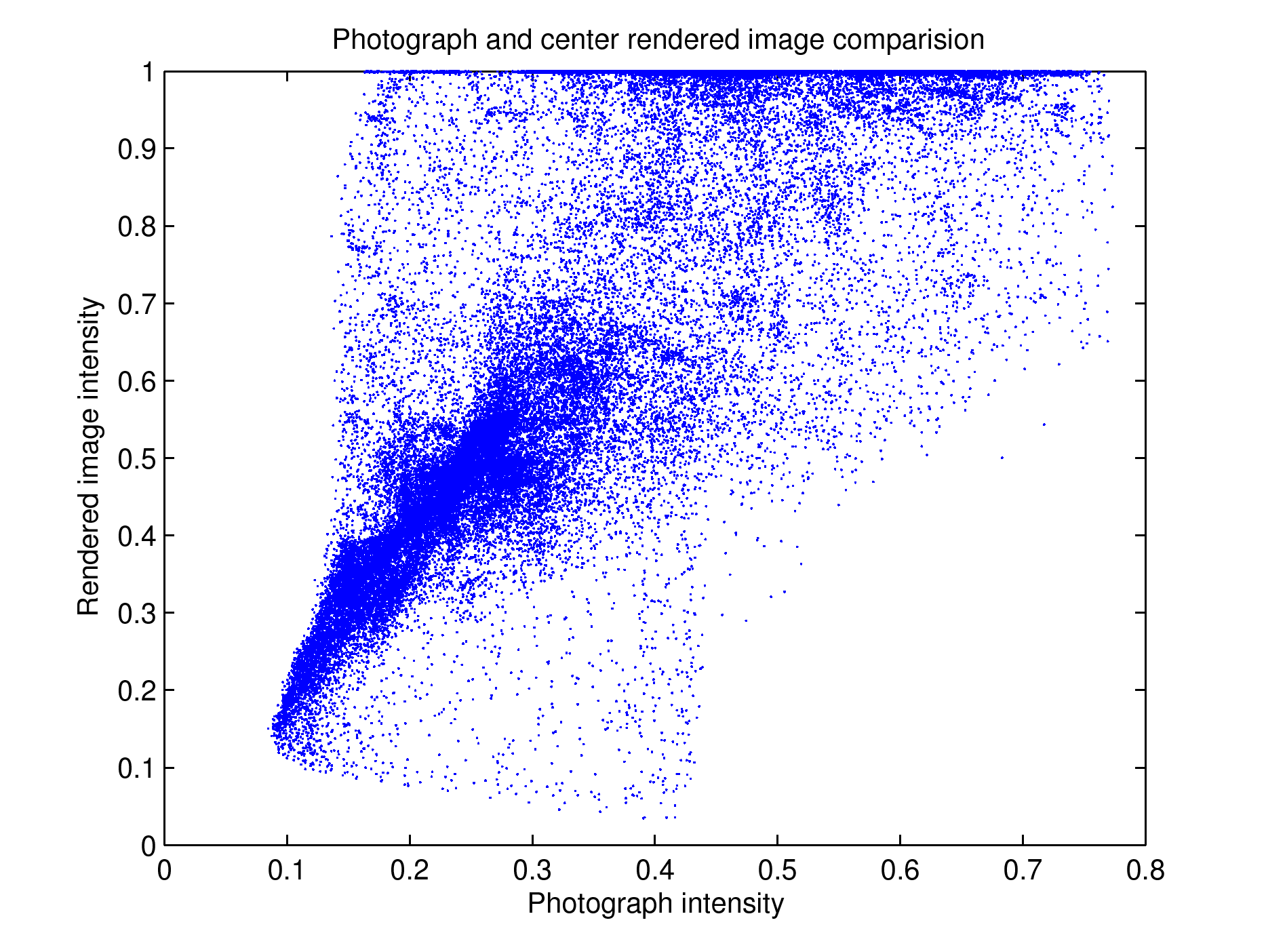}
\caption[Photograph intensity against rendered image intensity]{A plot of the intensity of each pixel on the photograph of the White Logo hologram plotted against the corresponding pixel on the rendered image shown in \Fref{fig:white_0360}.  The intensity scale is arbitrarily between 0 and 1 where 1 is a brighter, and 0 is darker.  The images were converted to greyscale and a Gaussian blur of 2 pixels in the horizontal direction applied first.}
\label{fig:photograph_against_rendered}
\end{figure}

\clearpage
Appendix~\ref{sec:whitelogo} gives the program listing for the code used to analyze the difference between the photograph and the rendered images.  The results are shown in \Fref{fig:photograph_against_rendered}.  There is a strong correlation for the low intensity pixels (the black areas of the image) and a much weaker correlation for the high intensity pixels.

It is the difference in intensities between the photograph (actual intensity) and the center rendered image (intended intensity) that we are interested in, as it is this difference that we need to compensate for.  This difference is a function, $f$, of the corresponding pixel on all of the rendered images.  Indicating the photograph intensity at a point (x,y) as $P(x,y)$, the intensity of the center rendered image as $I_\text{intended}(x,y)$, the intensity of a rendered image number '$i$' as $I_i$, we obtain:
\begin{align}
P(x,y) &\approx f\left( I_\text{intended}(x,y), I_1(x,y), I_2(x,y),\cdots \right)
\end{align}

The simplest non-trivial possible function is a simple summation of the intensities of a given pixel on all the rendered images (termed 'total intensity').  This function will be tried, as so:
\begin{align}
P(x,y) &\approx f\left( I_\text{intended}(x,y), \sum^i I_i(x,y) \right) \label{eq:general_correction_eq}
\end{align}

\begin{figure}[htp]
\centering
\includegraphics[width=\textwidth]{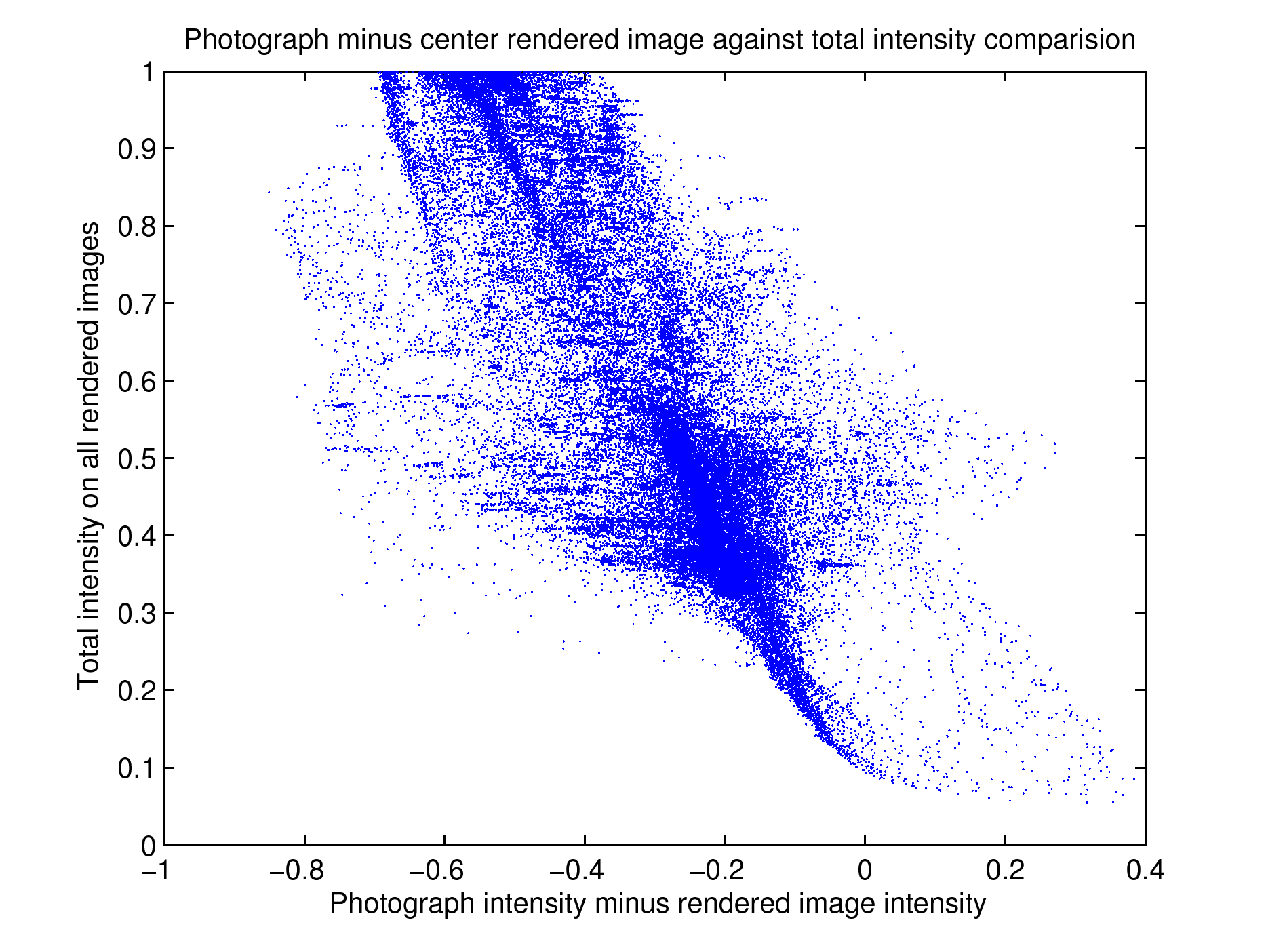}
\caption[Photograph minus image intensity against total image intensity]{A plot of the intensity of each pixel on the photograph minus the intensity of the corresponding pixel on the center rendered image of the White Logo hologram, plotted against the corresponding total intensity of that pixel.   The total intensity is a summation of the intensities of the given pixel for all of the rendered images.  The intensity scale is arbitrarily between 0 and 1 where 1 is a brighter, and 0 is darker.  The images were converted to greyscale and a Gaussian blur of 2 pixels in the horizontal direction applied first.}
\label{fig:photographdiff_against_total}
\end{figure}

A plot of the difference in actual and intended intensity against the total intensity is given in \Fref{fig:photographdiff_against_total}.  This shows a mostly-linear correlation of approximately $y=-1.8 x +0.04$  or, rearranging, $x=-0.54 y +0.02$.  Putting this into \Fref{eq:general_correction_eq}:
\begin{align}
P(x,y) &\approx -0.54 \times \sum^i I_i(x,y) + 0.02 + I_\text{intended}(x,y) \label{eq:correction}
\end{align}

\begin{figure}[htp]
\centering
\includegraphics[width=\textwidth]{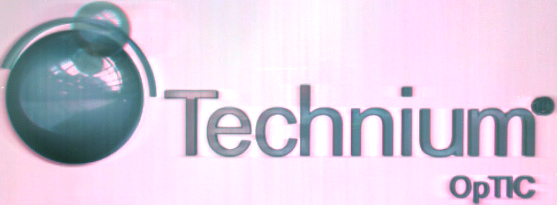}
\caption[Corrected image]{The photograph shown in \Fref{fig:photo_logo} with \Fref{eq:correction} applied to each pixel}
\label{fig:corrected_photograph}
\end{figure}

Using \Fref{eq:correction} the photograph can be modified to attempt to retrieve the intended image.  Applying the formula to the color photograph produced a few intensities that were outside of the range 0 to 1.  Where this happened, the value was rounded down to 1. The final results are shown in \Fref{fig:corrected_photograph}.  While this is a large improvement, it is clear that the separate colors need to be handled separately.  Since the hologram was printed with red (\unit[770]{nm}), green (\unit[532]{nm}) and blue (\unit[440]{nm}) laser light, the image was analyzed in the \ac{RGB} color space.

\clearpage

\begin{figure}
\centering
\subfloat[Red channel]{
\includegraphics[width=\textwidth*1/2]{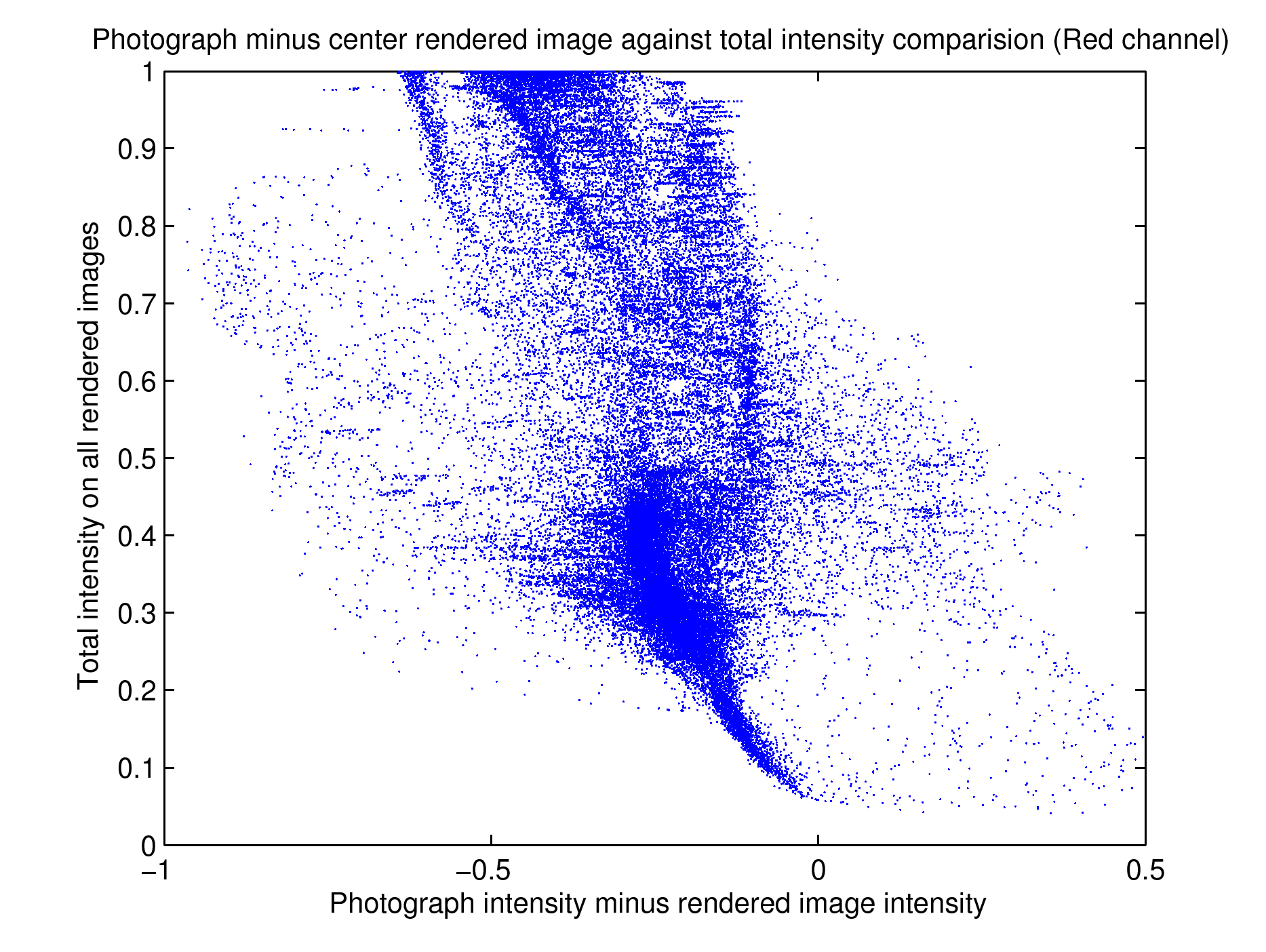}
\label{fig:redchannel}
}
\quad
\subfloat[Green channel] {
\includegraphics[width=\textwidth*1/2]{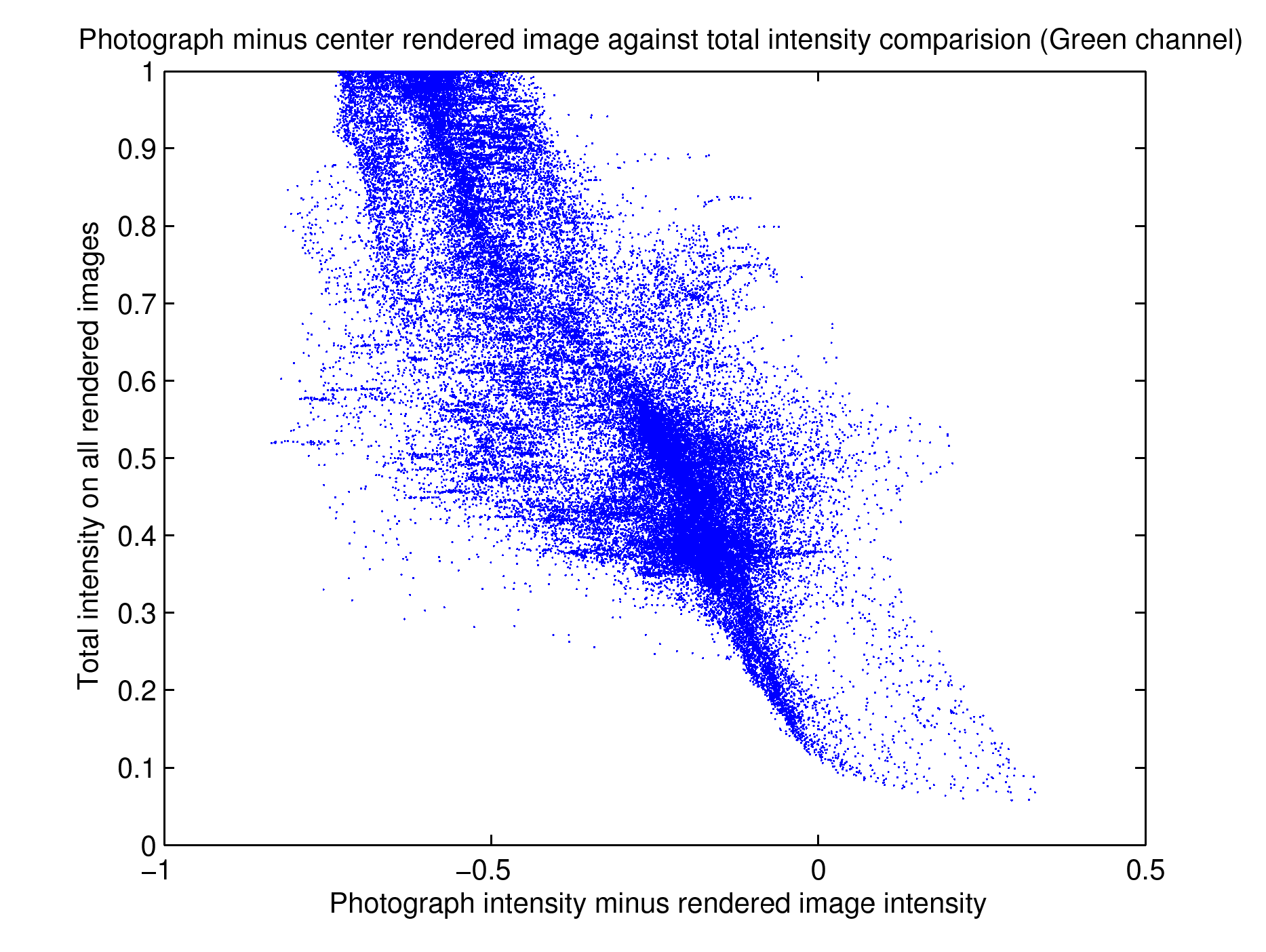}
\label{fig:greenchannel}
}
\quad
\subfloat[Blue channel] {
\includegraphics[width=\textwidth*1/2]{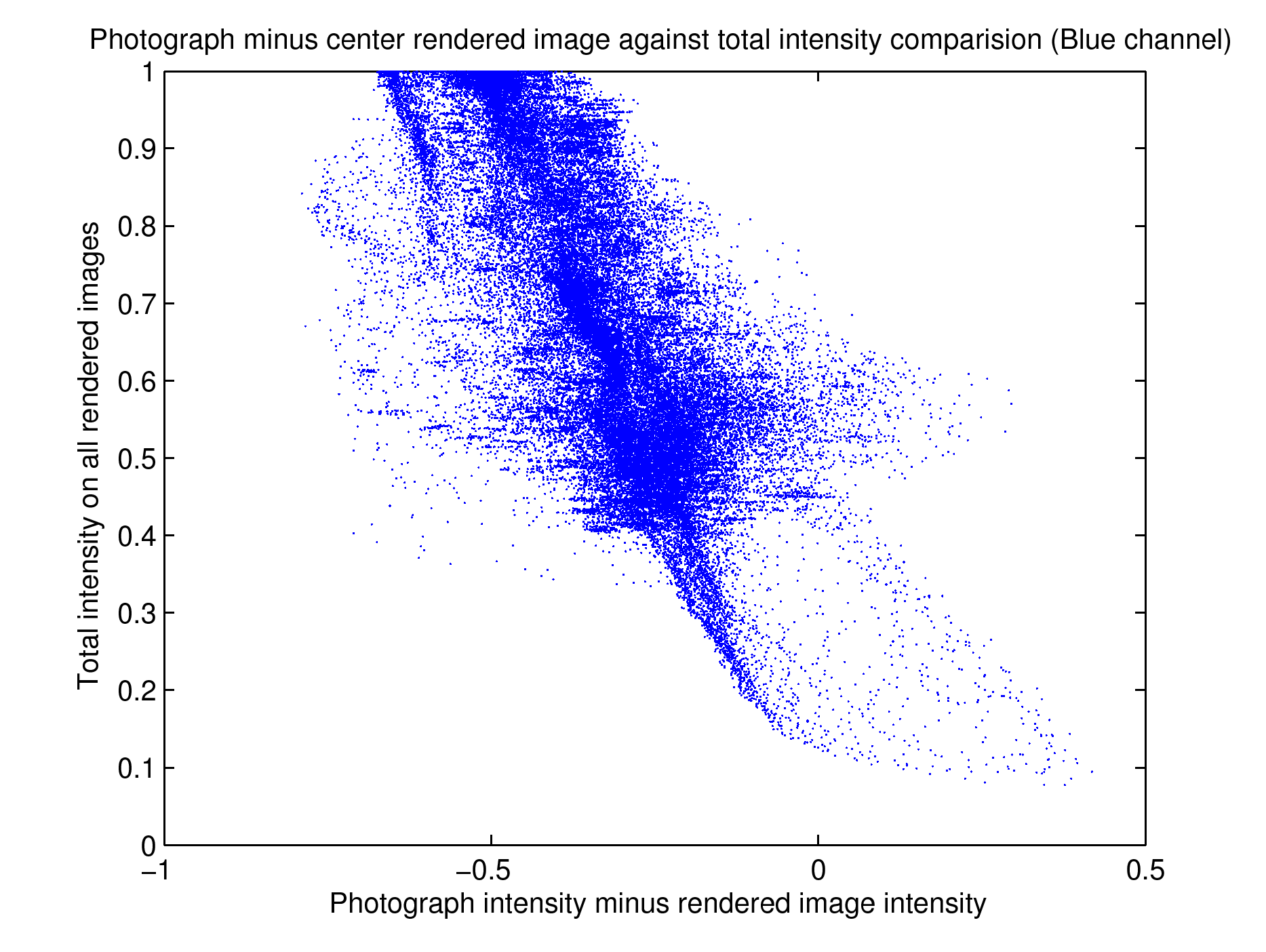}
\label{fig:bluechannel}
}
\caption[Photograph minus image against total (Individual channels)]{A plot of the intensity of each pixel on the photograph minus the intensity of the corresponding pixel on the center rendered image of the White Logo hologram, plotted against the corresponding total intensity of that pixel.  Each color channel has been plotted separately.}
\label{fig:individual_channels}
\end{figure}

To attempt to produce better results, each channel was processed separately in the \ac{RGB} color space, and the results are shown in \Fref{fig:individual_channels}.

The graph of the red channel, Figure~\ref{fig:redchannel}, has the least correlation, but attempting a correlation anyway we get a linear correlation of approximately $y=-1.5 x +0.2$  or, rearranging, $x=-0.67y +0.08$.

The green channel, Figure~\ref{fig:greenchannel}, has better correlation but strongly appears to be non-linear.  Attempting a linear correlation anyway, we get a correlation of approximately $y=-2.0 x$, or $x = -0.51 y$.

Finally, the blue channel, Figure~\ref{fig:bluechannel}, has a strong linear correlation of approximately $y=-2.8 x - 0.2$, or $x = -0.36 y - 0.07$.

\begin{figure}[htp]
\centering
\includegraphics[width=\textwidth]{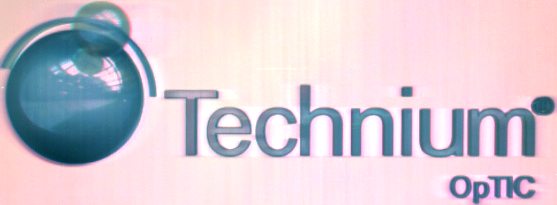}
\caption[Corrected image with each channel corrected separately]{The photograph shown in \Fref{fig:photo_logo} with the correction applied to each pixel.  Each color channel was corrected separately, in \ac{RGB} color space.}
\label{fig:corrected_photograph_separate}
\end{figure}

Using these corrections with \Fref{eq:general_correction_eq} but applying the correction separately to each channel, \Fref{fig:corrected_photograph_separate} is obtained.

This is subjectively better, but not entirely satisfactory.  It is still clearly brighter in between the letters.  Further work would be required on this.

To get a better analysis of the hologram, a spectrometer was used instead of a photograph.  This required a framework to support the spectrometer.  This is discussed in the next section.

\clearpage

\clearpage
\section{Spectrometer}\label{sec:spectrometer}

To measure the diffraction efficiency of the hologram, a framework was required to hold a hologram, a light source, and a spectrometer in a reliable and consistent way.  This was to allow the visible-light spectra and intensity of individual pixels to be compared between multiple points on a hologram, and between multiple prints of hologram.

\begin{figure}[ht]
\centering
\includegraphics[width=\textwidth]{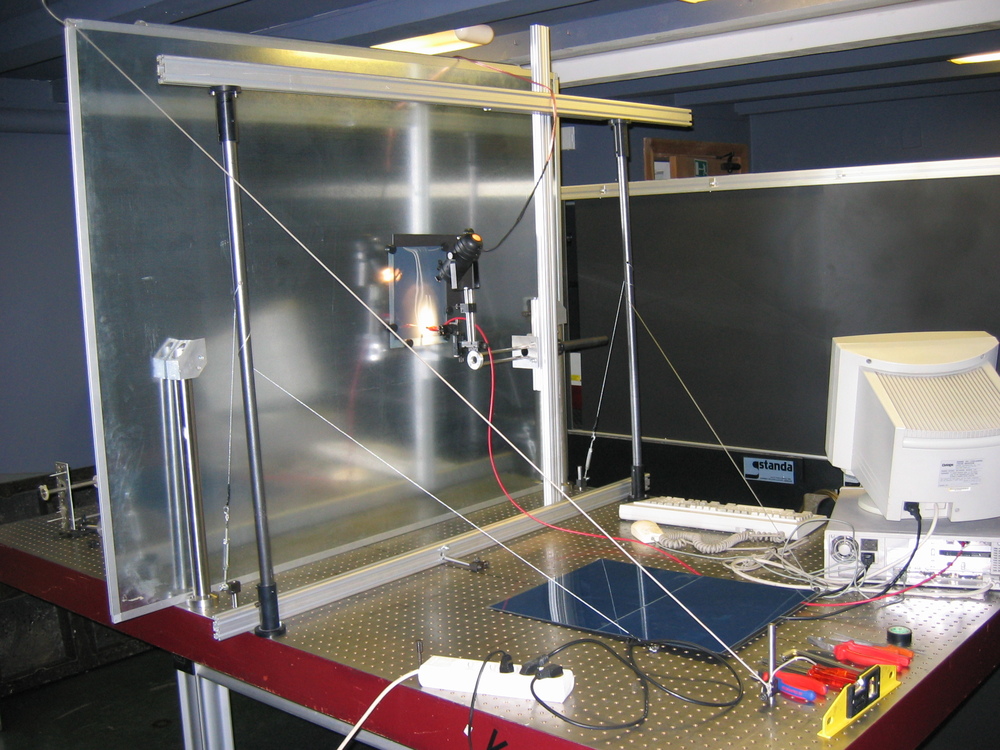}
\caption{Photograph of framework for spectrometer and halogen light}
\label{fig:spectrometer_overview_large}
\end{figure}

The framework shown in \Fref{fig:spectrometer_overview_large} was constructed for this purpose and is suitable for any white-light viewable visible-light reflection hologram. This framework consists of: a heavy metal screen which is securely mounted onto an optical table; a gantry system to allow two dimensional movement parallel to the screen; a fine-control mount with three degrees of freedom; a spectrometer and a light, mounted to the fine-control mount; and a \ac{PC}.

\begin{figure}[htp]
\centering
\includegraphics[width=\textwidth,angle=-90]{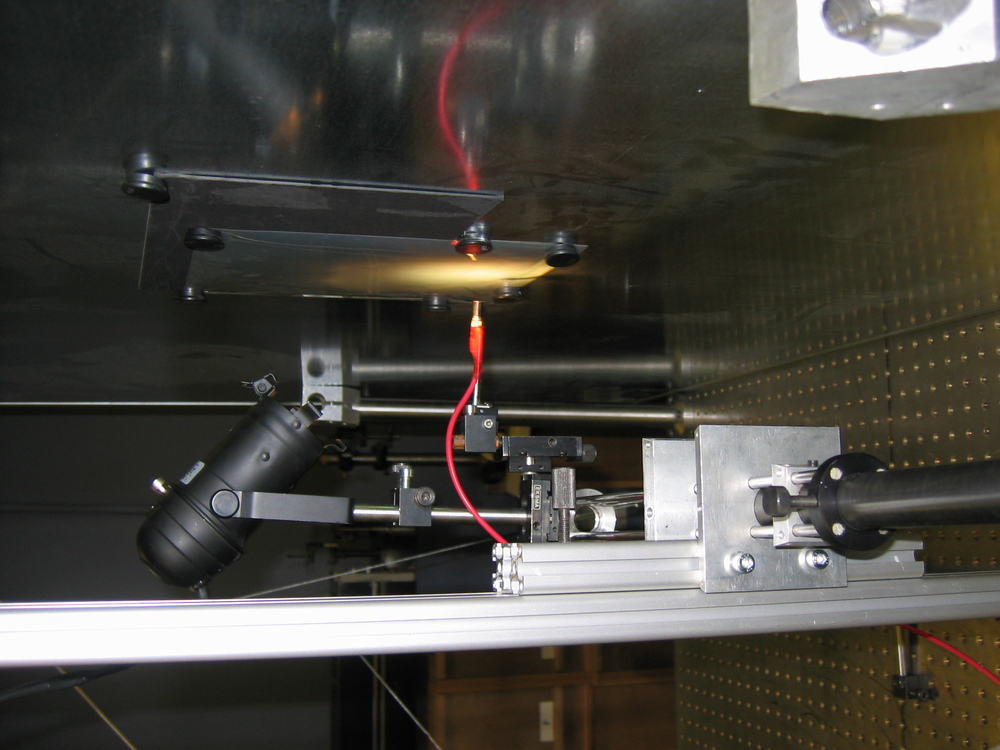}
\caption{Photograph of fine-control spectrometer mount}
\label{fig:spectrometer_sideview_large}
\end{figure}

\clearpage
The metal screen can hold a hologram of up to \unit[1.5]{m}$\times$\unit[1.5]{m} in size and is magnetic, allowing fragile holograms to be held in place with magnets.  The main \ac{2D} gantry system was greased for easy movement and can be locked in place with thumb-screws, to prevent accidental movement. The fine-control mount allows the 60W halogen light to be held at almost any angle and height relative to the spectrometer, and allows movement both parallel and normal to the screen. This can also be locked into place to prevent accidental movement.
The angle of the halogen was set to the designed replay-angle for the holograms -- equal to the reference beam angle, 37$\degree$ to the normal of the hologram, as shown in \Fref{fig:spectrometer_sideview_large}.

\begin{figure}[htp]
\centering
\includegraphics[width=\textwidth]{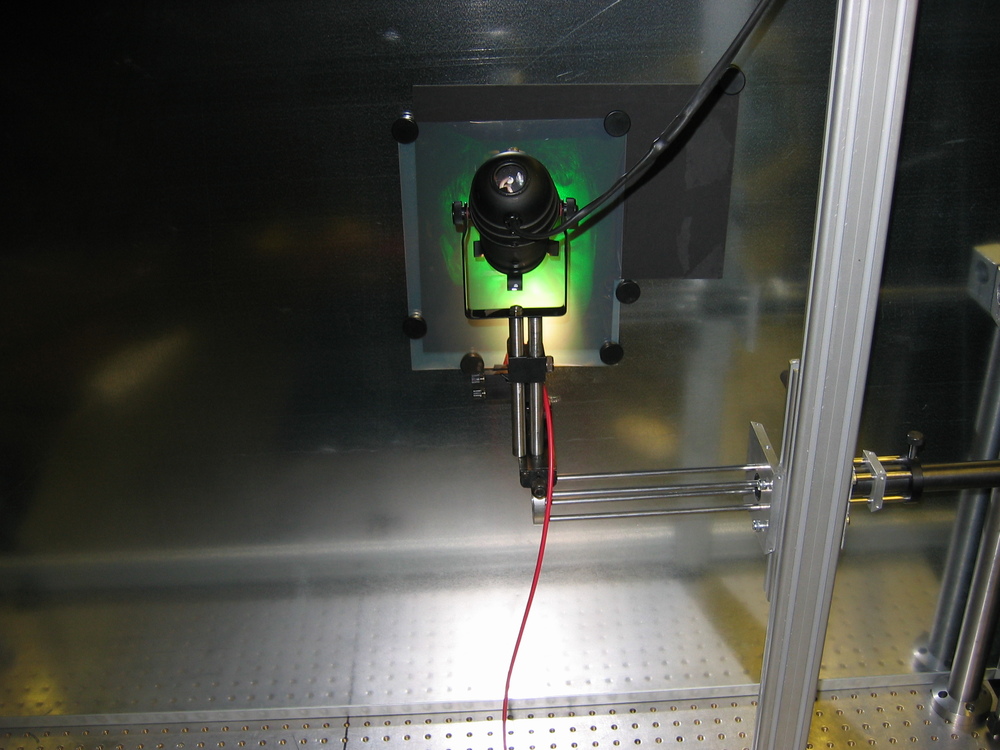}
\caption{Photograph of spectrometer in use}
\label{fig:spectrometer_zoomed_in_large}
\end{figure}

\clearpage
The details of the spectrometer are given in \Fref{tab:spectrometer}.  The spectrometer was connected to a \ac{PC} running \textit{OOIBase for Windows} and \textit{SpectraWin}, and was calibrated against a sheet of glossy white paper.  \Fref{fig:spectrometer_zoomed_in_large} shows the framework in use on a white-light reflection hologram.

\begin{table}[htp]
\centering
\begin{tabular}{ll}
\toprule
Company: & Ocean Optics Europe \\
Model:& PC2000-ISA \\
Description:& PC Plug-in Fiber Optic Spectrometer \\
Grating:& 600 lines blazed at \unit[500]{nm} \\
Bankwidth:& \unit[350-1000]{nm} \\
Options Installed:& \unit[25]{\textmu m} Slit \\
Serial: &PC2E637 \\
Sample time:& from \unit[3]{ms} to \unit[60]{s}\\
\bottomrule
\end{tabular}
\caption{Technical details of spectrometer}
\label{tab:spectrometer}
\end{table}

\clearpage

Using the spectrometer framework on the White Logo hologram, the results shown in \Fref{tab:spectrometer_results} were obtained.

\begin{table}[htp]
\centering
\begin{tabular}{lrrrr}
\toprule
Position & \multicolumn{2}{c}{Red Channel} & \multicolumn{2}{c}{Green Channel}\\
         & Wavelength & Intensity & Wavelength & Intensity\\
\midrule
White Region             & 625 & 33  & 505 & 28\\
In between the 'n'       & 627 & 145 & 508 & 106\\
Next to top of H         & 628 & 120 & 509 & 85\\
Next to the dot in 'i'   & 628 & 131 & 508 & 94\\
Next to floating sphere  & 629 & 181 & 509 & 150\\
In the middle of the 'c' & 628 & 110 & 509 & 73\\
In the middle top of 'c' & 629 & 194 & 509 & 133\\
\bottomrule
\end{tabular}
\caption[Spectrometer results]{Wavelength and counts per minute for peak counts per minute in red and green channel.  Wavelength is in nanometers, and intensity in counts per minute, as measured by the spectrometer detailed in \Fref{tab:spectrometer}.}
\label{tab:spectrometer_results}
\end{table}

\bigskip

\begin{figure}[htp]
\centering
\includegraphics[angle=-90,width=\textwidth]{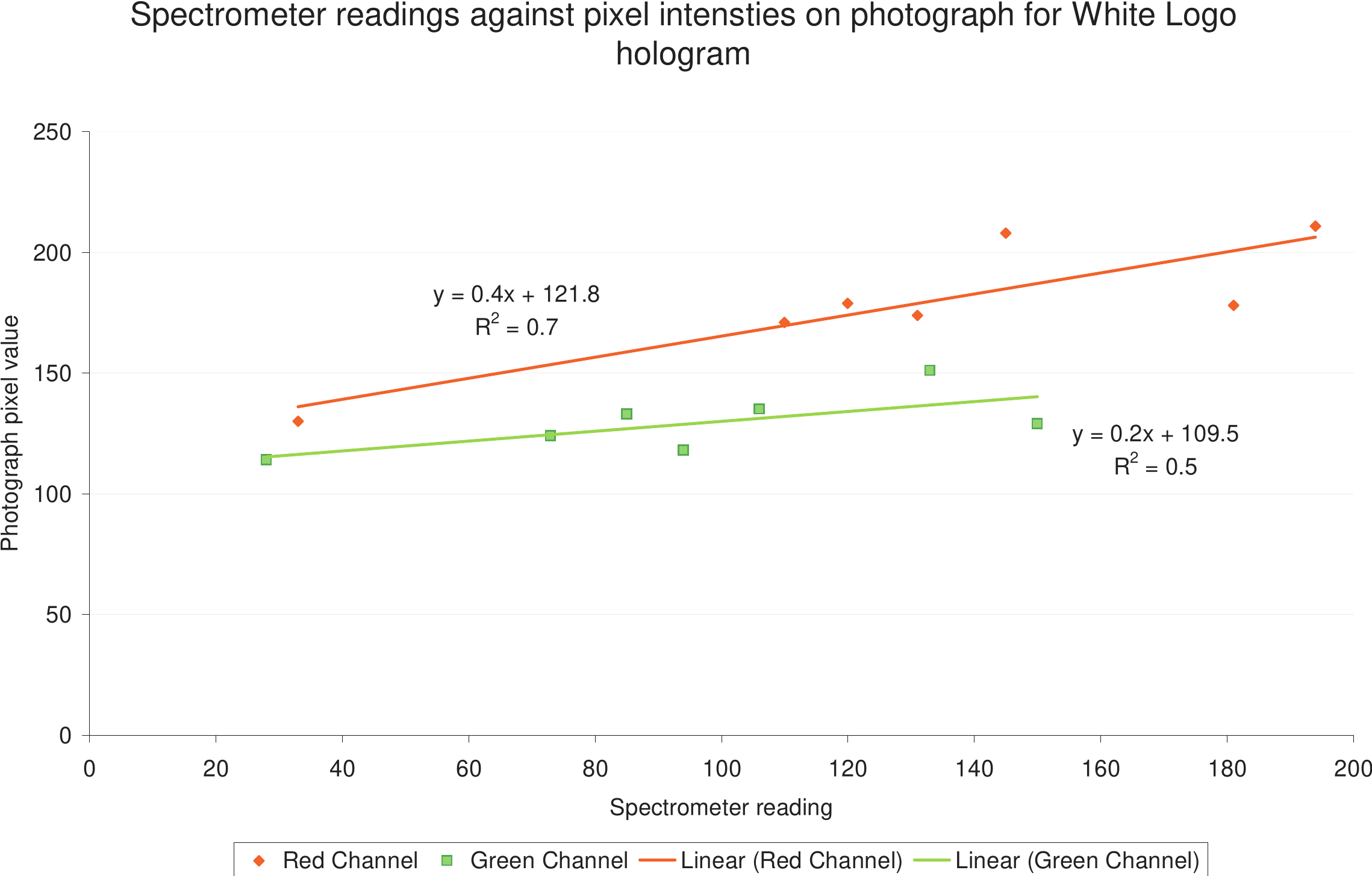}
\caption[Spectrometer readings against photograph readings]{Spectrometer readings (From \Fref{tab:spectrometer_results}) plotted against the intensity value recorded by the camera for the same position and same angle (From \Fref{tab:photograph_results}).  A linear-regression best-fit line is shown for the two color channels. The blue channel was not measured.}
\label{fig:spectrometer_against_photograph}
\end{figure}

To test the accuracy the photograph of the White Logo hologram (\Fref{fig:photo_logo}), the equivalent pixels in the photograph were compared against the intensity recorded by the spectrometer.  \Fref{tab:photograph_results} shows the \ac{RGB} values for the equivalent pixels in the photograph, and \Fref{fig:spectrometer_against_photograph} is a graph of the two sets of results plotted against each other, for each pixel and for each channel.

\bigskip

\begin{table}[htp]
\centering
\begin{tabular}{lrrrr}
\toprule
Position & Red Channel & Green Channel & Blue Channel\\
\midrule
White Region             & 130 & 114 & 125\\
In between the 'n'       & 208 & 135 & 164\\
Next to top of H         & 179 & 133 & 150\\
Next to the dot in 'i'   & 174 & 118 & 149\\
Next to floating sphere  & 178 & 129 & 151\\
In the middle of the 'c' & 171 & 124 & 149 \\
In the middle top of 'c' & 211 & 151 & 171\\
\bottomrule
\end{tabular}
\caption[Photograph intensity for spectrometer results]{Photograph intensity for the same position and angle as used by \Fref{tab:spectrometer_results}.  The values have the range 0 to 255, with 255 being the maximum intensity.}
\label{tab:photograph_results}
\end{table}

\clearpage

\Fref{fig:spectrometer_against_photograph} shows that the results from the camera do not match to sufficient precision to the results from the spectrometer.  This could be due to experimental error, or an indication that the chosen camera was not of sufficiently quality to capture a true-image of the scene.  If the image from the camera was not an accurate depiction of the light from the hologram, then this may explain some of the noise in the results found in \Fref{sec:photograph_analysis}.

\clearpage

\section{Future work}\label{sec:whitelogofuture}

A more accurate correction algorithm that could remove most of the ghosting problems would be a useful future work.  This work could be based on a theoretical approach to the problem, using coupled-wave theory to determine the exact correction algorithm required.  This algorithm can then be applied to the rendered images so that the final printed hologram appears without any ghosting effects.

\begin{figure}[htp]
\centering
\includegraphics[width=\textwidth]{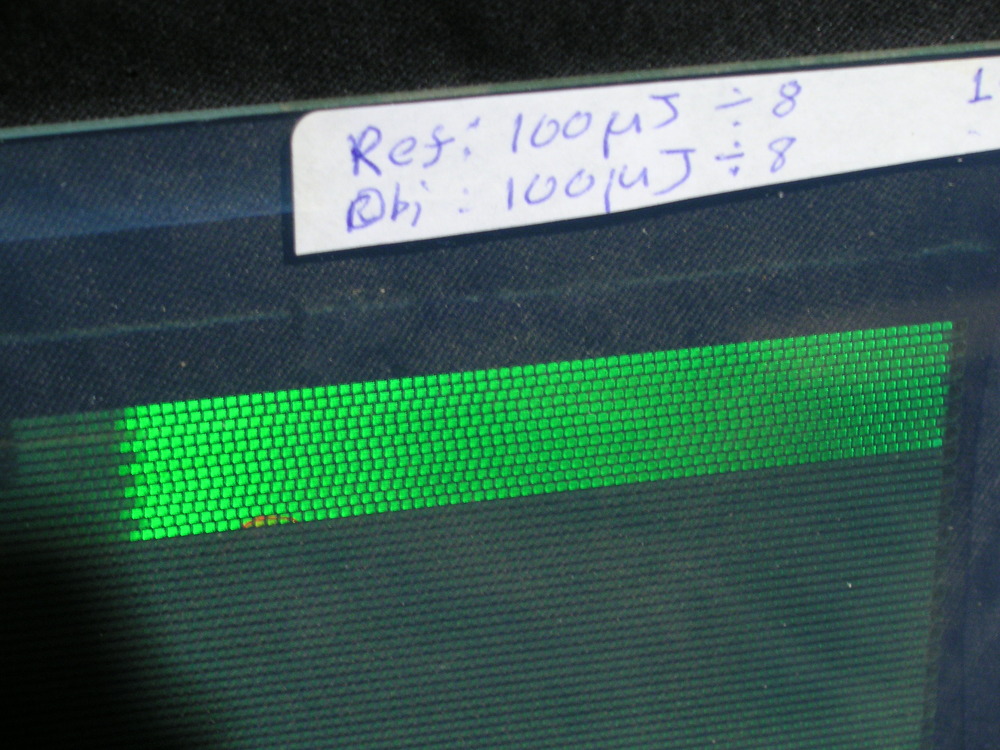}
\caption[Photograph of hologram with decreasing viewing window]{Photograph of hologram with decreasing viewing window.  The holopixels on left hand side of the photograph have the smallest viewing window.  Note that the label is indicating the laser energy was \unit[100]{\textmu J}, as measured after the wave plates and polarisers.  This was then decreased by a factor of eight through the use of \acl{NDF}s (NDF).  The '1' in the top right indicates that this was the first attempt, for the case that the hologram had to be reprinted. The pixels are square and \unit[1.0]{mm} wide with a \unit[2.0]{mm} separation between centers.  The photograph was taken with a f/5.0 aperture and 1/6 second exposure time.  One complete row is approximately \unit[7.5]{cm} across.}
\label{fig:window_size}
\end{figure}

To provide a better avenue for investigation, a hologram was printed such that the width of the viewing window is decreased for each holopixel, from left to right.  This was achieved by displaying a solid green rectangle of decreasing width on the \ac{LCOS} display.  The results are given in \Fref{fig:window_size}.  The twelve lines were all printed the same, to allow for a more accurate analysis.

\begin{figure}[htp]
\centering
\includegraphics[width=\textwidth]{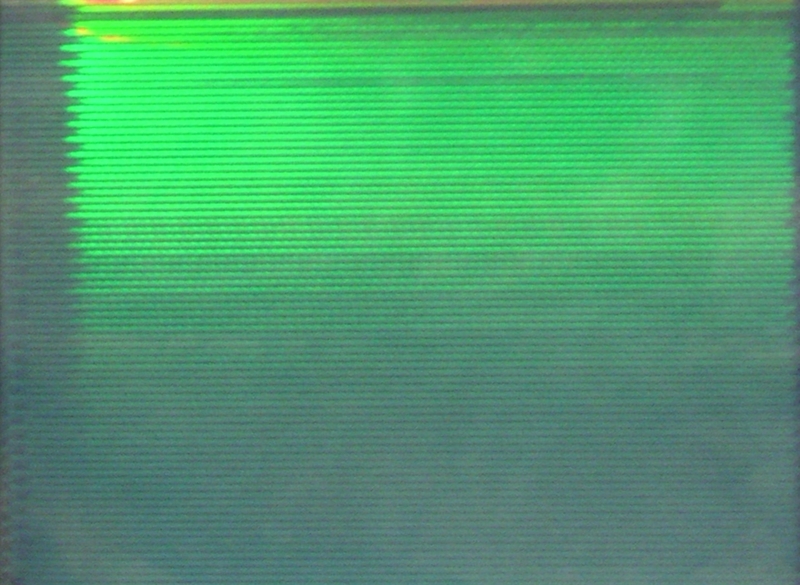}
\caption[Photograph of hologram with decreasing viewing window and energy]{Photograph of hologram with decreasing viewing window from left to right and decreasing reference beam energy top to bottom. The reference beam energy is decreased in steps, with 5 rows printed for each energy. The pixels are square and \unit[1.0]{mm} wide with a \unit[2.0]{mm} separation between centers.  The photograph was taken with a f/5.0 aperture and 1/3 second exposure time.}
\label{fig:window_energy}
\end{figure}

It is clear that the larger viewing window (to the left hand side of the photograph) results in a dimmer image in the direction normal to the hologram.  It would be useful future work to use the photographs to better analyze, and correct for, this problem.  To further help with the problem, several holograms were also printed with the object beam energy decreasing in steps.  This was achieved by changing the intensity of the rectangle display on the \ac{LCOS} after every five rows.  \Fref{fig:window_energy} shows a photograph of a such a hologram.  It is left as future work to analyze these.

\clearpage

\section{Summary}

A sample hologram was found to have visual `shadows' or `ghosting' where the size of the viewing window of a holopixel was determined to be affecting the apparent brightness of the holopixel in any given direction.  A photograph of the hologram was compared to the set of rendered images and the intended image, and correction algorithms were proposed.  The correction algorithms were applied to the photograph and found to improve the image.  These algorithms can be applied to rendered images before printing to correct for ghosting.

The algorithms can also be used to inform the artists where ghosting would occur, to allow them to adjust the model to remove any problem areas.

To better measure the hologram, a framework was built to hold and position a spectrometer and halogen light.  This gave good results and preliminary tests indicate that the camera was not accurately photographing the hologram intensity.

Future work is required to find a more optimal algorithm for correcting for ghosting, using a similar spectrometer framework and analysis code.

\chapter{Printing speed and resolution improvements}\label{chap:speed}

This chapter investigates methods to speed up the printing of digital holograms and ways to increase the holopixel resolution.  Sensitivity analysis methods are applied to determined the required mechanical stability of the hologram plate holder.  In addition to various other improvements, the power supply is replaced, the \ac{PC} software driving the printer is improved to run at a faster speed, and the \ac{LCD} display is replaced with an \ac{LCOS} display.

\section{Overview}
A monochromatic hologram printer similar to that described by \citet{ratcliffepatent} is investigated for possible speed increases.  The printer design described by Ratcliffe et al. is limited by both hardware and software to printing at four pixels per second.

\bigskip

This is an acceptable printing speed when printing \unit[1.0]{mm} by \unit[1.0]{mm} pixels on a small plate. A
small holographic plate produced, for example, by \citet{Slavich} is typically around \unit[127]{mm} by \unit[102]{mm} - a total of 12954 pixels at
\unit[1.0]{mm}$\,\times\,$\unit[1.0]{mm}. At 4 pixels per second, it would take just under an hour (54 minutes) to print.

\bigskip

This quickly becomes undesirable for a larger plate or for a higher resolution. It can 
take many hours or even days to print a large one meter squared hologram.  Additional problems 
can occur if the printing times are too long; the initial pixels `fade'
by the time the last pixels have been printed. Various techniques have been established 
with varying success in an attempt to compensate for this. One technique is to use a dim incoherent white-light source, such as a \unit[25]{W} lamp at a meter or so from the hologram.  This technique is known as latensification,
and its effect on green pulsed holography has been shown to be effective \citep{Vorzobova:04}.

\bigskip

With a high resolution print with pixels of size \unit[0.3]{mm} x \unit[0.3]{mm}, it would take
approximately 9~hours to print even the small plates mentioned.

To speed up the printing, both hardware and software adjustments were
required. 

\section{Power supply}

The laser used inside the hologram printer detailed by Ratcliffe et al. is flashlamp pumped, requiring the power supply to provide bursts of energy at precise regular intervals. The power
supply was fundamentally limited to \unit[10]{Hz}, thus limiting the number of
laser pulses to a maximum of 10 per second.  This was replaced with a newer design of power
supply provided by \citet{Geola}, capable of an operation of up to \unit[50]{Hz}.  The new power supply contains a larger charging unit and uses a standard simmer system to keep the flash lamp semi-powered continually. This also increases the lifetime of the flash lamp by two to five times longer.  The technical specification of the laser are given in Appendix~\ref{sec:laserspecs}.

\section{Stability}

With a faster operation, the motors provided more vibrations 
requiring the system to have a higher mechanical stability. Because the beam pulse is
short (approximately \unit[50]{ns}) compared to the vibrations, the system is essentially completely still
during the exposure {--} vibrations on the microscopic scale do not
affect the exposure or the print.  This is different to that of
\ac{CW} hologram printers that do suffer from even the sightest of vibrations.
At the macroscopic scale, however, small vibrations can move the point that the reference beam exposures 
laterally, causing the reference and object beams to overlap only partially, or even not overlap at all.
This effect is more pronounced at higher resolutions because of the smaller
overlap area.  At high translational velocities the stage motor vibrations can also compound this effect to further reduce the overlap area.

\begin{figure}[htp]
\centering
\includegraphics[angle=-90, width=\textwidth/2]{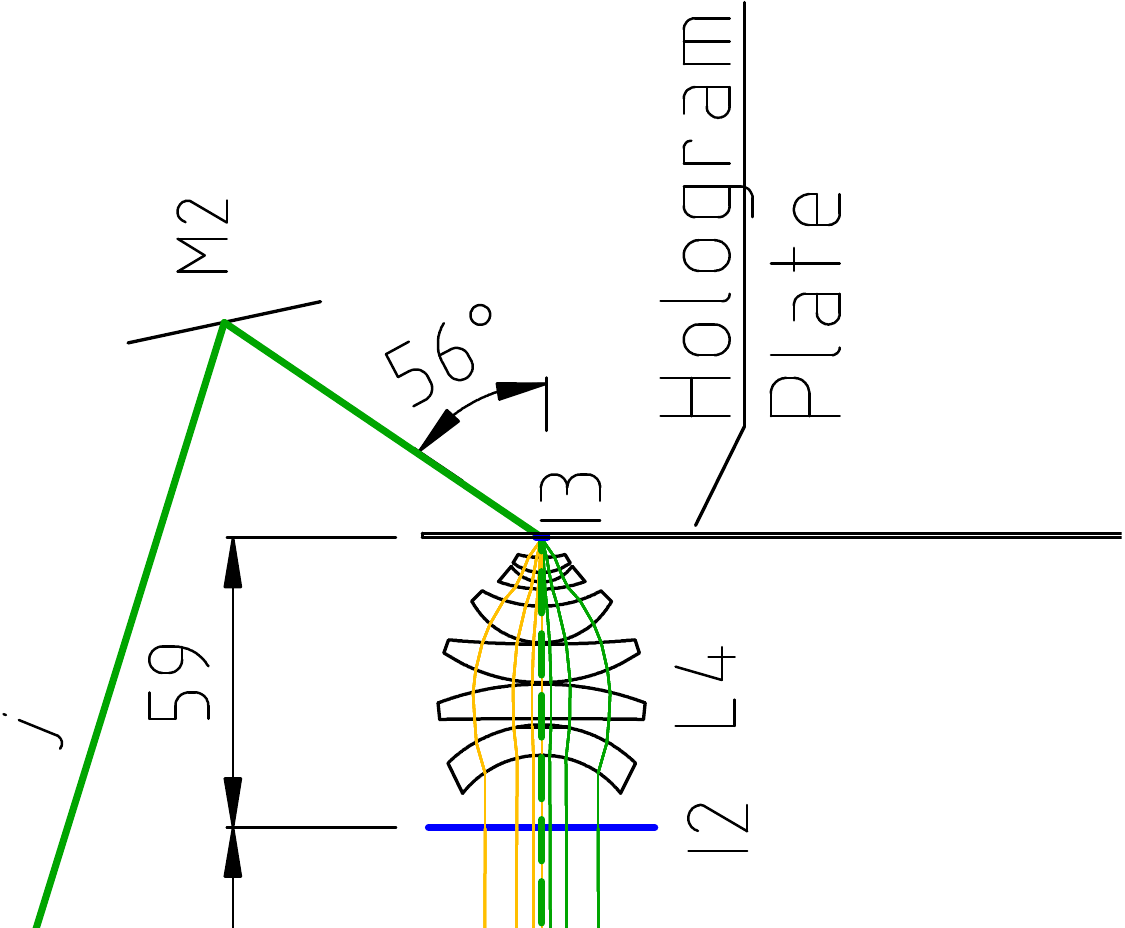}
\caption[Diagram of intersection plane of hologram plate and beams]{Diagram showing the plane of intersection of the hologram plate and the object and reference beams (shown in dashed and solid lines respectively).  The object beam is also shown raytraced through the objective lens for clarity.  A full diagram is shown in \Fref{fig:hrip_lcos_top_dimensions_big}.}
\label{fig:stagemovement}
\end{figure}

\bigskip

\begin{figure}[ht]
\centering
\def\firstcircle{(0,0) circle (1.5cm)}
\def\secondcircle{(2cm,0) circle (1.5cm)}

\begin{tikzpicture}
    \fill[gray!5!white] \firstcircle;
    \fill[gray!5!white] \secondcircle;
    \draw \firstcircle node[left] {$R$};
    \draw \secondcircle node  {$O$};


    \begin{scope}
      \clip \firstcircle;
      \fill[gray!30!white] \secondcircle;
      \draw \secondcircle;
    \end{scope}
    \begin{scope}
      \clip \secondcircle;
      \draw \firstcircle;
    \end{scope}
    \draw (1cm,0) node {$A$};
    \draw[<->,thick] (0,0) -- (0,-1.5) node [right,midway] {$r$};
    \draw[|<->|,thick] (0,-1.9) -- (2,-1.9) node [below,midway] {$c$};
\end{tikzpicture}
\caption[Object and reference beam misalignment]{The object ($O$) and reference ($R$) beam profiles, of radius $r$, at the plane of the hologram plate can become misaligned by a distance $c$.}
\label{fig:circleoverlap}
\end{figure}
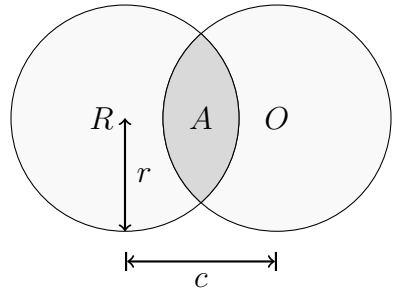

\Fref{fig:stagemovement} shows the intersection of the object and reference beams at the hologram plate. It is important to maintain some minimum overlap of the object and reference beams, $O$ and $R$ respectively, at the plane of the holographic plate (see \Fref{fig:circleoverlap}).

\clearpage

\nomenclature{$O$}{Object beam profile at the plane of the hologram plate}
\nomenclature{$R$}{Reference beam profile at the plane of the hologram plate}
\nomenclature{$A$}{Overlap area of reference and object beam profiles}
\nomenclature{$c$}{Misalignment distance between the centers of reference and object beam profiles at the hologram plate}
\nomenclature{$r$}{Radius of object and reference beam profiles at the hologram plate}

 Since the two beams are approximately circular they have reflectional symmetry about the chord that
crosses through their intersection points. Thus we can find the area
in one half of the intersection, and double that to obtain the total area of overlap.

The total area, $A$, of overlap of two circles of radius $r$ and distance $c$ between the 
centres of the two beams, is thus given by:
\begin{align}
A &=  2(q\cdot r^2/2-\sin(q)\cdot r^2/2)\\
&=r^2(q-\sin(q))
\intertext{where}
q&=2\cdot \cos^{-1}(c/(2r))
\end{align}

Since the area of a circle is $\pi\cdot r^2$, dividing through by this
obtains the percentage overlap, $P(c,r)$, of the two beams:
\begin{eqnarray}
P(c,r)=\frac{(q-\sin(q))}{\pi} &\text{where} &
q=2\cdot \cos^{-1}(c/(2r))
\end{eqnarray}
\nomenclature{$P(c,r)$}{Percentage overlap of reference and object beam profiles at hologram plate}

To get a numerical feel for these equations, two different beam sizes and two different minimum overlaps will be considered.  The design described by \citet{ratcliffepatent} uses circular pixels with a radius of \unit[0.5]{mm}.  For the purpose of setting a specific target, this will be compared to using round pixels with a radius of \unit[0.15]{mm} - a pixel that is one order smaller in area.

For a minimum overlap of 90\% of the two beams on the photographic plate, with the object and reference beams both having a radius of \unit[0.15]{mm} (and hence a height and width of \unit[0.3]{mm}), and by
solving the above equation numerically\footnote{Equation solved
numerically in Maxima with the command:
\Maxima{find\_root((2*acos(c/\ (2*r)) - sin(2*acos(c/(2*r)))) / \%pi - p, c, 0, 0.15), r=0.15, p=0.9;}}, it can be seen that the distance between the two centres must not be greater than
\unit[0.02]{mm}. 
The results shown in \Fref{tab:maximumbeamseparation} are obtained by 
solving the equation for the larger beam size of \unit[1.0]{mm} in radius, and also
repeating with a higher resolution beam separation of \unit[0.3]{mm}.

\bigskip
\begin{table}[ht]
\centering
\begin{tabular}{lrr}
\toprule
Overlap & Beam size & Separation\\
\midrule
70\% & \unit[1.0]{mm} & \unit[0.24]{mm}\\
             & \unit[0.3]{mm} & \unit[0.07]{mm}\\
90\% & \unit[1.0]{mm} & \unit[0.08]{mm}\\
             & \unit[0.3]{mm} & \unit[0.02]{mm}\\
\bottomrule
\end{tabular}
\caption[Maximum reference and object beam separation distance]{Maximum reference and object beam separation distance for certain beam diameters and a minimum beam overlap}
\label{tab:maximumbeamseparation}
\end{table}

\bigskip

To increase the resolution of a hologram by changing from a pixel size of \unit[1.0]{mm} diameter to pixels with a \unit[0.3]{mm} diameter, the alignment needs to be approximately four times better to achieve the same overlap area. A
sensitivity analysis shows that the relative sensitivity of the
percentage overlap, $P$, as the distance $c$ between the centers of the two beams is changed is\footnote{Equation
solved analytically in Maxima with the command:
\Maxima{P(c):=(\,2*acos(c/(2*r)) {}-
sin(2*acos(c/(2*r))))/\%pi; factor(diff(P(c),c) * c/P(c)); \% ,
2*acos(c/(2*r)) = q;}}:
\begin{eqnarray}
\text{Relative sensitivity of $P$ w.r.t. $c$} &=&\frac{c}{P}\cdot {\frac{\partial P}{\partial c}}\\
&=&\frac{2c(\cos(q)-1)}{(q-\sin(q))\sqrt{4r^2-c^2}}\label{eq:relativesensitivity}
\end{eqnarray}

\bigskip

\Fref{eq:relativesensitivity} shows that the sensitivity increases with the beam separation, $c$, and
inversely with the beam radius, $r$. However the sensitivity remains quite
low for the values we are considering. For example, for a beam radius
of \unit[0.15]{mm}, and a separation of \unit[0.07]{mm} (giving an overlap of 70\%), the
sensitivity of $P$ with respect to $c$ is only 40\%; the percentage overlap of the two beams will 
increase by only 40\% of the percentage change in the separation of their centres.

Likewise, the sensitivity of $P$ with respect to the beam radius
$r$ is\footnote{Equation solved analytically in Maxima with
the command: \Maxima{P(r):=(2*acos(c/(2*r)) {}-
sin(2*acos(c/(2*r))))/\%pi; factor(diff(P(r),r) * r/P(r)); \% ,
2*acos(c/(2*r)) = q;} }:
\begin{eqnarray}
\text{Relative sensitivity of $P$ w.r.t. $r$} &=&\frac{r}{P}\cdot {\frac{\partial P}{\partial
r}}\\
&=&-\frac{2c(\cos(q)-1)}{(q-\sin(q))\sqrt{4r^2-c^2}}
\end{eqnarray}

\bigskip

\bigskip

This is of the same magnitude as the relative sensitivity of $P$ with
respect to $c$. 

For a maximum beam movement of \unit[0.05]{mm} laterally, the reference beam
mirror (labelled M2 in \Fref{fig:stagemovement}) must also remain stationary by
approximately the same amount. To achieve this, the mirror was
mounted on a metal rod, which was secured to the table. Lateral
movement was restricted by an `A'{}-shaped structural support system.
This was found to be sufficient to mechanically damp vibrations from
the motors, and also withstand the occasional accidental knock without
knocking the system out of alignment.

\bigskip

Further mechanical stability adjustments are detailed in \Fref{sec:stagemovement} in the context of speeding up the movement of the stages.
The subject comes up again in \Fref{sec:increasingresolution} in the
context of decreasing the size of the pixels.

\clearpage
\section{Software}\label{sec:software}

The original software for the hologram printer was designed for low
speed printing. The basic algorithm of the code is listed in Algorithm~\ref{algorithm:printercode}.

\bigskip

\begin{algorithm}[ht]
\SetLine
Calculate the speed at which to move the plate vertically;

Move vertical stage to first pixel to print;

\ForEach{vertical row of pixels} {
	Start moving the vertical stage downwards at the calculated speed;

	\ForEach{pixel in the vertical row} {
		Display next pixel information on the \ac{LCD}/\ac{LCOS} screen;

		Call the windows function \textstyleHTMLCode{sleep(int milliseconds)} to wait until the stage has moved into position;

		Trigger the laser to fire a pulse;
	}

	Start moving the horizontal stage by the width of one pixel;

	Start moving the vertical stage back to the starting position;

	Wait until both stages are in position;
}
\caption{Printing algorithm suitable for printing at low speed}
\label{algorithm:printercode}
\end{algorithm}

There are several significant problems with this algorithm when trying
to scale it to faster speeds:
\begin{itemize}
\item { The granularity of the sleep function is by default 10 milliseconds.}
\item { There is no attempt to deal with `drift' {--} it assumes that the sleep
will be for a time specified, and that there will be no unforeseen
delays. Any problems will not be corrected for.}
\item { The \ac{LCD}/\ac{LCOS} takes a while to display the next image. At high speeds
we risk the \ac{LCD}/\ac{LCOS} not being updated by the time we print.}
\item { At faster speeds, the stage can have a significant inertia, resulting in
a non{}-negligible acceleration time.}
\item { At low speeds, moving the vertical stage back to the starting position
is many times faster than the printing. At fast speeds it starts to
take up a significant amount of time compared to the printing itself.}
\end{itemize}

Consider the first of these points.  The sleep time required will depend upon the speed at which
we print.  For printing at \unit[4]{Hz}, we will need the total time between
printing pixels to be approximately 250\,milliseconds. For printing at
\unit[30]{Hz}, we will need the total time to be approximately 33\,milliseconds.

\bigskip

{
From profiling the code, we found it takes approximately 20 milliseconds
to fetch the next pixel data and display it on the screen. The system
must then sleep for approximately 230\,milliseconds at \unit[4]{Hz}, and 
13\,milliseconds at \unit[30]{Hz}.}

\begin{figure}[htp]
\centering
\includegraphics[width=10cm]{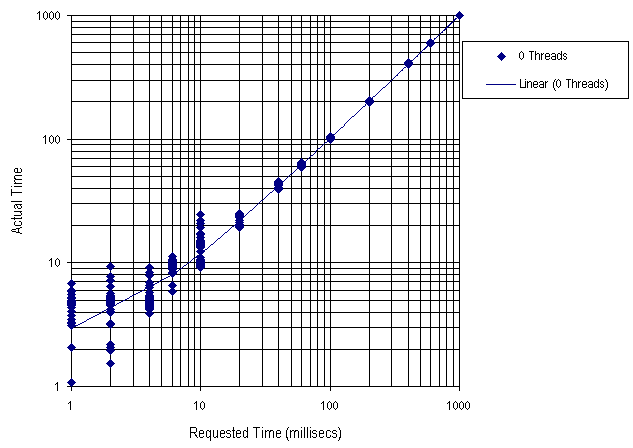}
\caption{Behavior of Microsoft Windows \Maxima{sleep()} function}
\label{fig:sleep}
\end{figure}

{
By default, the sleep function has a granularity of \unit[10]{ms}. So a call to
sleep for, say, just under \unit[225]{ms} will actually sleep for \unit[230]{ms} {--} an
error of 2\%. A sleep for \unit[13]{ms} will sleep for \unit[20]{ms}, however, an error
65\%. For short sleep times, the error is actually much worse than
this\footnote{Source \url{http://www.codeproject.com/system/sleepstudy.asp}} (See \Fref{fig:sleep}). This is because the kernel can take an arbitrary
amount of time in processing interrupts, I/O calls etc. While
handling interrupts etc, even a real time process cannot be scheduled.}

\bigskip

{
The first step to solving this is to increase the time granularity. We
can use the system call: \textstyleHTMLCode{timeBeginPeriod(1);}
which sets the granularity at around 1\,or 2\,milliseconds. This
reduces our average error at \unit[13]{ms} to \unit[2]{ms}$/$\unit[13]{ms} = 15\%. However the
situation is worse than this because it will always sleep for a minimum
of the time we give it i.e. the error is always in one direction, never
lower than the specified time. This means that the error will
accumulate much faster than the usual random walk $\log(n)$; we would
drift by a whole pixel after printing around 33 pixels, or \unit[1]{cm} of
printing. A drift of one pixel per centimeter is clearly not
acceptable.}

\bigskip

One possibility it to write our own sleep function that further
increases granularity by getting the CPU time and spin{}-lock until the
time desired is reached. However this is known to be unreliable under
high \ac{IO} activity, which we may have from reading the hard disk.

\bigskip

Instead we opted for using the one millisecond granularity sleeps combined with
continually adjusting the sleep time to compensate for drift. This should insure that
the system is never more than one millisecond out of step. The drift is tracked by
using the win32 system function \textstyleHTMLCode{timeGetTime()} and
comparing the returned result against the expected time. The delta is
then subtracted from the base sleep time. For better results, this
should ideally be done on a real time \ac{OS}, such as a real{}-time MS Windows
kernel, or a real{}-time Linux kernel. Alternatively the timing could be handled
in hardware.

\section{Display system refresh rates}

The design detailed by \citet{ratcliffepatent} utilizes an \ac{LCD} transmission display system,
with a native 800$\times$600 pixel resolution and capable of a refresh rate of \unit[20]{Hz}. This
means that with a simple implementation, the hologram printer can
print up to ten holopixels per second (so that you are guaranteed at
least one complete screen refresh before the pixel is printed.) With
a more careful implementation, synchronizing the laser pulses to the video \ac{VSYNC},
it would be possible to print at exactly twenty holopixels per second,
but with little flexibility. The timing would need to be much more
precise, and possible required dedicated timing hardware.

\bigskip

{
Instead a \acf{LCOS} display system was used. In particular the
Brillian BR768HC \ac{LCOS} which is capable of a refresh rate of \unit[60]{Hz} with
an \unit[18]{ms} response time. The \ac{LCOS} also has many other advantages, including a
higher contrast of 2000:1 and a higher native resolution (1280$\times$768).

\section{Stage movement}\label{sec:stagemovement}

The origin design's method of printing was to print a vertical column of pixels,
from top to bottom, then returning the vertical stage back to the top
while moving the horizontal stage across by one pixel.

\bigskip

There are multiple reasons for this method of operation:	
\begin{itemize}
\item {
Moving the vertical stage back to the top is much faster than moving the
vertical stage during printing. Thus as a percentage of time, it is
insignificant.}
\item {
Without the timing auto{}-correction system as described in \Fref{sec:software} (\nameref{sec:software}), the pixels would drift from their correct
positions. This can be crudely compensated for by always printing in
the same direction. In this manner the drift is approximately the same for
each column, and so does produce a noticeable effect.}
\item {
Simpler code. The pixel data can be loaded and processed sequentially.}
\item {
Mechanical stability. The horizontal stage needs time to move and
settle down; it is an order slower than the vertical stage as the load
is much higher (the horizontal stage carries the vertical stage).}
\end{itemize}

\bigskip

The first point is no longer valid at fast speeds. At \unit[30]{Hz}, the
vertical stage moves almost at its maximum speed while printing. The
second point will not be a problem when using the timing auto-correction
system previously mentioned. The third point is not that significant.

The mechanical stability was dealt with in two ways. The first step
was to mount the vertical stage off-center on the horizontal stage.
This added a frictional force which acted to stabilize the vertical
stage. The second step ties in with one of the original problems,
restated for convenience:

\begin{itemize}
\item {
At faster speeds, the stage can have a significant inertia, resulting in
a non{}-negligible acceleration time.}
\end{itemize}

\bigskip

By adding a small `buffer' space above and below the plate, such that the vertical stage
moves higher and lower than the bounds of the hologram plate, the
the vertical stage is giving time to accelerate and the horizontal stage
is given time to settle down. In practice, a buffer of 1cm on the top and bottom
was sufficient. 

\section{Increasing printed holopixel resolution}\label{sec:increasingresolution}

The previous section dealt with increasing the speed at which
`holopixels' could be printed. Many of the improvements detailed had
a secondary purpose in preparing the system for increasing the
resolution of the hologram i.e. decreasing the size of the holopixels
and decreasing the distance between their centres. The increased
mechanical stability, for example, paves the way for more precisely
placing the holopixels and hence allowing smaller holopixels.

\bigskip

Additional mechanical modifications were needed at higher resolutions,
and these are specified below.

\bigskip

The vertical stage is mounted on a horizontal stage. To move the plate
sideways, the horizontal stage is activated, thus moving collectively the vertical
stage and plate holder horizontally. Although this is obvious, it is not so
obvious that the angle of the axis that the horizontal stage
moves along is not critical. The angular placement of the horizontal stage is
not critical.

\bigskip

However it is critical that the hologram plate is held by the vertical
stage in the correct way; it must be parallel to the axis of motion of
both the horizontal and vertical stage. If the plate is not parallel
enough, then during printing the part of the plate that is being
exposed will move in the orthogonal direction, away or towards the
main lens (indicated as lens L4 in \Fref{fig:hrip_lens_system_big}).

\bigskip

For an idea of the accuracy required, consider the case that we wish to print a
hologram to a photosensitive plate with dimensions $W\!\times\!H$. Without loss of generality,
we consider the horizontal direction only. If the plate is held
at angles  $\theta_{x,\text{error}}$ 
and  $\theta_{y,\text{error}}$  to
horizontal and vertical motion respectively, then when the furthest
corner pixel is printed, the hologram plate will have moved a total
distance, in the direction normal to motion, of:
\begin{equation}
\sqrt{W^2\sin^2(\theta_{x,\text{error}})+H^2\sin^2(\theta_{y,\text{error}})}
\end{equation}
\nomenclature{$W$}{Width of hologram plate}
\nomenclature{$H$}{Height of hologram plate}
\nomenclature{$\theta_{x,\text{error}}$}{Angle of hologram plate to the horizontal motion}
\nomenclature{$\theta_{y,\text{error}}$}{Angle of hologram plate to the vertical motion}
Since $\theta_{x,\text{error}}$ and $\theta_{y,\text{error}}$ should be very close to zero, the 
approximation $\sin(x)\approx x$ is suitable. The situation can be simplified by
considering a square plate (such that $W=H$), and by assuming that 
$\theta_{x,\text{error}}$ and $\theta_{y,\text{error}}$ are approximately equal and both cause
normal motion in the same direction (the worst case scenario). Thus the maximum distance of
normal motion,  $z_{\text{max}}$, is:
\begin{equation}
z_\text{max}=\sqrt{2}\cdot W \theta_\text{error}
\end{equation}

\nomenclature{$\theta_\text{error}$}{Angle of hologram plate to motion}
\nomenclature{$z_{\text{max}}$}{Maximum distance of normal motion of hologram plate due to $\theta_\text{error}$}

This affects the hologram because the reference beam strikes the
hologram plate at an angle 
$\phi_{\text{reference}}$  to the
normal of motion of the hologram plate. ($\phi_{\text{reference}}$ will be
typically around 56\degree, close to the Brewster angle for glass).
As the exposed area of the plate moves away from the lens
L4 due to unwanted normal motion, the point at which the reference beam strikes will move laterally,
ultimately resulting in the object beam and reference beam no longer
fully overlapping.
\nomenclature{$\phi_\text{reference}$}{Angle of reference beam to the normal of the hologram plate}

\clearpage

The maximum distance, $c_\text{max}$, that the reference beam
will become misaligned due to $z_{\text{max}}$, is thus:
\begin{equation}
c_{\text{max}}=z_{\text{max}}\sin(\phi_\text{reference})=\sqrt{2}\cdot W{\theta}_{\text{error}}\sin(\phi_\text{reference})
\end{equation}
\nomenclature{$c_\text{max}$}{Maximum misalignment distance between the centers of the reference and object beam profiles at the hologram plate for a given $z_{\text{max}}$, $\phi_\text{reference}$ and $W$}

Rearranging:
\begin{equation}
\theta_{\text{error}}=\frac{c_\text{max}}{\sqrt{2}\cdot
W\sin(\phi_\text{reference})}
\end{equation}

\bigskip

If we assume that the beams are perfectly aligned in one corner, then
the maximum distance between the centres of the two beams, $c$,
will be equal to  $x_\text{max}$ . We found
previously that if we want to maintain at least a 70\%
overlap when both beams have a diameter of \unit[1.0]{mm}, then the distance
between the centres of the two beams must be at a maximum of \unit[1.0]{mm}.
Considering a plate of size, say, $W=\unit[0.5]{m}$, and angle 
$\phi_\text{reference}=56^\circ$
 we can get an idea of the maximum angles of error. These are listed
in \Fref{tab:maximumerror}. 

\begin{table}[ht]
\centering
\begin{tabular}[c]{lrr}
\toprule
Overlap & Beam Size& Separation\\
\midrule
70\% & \unit[1.0]{mm} & \unit[410]{\textmu rad}\\
             & \unit[0.3]{mm} & \unit[120]{\textmu rad}\\
90\% & \unit[1.0]{mm} & \unit[140]{\textmu rad}\\
             & \unit[0.3]{mm} & \unit[34]{\textmu rad}\\
\bottomrule
\end{tabular}
\caption[Maximum angle deviation for a photographic plate]{Maximum tolerated angle deviation from the plane of a photographic plate for certain beam sizes and minimum overlap, at $\phi_\text{reference}=56^\circ$
and $W=\unit[0.5]{m}$}
\label{tab:maximumerror}
\end{table}

To achieve the high level of precision needed, an adjustable holding frame
was made for the glass plate to allow adjustment of the angle of the
plate. This was made from a Standa lens holder that allowed for two
dimensional angle adjustments with an accuracy of between 1 to 2 {\textmu}rad,
sufficient for this application.

\bigskip

Alignment was done by eye. The beam energy was reduced to a few percent of the
normal energy level (for both safety reasons and because it is easier to align when the
beam is dim). A blank glass plate can then be covered in white
masking tape, and placed in the holder, with the masking tape closest
to the main lens (where the emulsion would be located). It is best to
use an actual hologram plate, as it is important that the glass plate
is of the same thickness as the hologram plate that will be employed.

\bigskip

The plate was then moved (using computer control or otherwise) to one
corner (meaning that the glass plate is moved so that the object and
reference beams are illuminating one of the corners of the plate). The
reference beam mirror (labelled M2 in \Fref{fig:stagemovement}) can then be adjusted until
the beams overlap, judging this as best as possible by eye.

The glass plate is then moved (by computer control) so that the other
corner is exposed. If the beams no longer overlap completely, the
angles are adjusted, and the process repeated.

Although somewhat crude, this actually works fairly well in practice.

\section{3D model}

The produced holograms were based upon two different 3D models.  The low resolution holograms used a 3D model of the KDE mascot, Konqi.  For the higher resolution holograms, a greyscale, high contrast and high detail model was required.  For this, the kind permission from Doug Ollivier was granted for the non-commercial use of his model of a futuristic tank model.

The increase in holopixel resolution meant that higher resolution images of the model are required.  Additionally, the increase in pixels on the LCOS (1280$\times$768 on the \ac{LCOS} compared to 800$\times$600 on the \ac{LCD}) potentially means that more rendered images can be used, to provide a greater depth of the field.  The \ac{LCOS} is used in landscape orientation in the hologram printer, meaning that for single parallax holograms, up to 1280 separate images are required, each rendered from a slightly different angle.  For full parallax, up to 1280$\times$768 $\approx$ 1 million images would be required.

To determine how to render the images from the \ac{3D} model for the horizontal parallax case only, an intuitive approach can be taken.  The scene needs to be rendered from many different angles, producing an image for some discrete set of angles.  By moving the virtual camera along a virtual track, keeping the camera pointed at the center of the scene, we can produce the required series of images.  However these images will need to be algorithmically distorted as each image is taken at a different angle.

Instead, the camera was set to capture a view that was twice as wide as the scene.  The virtual camera was first calibrated such that its field of view was equal to the field of view produced by the imaging lens system of the \ac{LCOS} onto the hologram plate.  The camera was then placed at a distance such that the scene took up exactly the whole of the right half of the view (see the scene setup in Blender in \Fref{fig:blender_tank}).  This established the start point of the camera track (see the rendered image in Figure~\ref{fig:tank_left}).  The camera was then moved parallel to the scene into a position such that the scene took up exactly the whole of the left half of the view.  This established the last point of the camera track (see Figure~\ref{fig:tank_right}).

\begin{figure}[htp]
\centering
\subfloat[Setting up the scene in Blender]{
\includegraphics[width=\textwidth *35/40]{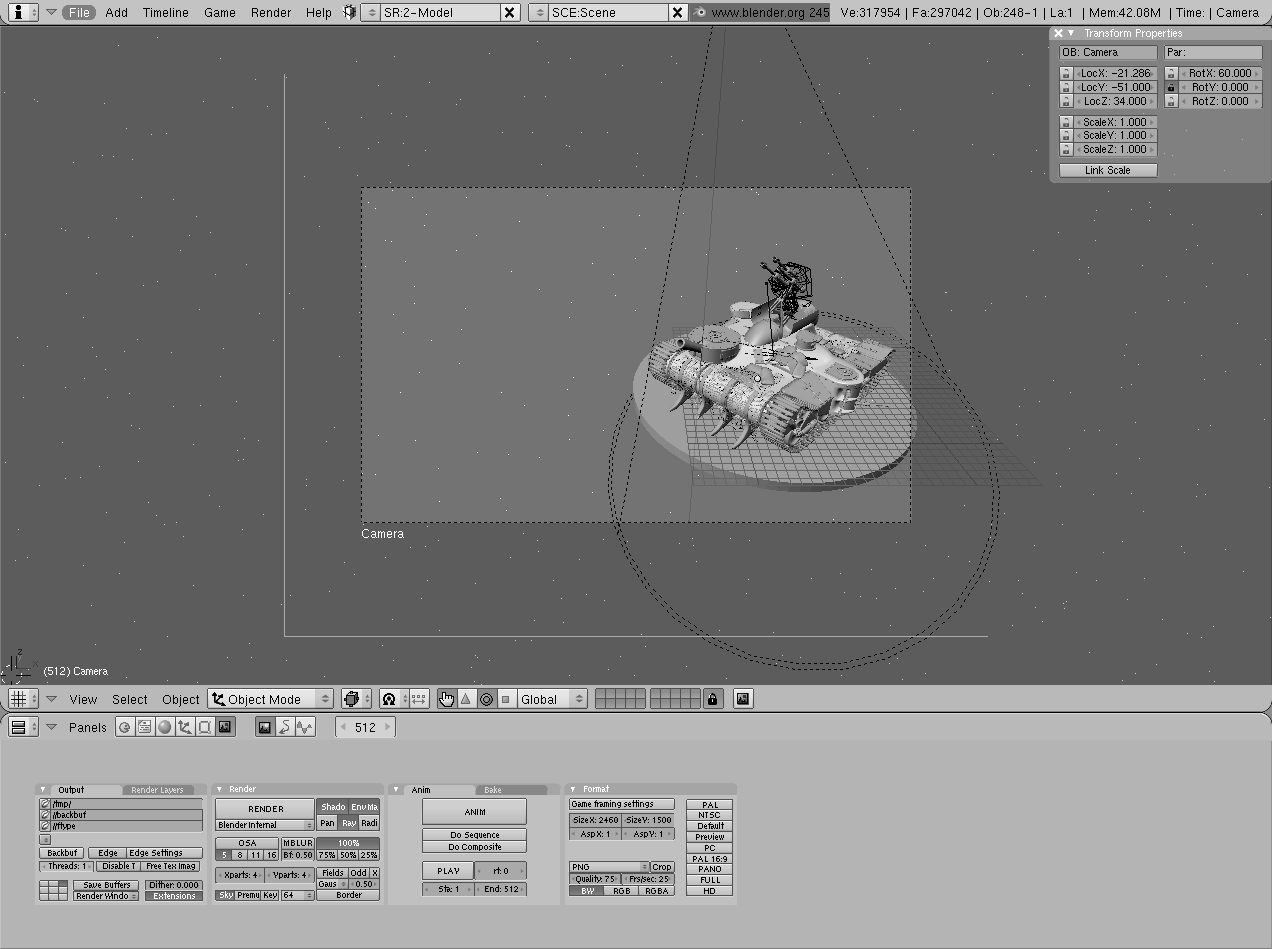}
\label{fig:blender_tank}
}
\qquad
\qquad
\subfloat[Rendered left-hand side]{
\includegraphics[width=\textwidth *4/10]{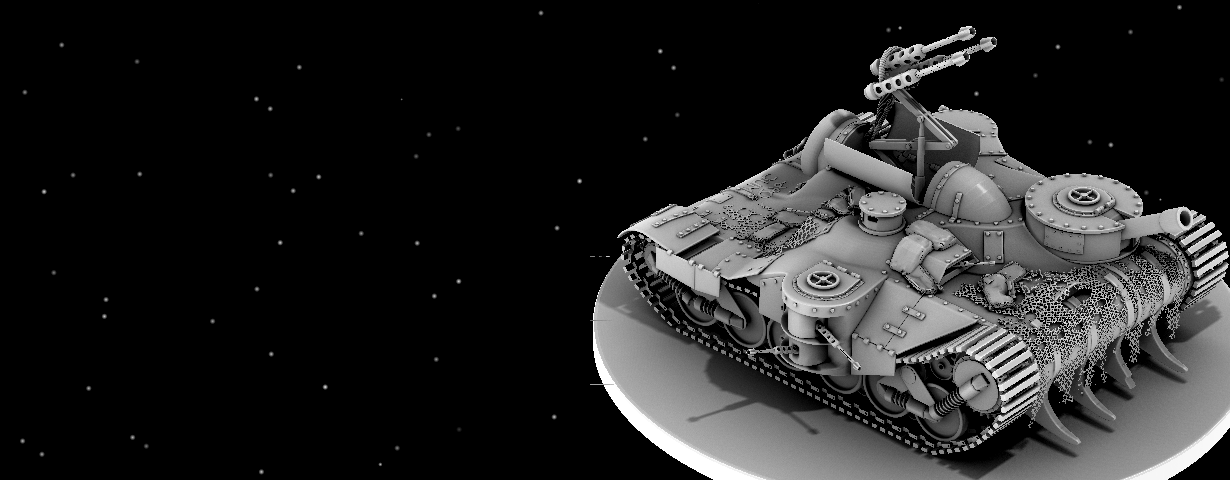}
\label{fig:tank_left}
}
\qquad
\subfloat[Rendered right-hand side]{
\includegraphics[width=\textwidth *4/10]{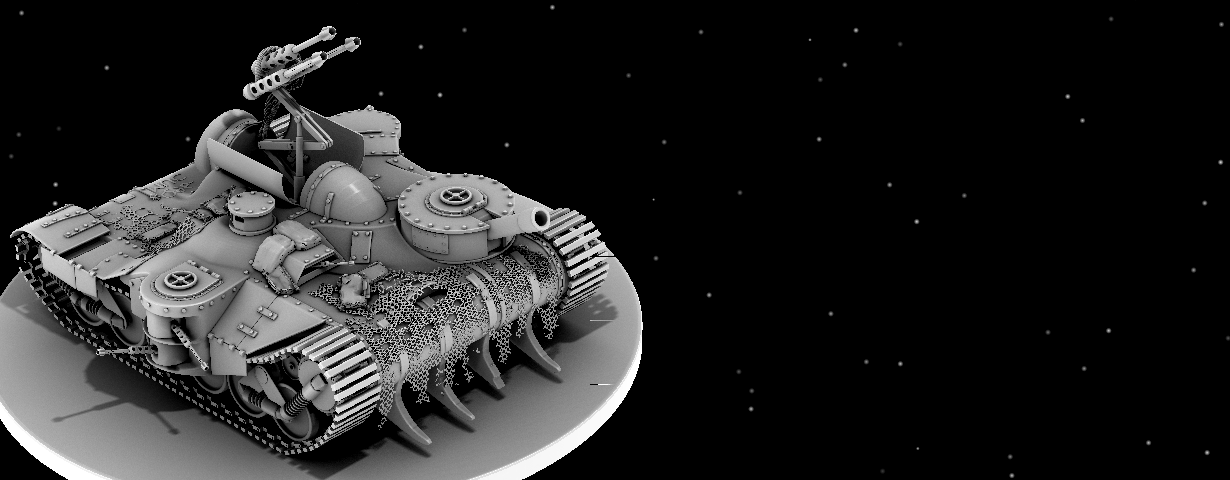}
\label{fig:tank_right}
}
\qquad
\subfloat[Cropped left]{
\includegraphics[width=\textwidth *2/10]{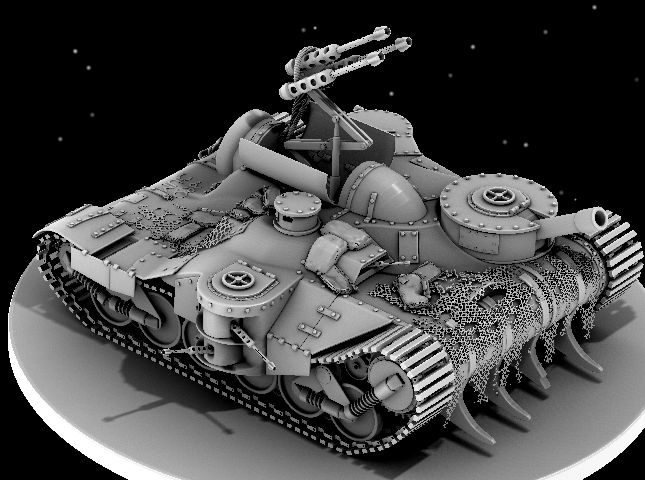}
\label{fig:tank_left_cropped}
}
\qquad
\subfloat[Cropped right]{
\includegraphics[width=\textwidth *2/10]{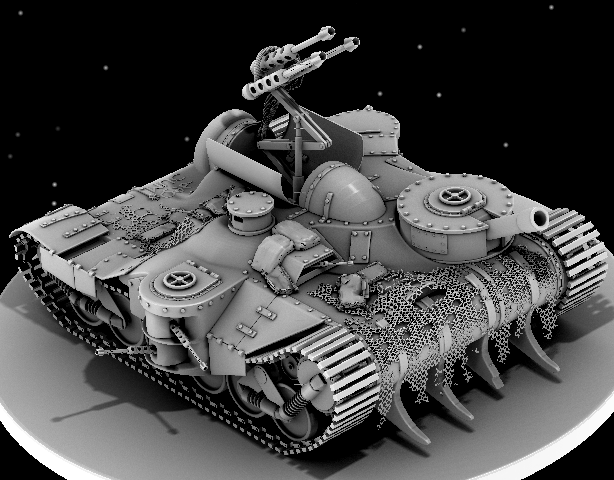}
\label{fig:tank_right_cropped}
}
\caption{Rendered model images}
\label{fig:leftandrighttank}
\end{figure}

A linear track was then made between these two points, discretized such that $N_\text{images}$ number of images was produced.  These images are then cropped with a sliding window such that the new image has the same height, but half the width, resulting in a series of $N_\text{images}$ images, all with the scene centered in each image (see Figure~\ref{fig:tank_left_cropped} and Figure~\ref{fig:tank_right_cropped}).
\nomenclature{$N_\text{images}$}{Number of rendered images of 3D computer model}
Since these images need to be cropped, it is advantageous to be able to crop by an integer number of pixels 
to avoid the need for interpolation. This means that the sliding cropping window must move by an integer number, $k$, of pixels each time.  Writing the cropped image width as $W_\text{image}$ pixels, the uncropped image width is thus $2\,W_\text{image}$. This gives us the restriction that $2\,W_\text{image} / k \in \mathbb{N}^*$ and that it is equal to the number of images, $N_\text{images}$.
\nomenclature{$W_\text{image}$}{Width of uncropped rendered image in pixels}
\nomenclature{$H_\text{image}$}{Height of uncropped rendered image in pixels}
\nomenclature{$k$}{Step size of sliding crop window for rendered images}
\begin{figure}[tp]
\centering
\includegraphics[width=\textwidth]{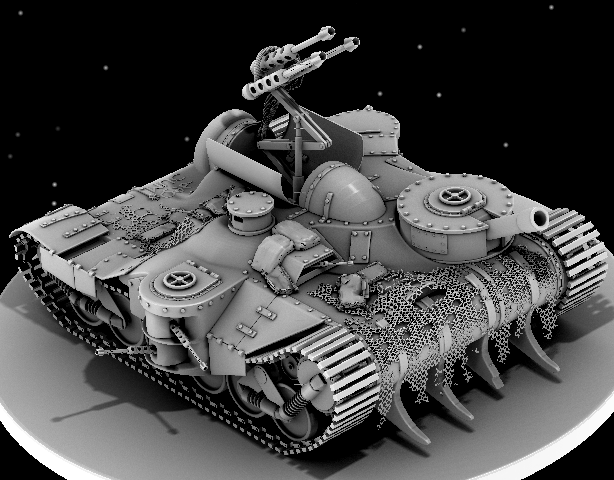}
\caption{Large rendered model image}
\label{fig:highresolutionrendered}
\end{figure}

So the width of the uncropped images, $2\,W_\text{image}$, must be some integer multiple of number of images rendered.  The width of a rendered and cropped image, in pixels, also corresponds to the width, in holopixels, of the final printed hologram.  Likewise, the height of the rendered image corresponds to the height, in holopixels, of the final printed hologram.

A further restriction can be made from the \ac{LCOS} resolution.  Consider the holopixel at coordinate $(x,y)$ being printed, with $x$ and $y$ in units of pixels.  The \ac{LCOS} must display an image such that the same $(x,y)$ coordinate pixel from each image is mapped to a pixel on the \ac{LCOS}.  The solution can be simplified by imposing the restriction that the effective\footnote{Sometimes it is advantageous to only use a portion of the LCOS screen, hence the prepended qualifier 'effective'.} horizontal resolution of the \ac{LCOS} is $W_\text{lcos} = m\,N_\text{images}$ where $m \in \mathbb{N}^*$.

To summarize, the following restrictions were imposed in order to optimize the code:
\begin{align}
N_\text{images} &= \frac{2\,W_\text{image}}{k}\label{eq:nimages1}\\
N_\text{images} &= \frac{W_\text{lcos}}{m}\label{eq:nimages2}\\
H_\text{image} &= H_\text{hologram}\\
W_\text{image} &= W_\text{hologram}
\end{align}
\nomenclature{$W_\text{hologram}$}{Width of hologram in holopixels}
\nomenclature{$H_\text{hologram}$}{Height of hologram in holopixels}
\nomenclature{$W_\text{lcos}$}{Width of LCOS display in pixels}
\nomenclature{$H_\text{lcos}$}{Height of LCOS display in pixels}
\nomenclature{$\mathbb{N^*}$}{Set of positive natural numbers}
Where $N_\text{images}$ is the number of rendered images, $W_\text{image}$ and $H_\text{image}$ are the width and height respectively of the cropped image, $W_\text{lcos}$ is the effective width of the \ac{LCOS}, $W_\text{hologram}$ and $H_\text{hologram}$ are the effective width and height in pixels respectively of the hologram, and $m$ and $k$ are arbitrary non-zero natural numbers.}
\nomenclature{$m$}{Parameter for image cropping -- non-zero natural number}
\bigskip

To print the test model for the purpose of evaluating the printer, a holopixel size of approximately \unit[0.3]{mm} was used. A resolution of \unit[0.3]{mm} between pixel centers is approximately 85 \acf{DPI}, similar to that of the resolution of a computer screen.  To aid in the visual evaluation of the completed hologram, however, the distance between the pixels was increased to \unit[0.5]{mm}, leaving a gap between each pixel (See \Fref{fig:highresolutiontankzoomed}).  

The hologram was printed on a large Slavich VRP-M glass plate of dimensions \unit[60]{cm}$\times$\unit[40]{cm} in landscape. With a \unit[0.5]{mm} inter-pixel spacing, the maximum width and height of the hologram in pixels is thus:
\begin{align}
W_\text{hologram} &\le \unit[60]{cm} \div \unit[0.5]{mm/pixel} = \unit[1200]{pixels}\\
H_\text{hologram} &\le \unit[40]{cm} \div \unit[0.5]{mm/pixel} \approx \unit[800]{pixels}
\end{align}

For the Brillian \ac{LCOS} BR1080HC display system with a resolution of 1280$\times$768 used in landscape format, $W_\text{lcos} = 1280$.  To produce a slightly brighter hologram, the hologram replay viewing window was reduced by reducing the effective width of the \ac{LCOS} to \unit[984]{pixels}.  This particular size was chosen to allow the image size to be nice resolution.

Rearranging Equation~(\ref{eq:nimages1}) and \Fref{eq:nimages2}:
\begin{align}
W_\text{image} &= \frac{k}{2m}W_\text{lcos}\\
\intertext{Setting $m=2$ \& $k=5$:}
1230 &= \frac{5}{4} \times 984\\
N_\text{images} &= \frac{2\times1230}{5} = 492
\end{align}

An image height of 960 was chosen.  492 images were thus rendered at 2460$\times$960 and cropped to 1230$\times$960 using the program listed in Appendix~\ref{sec:imagecropping}.

A single rendering of the test model at this resolution took approximately five hours to do on a single CPU \ac{PC}.  Extrapolating, this would take a total rendering time of 1280$\times$\unit[5]{hours} = \unit[1]{month}.  To reduce this time, the rendering farm company ResPower was used, completing the render in just under 24 hours.  If maximum resolution and maximum depth was used instead, the rendering would have taken an estimated \unit[1.5]{years} of rendering time - around a week on the ResPower rendering farm.

\section{Analysis}

After several attempts, a successful bright hologram was produced.  With just over a million pixels, the hologram took around 11 hours to print at rate of 30 pixels per second.  This speed was experimentally pushed up to 42 pixels per second for smaller holograms without noticeable problems.

To protect and display the hologram, the hologram was printed back to front and then the emulsion side spray-painted black.  A few different brands of spray paint were tested to find one that not harm the emulsion.  The best performing paint was a cheap quick-dry car matt-black spray-paint.  This caused the hologram to color shift slightly, but had no noticeable adverse affect on the brightness.

\begin{figure}[htp]
\centering
\includegraphics[width=\textwidth]{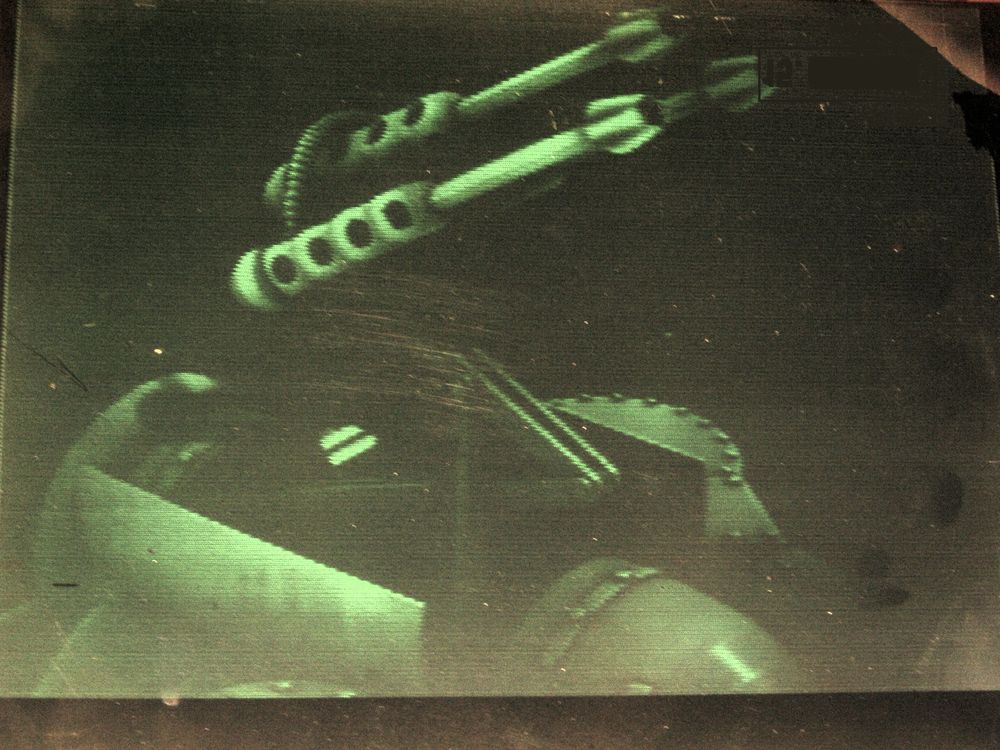}
\caption[High resolution hologram of a futuristic tank]{High resolution (\unit[0.3]{mm} in diameter pixels) hologram of futuristic tank.  Some of the low contrast and low detail are due to the difficulty in capturing the image with the camera.}
\label{fig:highresolutiontank}
\end{figure}

\Fref{fig:highresolutiontank} shows a photograph of the final hologram.  This was taken with a 0.625 second exposure and a $f/4.2$ aperture.  The ghost-like artifacts are due to reflections from the glass, and are not prominent when viewed by eye.  

\bigskip

A single attempt was made to try to print with an inter-pixel size of \unit[0.28]{mm}.  This produced a dim hologram that is visible to the eye but too dim to photograph with a standard digital camera.  This failure was most likely due to a bad alignment of the object and reference beams on the holographic emulsion, and further experimentation is required.

\begin{figure}[htp]
\centering
\includegraphics[width=\textwidth]{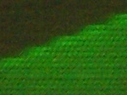}
\caption[Zoomed in on a small section of the high resolution hologram]{Zoomed in on a small section of the high resolution hologram.  Blurriness and artifacts are due to the difficulty in capturing the image with the camera.}
\label{fig:highresolutiontankzoomed}
\end{figure}

\Fref{fig:highresolutiontankzoomed} shows an enlarged portion of the hologram.  Notice that the pixels are perfectly aligned in both axes (The tilt of the y-axis is just due to the angle of the hologram when the photograph was taken.  The axes are orthogonal.)  The spacing between the pixels has been left slightly larger than the actual pixel size (Approximately \unit[0.5]{mm} spacing) to make such analysis of the final hologram easier.  

\clearpage
\section{Summary}
Various printing speed and resolution improvements to the digital direct-write monochromatic hologram printer described in Chapter~\ref{chap:design} were investigated and implemented.  The printing speed was successfully increased from 4 holopixels per second to 40 holopixels per second through a variety of software and mechanical improvements along with a newer design of a laser power supply.  The printing resolution of holopixels was increased, with the holopixel diameter reduced from \unit[1.0]{mm} to \unit[0.3]{mm} - a factor of ten increase in resolution.  The rendering of the input images was also discussed along with the restrictions required for optimal holograms.

It was found that achieving the required high precision timing on a normal \ac{PC} for printing at 40 holopixels per second is possible but that further improvement to printing speed would require a realtime operating system with high performance hardware timers.  

A demonstration of the final results is given in \Fref{fig:blender_tank}.
\chapter{Temperature-energy feedback}\label{chap:temperature_energy_feedback}

This chapter details the custom energy meter and heating system that was built to improve the pulse to pulse laser energy stability by reducing laser mode beating.  A computer controlled heating system was created for adjusting the position of the laser cavity rear mirror and was driven by a temperature-energy feedback algorithm.  The heating system was additionally used for maintaining a constant temperature within the laser, to further help with stability. The effectiveness of these improvements is determined.

\section{Problems with laser instability}
The particular laser investigated for use in the hologram printer is an
infrared \unit[1064]{nm} passively Q-switched long-cavity
pulsed laser. The fundamental beam has a frequency of \unit[1064]{nm} beam which is frequency doubled to
\unit[532]{nm} (visible green). The laser produces up to 50 pulses a second,
each with \unit[50]{ns} pulse width, and has a single 
$\text{TEM}_{00}$ beam mode.  The full technical specifications are given in Appendix~\ref{sec:laserspecs}.

\begin{figure}[ht]
\centering
\begin{tikzpicture}
	\begin{axis}[
	        xlabel=Exposure (microJoules/cm$^2$),
	        ylabel=Diffraction efficiency,
		tickpos=left,
		xmin=25,
		xmax=125,
		ymin=30,
		ymax=50,
		yticklabel={$\tick \%$},
		xtick={50,75,...,125},
		ytick={30,35,40,45,50}]
	
	\draw[gray!25!white, very thin,xstep=25,ystep=5] (27,30.45) grid (124.8,49.9);
	
    	\addplot[smooth, line width=5pt,cap=round] plot coordinates {
	(50,36) (58,40) (69,45) (75,46.75) (82.7,47.5) (100,46.75) (120.5,45.75)
    	};
	\end{axis}
\end{tikzpicture}
\caption[Diffraction efficiency curve for VRP-M emulsion]{Diffraction efficiency curve for Slavich VRP-M monochromatic emulsion suitable for holography at \unit[532]{nm} (Data source: \citep{Zacharovas2001}). Diffraction efficiency curves for other emulsions are similar\citep{Zacharovas2001,zacharovas:73}.}
\label{fig:diffractionefficiency}
\end{figure}
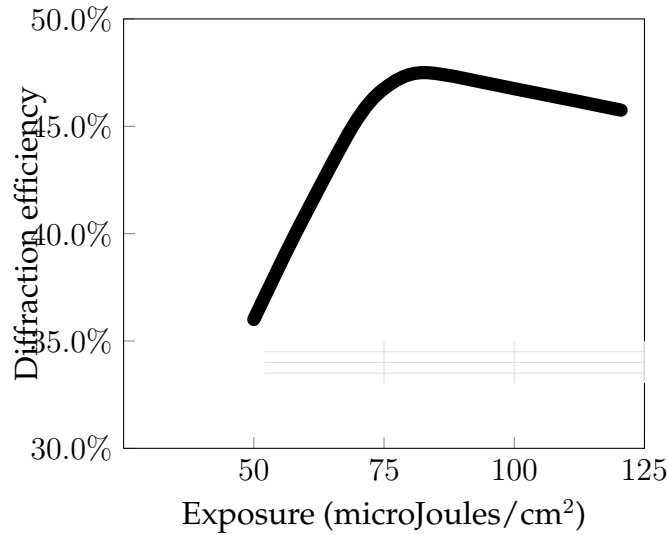

The pulse energy was found to be unstable with between a 10\% to 30\%
pulse-to-pulse standard deviation in energy. This produces
noticeable defects in the replay of the printed hologram (as seen in \Fref{fig:greendragon}), as the pixels
are printed with different energy densities.  This in turn affects the hologram
because the maximum diffraction efficiency of the final developed hologram is
 a function of the incident energy density \citep{zacharovas:73}.
The exact nature of this function varies between emulsions, but typically has a 
peak diffraction efficiency at some particular energy density with a gradual decrease in  
diffraction efficiency with increasing energy density, as shown in \Fref{fig:diffractionefficiency}.  The energy stability required from the hologram printer laser is thus dependant on the slope of this curve.
By inspection of the relationship between energy density and diffraction efficiency, and
from experience of printing holograms, we found that for the Slavich monochromatic emulsion Slavich VRP-M and the panchromatic Slavich PFG-03C, a pulse-to-pulse standard deviation of less than 10\% is not noticeable under normal viewing conditions. 

It is worth mentioning that the contrast of the image is related to the interference term which relies
on the relative intensities of the reference and object beams.  Since the two beams originate from the same
laser source, the relative intensities remain unaffected by laser instability.

\bigskip

The pulse-to-pulse standard deviation changes due to uncontrolled changes to 
optical length of the laser resonance cavity due to temperature
fluctuations, changes in moisture, etc\citep{patent:EP0109254}. The laser is designed to be locked to a certain mode
by the combination use of a thin and thick etalon.  However if the cavity optical length changes sufficiently, a second mode
frequency can be selected.  If multiple mode frequencies can coexist in the resonance cavity, they compete, producing mode beating which causes a drop in the beam intensity. It is this drop in the beam
intensity that can cause some
pixels on the final hologram image to appear much dimmer than others in
replay.  Severe mode beating can often be observed to the naked eye as an occasional flickering
of the beam.

We have argued for the need for improved energy stability in order to produce high quality holograms, and that   
the reason that the laser is unstable is because it uses a
long resonance cavity that can change its effective optical size due to
temperature, moisture, etc. fluctuations.

This was approached on two fronts.  The first was to try to limit the temperature
fluctuations. The laser was enclosed in a case with a heated
breadboard. The breadboard was heated with a temperature feedback
system that continually measures and heats the breadboard to maintain a
stable temperature.

The second step is to have an active way to adjust the cavity length
size. By implementing a system that can change the physical cavity
length, we can counter changing air densities, and other changes, in an attempt to keep the
optical cavity length constant.

To achieve this, the rear mirror of the cavity was mounted onto one end
of a metal block. The other end was mounted in a holder for mounting in the laser. Two
resistors were affixed to the metal block to provide the ability to heat the block, along with a
thermiresistor to measure the temperature of the block.  When the metal block is
heated the block expands, adjusting the position of the rear mirror.  In
this way the position of the rear mirror can be adjusted to high precision (see \Fref{eq:thermalexpansion})
to compensate for uncontrolled resonance length variations.

Aluminum was chosen for the material of the metal block.  Aluminum has a coefficient of linear thermal expansion of $\unit[23\times10^{-6}]{K^{-1}}$.  As we will see, the temperature of the block can be controlled to within an accuracy of $\unit[0.01]{\degree C}$.  For an aluminum block of length $\unit[1.5]{cm}$, the thermal expansion for a change in temperature of $\unit[0.01]{\degree C}$ is:
\begin{eqnarray}
\Delta L &=& \unit[1.5\times10^{-3}]{m} \times \unit[23\times10^{-6}]{K^{-1}} \times \unit[0.01]{K}\\
&\approx&\unit[0.35]{nm}
\label{eq:thermalexpansion}
\end{eqnarray}

This is of sufficient accuracy for a fundamental mode of \unit[1064]{nm}.

\bigskip

Rather than attempting to determine the optical cavity length at any given time,
in order to compensate for its changes, the energy can measured instead and used as the basis 
for adjusting the position of the rear mirror. A feedback system was implemented
to monitor the beam energy.  If temperature/moisture/etc fluctuations cause the optical cavity length to change, the beam
energy changes, and the feedback system can adjust the position of the rear block to compensate.

Both the energy-feedback system and the breadboard heating system require a heating and monitoring component, with its
own feedback system, to heat to a given temperature.  The same system was used for both of these needs, and will be hereby
referred to as the \acf{DTC}.

\section{Custom energy meter}

To control the energy stability,  a computerized
system is required to measure the energy in each beam pulse.  For this task, a custom energy meter was built.  Although off-the-shelf
energy meters exist, a simple and cost effective device was required that worked at low energies (of the order of micro Joules).  Since it
was to be fixed in place and communicate only to a computer, a display etc was not required.  Since the energy stability is the key piece of information
that is required, overly accurate calibration and preciseness was also not required.

\bigskip

A general overview of the custom energy meter is discussed below, followed by a discussion of methods for diverting the energy beam into the custom energy meter.  Next the
calibration of energy meter is explained, concluding with a comparison of energy readings with an off-the-shelf energy meter from the company Ophir.

\subsection{Overview of custom energy meter}

A small portion of the beam to be measured is diverted onto the surface of an awaiting photodiode. 
The \unit[50]{ns} laser pulse used is too short to measure directly, so current
from the photodiode is collected onto the plates of a capacitor. The
capacitor is then allowed to leak to ground through a resistor. The
capacitor leaks in a consistent way, with the voltage across it following a very specific profile
which is determined by the initial charge collected. By using an
\ac{A-D} converter to measure the voltage across the capacitor with time
and timing how long it takes for the capacitor to discharge, the
initial collected energy can be deduced. From that, with calibration,
we can deduce the energy that was in the incident laser pulse.

\subsection{Methods for diverting beam}

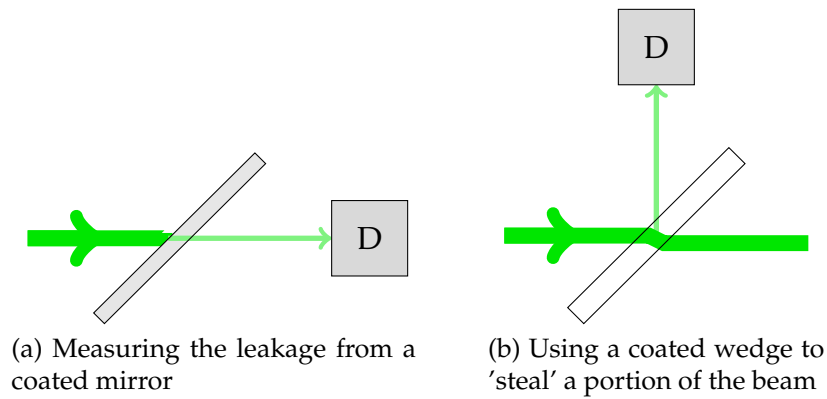
\begin{figure}[ht]
\centering
\subfloat[Measuring the leakage from a coated mirror] {
	\begin{tikzpicture}
  	\draw[->,line width=6pt, green!90!black] (-2,0) -- (-1,0);
  	\draw[-,line width=6pt, green!90!black,join=bevel] (-2,0) -- (-0.2,0) -- (-0.2,2pt);
	\fill[gray!20!white, rotate=-45] (-1mm,-1.5cm) rectangle (1mm,1.5cm);
  	\draw[->,line width=2pt, green!90!black!50!white] (-0.15,0) -- (2,0);
  	\draw[rotate=-45] (-1mm,-1.5cm) rectangle (1mm,1.5cm);
  	\filldraw[fill=gray!30!white,draw] (2,-0.5) rectangle node {D} (3,0.5) ;
	\end{tikzpicture}
	\label{fig:energydectectormirror}
}
\qquad
\subfloat[Using a coated wedge to 'steal' a portion of the beam] {
	\begin{tikzpicture}
  	\draw[->,line width=2pt, green!90!black!50!white] (0,0) -- (0,2);
  	\draw[->,line width=6pt, green!90!black] (-2,0) -- (-1,0);
  	\draw[-,line width=6pt, green!90!black] (-2,0) -- (-0.1,0) -- (0.1, -0.1) -- (2,-0.1);
  	\draw[rotate=-45] (-1.5mm,-1.5cm) rectangle (1.5mm,1.5cm);
  	\filldraw[fill=gray!30!white,draw] (-0.5,2) rectangle node {D} (0.5,3) ;
	\end{tikzpicture}
	\label{fig:energydectectorwedge}
}
\caption[Two possible layouts to deflect energy to the detector]{Two possible layouts to deflect energy to the detector, D.}
\label{fig:energydector}
\end{figure}

Since the feedback system needs to be working continually, a 
real-time monitoring system needs to be put in
place; blocking the beam in order to measure its energy is clearly
not desirable.  As stated before, the laser has a fundamental frequency of 
\unit[1064]{nm} which is frequency doubled to \unit[532]{nm}.  Either of these are
suitable for monitoring the energy, but the visible second-harmonic (\unit[532]{nm}) is the more 
obvious candidate.
There are various methods for diverting a small amount of the beam in order to measure
the energy. Two different approaches were evaluated.  The first approach tried was to 
measure the back leakage from one of the various 90\degree reflective mirrors used (\Fref{fig:energydectectormirror}).

The mirrors, although coated, are not 100\% perfect {--} a small amount of
the beam energy transmits (`leaks') through the mirror and would be
normally discarded.

The second approach tried used a thin glass wedge coated on
one side, placed in the beam. The wedge reflects around 3\% of the
beam energy to the waiting photo-diode (\Fref{fig:energydectectorwedge}) \footnote{In practice, two right-angled coated wedges were used, with one rotated by 180\degree, to act as a single wedge.}.

The second approach used a thin glass wedge two glass right-angled coated wedges, placed next to each other in the beam.
The wedge reflects around 3\% of the
beam energy to the waiting photo-diode (\Fref{fig:energydectectorwedge}).

The calibration of the energy meter is discussed later, but
the results regarding the method for deflection are mentioned here, for the sake of continuity.

\begin{figure}[htp]
\centering
\includegraphics[angle=-90,width=\textwidth]{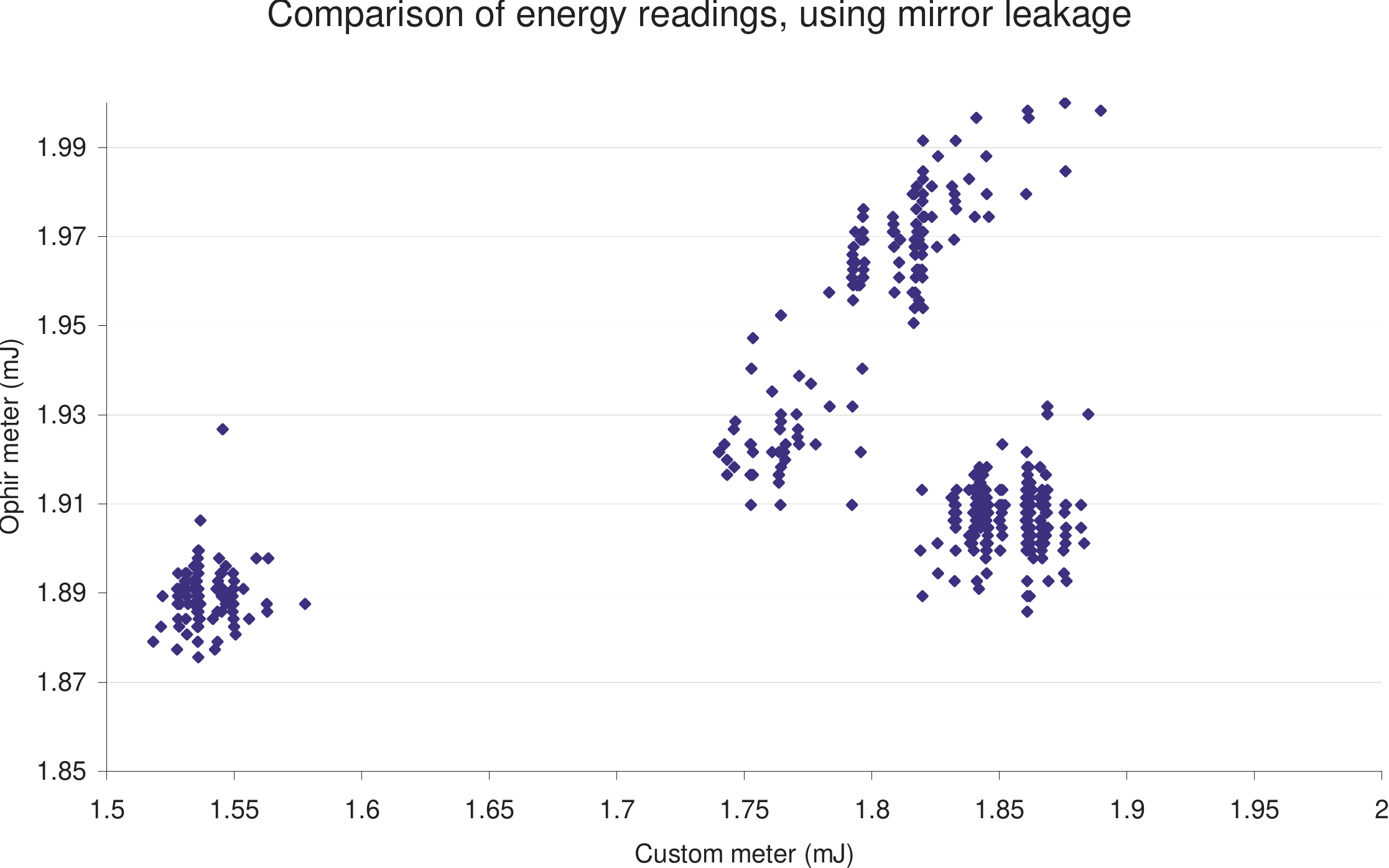}
\caption[Comparison of energy readings for mirror-leakage method]{Comparison of energy readings from the Ophir energy meter and the custom energy meter, obtained for mirror-leakage.}
\label{fig:comparisons}
\end{figure}

The first method was found to give unreliable results.  The percentage of
leakage through the mirror was not predictable with time.  This can be demonstrated
by a plot of the energy readings taken from the off-the-shelf pre-calibrated Ophir meter
and comparing them against the readings from the custom energy detector.  There should be a
one-to-one mapping between the two sets of readings with some small margin of error.  Instead there
appears to be two distinct clusters, as can be see in \Fref{fig:comparisons}\footnote{Note that the scale for the readings from the custom energy meter are somewhat arbitrary, given that no sensible calibration can be obtained}. 

The second method, to use a thin coated wedge, was found to give more reliable results.  Testing shows a strong, and time-constant, correlation between the energy reading given by the custom energy meter and the energy reading given by the Ophir energy meter.  All future graphs and values are given for this second method, unless otherwise noted.

\subsection{Calibrating custom energy meter}

The \unit[50]{ns} laser pulse used is too short to measure directly, so current
from the photodiode is collected onto the plates of a capacitor. The
capacitor is then allowed to leak to ground through a resistor. The
capacitor leaks in a consistent way, following a very specific profile
which is determined by the initial charge collected. By using an
\ac{A-D} converter to measure the voltage across the capacitor with time
and timing how long it takes for the capacitor to discharge, the
initial collected energy can be deduced. From that, with calibration,
we can deduce the energy that was in the incident laser pulse.

\bigskip

\begin{figure}[htp]
\centering
\includegraphics[angle=-90,width=\textwidth]{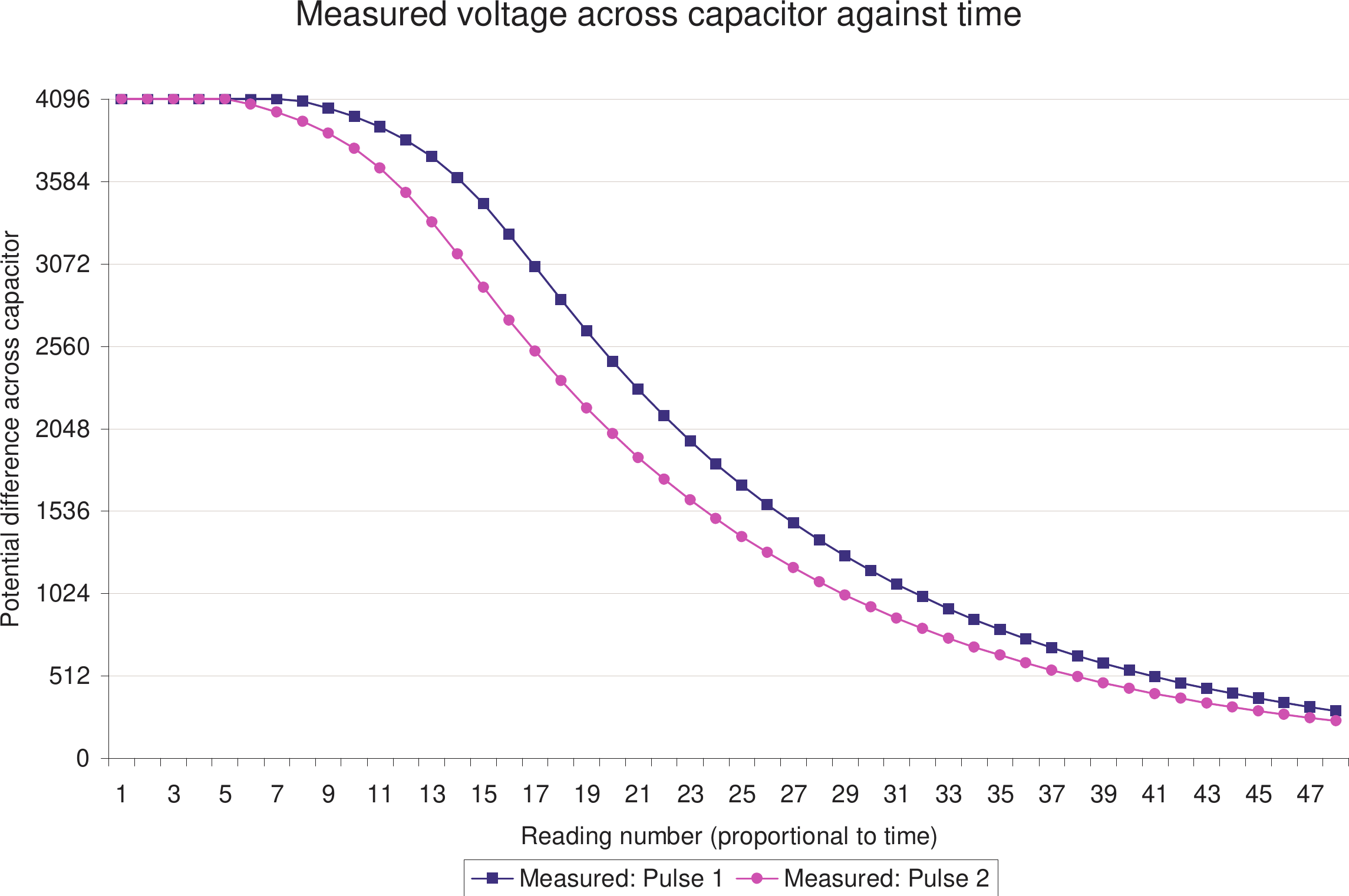}
\caption[Measured voltage on custom energy meter]{Measured voltage across the energy meter's capacitor against time, after a pulse (Arbitrary units)}
\label{fig:voltage}
\end{figure}

{
To calibrate the photo{}-detector, an off{}-the{}-shelf calibrated
energy meter was placed in the optical path of the \unit[532]{nm} beam and the
beam energy adjusted to calibrate and check the photo{}-detector
software. The energy meter triggers sending the voltage reading data
to the \ac{PC} when the voltage suddenly spikes {--} i.e. at the start of a
pulse. The first few readings saturate the \ac{A-D} converter and just
return the maximum value {--} 4096. Then as the voltage drops we
measure the decay curve. Two example pulses with different energies
are shown in \Fref{fig:voltage}.}

To determine the energy from this curve, the values between 2400 and
1000 were considered only.

\bigskip

{
The potential difference across the capacitor curve follows a
characteristic capacitor decay curve of:
\begin{equation}
V_t=V_0 e^{-t/RC}
\end{equation} 
}

The energy in capacitor is related to the potential difference across
the capacitor by the equation:
\begin{equation}
E=\frac{1}{2}CV^2
\end{equation}

So indicating initial energy as  $E_0$ we obtain:
\begin{equation}
V_t=\sqrt{\frac{2E_0}{C}}e^{-t/RC}
\end{equation}

Using variable substitution of  ${A=\frac{2}{C}}$  and 
$k=\frac{2}{RC}$ :
\begin{equation}
V_t=\sqrt{AE_0}e^{-kt/2}  \label{eq:potentialdifference}
\end{equation}

\begin{figure}[htp]
\centering
\includegraphics[angle=-90,width=\textwidth]{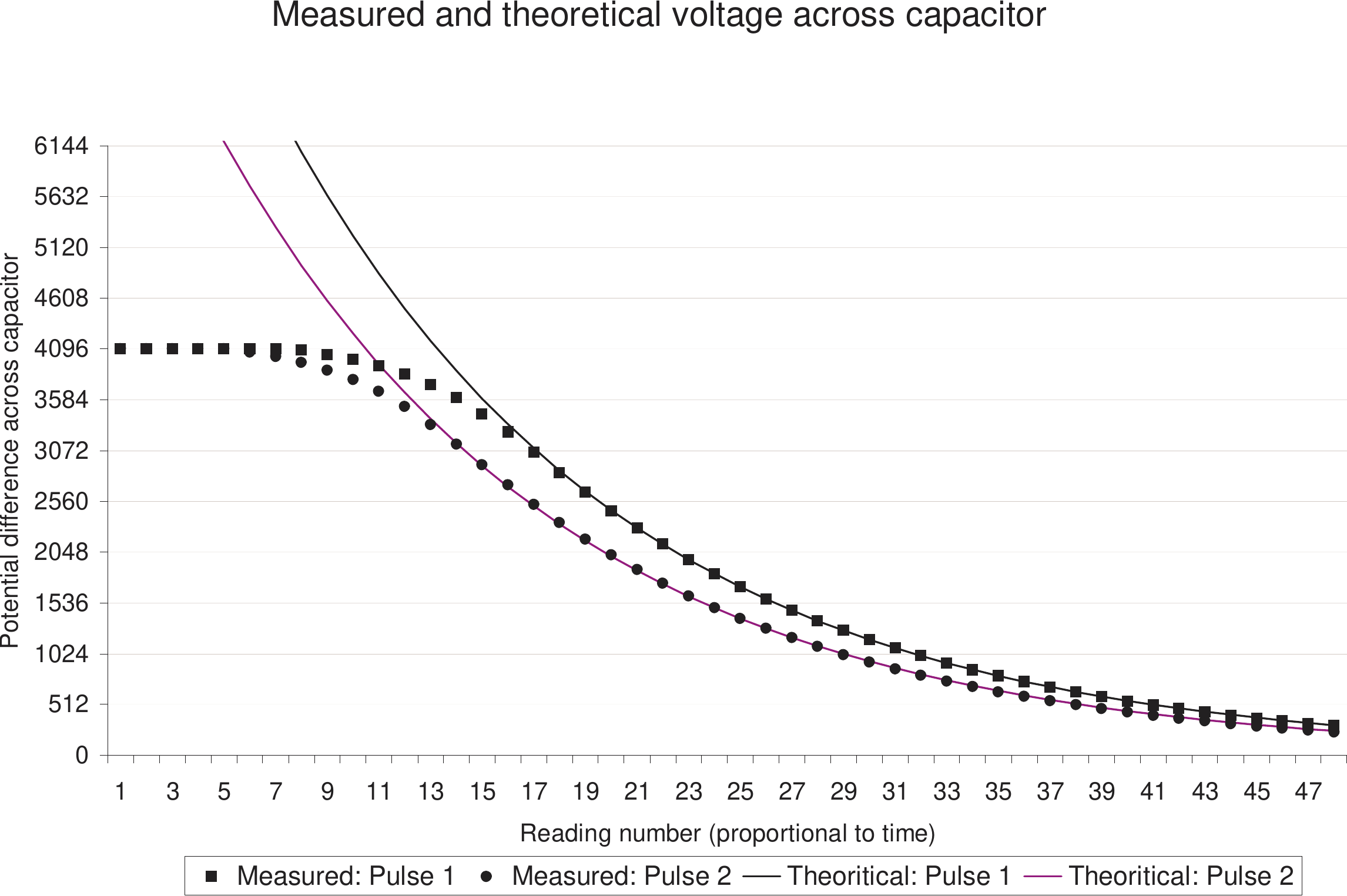}
\caption[Theoretical voltage on custom energy meter]{Measured and theoretical voltage curve of capacitor for energy meter}
\label{fig:theoriticalvoltage}
\end{figure}

By using the data measured for the voltage across the capacitor after a single pulse and setting 
$E_0$ to one we can use the least squares algorithm to determine
the best fitting $A$ and $k$ constants. Repeating this process for many pulses, different values $A$ will 
be obtained, but the $k$ values will be very similar. The average value of $k$ thus
determines the calibration $k$ constant. An arbitrary value for
$A$ can be chosen as $A$ simply sets the calibration scale of the
energy.  After calibration, we obtain a fit as shown in \Fref{fig:theoriticalvoltage}.

With the now determined values of $A$ and $k$ in hand, it is useful to rearrange the \Fref{eq:potentialdifference} to obtain the energy in the pulse in terms of the measured potential differences, as so:
\begin{equation}
E_0=V_t^2 \cdot A^{-1}e^{kt} 
\end{equation}

So using the measured values for the voltage across the capacitor, $V_t$,
and our determined values for $k$ and $A$, the initial
energy (measured on some arbitrary scale) of the pulse, $E_0$, can be found.

An estimate for the energy for any pulse can now be determined. Using
an energy meter that has already been calibrated, the value of $A$ 
can be determined such that our energy is measured in Joules.

\begin{figure}[htp]
\centering
\includegraphics[angle=-90,width=\textwidth]{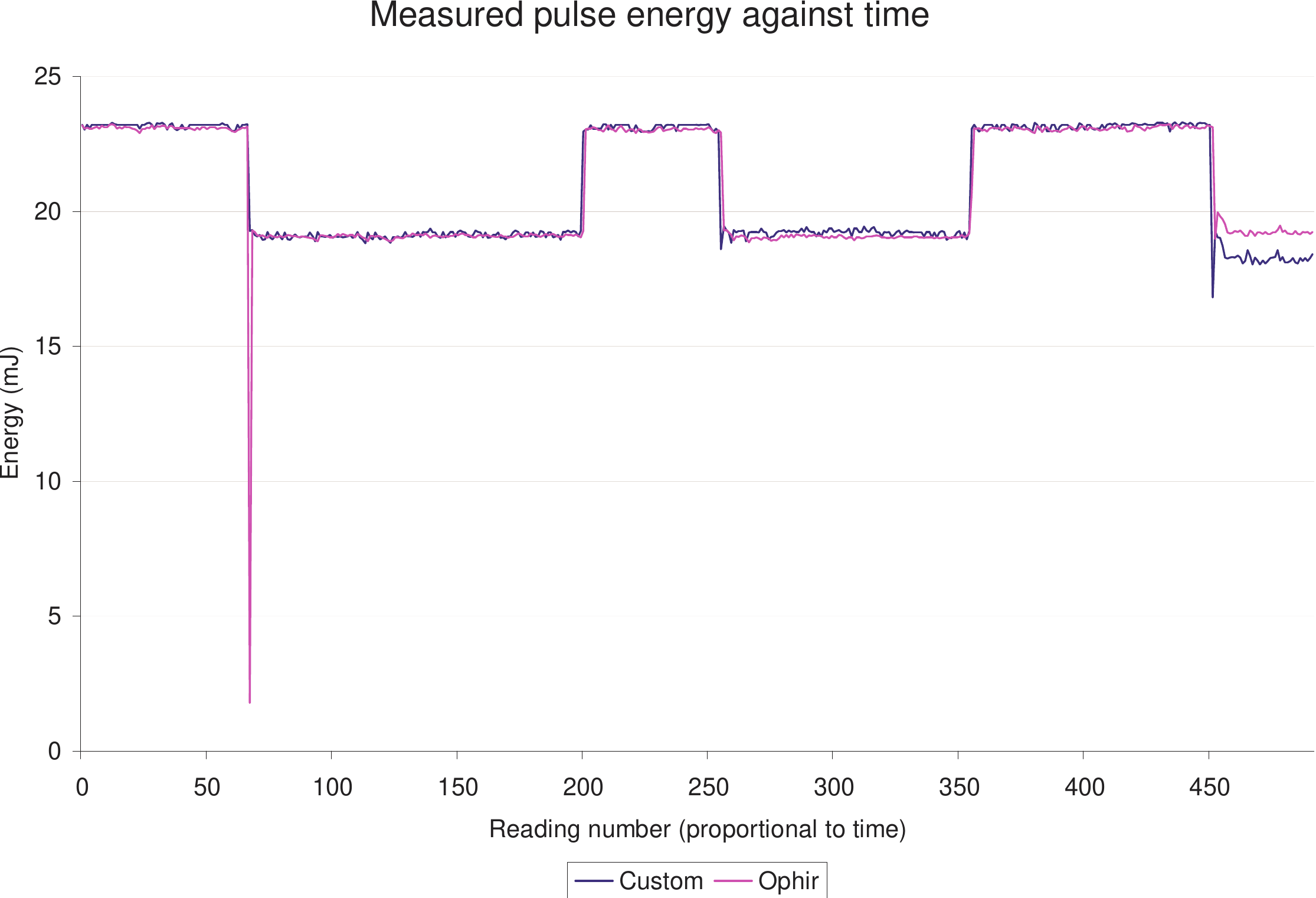}
\caption[Comparison of custom energy meter vs. Ophir's]{Comparison of energy meter and the Ophir energy meter measuring energy against time}
\label{fig:comparisonEnergyMeterOrphir}
\end{figure}

A comparison of the energy measured by this device against
the energy measured by an off-the-shelf energy meter by the company
Ophir is given in \Fref{fig:comparisonEnergyMeterOrphir}. For 
this graph the photodiode was using
the energy leakage from a mirror. The two graphs do not fit
perfectly, although it is adequate for the energy-temperature
feedback system.

\clearpage

\section{Digital temperature controller}

As mentioned previously, two resistors were affixed to a metal block that holds the rear mirror, along with a
thermiresistor to measure the temperature of the block.  To control these components, the PIC18F876A microprocessor was used.  This microprocessor contains 8k x 14-bit internal non-volatile EEPROM memory, and a 20MHz clock.

\begin{figure}[htp]
\centering
\includegraphics[width=\textwidth/3*2]{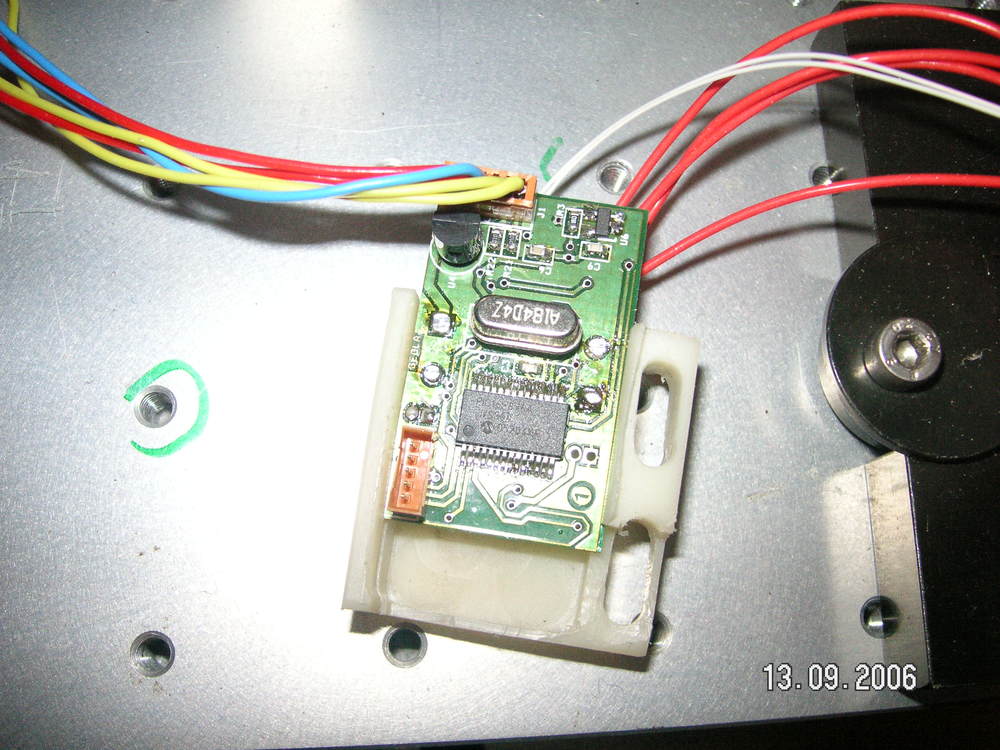}
\caption[Photograph of DTC]{Photograph of the final \ac{DTC} board.  The two white wires on the right lead to the thermiresistor, the four red wires on the right lead to a pair of resistors for heating, and the rest of the wires on the left provide power and communication lines to the \ac{PC}.}
\label{fig:dtc}
\end{figure}

These hardware components are mounted on a small custom made controller
\ac{PCB} board as shown in \Fref{fig:dtc}.  Several of these \ac{DTC} boards were mounted on the underside of the laser breadboard as well to carefully control the temperature of the breadboard and components on it. The \ac{PCB} board, named the \ac{DTC}, communicates via the serial communications protocol RS485.

\section{Temperature feedback algorithm}

The Digital Temperature Controller has one input (the current measured
temperature) and one output (whether to heat or not). There is a time
delay between switching on the heating and the new temperature
registering on the thermometers. This is because it takes time for
the resistors to heat up and time for the heat to conducted through the
material. This means that if we simply heat until we measure the
required temperature and then switch of the heating, we will find that
the measured temperature continues to climb, overshooting the target
temperature. If we are too cautious in heating then we risk taking a
much longer time to heat than necessary.

\bigskip

This can be solved by using a \ac{PID} controller
algorithm \citep{andras}.
This is a general purpose control-loop fedback algorithm that is
widely used in industrial control systems.

The  \ac{PID} controller attempts to correct the error between a measured
input variable (the temperature) and a desired target (the temperature
we want to heat to) by calculating and then outputting a corrective
action that can adjust the process accordingly (by heating). 

\bigskip

There are 3 steps to this algorithm. 

The first step is a \textbf{P}roportional response that changes the
output proportionally to the current temperature error. The constant
of proportionality is labelled $Kp$.

The second step is an \textbf{I}ntegral response that changes the output
based on the accumulated instantaneous temperature readings over time.
The integral gain constant is labelled $Ki$.

The final step is a \textbf{D}ifferential response that changes the
output proportionally to the rate of change of the temperature error.
The differential gain constant is labelled $Kd$.

\bigskip

Using a fairly standard adaptation, the integral response was changed to
limit the time period over which the integral is done. This is
omitted from the pseudo code for sake of clarity.

This algorithm was implemented in the computer language C and run on the
Digital Temperature Controller boards. The pseudo code is shown in Algorithm~\ref{algorithm:picfeedback}.

\begin{algorithm}[ht]
\SetLine
$\Delta T_0=0$\;
$t=1$\;
\Repeat{finished}{
$\Delta T_t=T^\text{Target}-T_t$\;
$P_t=Kp \cdot \Delta T_t$\;
$I_t=I_t-1+Ki \cdot \Delta T_t \cdot \Delta t$\;
$D_t=Kd \cdot \Delta T_t-\Delta T_{t-1}/ {\Delta t}$\;
$\text{output}=P_t+I_t+D_t$\;
heat output\;
wait for time $\Delta t$\;
$t=t+1$\;
}
\caption[DTC feedback using PID algorithm]{\acl{DTC} feedback using \ac{PID} algorithm}
\label{algorithm:picfeedback}
\end{algorithm}

Calibration of the parameters $Kp$, $Ki$ and $Kd$ was done manually. They are
dependant upon many factors {--} mechanical setup, size of the mechanical
blocks, thermal insulation, heating resistor used, etc. The resistors were tested against various
target temperature functions and the optimal set of parameters chosen.
The time period across which to integrate was limited to the previous
32 seconds.

The final values decided on were: 
\begin{align}
\begin{split}
Kp &= 0.7\\
Ki &= 0.02\\
Kd &= 0.7
\end{split}
\end{align}

This results in the fit shown in \Fref{fig:comparisonEnergyMeterOrphir}.

\section{Temperature feedback system on rear mirror}

As mentioned before, the rear mirror in the laser cavity was mounted on
one end of a metal block whose other end is mounted in a holder. One
of these \ac{DTC} boards was used to control the temperature of this rear
mirror metal block.

\bigskip

\begin{figure}[htp]
\centering
\includegraphics[angle=-90,width=\textwidth]{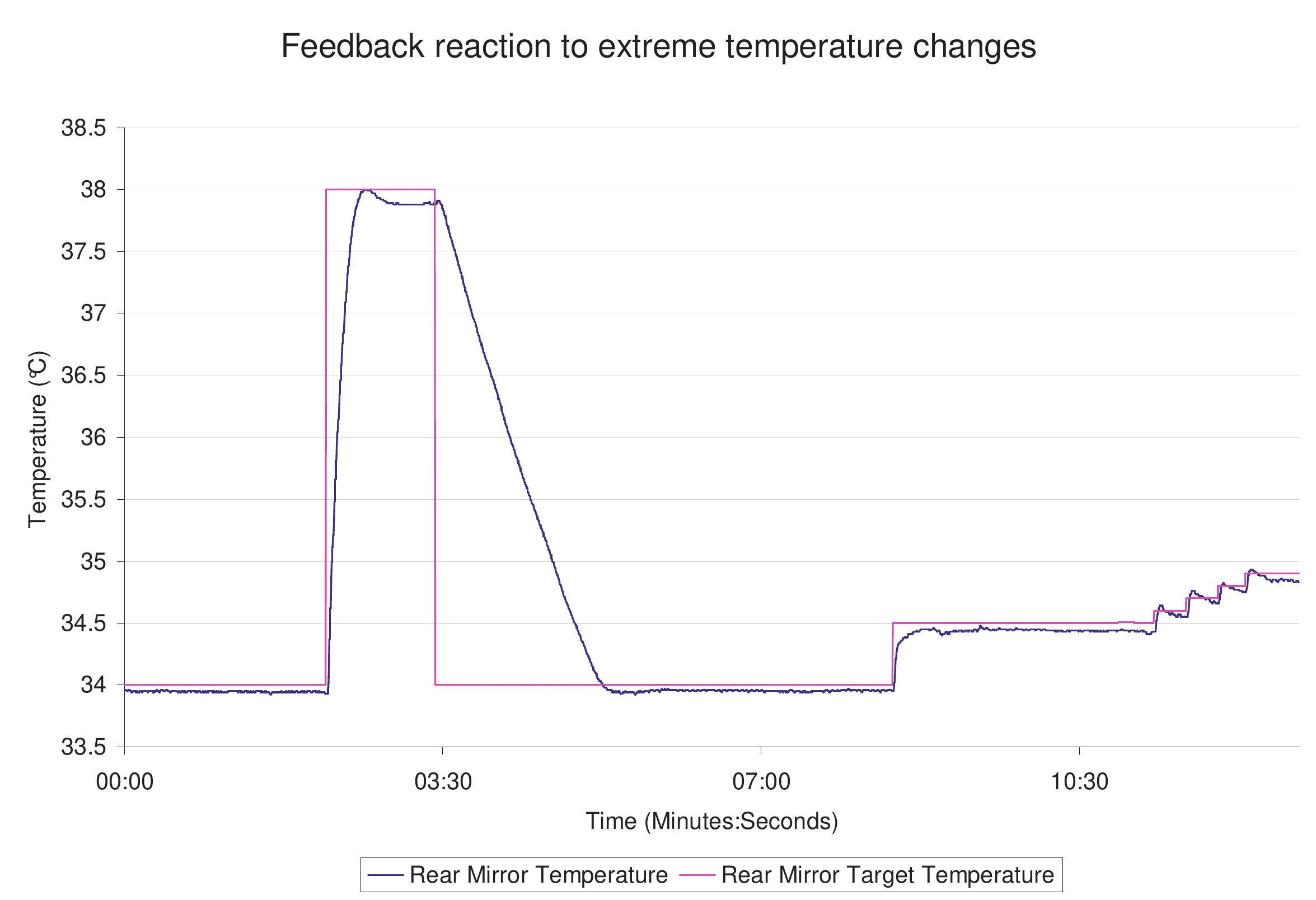}
\caption[DTC response to target temperature changes]{Digital Temperature Controller \ac{PID} response to target temperature changes}
\label{fig:response}
\end{figure}

This gives a response such as that shown in \Fref{fig:response}. The light gray
line indicates the target temperature, $T^\text{Target}$, as a function of time.
The black line shows the measured temperature  $T_t$  as
a function of time. To increase the temperature, the system responds
rapidly. It takes approximately 20 seconds to increase the
temperature by \unit[4]{\textcelsius}. Cooling down requires simply
waiting for the excess heat to be carried away, requiring about one and
a half minutes to decrease the temperature by \unit[4]{\textcelsius}.

Such large temperature changes are not, however, useful to us for any
normal operation. It is shown simply to check the behavior of the
system.

\bigskip

We can use the same temperature system to control the temperature of the
bread board.

\bigskip

\begin{figure}[htp]
\centering
\includegraphics[angle=-90,width=\textwidth]{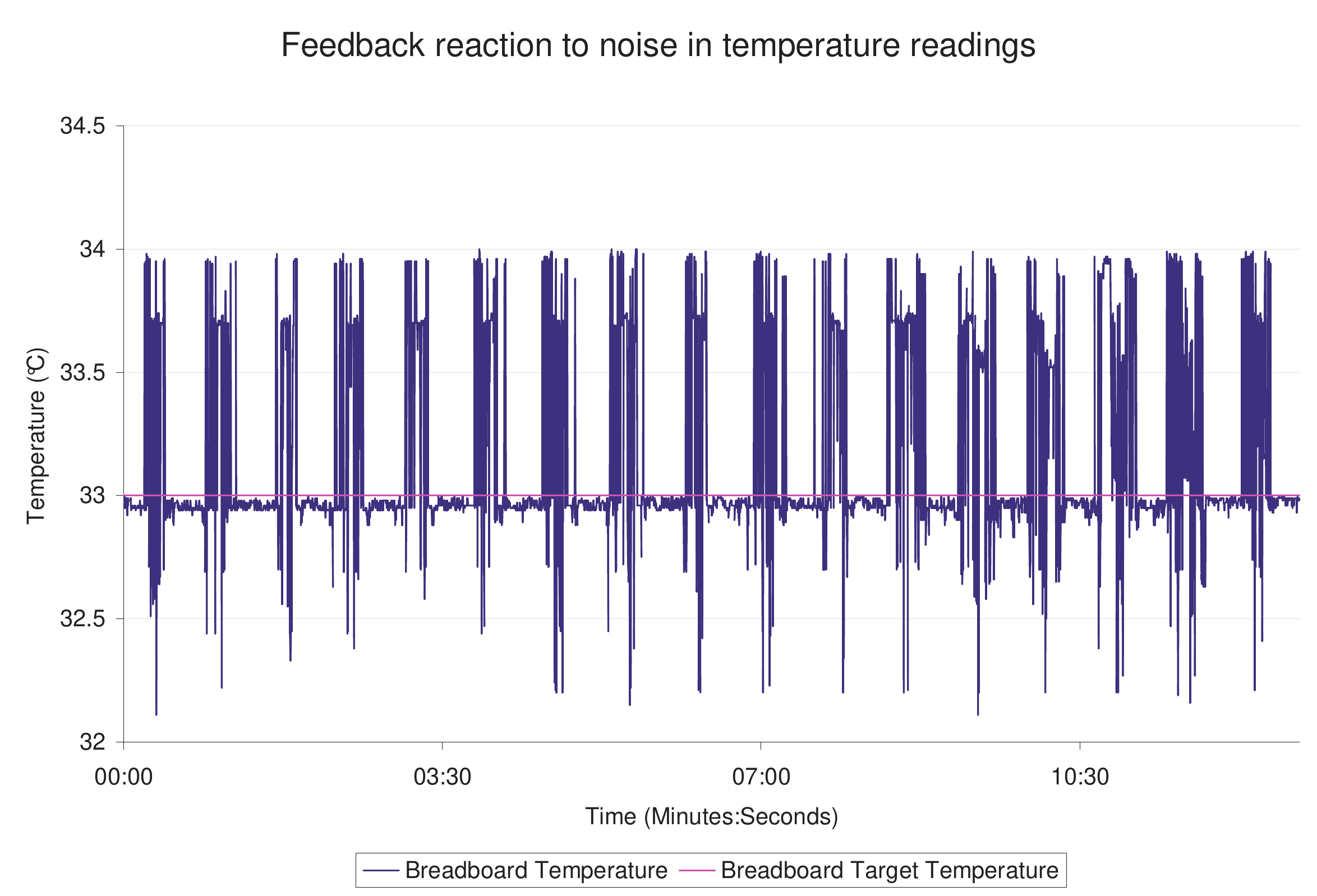}
\caption[DTC response to noise in temperature readings]{Digital Temperature Controller \ac{PID} response to noise in temperature readings}
\label{fig:breadboardunstable}
\end{figure}

\begin{figure}[htp]
\centering
\includegraphics[angle=-90,width=\textwidth]{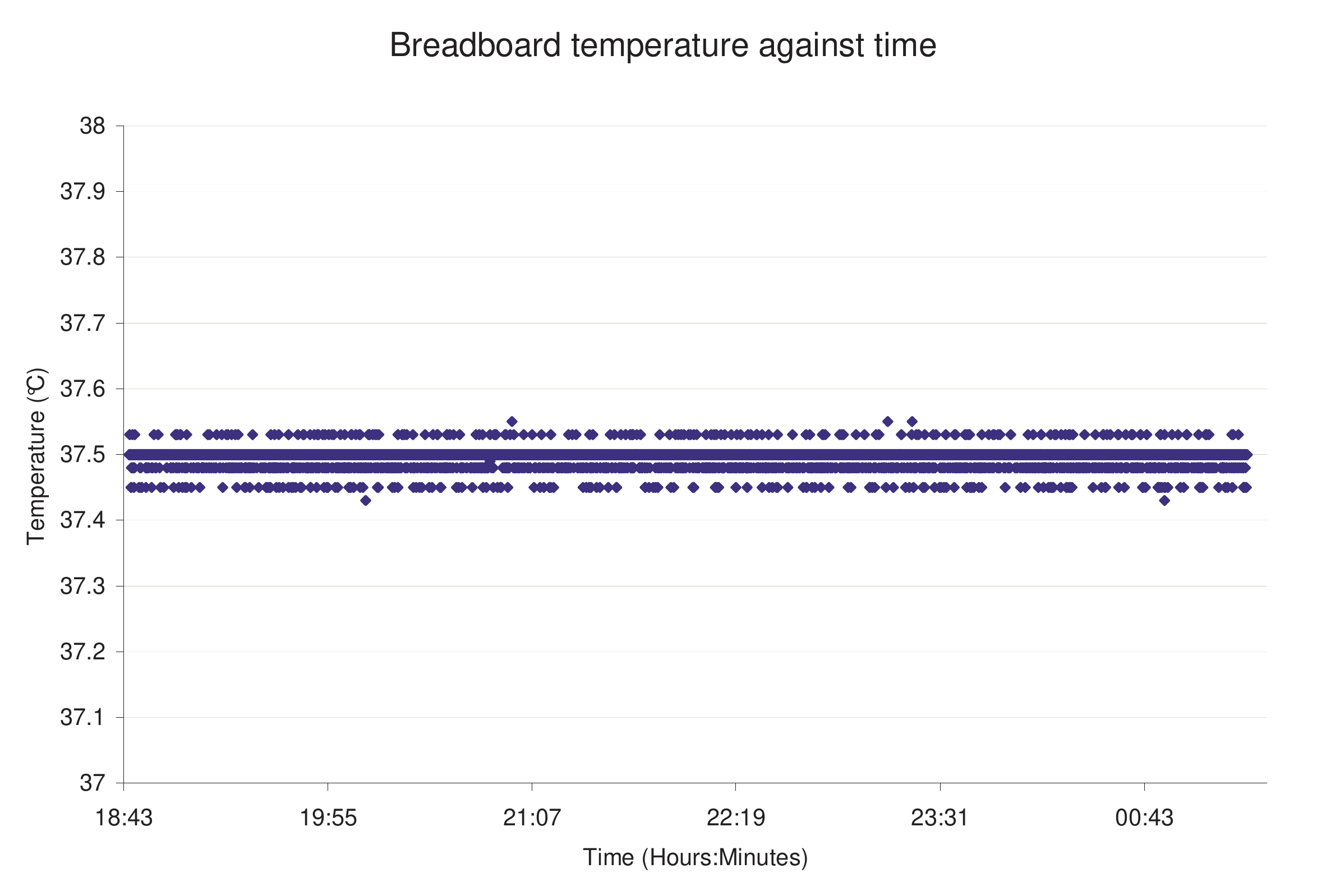}
\caption{Breadboard temperature against time}
\label{fig:breadboardstable}
\end{figure}

\bigskip

\Fref{fig:breadboardunstable} shows the same system working for heating the breadboard. The
target temperature is set here at \unit[33]{\textcelsius}, and kept constant. This shows
that there was a large source of noise arising from somewhere,
but despite the noise the feedback system still managed to cope
exceptionally well. After fixing the various noise problems, the
temperature looks like that shown in \Fref{fig:breadboardstable}.
The standard deviation of the temperature is a lot better than it might
appear from the graph {--} around 0.02\%. For the holographic printer, the temperature of the 
breadboard is kept at \unit[37.50{\textpm}0.02]{\textcelsius}.

\section{DTC-PC communication}

So now we have a way of maintaining a given temperature. The \acf{DTC}
board works independently to maintain
the set target temperature. To set the target temperature, the \ac{DTC}
board talks to the computer via a serial communication protocol. The
\ac{DTC} boards use the RS485 serial protocol, produced by \ac{EIA}. This protocol uses a
voltage differential signal increasing its immunity against signal
noise and suitable for a multi-point communications network.  It also offers protection
against data collision (multiple nodes accidentally talking simultaneously) and bus fault problems. 
The protection against signal noise is particularly important for the application of a \ac{DTC} in a hologram
printer because of the high
voltage power supplies, flash lamps and other high voltage equipment that is present in the environment around
the \ac{DTC} communication wires.

RS485 also allows for multiple systems on a single wire. This allowed
us to have multiple \ac{DTC}s as well as other subsystems all connected via
the same wires. For our purposes it was convenient to have all the
data wires join together and feed into a \acs{USB}-to-serial converter,
and then plug the \ac{USB} converter into a \ac{PC}.

\bigskip

For the higher level communication protocol, a custom protocol was developed.
Each  \ac{DTC}, as well as other subsystems, were considered as separate
nodes. Each node was given a unique number from 0 to 254. The node
number 255 was reserved to indicate all nodes. Each node thus
responded only to commands addressed to their own node number, or to
the node number 255.

RS485 is half duplex and only allows one device to be talking at a time
-- to prevent cross talk.  The protocol is robust enough to recover from accidentally cross talk.

To make the protocol simple, the \ac{PC} acted as a master, and the nodes act
as slaves {--} only communicating in reply to a message. At start up,
the \ac{PC} can send a `ping' message to each node  \ac{ID} number in turn. If a
node sees a ping message with its  \ac{ID} on it, it replies. If there is
no reply within some timeout period then the \ac{PC} knows that there is no
node connected with that \ac{ID}. In this way the \ac{PC} can `scan' the bus
looking for devices.

\bigskip

{
Every second, or so, the \ac{PC} sends a message to the first device that it
knows is connected asking for the current temperature. It then waits
for a reply (or a timeout in case of an error) before asking the next
device for its current temperature.}

{
In this fashion the \ac{PC} can be kept updated about the current
temperatures of all the parts of the laser. At any time the \ac{PC} can
send a message to set the temperature.}

\bigskip

{
To set a node \ac{ID} on a  \ac{DTC} or to reprogram a node, a single node must be
connected to the computer by itself. Then the node  \ac{ID} 255 can be used
to communicate to it and to reprogram it.}

\section{Temperature-energy feedback}

We now have the components required to measure the current beam energy
and to adjust the cavity length by heating and thus expanding a small
block that the rear mirror is attached to.

The aim now is to adjust the rear mirror as needed in order to try to
maximise the beam energy and minimize the point{}-to{}-point standard
deviation of the beam energy.

\bigskip

\begin{figure}[htp]
\centering
\includegraphics[angle=-90,width=\textwidth]{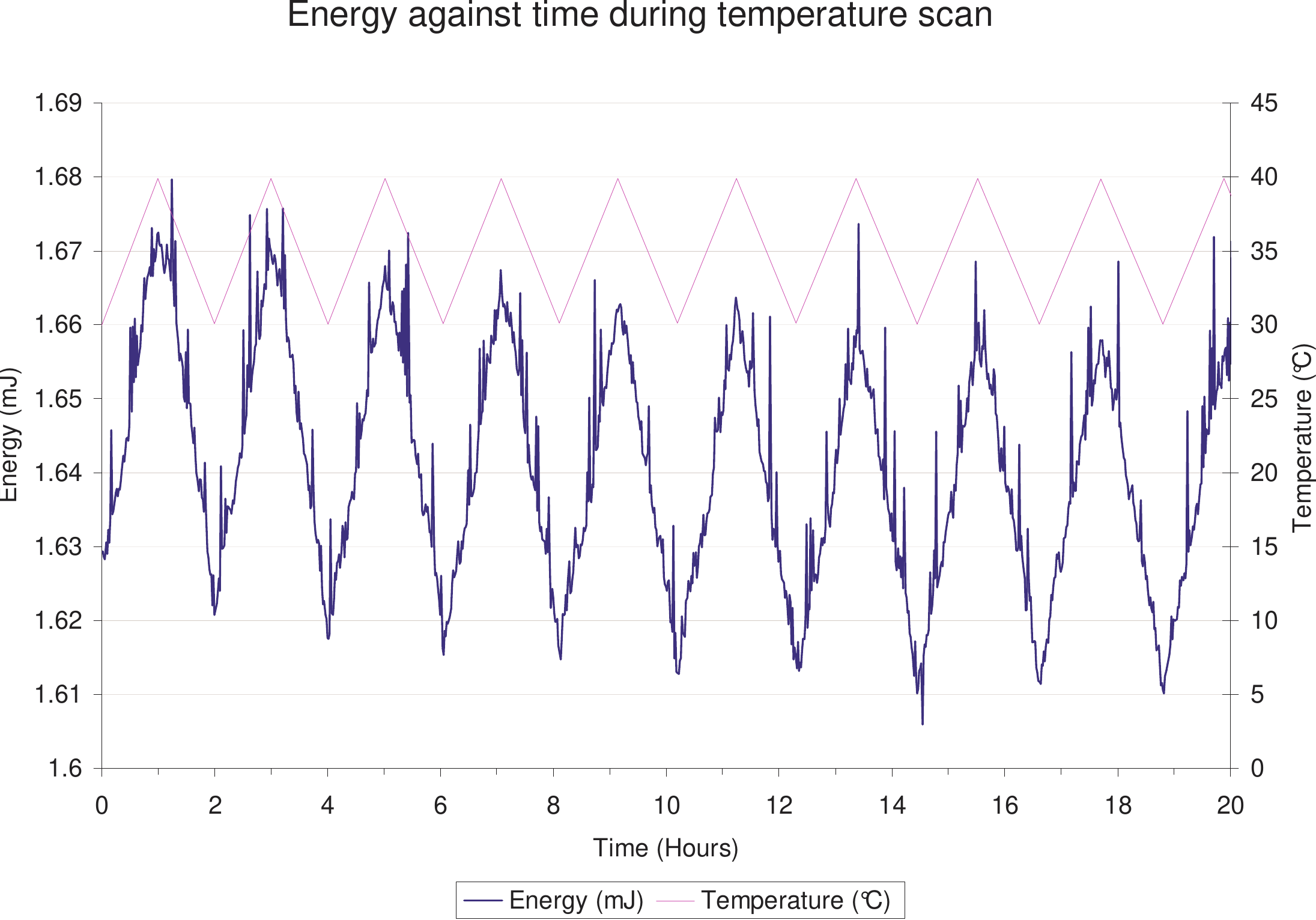}
\caption{Energy against time in rear mirror position scan over 20 hours}
\label{fig:energytime}
\end{figure}

To first get a qualitative feel for how the energy depends on the cavity length, a
temperature scan was performed. This changes the temperature of the
metal that the rear mirror is mounted on, and hence changes the cavity
size. The first attempt of an energy scan was run for just over twenty
hours, with the results shown in \Fref{fig:energytime}. 

This shows that something has gone wrong. The energy increases with
the temperature (and hence shortening of the cavity) to within the
limits of the heating system. This means that the cavity is
misaligned, and the rear mirror is far away from the optimal position.

{
To fix, the rear mirror temperature is set to a
middle{}-of{}-the{}-range temperature. The laser was opened up and
the rear mirror's position mechanically adjusted, using an energy meter
to recalibrate it.}

{
Despite this being a laser alignment mistake, we mention it here because
this is a useful way to tell if the laser is correctly aligned.}

\bigskip

{
To control the energy, we must control the rear mirror position, and
thus the temperature of the block it is mounted on. However how can
we know whether to heat or cool the block in order to increase the
energy? }

{
Initially a simple discrete algorithm was used to decide whether to heat
or cool. It was heated initially, say, until the energy decreased.
It then cooled (i.e. not heated), expecting that the energy would
thus increase again. It continued to cool until the energy decreased
again.}

{
This was found to be too sensitive to point{}-to{}-point variations in
the energy, even if some average was taken. It would oscillate
between heating and cooling, getting stuck in local minima and
responding to insignificant noise.}

\bigskip

\begin{figure}[ht]
\centering
\begin{tikzpicture}[>=stealth, shorten >=1pt,auto,node distance=2.8cm,line width=2pt]
   \matrix[matrix of nodes,
        column sep={7cm,between origins},
        row sep={5cm,between origins},
        nodes={circle, draw, minimum size=7.5mm, fill=gray!5!white, semithick}]
        {
            |(H1)| Heat & |(C1)| Cool \\
	    |(H2)| Heat & |(C2)| Cool \\
        };
    \draw[->,color=red] (C1) to [bend right, looseness=0.3] (H1) node [midway,above] {$\Delta E < -\tau $};
    \draw[->,color=blue] (H1) to [bend right, looseness=0.3] (C1) node [midway,below] {$\Delta E > \tau $};

    \draw[->,color=blue] (H2) to [bend right, looseness=0.3] (C2) node [midway,below] {$\Delta E < -\tau $};
    \draw[->,color=red] (C2) to [bend right, looseness=0.3] (H2) node [midway,above] {$\Delta E > \tau $};

    \draw[->,color=red] (H1) to [bend right, looseness=0.3] (H2) node [midway,left] {$\Delta E < -\tau $};
    \draw[->,color=red] (H2) to [bend right, looseness=0.3] (H1) node [midway,right] {$\Delta E > \tau $};

    \draw[->,color=blue] (C2) to [bend right, looseness=0.3] (C1) node [midway,right] {$\Delta E < -\tau $};
    \draw[->,color=blue] (C1) to [bend right, looseness=0.3] (C2) node [midway,left] {$\Delta E > \tau $};

    \draw[->,color=blue] (C1) to [in=30,out=65,loop] (C1) node [midway,right] {\,$|\Delta E| < \tau $\,};
    \draw[->,color=red] (H1) to [in=150,out=115,loop] (H1) node [midway,left] {\,$|\Delta E| < \tau $\,};
    \draw[->,color=blue] (C2) to [in=-30,out=-65,loop] (C2) node [midway,right] {\,$|\Delta E| < \tau $\,};
    \draw[->,color=red] (H2) to [in=-150,out=-115,loop] (H2) node [midway,left] {\,$|\Delta E| < \tau $\,};
\end{tikzpicture}
\caption[Energy temperature feedback algorithm]{Energy temperature feedback algorithm for change of energy $\Delta E$ and threshold $\tau$.}
\label{fig:energytemperature}
\end{figure}
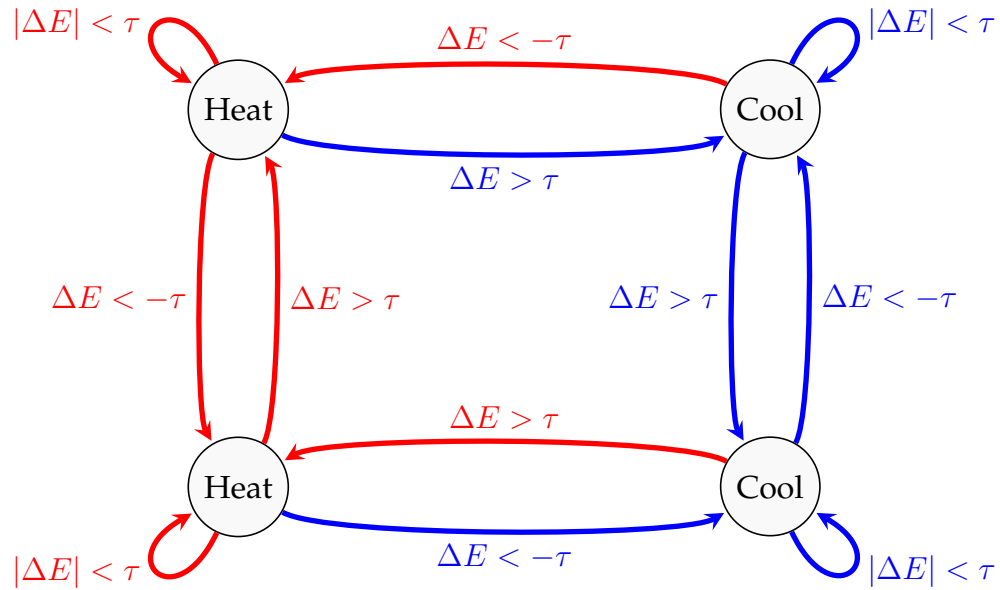

{
Instead, a more robust, but still discrete, algorithm based on that
indicated in \Fref{fig:energytemperature} was used. This algorithm requires the energy to
drop twice in succession in order to cause the temperature to change.}

\bigskip
\usepgflibrary{arrows}

\bigskip

To improve the algorithm, the energy standard deviation was also looked
at. As the standard deviation increases, so does the amount that the
temperature is increased or decreased by. This increases its ability
to get out of local minima. When the energy has dropped a lot and
cavity is mode beating, the energy can become highly unstable, meaning
that algorithm in \Fref{fig:energytemperature} can no longer work. By having a high
standard deviation causing a large change in temperature, it allows us
to `jump out' of the bad mode beating region.

\begin{figure}[ht]
\centering
\includegraphics[angle=-90,width=\textwidth]{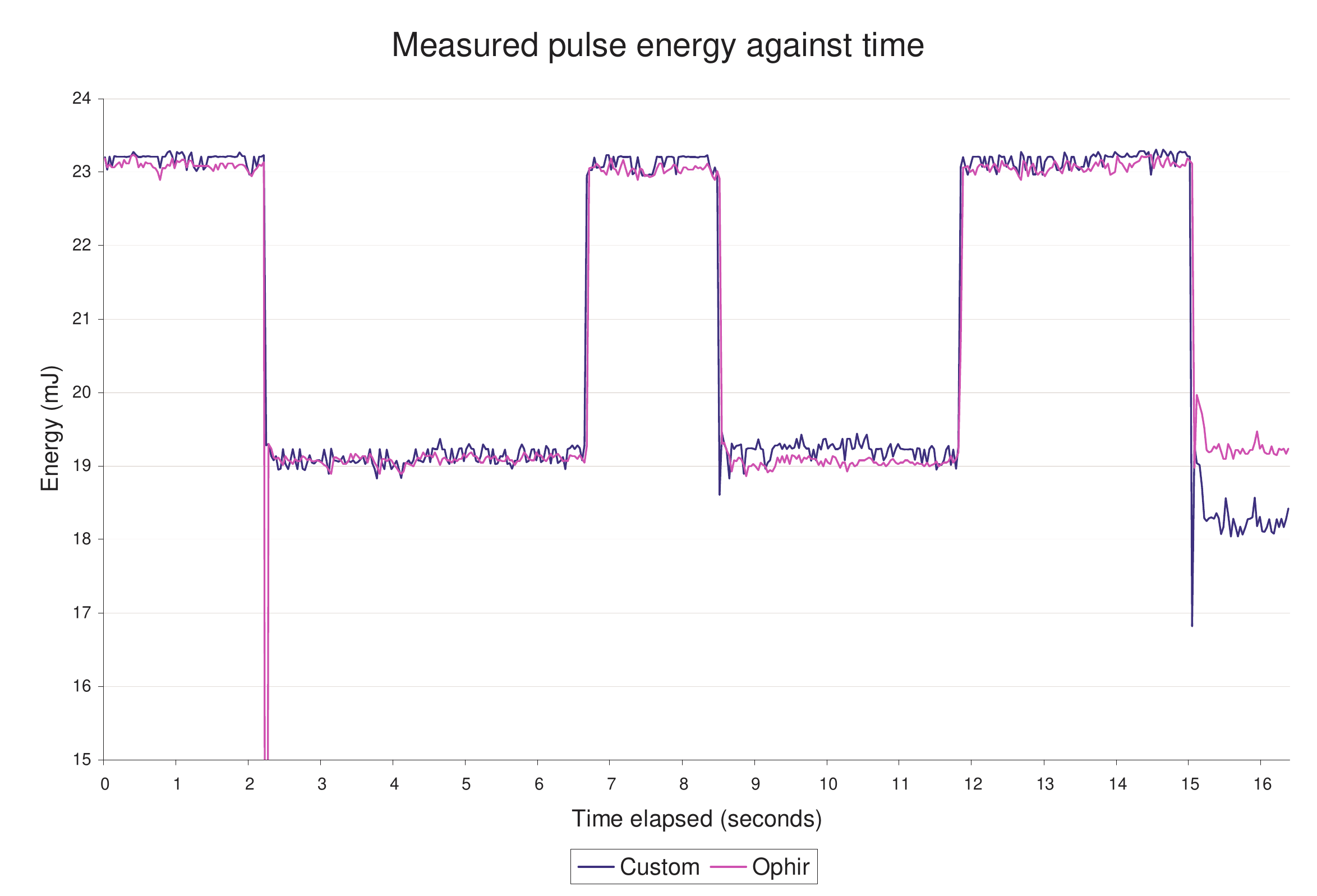}
\caption[Pulse energy with active feedback]{Second harmonic (\unit[532]{nm}) pulse energy with active feedback}
\label{fig:secondharmonic}
\end{figure}

\begin{figure}[ht]
\centering
\includegraphics[angle=-90,width=\textwidth]{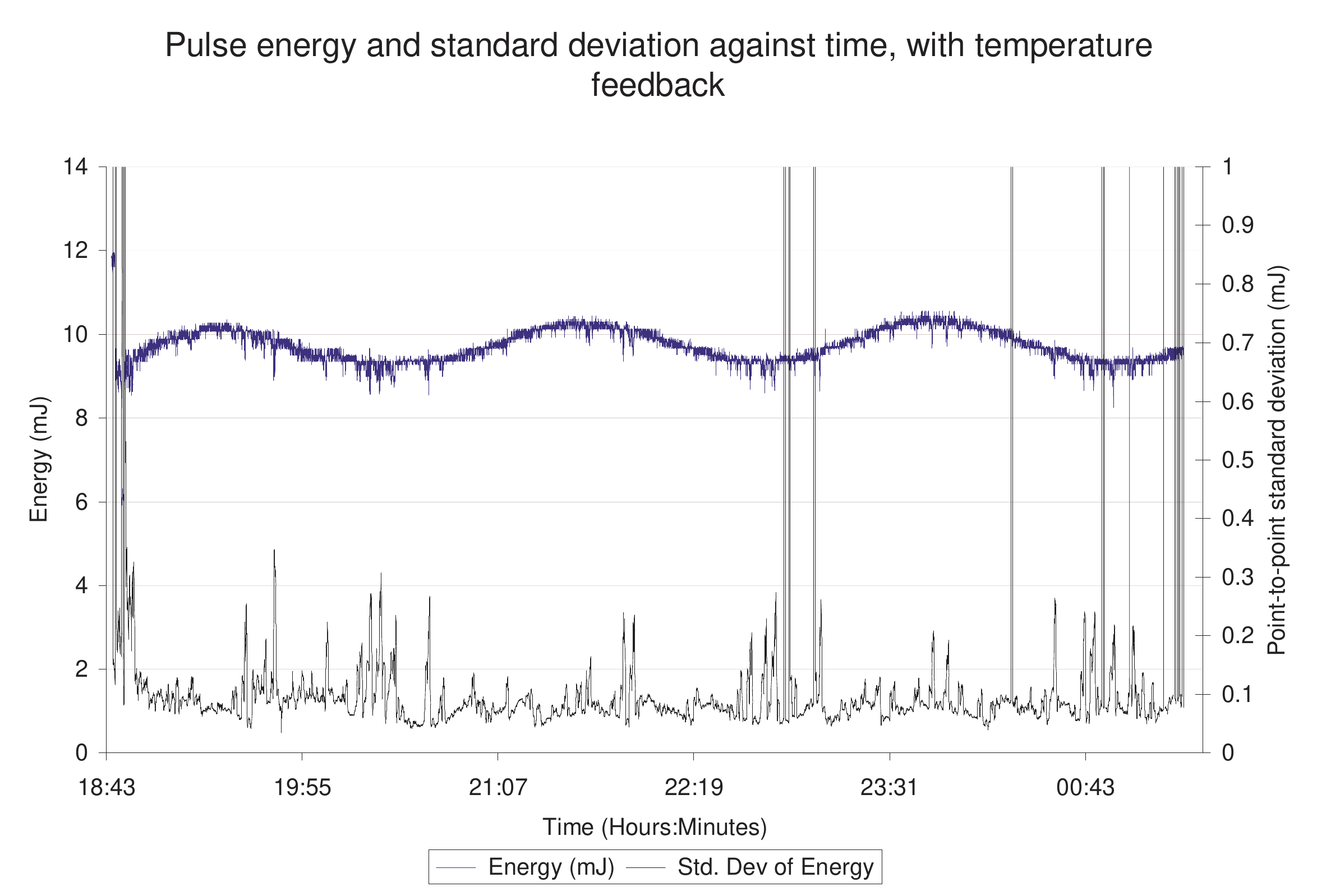}
\caption[Pulse std. dev. with active feedback]{Second harmonic (\unit[532]{nm}) pulse energy with active feedback showing standard deviation}
\label{fig:secondharmonicstddev}
\end{figure}

This algorithm was found to perform very well, as shown by \Fref{fig:secondharmonic}. The
point-to-point standard deviation remains less than \unit[0.3]{mJ} (approx
3\% of the beam energy) for the vast majority of the time, well within acceptable limits
for hologram printing, as \Fref{fig:secondharmonicstddev} shows.  The figure shows peaks of high standard deviation, but these were all below 10\% standard deviation, producing no noticeable effects to the naked eye on the final hologram.  Note that the square wave appearance of the beam energy is just due to a Neutral  Density Filter being placed in the beam at arbitrary times, just to check that the beam energy is monitored over a range of energies and that no hysteresis effects are present.  Also note that the graph indicates that Ophir meter has a glitch at around reading number 60 that was not picked up on the custom energy meter.  The reason for this was not determined.

\clearpage
\section{Summary}

Instability in the laser energy output resulted in printed holograms having dim or missing pixels upon replay. A temperature control system was developed, and used to control the laser breadboard temperature.  In addition, the temperature control system was used to adjust the position of the laser resonance cavity's rear mirror.  This was coupled with an energy monitoring system and a \ac{PC} to provide an active feedback system to stabilize the energy.  
Two different methods were investigated for diverting the beam energy into energy monitoring system, with the glass beam wedge being the resulting preferred method.  The energy monitoring system was designed with the goal of being unobtrusive, cheap, and accurate enough for use in the feedback system.  It succeeded at achieving all of these aims.

Using the active feedback for adjusting the position of the rear mirror based on the
energy gave us a much more stable beam. The pulse{}-to{}-pulse
standard deviation was reduced from over 10\% to less than 3\%,
removing visual artifacts from the printed holograms.
The \acl{DTC} was found to react fast enough for the requirements of a digital hologram printer, and performed satisfactory.

\chapter{Conclusion and future work}

\section{Conclusion}

This thesis researched and developed a direct-write pulse-laser hologram printer design suitable for recording white-light viewable reflection holograms in an off-axis geometry.

Chapter~3 examined the design of a one-step monochromatic hologram printer capable of producing white-light viewable transmission holograms created with the aid of an LCOS display system and printed in a dot-matrix sequence. The lens system was analyzed and documented in detail, with a particular focus on the microlens array system.  The magnification of the object beam aperture due to the afocal telescopic reversing lens system, combined with an upstream microlens array system, was determined based upon a matrix method of lens analysis.

An optimal range of positions was analytically determined for each of the three rectangularly-packed microlens array of differing lenslet size and lenslet curvature.  While the overall fidelity of the hologram was demonstrated to be sensitive to the microlens array position, in their respective optimal positions there was no significant difference between the three different microlens arrays tested in terms of image fidelity, contrast and speckle.  This implies that the contrast, diffraction efficiency and depth of view of the hologram's replay image will not be significantly affected by the choice in the microlens array employed, to within the range tested.

Chapter~4 analyzed the unwanted side effects of the angular intensity distribution of a hologram pixel, using a 'White Logo' hologram for case study.  A visible 'ghosting' effect was demonstrated to be significantly reduced by applying a linear formula whose parameters were determined from either a grayscale image of the hologram or by considering each color channel (in \ac{RGB} color space) separately.  This correction can be applied as a pre-processing operation before printing holograms in order to reduce ghosting in the final image.  Additionally, the artists can be warned when such effects may occur.

Chapter~5 examined methods for increasing both the printing speed and resolution of the hologram printer based upon increasing mechanical stability, improving hologram printer software, upgrading the laser power supply and using an improved display system.  These improvements enabled the pixel size to be reduced from \unit[1.0]{mm}$\times$\unit[1.0]{mm} to \unit[0.3]{mm}$\times$\unit[0.3]{mm} and the printing speed to be increased from approximately \unit[10]{holopixels per second} to approximately \unit[42]{holopixels per second}.

 Chapter~6 described the analysis and design of a temperature-energy feedback system to correct for pulsed laser instabilities arising from mode beating due to temperature variations.  A small feedback heating system was used to control the optical length of the laser cavity with an accuracy of $\approx$\unit[0.35]{nm} by heating a metal block holding the rear cavity mirror.  Combined with a custom-made energy meter and a feedback \ac{PID} algorithm to adjust the temperature based on the energy stability, the pulse{}-to{}-pulse standard deviation was reduced from over 10\% to less than 3\%.  This improvement was shown to reduce visual artifacts from the printed holograms, improving their overall quality.  The \acl{DTC} was found to react fast enough for the requirements of a digital hologram printer, and performed satisfactory.

\clearpage

\begin{figure}[htp!]
\centering
\includegraphics[height=\vsize/32*10]{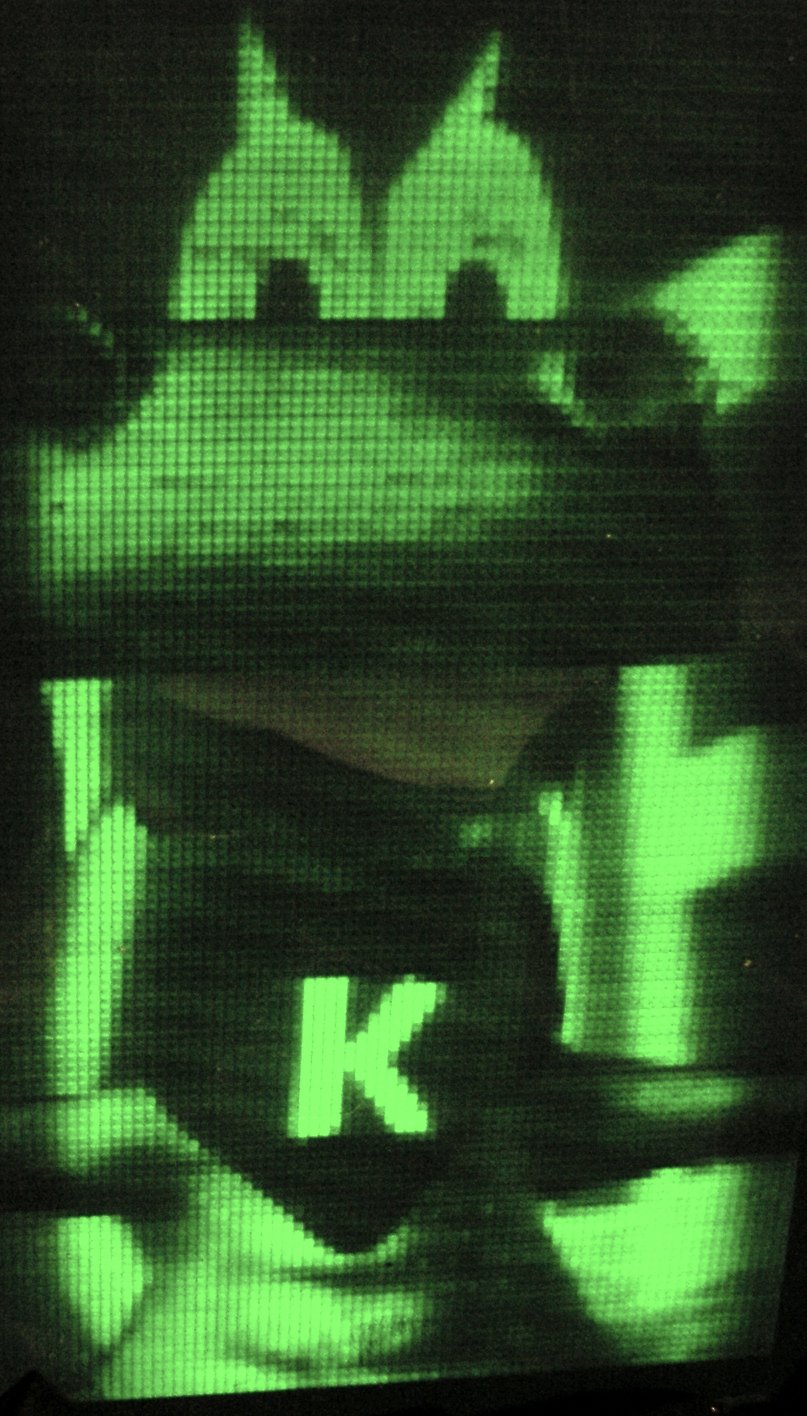}
\caption[Photograph of dragon hologram]{Photograph of green dragon hologram with \unit[1]{mm}$\times$\unit[1]{mm} pixels printed on the hologram printer before improvements, printed at 4 holopixels per second.  A larger reproduction of photograph in \Fref{fig:greendragon}.}
\label{fig:greendragon_2}
\end{figure}

\begin{figure}[hbp!]
\centering
\includegraphics[height=\vsize/32*10]{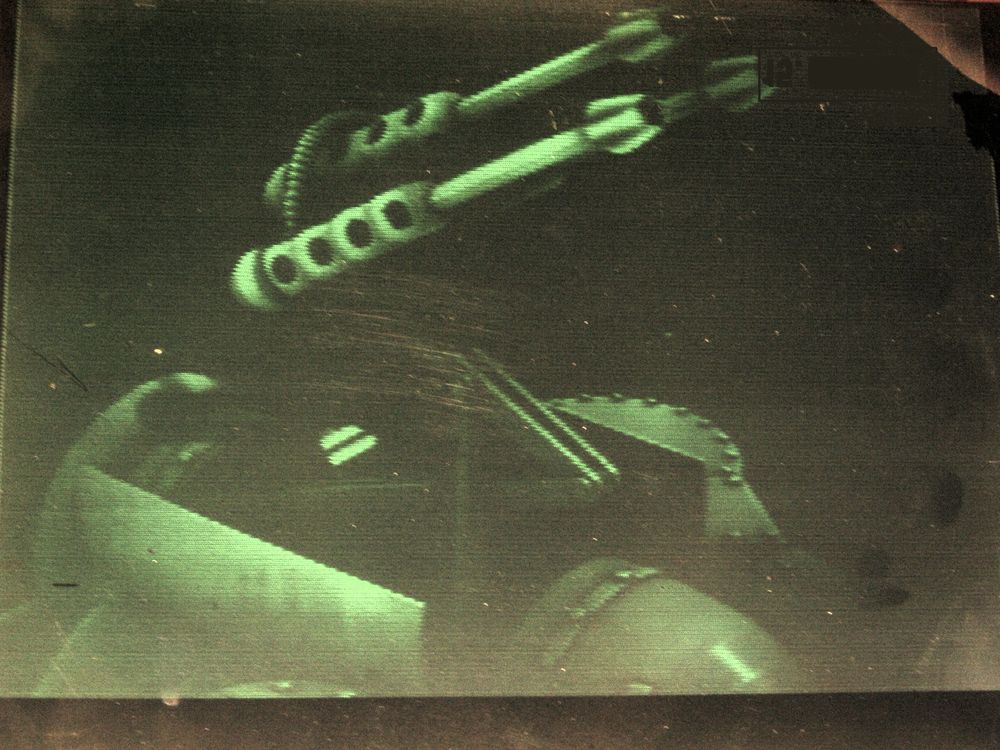}
\caption[High resolution hologram of a futuristic tank]{High resolution (\unit[0.3]{mm} in diameter pixels) hologram of futuristic tank, printed at 40 holopixels per second.  Some of the low contrast and low detail are due to the difficulty in capturing the image with the camera. A larger reproduction of photograph in \Fref{fig:highresolutiontank}.}
\label{fig:highresolutiontank_2}
\end{figure}

\Fref{fig:greendragon_2} and \Fref{fig:highresolutiontank_2} show a visual `before and after' for the improvements to the digital hologram printer mention in this thesis.  In particular note the occasional dim and missing pixels in the first hologram due to laser mode beating, and note the higher resolution of the second hologram.

\clearpage

\section{Future work}

The increased resolution changes could be adopted on a hologram printer that uses an \ac{RGB} laser source, such as the digital hologram printer by \citet{rodin2007dwh}.  \citet{bjelkhagen2008chp} showed that while a minimum of three monochromatic colors are required for color holograms, four or more covers more of the color gamut observable by the human eye, allowing for more realistic colors to be reproduced.  

For the beam illumination on the \ac{LCOS}, \Fref{sec:analysis_of_lens_system} assumed that an even spatial beam intensity profile would be preferable, in order to produce a replay viewing window with an even intensity.  However non-even beam distributions, such as a Gaussian profile, could offer the advantage of having a brighter optimal replay angle at the sacrifice of dimmer non-optimal replay angles.  This could also allow for a much larger replay window while keeping the image bright for viewers at some optimal angle (for example, at the normal to the hologram).

A modern laser printer typically prints at \unit[600]{DPI} resolution.  This translates to a pixel size of \unit[0.04]{mm} in diameter.  This makes for an interesting goal, requiring the holopixel to be reduced by an order of magnitude.  This is an order of magnitude smaller than the high resolution holograms produced for this thesis. Achieving this would significantly increase the problems with beam alignment and printing speed.  For an A4 sized hologram, approximately 35 million holopixels would need to be printed, requiring 8 days to print at \unit[50]{Hz}.

The hologram printer could be adapted to print full parallax holograms, similar to the work done by \citet{hrynkiw2008mwc}, although this can introduce problems if printed in three of more colors. The printer could also be adapted to produce transmission holograms.  This would require the reference beam to strike the hologram plate from the same side as the object beam.  The only difficulty currently is that the objective and its mount block the path of the beam.  This should be easily solvable, however.

Making the printer physically smaller and more compact would be useful, ideally reducing the size down to the size of a typical office laser printer.  This could possibly be achieved by using a pulsed laser diode, and by shrinking the afocal telescopic reversing lens system.

Building and maintaining the laser requires a high degree of patience and skill.  It would be interesting to investigate an automatic alignment system of the mirrors and lenses, possibly based on piezoelectric actuators to provide between three to six degrees of freedom as required \citep{chang2002pmd}.  This could use the energy meter demonstrated with a feedback system to scan the parameter space for an optimal alignment, maximizing both beam energy and energy stability.  Such a system, combined with automatic re-alignment as needed, could drastically reduce printer downtime and reduce the requirement for a full-time laser specialist.

The use of a diode pumped laser instead of a flash-lamp pumped laser could also increase reliability.  This would reduce the continual maintenance requirement of replacing the flash lamp.  Diode pumping could also drastically reduce the unwanted acoustic noise of the printer - by eliminating the noise produced by the flash lamp and by reducing the cooling requirements, allowing for a quieter but less powerful cooling system.  

The temperature-energy feedback system used to compensate for laser cavity optical-length fluctuations worked by heating a metal block which thermally expanded, adjusting the position of the rear mirror.
This approach was found to work well with a long cavity laser, but did not work well when extended to a short-cavity laser.  Using a piezoelectric actuator instead would allow extension of the control scheme to a short-cavity laser.

On the software side, the rendered images could be preprocessed to correct for the 'ghosting' effect in holograms, using the parameters determined.

A more fundamental change would be to remove the reliance of wet processing.  This could be achieved through the use of photopolymers or by direct abrasion or ablation of a material.  This would truly open up the realization of an office hologram printer.

\renewcommand\chapterheadstartvskip{\vspace*{2\baselineskip}}

\bibliographystyle{unsrtnat}
\bibliography{hologram}

\begin{thebibliography}{84}
\providecommand{\natexlab}[1]{#1}
\providecommand{\url}[1]{\texttt{#1}}
\expandafter\ifx\csname urlstyle\endcsname\relax
  \providecommand{\doi}[1]{doi: #1}\else
  \providecommand{\doi}{doi: \begingroup \urlstyle{rm}\Url}\fi

\bibitem[Chartier(2005)]{IntroOptics}
Germain Chartier.
\newblock \emph{Introduction to Optics}.
\newblock Springer, May 2005.

\bibitem[Gabor et~al.(1965)Gabor, Stroke, Brumm, FunkHouser, and
  Labeyrie]{gabor1965rpo}
D.~Gabor, G.~W. Stroke, D.~Brumm, A.~FunkHouser, and A.~Labeyrie.
\newblock {Reconstruction of Phase Objects by Holography}.
\newblock \emph{Nature}, 208\penalty0 (5016):\penalty0 1159--1162, 1965.

\bibitem[Gabor(1972)]{gabor1972h}
D.~Gabor.
\newblock {Holography, 1948-1971}.
\newblock \emph{Science}, 177\penalty0 (4046):\penalty0 299--313, 1972.

\bibitem[Gabor and Stroke(1968)]{gabor1968tdh}
D.~Gabor and GW~Stroke.
\newblock {The Theory of Deep Holograms}.
\newblock \emph{Proceedings of the Royal Society of London. Series A,
  Mathematical and Physical Sciences}, 304\penalty0 (1478):\penalty0 275--289,
  1968.

\bibitem[Talbot(1841)]{talbot1841aca}
H.F. Talbot.
\newblock An account of some recent improvements in photography.
\newblock 1841.

\bibitem[Archer(1852)]{archer1852cp}
F.S. Archer.
\newblock {Collodion Process}.
\newblock \emph{Notes and Queries}, \penalty0 (165):\penalty0 612--612, 1852.

\bibitem[Eastman et~al.(1884)]{1884pf}
Eastman et~al.
\newblock Photographic film, Oct 1884.
\newblock US Patent 306,470.

\bibitem[Dillon et~al.(1976)Dillon, Brault, Horak, Garcia, Martin, and
  Light]{dillon1976icf}
PLP Dillon, AT~Brault, JR~Horak, E.~Garcia, TW~Martin, and WA~Light.
\newblock {Integral color filter arrays for solid state imagers}.
\newblock \emph{1976 International Electron Devices Meeting}, 22, 1976.

\bibitem[Thomas B.~Greenslade(2004)]{greenslade:76}
Jr. Thomas B.~Greenslade.
\newblock Wire diffraction gratings.
\newblock \emph{The Physics Teacher}, 42\penalty0 (2):\penalty0 76--77, 2004.

\bibitem[Gabor et~al.(1946)]{gabor1946tc}
D.~Gabor et~al.
\newblock Theory of communication.
\newblock \emph{Institution of Electrical Engineering}, 93\penalty0
  (4):\penalty0 29--4, 1946.

\bibitem[Gabor(1944)]{gabor1944spa}
D.~Gabor.
\newblock System of photography and projection in relief, 1944.
\newblock US Patent 2,351,032.

\bibitem[Gabor(1965)]{gabor1965ite}
D.~Gabor.
\newblock Information theory in electron microscopy.
\newblock \emph{Lab Invest}, 14:\penalty0 801--7, 1965.

\bibitem[Gabor(1949)]{gabor1949mrw}
D.~Gabor.
\newblock {Microscopy by Reconstructed Wave-Fronts}.
\newblock \emph{Proceedings of the Royal Society of London. Series A,
  Mathematical and Physical Sciences}, 197\penalty0 (1051):\penalty0 454--487,
  1949.

\bibitem[Gabor(1948)]{gabor1948nmp}
D.~Gabor.
\newblock A new microscopic principle.
\newblock \emph{Nature}, 161\penalty0 (4098):\penalty0 777--778, 1948.

\bibitem[Gabor(1969{\natexlab{a}})]{gabor1969ph}
D.~Gabor.
\newblock Progress in holography.
\newblock \emph{Reports on Progress in Physics}, 32\penalty0 (1):\penalty0
  395--404, 1969{\natexlab{a}}.

\bibitem[Gabor(1969{\natexlab{b}})]{gabor1969tdp}
D.~Gabor.
\newblock Three-dimensional picture projection, 1969{\natexlab{b}}.
\newblock US Patent 3,479,111.

\bibitem[Leith and Upatnieks(1961)]{leith1961ntw}
E.N. Leith and J.~Upatnieks.
\newblock New techniques in wavefront reconstruction.
\newblock \emph{J. Opt. Soc. Am}, 51:\penalty0 1469--1469, 1961.

\bibitem[Maiman(1960)]{maiman1960sor}
TH~Maiman.
\newblock {Stimulated optical radiation in ruby masers}.
\newblock \emph{Nature}, 187\penalty0 (4736):\penalty0 493--494, 1960.

\bibitem[Maiman et~al.(1961)Maiman, Hoskins, D'Haenens, Asawa, and
  Evtuhov]{maiman1961soe}
TH~Maiman, RH~Hoskins, IJ~D'Haenens, CK~Asawa, and V.~Evtuhov.
\newblock {Stimulated Optical Emission in Fluorescent Solids. II. Spectroscopy
  and Stimulated Emission in Ruby}.
\newblock \emph{Physical Review}, 123\penalty0 (4):\penalty0 1151--1157, 1961.

\bibitem[Leith and Upatnieks(1962)]{Leith:62}
Emmett~N. Leith and Juris Upatnieks.
\newblock Reconstructed wavefronts and communication theory.
\newblock \emph{J. Opt. Soc. Am.}, 52\penalty0 (10):\penalty0 1123--1130, 1962.

\bibitem[Leith and Upatnieks(1964)]{leith1964wrd}
E.N. Leith and J.~Upatnieks.
\newblock {Wavefront reconstruction with diffused illumination and
  three-dimensional objects}.
\newblock \emph{J. Opt. Soc. Am}, 54\penalty0 (11):\penalty0 1295--1301, 1964.

\bibitem[Benton(1977)]{benton1977wlt}
S.A. Benton.
\newblock White light transmission/reflection holographic imaging.
\newblock \emph{Applications of Holography and Optical Data Processing,
  Pergamon Press, Oxford}, 1977.

\bibitem[Benton et~al.(1980)Benton, Mingace~Jr, and Walter]{benton1980osw}
SA~Benton, HS~Mingace~Jr, and WR~Walter.
\newblock {One-step white-light transmission holography}.
\newblock \emph{Recent advances in holography; Proceedings of the Seminar, Los
  Angeles, Calif., February 4, 5, 1980.(A81-32403 14-35) Bellingham, Wash.,
  Society of Photo-Optical Instrumentation Engineers, 1980, p. 156-161.}, 1980.

\bibitem[Bjelkhagen(1992)]{bjelkhagen1992hpm}
H.I. Bjelkhagen.
\newblock {Holographic Portraits Made by Pulse Lasers}.
\newblock \emph{Leonardo}, 25\penalty0 (5):\penalty0 443--448, 1992.

\bibitem[De~Bitetto(1970)]{debitetto1970b}
D.J. De~Bitetto.
\newblock Holographic image system and method employing narrow strip holograms,
  1970.
\newblock US Patent 3,547,510.

\bibitem[{De Bitetto}(1968)]{1968ApPhL..12..176D}
D.~J. {De Bitetto}.
\newblock {Bandwidth Reduction of Hologram Transmission Systems by Elimination
  of Vertical Parallax}.
\newblock \emph{Applied Physics Letters}, 12:\penalty0 176--178, March 1968.

\bibitem[{Bell Telephone Labor Inc}(1974)]{patent:3832027}
{Bell Telephone Labor Inc}.
\newblock Synthetic hologram generation from a pluarlity of two-dimensional
  views, August 1974.
\newblock US Patent 3,832,027.

\bibitem[McGrew(1980)]{patent:4206965}
Stephen~P. McGrew.
\newblock System for synthesizing strip-multiplexed holograms, June 1980.
\newblock US Patent 4,206,965.

\bibitem[Grichine et~al.(1998)Grichine, Ratcliffe, and Skokov]{grichine:203}
Mikhail~V. Grichine, David~B. Ratcliffe, and Gleb~R. Skokov.
\newblock Integrated pulsed holography system for mastering and transferring
  onto agfa or vr-p emulsions.
\newblock volume 3358, pages 203--210. SPIE, 1998.

\bibitem[Chen and Yu(1978)]{chen1978osr}
H.~Chen and F.T.S. Yu.
\newblock One-step rainbow hologram.
\newblock \emph{Opt. Lett}, 2\penalty0 (4):\penalty0 85, 1978.

\bibitem[Ohe et~al.(1998)Ohe, Ito, and Watanabe]{patent:EP0697631}
Yasushi Ohe, Hiromitsu Ito, and Niro Watanabe.
\newblock Photosensitive recording material, photosensitive recording medium,
  and process for producing hologram using this photosensitive recording
  medium, June 1998.
\newblock European Patent EP0697631.

\bibitem[King et~al.(1970)King, Noll, and Berry]{King:70}
Marvin~C. King, A.M. Noll, and D.H. Berry.
\newblock A new approach to computer-generated holography.
\newblock \emph{Appl. Opt.}, 9\penalty0 (2):\penalty0 471, 1970.

\bibitem[Kock(1974)]{patent:3843225}
Manfred Kock.
\newblock Method of holographically forming a three-dimensional image from a
  sequence of two-dimensional images of different perspective., October 1974.
\newblock US Patent 3,843,225.

\bibitem[Gale and Kane(1976)]{gale1976gpp}
M.T. Gale and J.~Kane.
\newblock {Generation of permanent phase holograms and relief patterns in
  durable media by chemical etching}, March~16 1976.
\newblock US Patent 3,944,420.

\bibitem[McGrew(1983)]{mcgrew1983fch}
S.P. McGrew.
\newblock {Full-color hologram}, December~20 1983.
\newblock US Patent 4,421,380.

\bibitem[Pizzanelli()]{pizzanelli:ddw}
D.~Pizzanelli.
\newblock {The development of direct-write digital holography}.
\newblock \emph{Digital Holography}.

\bibitem[Davis(1998)]{patent:5822092}
Frank Davis.
\newblock System for making a hologram of an image by manipulating object beam
  characteristics to reflect image data, October 1998.
\newblock US Patent 5,822,092.

\bibitem[Newswanger(1994)]{patent:5291317}
Craig Newswanger.
\newblock Holographic diffraction grating patterns and methods for creating the
  same, March 1994.
\newblock US Patent 5,291,317.

\bibitem[Spierings and Nuland(1992)]{spierings:52}
Walter~C. Spierings and Eric~Van Nuland.
\newblock Development of an office holoprinter ii.
\newblock volume 1667, pages 52--62. SPIE, 1992.

\bibitem[Nuland and Spierings(1993)]{nuland:9}
Eric~Van Nuland and Walter~C. Spierings.
\newblock Development of an office holoprinter iii.
\newblock volume 1914, pages 9--14. SPIE, 1993.

\bibitem[Yamaguchi et~al.(1994)Yamaguchi, Endoh, Honda, and
  Ohyama]{Yamaguchi:94}
Masahiro Yamaguchi, Hideaki Endoh, Toshio Honda, and Nagaaki Ohyama.
\newblock High-quality recording of a full-parallax holographic sterogram with
  a digital diffuser.
\newblock \emph{Opt. Lett.}, 19\penalty0 (2):\penalty0 135, 1994.

\bibitem[Yamaguchi et~al.(1995)Yamaguchi, Koyama, Endoh, Ohyama, Takahashi, and
  Iwata]{yamaguchi:50}
Masahiro Yamaguchi, Takahiro Koyama, Hideaki Endoh, Nagaaki Ohyama, Susumu
  Takahashi, and Fujio Iwata.
\newblock Development of a prototype full-parallax holoprinter.
\newblock volume 2406, pages 50--56. SPIE, 1995.

\bibitem[Kodera(1992)]{kodera1992hj}
M.~Kodera.
\newblock {Holography in Japan}.
\newblock \emph{Leonardo}, 25\penalty0 (5):\penalty0 451--455, 1992.

\bibitem[Ohnuma et~al.(1989)Ohnuma, Nishihara, and Iwata]{ohnuma1989fcr}
K.~Ohnuma, T.~Nishihara, and F.~Iwata.
\newblock {Full color rainbow hologram using a photoresist plate}.
\newblock \emph{Proc. SPIE}, 1051, 1989.

\bibitem[Perlmutter(1987)]{patent:4701006}
Robert~J. Perlmutter.
\newblock Optical-digital hologram recording, October 1987.
\newblock US Patent 4,701,006.

\bibitem[Hamano and Yoshikawa(1998)]{hamano:2}
Tomohisa Hamano and Hiroshi Yoshikawa.
\newblock Image-type cgh by means of e-beam printing.
\newblock volume 3293, pages 2--14. SPIE, 1998.

\bibitem[Ratcliffe et~al.(2008)Ratcliffe, Vergnes, Rodin, and
  Grichine]{ratcliffepatent}
David~Brotherton Ratcliffe, Florian Michel~Robert Vergnes, Alexey Rodin, and
  Mikhail Grichine.
\newblock Holographic printer.
\newblock Patent 7324248, Jan 2008.

\bibitem[Hecht(1990)]{Hecht1990}
Eugene Hecht.
\newblock \emph{Optics}, chapter~8.
\newblock Addison Wesley, 2nd edition, 1990.

\bibitem[{ZEMAX Development Corporation}()]{zemax}
{ZEMAX Development Corporation}.
\newblock 3001 112th avenue ne, suite 202, bellevue, wa 98004-8017 usa.

\bibitem[Carson(1969)]{BasicOptics}
Fred~A. Carson.
\newblock \emph{Basic Optics and Optical Instruments}, chapter~4.
\newblock Courier Dover Publications, 1969.

\bibitem[Oliver(1963)]{oliver1963ssa}
BM~Oliver.
\newblock {Sparkling spots and random diffraction}.
\newblock \emph{Proceedings of the IEEE}, 51\penalty0 (1):\penalty0 220--221,
  1963.

\bibitem[Goodman(1976)]{goodman1976sfp}
JW~Goodman.
\newblock {Some fundamental properties of speckle}.
\newblock \emph{J. Opt. Soc. Am}, 66\penalty0 (11):\penalty0 1145--1150, 1976.

\bibitem[Iwai and Asakura(1996)]{iwai1996src}
T.~Iwai and T.~Asakura.
\newblock {Speckle reduction in coherent information processing}.
\newblock \emph{Proceedings of the IEEE}, 84\penalty0 (5):\penalty0 765--781,
  1996.

\bibitem[Yu and Wang(1973)]{yu1973srh}
FTS Yu and EY~Wang.
\newblock {Speckle reduction in holography by means of random spatial
  sampling}.
\newblock \emph{Applied Optics}, 12\penalty0 (7):\penalty0 1656, 1973.

\bibitem[Kato and Okino(1973)]{kato1973srd}
M.~Kato and Y.~Okino.
\newblock {Speckle reduction by double recorded holograms}.
\newblock \emph{Appl. Opt}, 12:\penalty0 1199--1201, 1973.

\bibitem[KATO et~al.(1975)KATO, NAKAYAMA, and SUZUKI]{kato1975srh}
M.~KATO, Y.~NAKAYAMA, and T.~SUZUKI.
\newblock {Speckle reduction in holography with a spatially incoherent source}.
\newblock \emph{Applied Optics}, 14:\penalty0 1093--1099, 1975.

\bibitem[Trahey et~al.(1986)Trahey, Allison, Smith, and
  Van~Ramm]{trahey1986qas}
GE~Trahey, JW~Allison, SW~Smith, and OT~Van~Ramm.
\newblock {A quantitative approach to speckle reduction via frequency
  compounding}.
\newblock \emph{Ultrasonic imaging(Print)}, 8\penalty0 (3):\penalty0 151--164,
  1986.

\bibitem[Shin and Javidi(2002)]{Shin:02}
Seung-Ho Shin and Bahram Javidi.
\newblock Speckle-reduced three-dimensional volume holographic display by use
  of integral imaging.
\newblock \emph{Appl. Opt.}, 41\penalty0 (14):\penalty0 2644--2649, 2002.

\bibitem[Ambar et~al.(1985)Ambar, Aoki, Takai, and Asakura]{ambar1985msr}
H.~Ambar, Y.~Aoki, N.~Takai, and T.~Asakura.
\newblock {Mechanism of speckle reduction in laser-microscope images using a
  rotating optical fiber}.
\newblock \emph{Applied Physics B: Lasers and Optics}, 38\penalty0
  (1):\penalty0 71--78, 1985.

\bibitem[AMBAR et~al.(1986)AMBAR, AOKI, TAKAI, and ASAKURA]{ambar1986rss}
H.~AMBAR, Y.~AOKI, N.~TAKAI, and T.~ASAKURA.
\newblock {Relationship of speckle size to the effectiveness of speckle
  reduction in laser microscope images using rotating optical fiber}.
\newblock \emph{Optik(Stuttgart)}, 74\penalty0 (1):\penalty0 22--26, 1986.

\bibitem[Ambar et~al.(1986)Ambar, Aoki, Takai, and Asakura]{ambar1986fci}
H.~Ambar, Y.~Aoki, N.~Takai, and T.~Asakura.
\newblock {Fringe contrast improvement in speckle photography by means of
  speckle reduction using vibrating optical fiber}.
\newblock \emph{Optik(Stuttgart)}, 74\penalty0 (2):\penalty0 60--64, 1986.

\bibitem[Wang et~al.(1998)Wang, Tschudi, Halldorsson, and
  Petursson]{wang1998srl}
L.~Wang, T.~Tschudi, T.~Halldorsson, and P.R. Petursson.
\newblock {Speckle reduction in laser projection systems by diffractive optical
  elements}.
\newblock \emph{Applied Optics}, 37\penalty0 (10):\penalty0 1770--1775, 1998.

\bibitem[Michaloski and Stone(2001)]{curtis2001lis}
P.F. Michaloski and B.D. Stone.
\newblock {Laser illumination with speckle reduction}, February~20 2001.
\newblock US Patent 6,191,887.

\bibitem[Zacharovas et~al.(2001)Zacharovas, Rodin, Ratcliffe, and
  Vergnes]{Zacharovas2001}
Stanislovas~J. Zacharovas, Alexey~M. Rodin, David~B. Ratcliffe, and Florian~R.
  Vergnes.
\newblock Holographic materials available from geola.
\newblock \emph{Practical Holography XV and Conference 4296B: Holographic
  Materials VII (Proceedings of SPIE Volume 4296)}, 4296:\penalty0 206--212,
  June 2001.

\bibitem[Slavich()]{Slavich}
Slavich.
\newblock Joint stock co., micron branch co., 2 pl. mendeleeva, 152140
  pereslavl-zalessky, russia.

\bibitem[Markov()]{markov:33}
Vladimir~B. Markov.
\newblock Some characteristics of a single-layer color hologram.

\bibitem[Markov and Khizhnyak()]{markov:304}
Vladimir~B. Markov and Anatoliy~I. Khizhnyak.
\newblock Selective characteristics of single-layer color holograms.

\bibitem[Bjelkhagen et~al.()Bjelkhagen, Huang, and Jeong]{bjelkhagen:104}
Hans~I. Bjelkhagen, Qiang Huang, and Tung~H. Jeong.
\newblock Progress in color reflection holography.

\bibitem[Liebling(2004)]{liebling2004fif}
M.~Liebling.
\newblock \emph{{On Fresnelets, Interference fringes, and digital holography}}.
\newblock PhD thesis, {\'E}COLE POLYTECHNIQUE F{\'E}D{\'E}RALE DE LAUSANNE,
  2004.

\bibitem[Sandoz(1997)]{sandoz1997wtp}
P.~Sandoz.
\newblock {Wavelet transform as a processing tool in white-light
  interferometry}.
\newblock \emph{Optics Letters}, 22\penalty0 (14):\penalty0 1065--1067, 1997.

\bibitem[Recknagel and Notni(1998)]{recknagel1998awl}
R.J. Recknagel and G.~Notni.
\newblock {Analysis of white light interferograms using wavelet methods}.
\newblock \emph{Optics Communications}, 148\penalty0 (1-3):\penalty0 122--128,
  1998.

\bibitem[Sotthivirat and Fessler(2004)]{sotthivirat2004pli}
S.~Sotthivirat and J.A. Fessler.
\newblock {Penalized-likelihood image reconstruction for digital holography}.
\newblock \emph{Journal of the Optical Society of America A}, 21\penalty0
  (5):\penalty0 737--750, 2004.

\bibitem[Allebach et~al.(1976)Allebach, Liu, and Gallagher]{allebach1976aed}
JP~Allebach, B.~Liu, and NC~Gallagher.
\newblock {Aliasing error in digital holography}.
\newblock \emph{Applied Optics}, 15\penalty0 (9):\penalty0 2183--2188, 1976.

\bibitem[Evans(2003)]{evans2003rla}
A.K. Evans.
\newblock {Resolution limits and noise reduction in digital holographic
  microscopy}.
\newblock \emph{Proceedings of SPIE}, 4659:\penalty0 35, 2003.

\bibitem[Jacquot et~al.(2001)Jacquot, Sandoz, and Tribillon]{jacquot2001hrd}
M.~Jacquot, P.~Sandoz, and G.~Tribillon.
\newblock {High resolution digital holography}.
\newblock \emph{Optics Communications}, 190\penalty0 (1-6):\penalty0 87--94,
  2001.

\bibitem[Vorzobova(2004)]{Vorzobova:04}
N.~D. Vorzobova.
\newblock Pulsed holographic recording under various external lighting
  conditions.
\newblock \emph{J. Opt. Technol.}, 71\penalty0 (10):\penalty0 682--684, 2004.

\bibitem[{Geola Technologies Ltd}()]{Geola}
{Geola Technologies Ltd}.
\newblock Geola uab., 2006 vilnius, lithuania.

\bibitem[Zacharovas et~al.(2000)Zacharovas, Ratcliffe, Skokov, Vorobyov,
  Kumonko, and Sazonov]{zacharovas:73}
Stanislovas~J. Zacharovas, David~B. Ratcliffe, Gleb~R. Skokov, Sergey~P.
  Vorobyov, Petr~I. Kumonko, and Yury~A. Sazonov.
\newblock Recent advances in holographic materials from slavich.
\newblock volume 4149, pages 73--80. SPIE, 2000.

\bibitem[Koo(1984)]{patent:EP0109254}
Yuk Wah~Joseph Koo.
\newblock Single mode pulsed laser., May 1984.
\newblock European Patent EP0109254.

\bibitem[B\'alint(2005)]{andras}
Andr\'as B\'alint.
\newblock \emph{Instrument Engineers Handbook Process Control and
  Optimization}.
\newblock 2005.

\bibitem[Rodin et~al.(2007)Rodin, Vergnes, and
  Brotherton-Ratcliffe]{rodin2007dwh}
A.~Rodin, F.M.R. Vergnes, and D.~Brotherton-Ratcliffe.
\newblock {Direct write holographic printer}, April~3 2007.
\newblock US Patent 7,199,814.

\bibitem[Bjelkhagen and Mirlis(2008)]{bjelkhagen2008chp}
H.I. Bjelkhagen and E.~Mirlis.
\newblock {Color holography to produce highly realistic three-dimensional
  images}.
\newblock \emph{Applied Optics}, 47\penalty0 (4):\penalty0 123--133, 2008.

\bibitem[Hrynkiw et~al.(2008)Hrynkiw, Rodin, and
  Brotherton-Ratcliffe]{hrynkiw2008mwc}
L.~Hrynkiw, A.~Rodin, and D.~Brotherton-Ratcliffe.
\newblock {Method of writing a composite 1-step hologram}, February~19 2008.
\newblock US Patent 7,333,252.

\bibitem[Chang(2002)]{chang2002pmd}
T.N. Chang.
\newblock {Piezoelectric multiple degree of freedom actuator}, March~19 2002.
\newblock US Patent 6,359,370.

\end{thebibliography}
\renewcommand\chapterheadstartvskip{\vspace*{0\baselineskip}}
\appendix
\chapter{Program listing for image cropping }\label{sec:imagecropping}

This is an example perl program to take a folder of files with the image type 'png' and crops each image with a sliding window.
The code assumes that there are 492 images named 0001.png, 0002.png, etc, each of size 2460$\times$960 pixels that need to be cropped to the size 1230$\times$960.  It utilizes the ImageMagick 'convert' program.

\lstset{language=perl, basicstyle=\ttfamily,
    keywordstyle=\color{blue},commentstyle=\color{red},
    stringstyle=\color{dkgreen},
    numbers=left,
    numberstyle=\tiny\color{gray},
    stepnumber=1,
    numbersep=10pt,
    backgroundcolor=\color{white},
    tabsize=4,
    showspaces=false,
    showstringspaces=false }
\begin{figure}[htp]
\begin{lstlisting}
#!/usr/bin/perl 
for($f=0;$f <= 492; $f++) {
  system "echo $f";
  $command= "echo convert -crop 1230x960+" . (($f-1)*5) . \
     "x0 frame" . sprintf("%04d", $f) . ".png cropped_" . \
      sprintf("%04d", $f) . ".png";
  print $command;
}
\end{lstlisting}
\caption[Program for cropping images with a sliding window]{Program listing for cropping a series of images with a sliding window}
\end{figure}

\chapter{DTC commands}
The communication protocol between the \acl{DTC} boards underwent many revisions.  The final revision of the command set is given in \Fref{tab:communicationProtocol}.  Missing identification numbers are due to deprecated commands due to the revision process.  For completeness, the commands for the shutter system have been left in.  The \ac{DTC} boards were extended in their use to also control shutters in the beam to block the beam as needed.

'D1' indicates an 1 byte number.  'D1 D2' indicates a 16 byte number with most signifance bit first.

\begin{landscape}
\begin{table}
\centering
\begin{tabular}[l]{p{0.6cm}llll}
\toprule
{\bfseries ID} & {\bfseries Returns} & {\bfseries Command} & {\bfseries Arguments} & {\bfseries Description} \\
\midrule
1 & D1 D2 & get\_temp1 & & Get current temperature \\
2 & D1 D2 & get\_temp2 & & Get internal temperature value \\
6 & D1 D2 & get\_status & 1 -- DTC, 2 -- Shutter & Get status of the specified component. \\
8 & & set\_tset & D1 D2 & Set target temperature to heat to \\
9 & D1 D2 & get\_tset & & Get target temperature to heat to \\
10& & set\_kp & D1 & Set Kp -- the proportional coefficient in  PID algorithm \\
11& & set\_ki & D1 & Set Ki -- the integrating coefficient in PID algorithm\\
12&&set\_kd& D1 &Set Kd -- the differential coefficient in  PID algorithm\\
13&&set\_eilimit& D1 &Set maximum number of points to integrate in PID algorithm\\
16&&set\_status&&Reset the error status\\
17&&set\_alltoeeprom&&Save all the parameters to non-volatile EEPROM\\
18&&get\_allfromeeprom&&Load all the parameters to non-volatile EEPROM\\
20&&set\_address& D1 &Set the node address ID. There must only be one node connected\\
21&&set\_shutters& 0 -- Close, 1 -- Open &Open/Close shutter, if attached\\
22&D1&get\_shutters& &Get open/close status from shutter, if attached\\
28&D1&get\_version&&Get protocol version -- to allow for future extensions\\
\bottomrule
\end{tabular}
\caption{Communication protocol for Digital Temperature Controller}
\label{tab:communicationProtocol}
\end{table}
\end{landscape}

\chapter{Hologram printer diagrams}\label{chap:hologramprinterdiagrams}

\Fref{fig:hrip_lcos_big} illustrates the final design of the hologram printer.  For comparison, the hologram printer as detailed by \citet{ratcliffepatent} is illustrated in \Fref{fig:hrip_lcd_big}.  The main difference is that the display system has been changed from a transmissive \ac{LCD} to a reflective \ac{LCOS} and moved away from the objective.  An afocal relay lens system transports the \ac{LCOS} image into the position that the \ac{LCD} original was.

The final objective lens piece is now mounted on a heavy metal board, rather than held by mount.

The final reference beam mirror is now mounted to the optical table via an `A' shaped support.  This isolates it from the vibrations from the translation stages.

The hologram plate holder has been modified to allow for fine adjustment, and the power supply and cooling unit upgraded.

The components labelled in the figures are detailed in the Nomenclature section on page~\pageref{sec:printercomponents}.

\begin{landscape}

\begin{figure}[htp]
\centering
\includegraphics[width=\textwidth-28.9pt, angle=90]{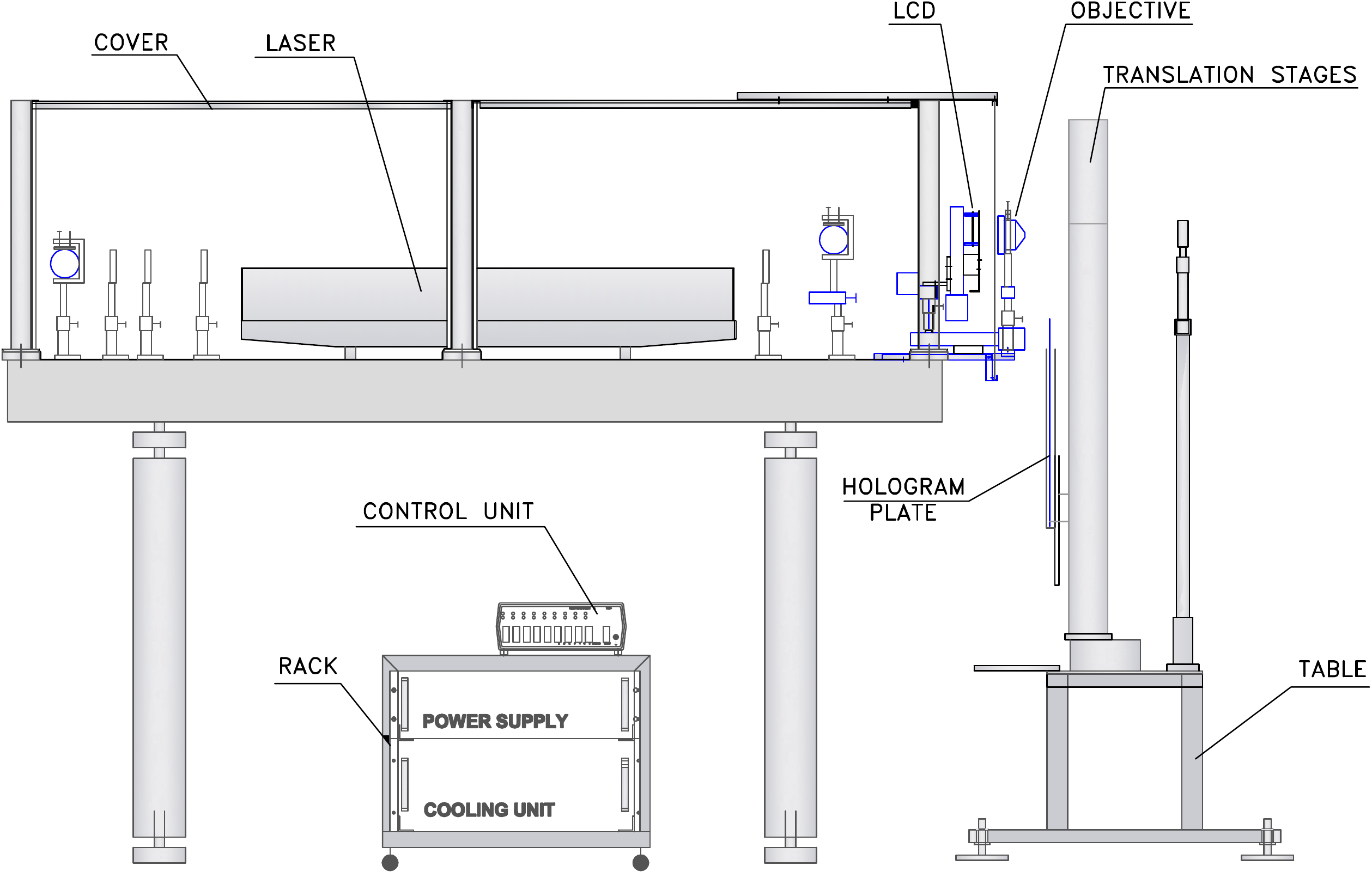}
\caption[Original hologram printer with LCD]{Side-on orthographic projection of original hologram printer with \ac{LCD} display system as detailed by \citet{ratcliffepatent}}
\label{fig:hrip_lcd_big}
\end{figure}

\begin{figure}[htp]
\centering
\includegraphics[width=\textwidth-28.9pt, angle=90]{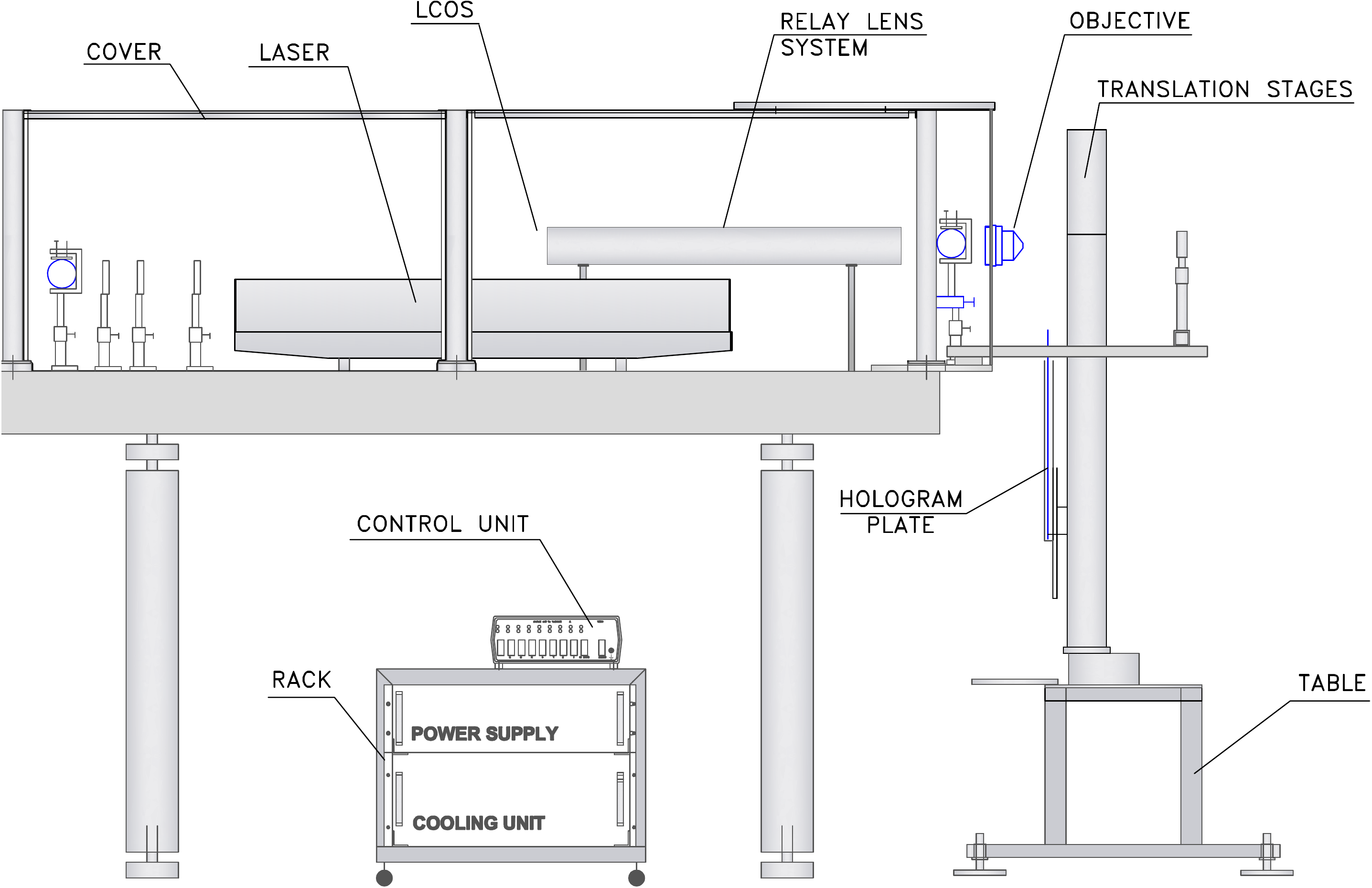}
\caption[Hologram printer with LCOS]{Side-on orthographic projection of Hologram printer with \ac{LCOS} display system}
\label{fig:hrip_lcos_big}
\end{figure}

\begin{figure}[htp]
\centering
\includegraphics[angle=90, width=\hsize]{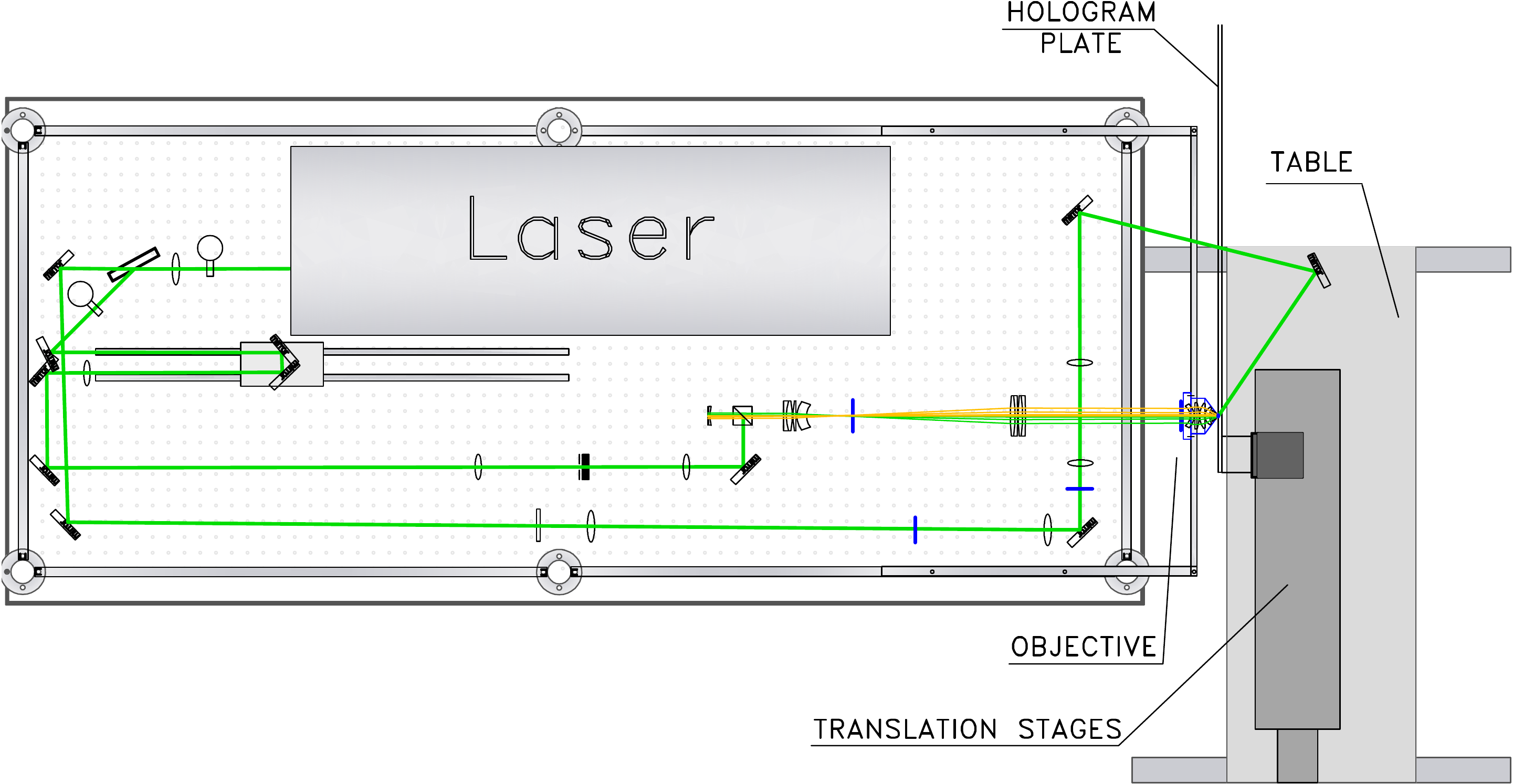}
\caption[Layout of Hologram Printer]{Top-down orthographic projection of Hologram Printer}
\label{fig:hrip_lcos_top_big}
\end{figure}

\begin{figure}[htp]
\centering
\includegraphics[angle=90, width=\hsize]{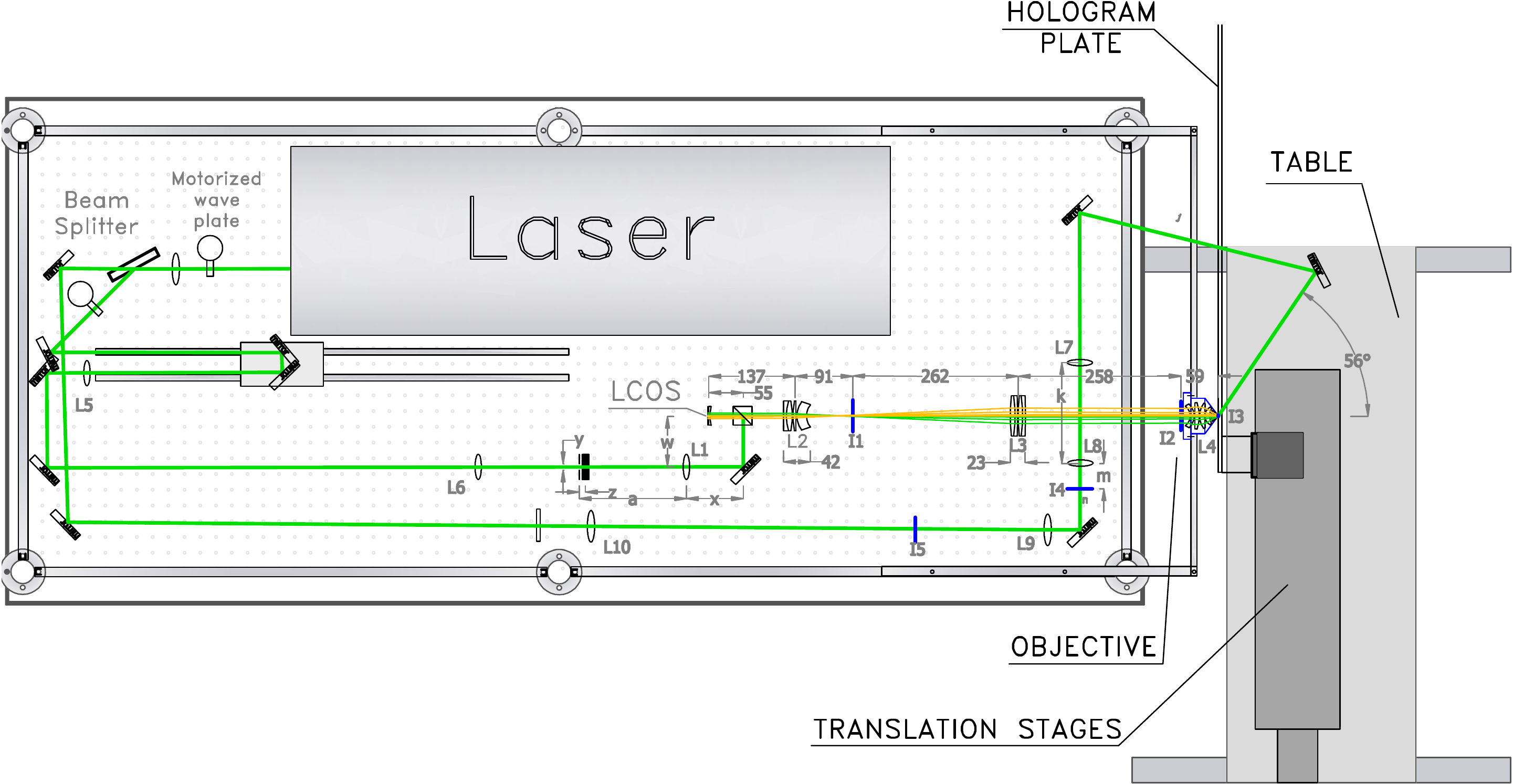}
\caption[Layout of Hologram Printer with dimensions]{Top-down orthographic projection of Hologram Printer with dimensions labelled}
\label{fig:hrip_lcos_top_dimensions_big}
\end{figure}

\begin{figure}[htp]
\centering
\includegraphics[width=\textwidth-28.9pt, angle=90]{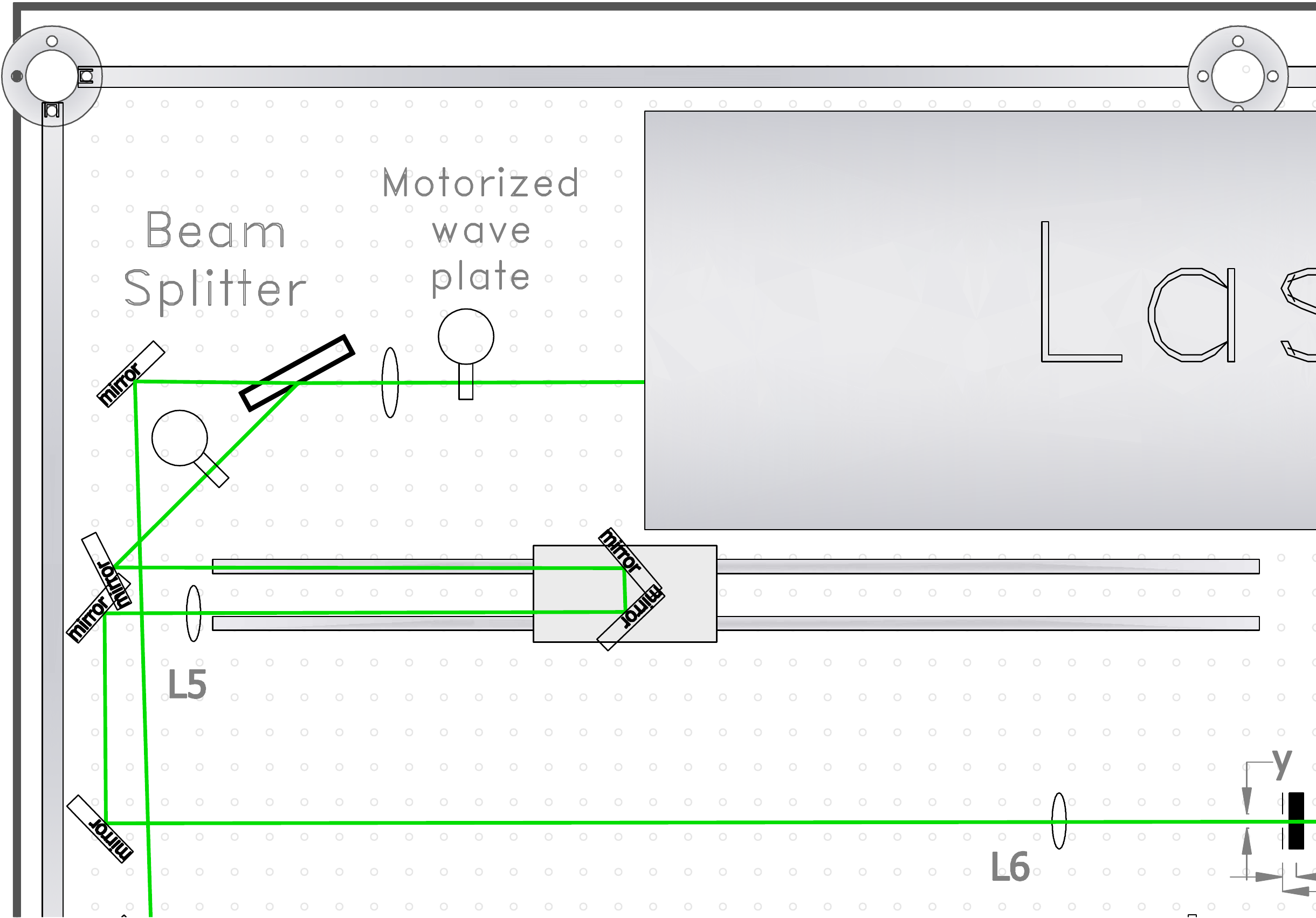}
\caption[Length adjustment system]{Top-down orthographic projection of system to adjust optical-path distance of object beam}
\label{fig:hrip_rails_big}
\end{figure}

\begin{figure}[htp]
\centering
\includegraphics[width=\textwidth-28.9pt, angle=90]{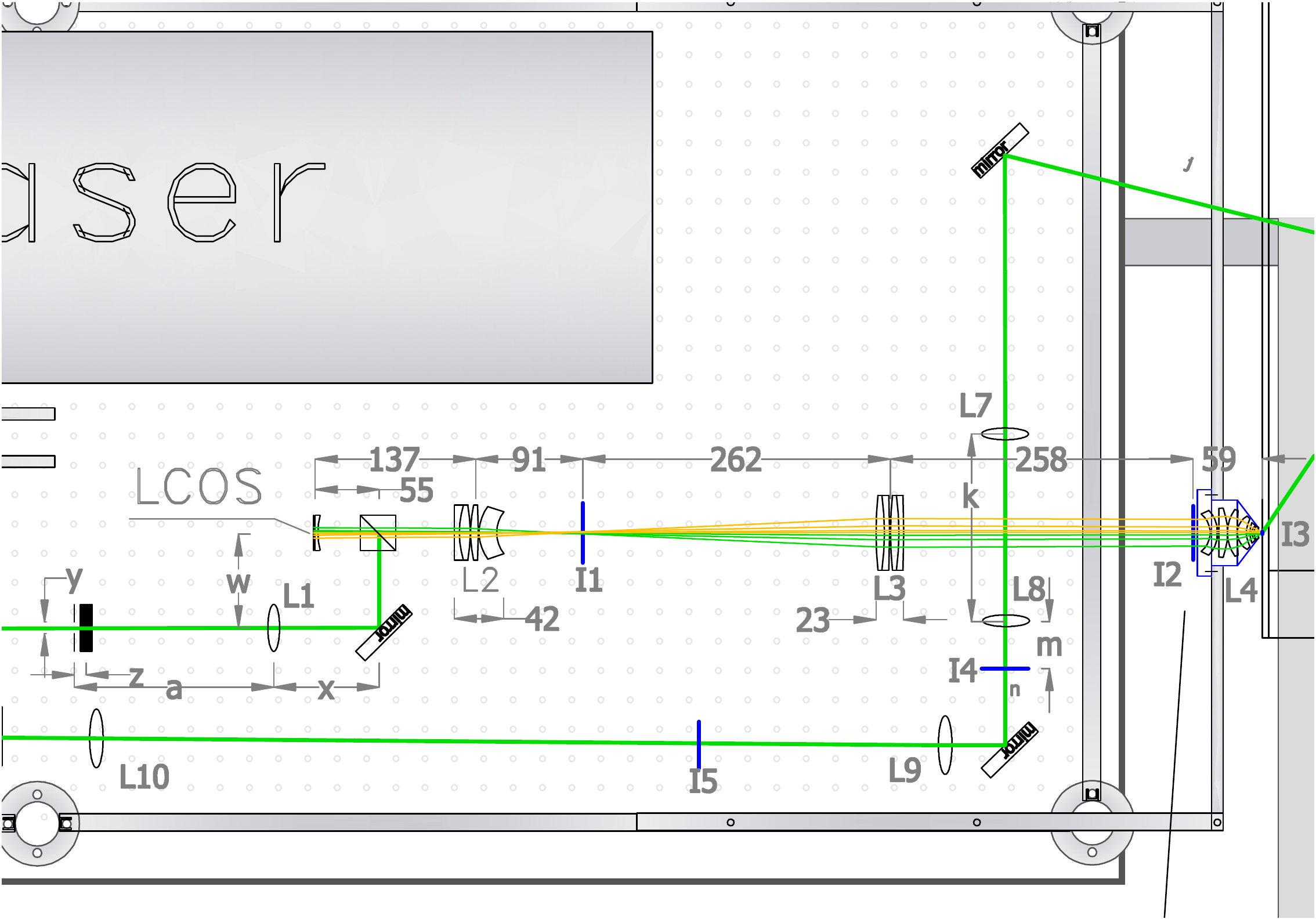}
\caption[Object lens system on printer]{Top-down orthographic projection of object lens system in hologram printer}
\label{fig:hrip_lens_system_big}
\end{figure}

\end{landscape}

\chapter{Program listing for image analysis}\label{sec:whitelogo}

Chapter~\ref{chap:whitelogo} analysed the White Logo hologram with various techniques.  Program listing~\ref{lst:whitelogo} gives a matlab program for the simple analysis of a set of images.  A Guassian-blur filter is optionally applied to each image.

\lstset{language=matlab, basicstyle=\ttfamily,
    keywordstyle=\color{blue},commentstyle=\color{red},
    stringstyle=\color{dkgreen},
    numbers=left,
    numberstyle=\tiny\color{gray},
    stepnumber=1,
    numbersep=10pt,
    backgroundcolor=\color{white},
    tabsize=4,
    showspaces=false,
    showstringspaces=false }
\begin{figure}[htp]
\begin{lstlisting}
% Copyright (C)  John Tapsell  2007
% Whether to apply a Guassian-blur 
% filter to the rendered images:
blur_input_images=true;

% Analysing all the rendered images can take a few minutes. 
% Setting this reuses the values obtained by the last run.
reuse=false;

% Read in the photograph, and convert it to grayscale.
photo = rgb2gray(imread('photo.png'));

% Read in the closest rendered image to the photograph.
% The photograph was taken at close to the normal angle.
closest_match = rgb2gray(imread('white_0360.png'));

% Apply a guassian blur filter to the photograph.
h = fspecial('gaussian',5,1);
blur = imfilter(photo,h);

if ~reuse
  total_intensity = 0;
  for i=0:663 % Read image white_0000.png to white_0663.png
      image_filename = ['white_',num2str(i,'%04d'),'.png'];
      new_image = rgb2gray(imread(image_filename));
      if blur_input_images
          new_image = imfilter(new_image,h);
      end
      total_intensity=total_intensity+im2double(new_image);
  end
end
plot(blur(:), total_intensity(:),'.', 'MarkerSize', 5);

\end{lstlisting}
\caption[Program for White Logo analysis]{Program listing for comparing the photograph of the White Logo hologram and the rendered images that were used to construct the said hologram}
\label{lst:whitelogo}
\end{figure}

\chapter{Optical fourier transform lens system}\label{sec:opticalfourier}

The telecentric afocal reversing lens system for the objective beam of a direct write digital hologram printer is given below in Figures \ref{fig:lenssystem2}, \ref{fig:lenssystem3} and \ref{fig:lenssystem4}.  The lens system shown is suitable for monochromatic light of wavelength \unit[532]{nm} (visible green), entering from the polarizing beam splitter cube.  The rays reflected from the beam splitter cube are not indicated on the diagrams.

\bigskip

The two focal planes, I1 and I2, are also indicated\footnote{Drawn in blue, if in color} on the diagrams.

\bigskip

Figure~\ref{fig:qcad_lenses} shows the lens system in in the direct-write hologram printer.

\begin{landscape}
\begin{figure}[p]
\centering
\includegraphics[angle=-90, width=\hsize]{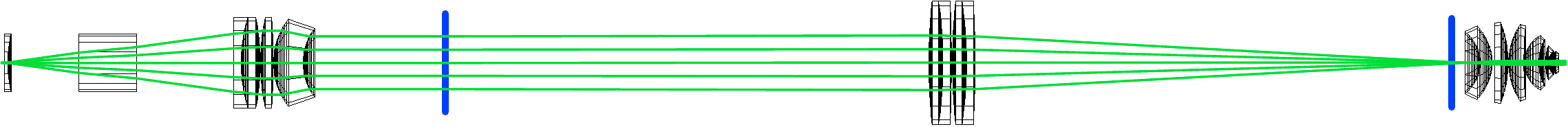}
\caption[Lens system for imaging LCOS Fourier image]{Lens system for imaging the Fourier transform of the \ac{LCOS} (shown on the left) onto the the photosensitive emulsion, producing a holopixel (shown on the right).  Figures \ref{fig:lenssystem2}, \ref{fig:lenssystem3} and \ref{fig:lenssystem4} below provide an enlarged view.  The indicated rays trace out a single point on the \ac{LCOS} display.}
\label{fig:lenssystem_overall}
\end{figure}

\begin{figure}[p]
\centering
\includegraphics[angle=-90, width=\hsize]{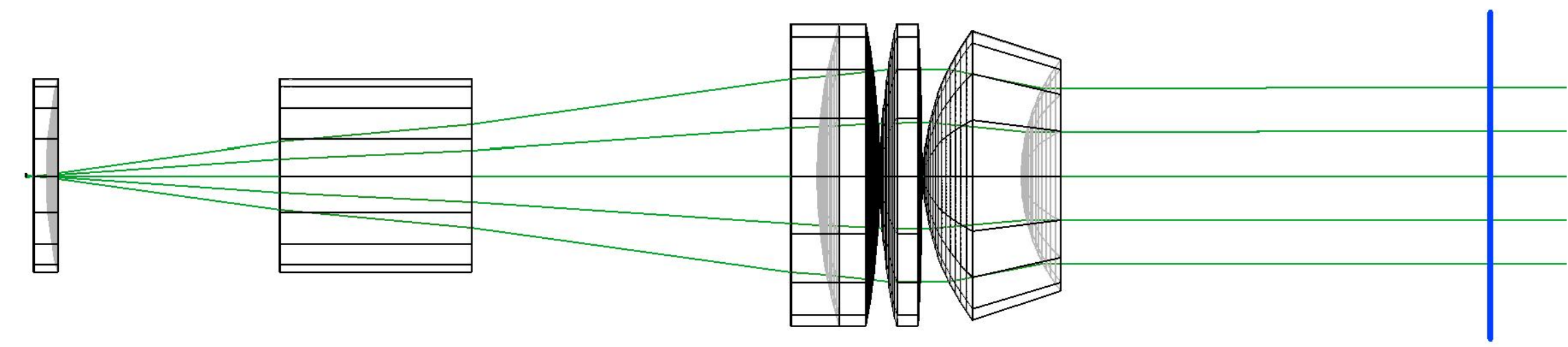}
\caption[Compound Lens L2]{Compound Lens L2 - First component of telecentric afocal reversing lens system.  Images the Fourier transform of the \ac{LCOS} plane onto the real image I2.}
\label{fig:lenssystem2}
\end{figure}

\begin{figure}[ht]
  \centering
  \hfill
  \begin{minipage}[t]{5.5cm}
    \begin{center}  
      \includegraphics[angle=-90, totalheight=10cm]{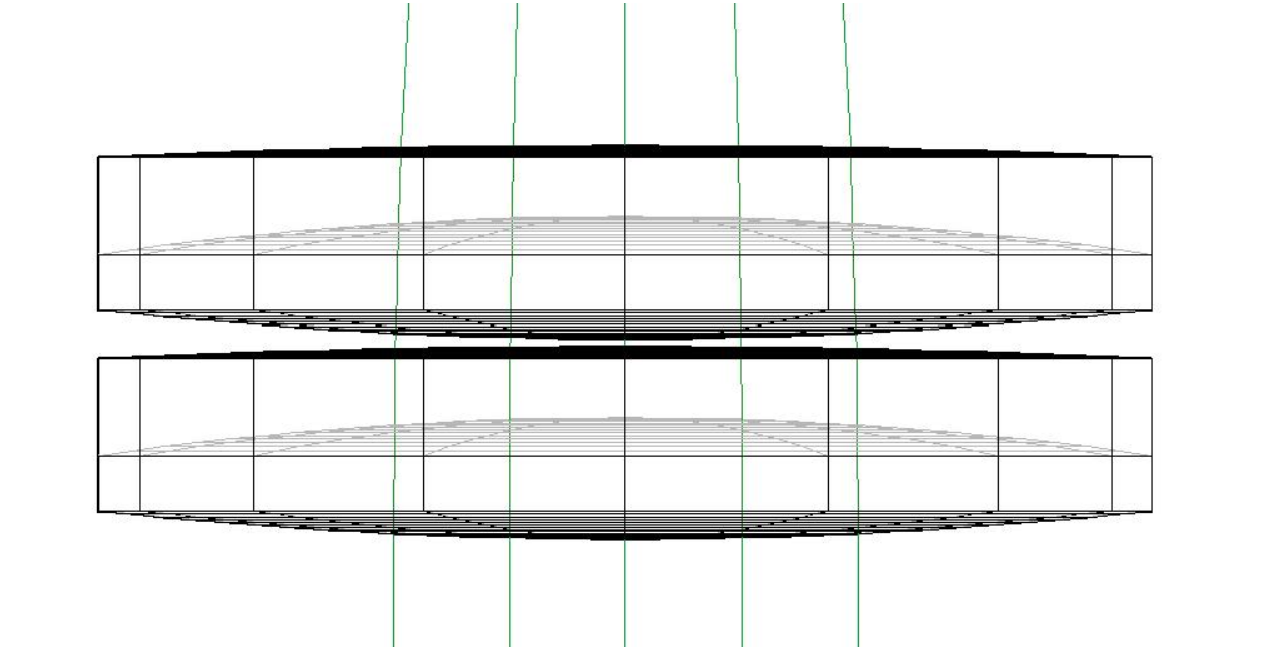}
      \caption[Compound Lens L3]{Compound Lens L3 -- Second component of telecentric lens system}
      \label{fig:lenssystem3}
    \end{center}
  \end{minipage}
  \hfill
  \begin{minipage}[t]{14.6cm}
    \begin{center}  
      \includegraphics[angle=-90, totalheight=10cm]{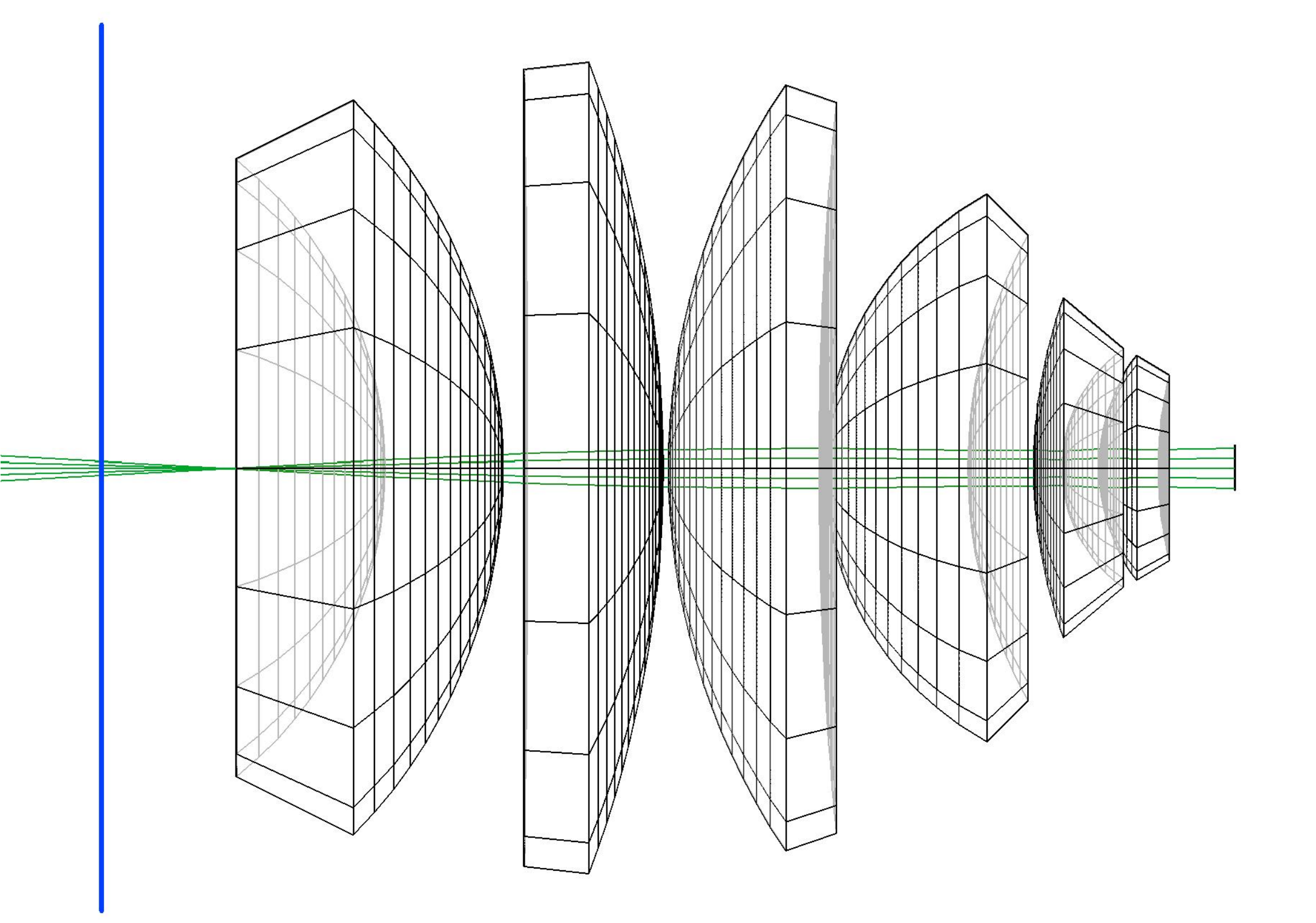}
      \caption[Compound Lens L4]{Compound Lens L4 -- Fourier transforms the real image plane (I2) on the left to the holopixel plane on the right (I3)}
      \label{fig:lenssystem4}
    \end{center}
  \end{minipage}
\hfill
\end{figure}

\begin{figure}[p]
\centering
\includegraphics[angle=-90, width=\hsize]{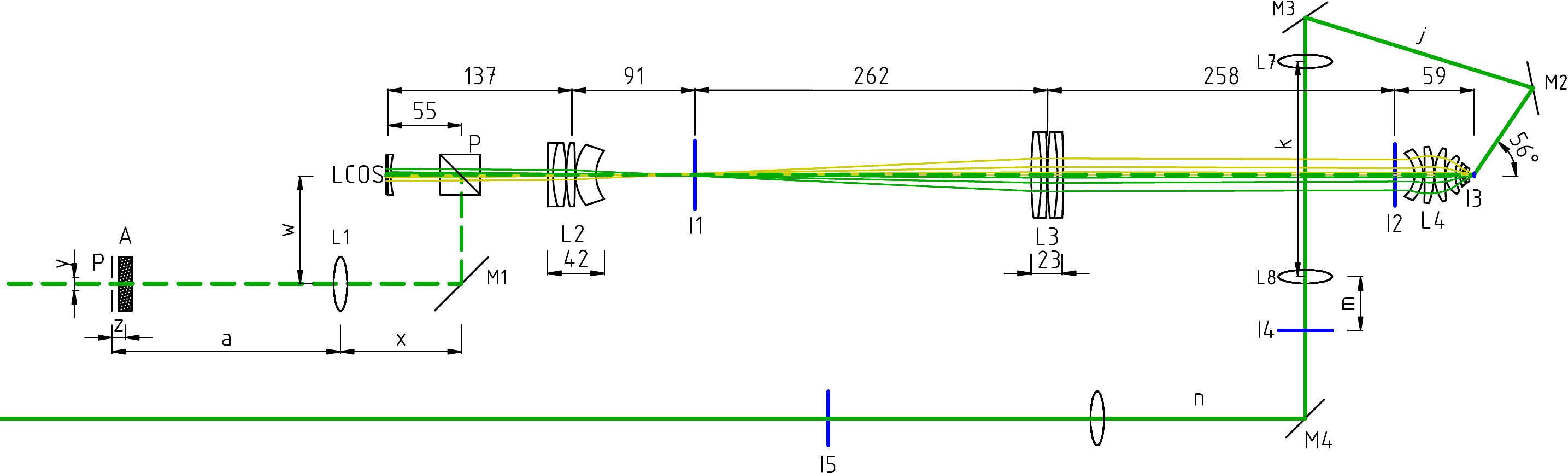}
\caption[Objective Lens system, ray traced]{Top down view of design suitable for direct-write holography of reflection holograms.  Sample light rays are shown for the telescopic system.  These originate parallel and spatially distinct points on the \ac{LCOS} display and focus to single point with distinct angles at the final hologram plane I3. Rays below the axial ray are indicated in yellow and rays above the axial ray are indicated in green, for visual confirmation that the afocal lens system is indeed reversing.}
\label{fig:qcad_lenses}
\end{figure}

\clearpage
\end{landscape}

\chapter{LCOS mechanical mount}\label{chap:lcos_mount}

\begin{figure}[hp]
\centering
\includegraphics[width=\textwidth]{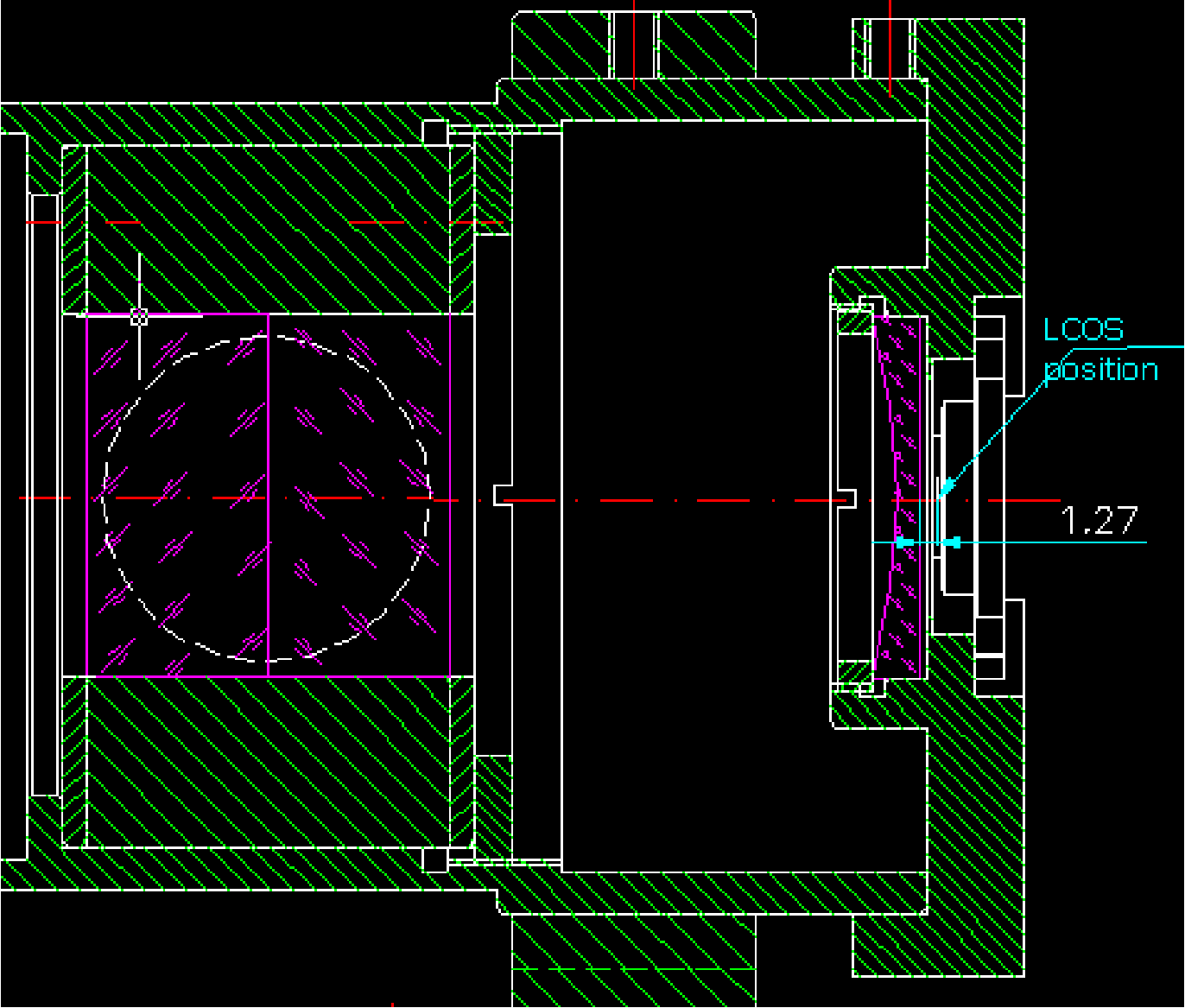}
\caption{Side-on orthogonal view of mechanical mount for LCOS and split beam polariser}
\label{fig:lcos_mount}
\end{figure}

\Fref{fig:lcos_mount} shows that the \ac{LCOS} was mounted in place with a weak corrective lens in front.  This corrective lens ensures that the final projected image has a flat focal plane.  The cube split-beam polariser is mounted in the same holder as shown.  The laser light enters into the page as a plane wave into the polariser cube.  Dimension shown is in millimeters.

\chapter{Lens system}\label{chap:lenscomponents}

The parameters of the lens components in the \ac{LCOS}-based hologram printer's relay lens system are given in \Fref{tab:opticalcomponents}.  Each surface is given with the material and separation distance to the next surface.  All distances are given in millimeters.  Note that surfaces 14 and 21 indicate the focal image planes I2 and I1 respectively.

\clearpage

Optical Component Values - Green Channel

Effective Focal Length = \unit[-7.669]{mm}

Total distance from \ac{LCOS} to image plane I3 = \unit[807.707]{mm}

\begin{table}[ht]
\centering
\begin{tabular}{rrrrll}
\toprule
 No. &    Radius  & Clear Diameter  &         Separation &      Material & Lens\\
\midrule
 1 &     PLANE    &           2.301 &            4.00000 &      Air      &    \\
 2 &    -20.34000 &           9.562 &            3.07000 &      S-SF6    & L4 \\
 3 &     -9.61600 &          11.630 &            1.93000 &      Air      & L4 \\
 4 &     -7.60000 &          12.360 &            1.45000 &      S-SF6    & L4 \\
 5 &    -26.03000 &          17.598 &            3.45000 &      Air      & L4 \\
 6 &    -25.02700 &          24.134 &            7.45000 &      S-SF6    & L4 \\
 7 &    -16.14400 &          28.390 &            0.30000 &      Air      & L4 \\
 8 &   -201.01914 &          37.924 &            7.80000 &      S-SF6    & L4 \\
 9 &    -35.57000 &          39.759 &            0.30000 &      Air      & L4 \\
10 &     59.70000 &          42.100 &            7.03000 &      S-SF6    & L4 \\
11 &   1310.14201 &          41.384 &            1.27000 &      Air      & L4 \\
12 &     27.27000 &          38.142 &            6.15000 &      S-SF6    & L4 \\
13 &     20.51000 &          32.078 &           14.72777 &      Air      & \\
14 &     PLANE    &          46.000 &          246.59396 &      Air      & \\
15 &    693.00000 &          64.000 &            4.33000 &      S-SF5    & L3 \\
16 &    224.90000 &          64.000 &            7.33000 &      S-BK7    & L3 \\
17 &   -304.80000 &          64.000 &            0.50000 &      Air      & L3 \\
18 &    693.00000 &          64.000 &            4.33000 &      S-SF5    & L3 \\
19 &    224.90000 &          64.000 &            7.33000 &      S-BK7    & L3 \\
20 &   -304.80000 &          64.000 &          250.10000 &      Air      & \\
21 &     PLANE    &          50.895 &           73.58000 &      Air      & \\
22 &    -29.51000 &          35.980 &           15.20000 &      J-BAF7   & L2 \\
23 &    -37.10000 &          45.500 &            0.40000 &      Air      & L2 \\
24 &    483.10001 &          47.000 &            6.40000 &      J-SK4    & L2 \\
25 &   -110.15000 &          47.000 &            0.20000 &      Air      & L2 \\
26 &    129.42000 &          47.000 &            9.80000 &      J-SK12   & L2 \\
27 &    -80.91000 &          47.000 &            4.00000 &      J-SF14   & L2 \\
28 &     PLANE    &          47.000 &           50.00000 &      Air      & \\
29 &     PLANE    &          30.000 &           30.00000 &      S-BK7    & Beam cube \\
30 &     PLANE    &          30.000 &           36.68504 &      Air      & \\
31 &    -65.84500 &          30.000 &            2.00000 &      S-SF5    & LCOS lens \\
32 &     PLANE    &          30.000 &            1.15834 &      Air      & LCOS\\
\bottomrule
\end{tabular}
\caption[Lens system optical components]{Optical components in afocal reversing telescopic lens system used in \ac{LCOS} based hologram printer, for green channel.  All distances are given in millimeters.}
\label{tab:opticalcomponents}
\end{table}

\chapter{Laser specifications}\label{sec:laserspecs}
The laser used for the digital hologram printer is a compact single oscillator flashlamp pumped Nd:YAG laser producing emissions in the nanosecond regime.  The laser produces TEM$_{00}$ near-diffraction limited
radiation at \unit[532]{nm}. Single Longitudinal Mode operation is provided with a built-in SLM option. Design features include highly stable passive Q-switched linear oscillator and harmonic generation.  Fundamental lasing is at 1064 nm.

\begin{figure}[ht]
\centering
\includegraphics[width=\hsize*2/3]{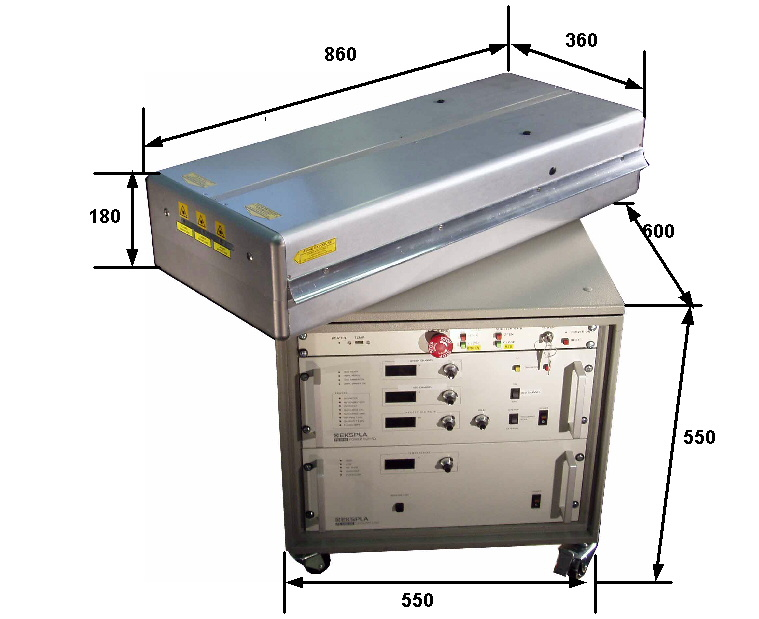}
\caption[Photograph of laser, powersupply and cooling unit]{Photograph of laser, powersupply and cooling unit.  Dimensions shown in mm.}
\label{fig:laser_dim}
\end{figure}

\begin{figure}[ht]
\centering
\includegraphics[width=\hsize]{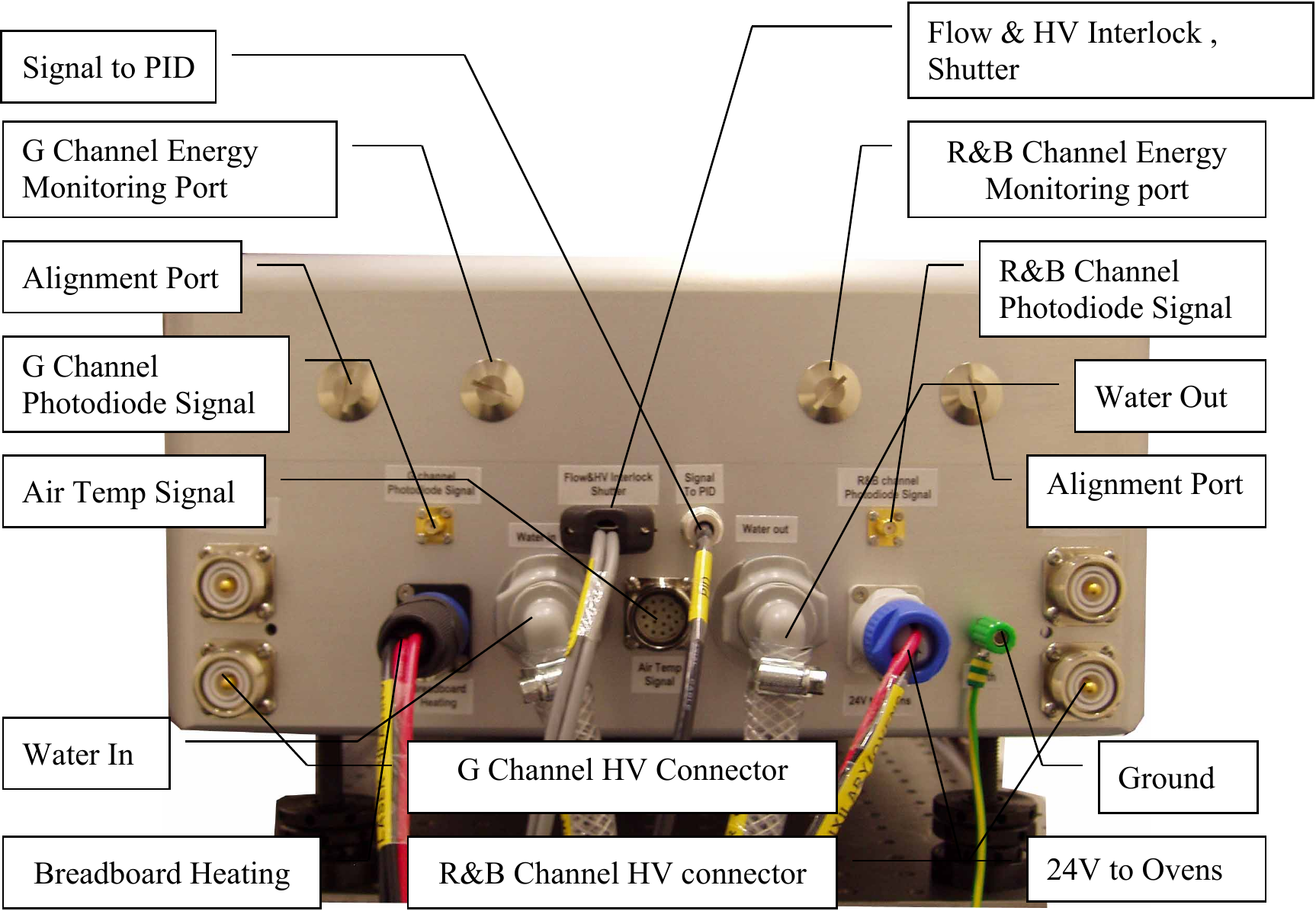}
\caption[Photograph of laser case]{Labelled photograph of the laser case.  Laser shown is the Red-Green-Blue but only the green channel was used for this thesis.  The photodiode and heating connection wires for the temperature-feedback system can be seen.}
\label{fig:laser_sideview}
\end{figure}

\begin{table}[ht]
\centering
\begin{tabular}{ll}
\toprule
Fundamenta Wavelength & \unit[1064]{nm} \\
Output Wavelength & \unit[532]{nm} \\ 
Output Energy & \unit[5]{mJ} \\ 
Shot-to-Shot Energy Stability & \unit[3]{\%} \\ 
Pulsewidth (FHM) & \unit[50]{ns} \\ 
Beam Divergence (1/e$^2$) & \unit[5-7]{mm} \\ 
Beam Divergence & Near diffraction limit \\ 
Linewidth & < \unit[0.002]{cm$^{-1}$}\\ 
Repetition Rate & \unit[50]{Hz} \\ 
Spatial profile & TEM$_{00}$ \\ 
Laser Head Size & 860$\times$360$\times$\unit[180]{mm$^3$} \\ 
Resonator Base Plate & Al-alloy \\ 
Q-switching & Passive \\ 
Flash Tube Lifetime & > 107 shots \\ 
Advised Operating Temperature & \unit[19 - 21]{\degree C} \\ 
Advised Operating Humidity & < \unit[70]{\%} non-condensing \\ 
Powering & \unit[220]{V}, 1 phase \\ 
Warranty Period & 12 months \\
\bottomrule
\end{tabular}
\caption[Specifications of laser used]{Specifications of laser used.  Energy stability figure given is with temperature-energy feedback system described in \Fref{chap:temperature_energy_feedback}.}
\end{table}

\chapter{Microlens array analysis}\label{sec:lensarrayappendix}

The following three tables give the raw data from an experiment with three different microlens arrays.  In order to reduce speckle on the projected image, the three different microlens arrays of were tested at different positions and the results recorded.

The mcrolens arrays are composed of thousands of small lenses (aka \textit{lenslets}) made from BK7 glass.  The lenslets are spherical lenses with a radius of curvature of the order of a millimeter.  The lenslets are rectangularly packed with an aspect ratio close or equal to that of the \ac{LCOS} display system.
	
In the tables, '\textit{Dist. A}' is the distance of the lens array from the closest lens with an uncertainty of $\pm$\unit[0.1]{cm}.  `\textit{Dist. B}' is the distance of the interference pattern from the aperture with an uncertainty of $\pm$\unit[0.5]{cm}, discussed in more detail in \Fref{sec:experimental_evaluation_lens_array}.

\bigskip

The important data to note in the tables is the 'Noise pattern' column.  The image projected from the \ac{LCOS} has speckle.  This image was projected onto a white board \unit[1.2]{m} away from the final objective lens and the repeating patterns in the speckle measured with a ruler, averaging over several.
The overall contrast of the speckle was also noted (lower is better) as well as the overall fidelity of the image.  This was judged by eye.

When the lens array focal point is imaged directly onto the \ac{LCOS}, the energy density on the \ac{LCOS} is sufficiently high to risk breaking the \ac{LCOS}.  This was avoided and noted in the tables were applicable.

\begin{landscape}
\section{First microlens array}
Each lenslet is rectangular with a size of \unit[0.40]{mm}$\times$\unit[0.46]{mm}.  The lenslets are spherical lenses with a radius of \unit[1.3]{mm} 
\begin{center}
\singlespacing
\begin{longtable}{R{1.3cm}R{1.2cm}L{3.5cm}L{5cm}p{9.9cm}}
\multicolumn{5}{A}{\begin{small}Raw data for first lens array continued\end{small}}
\endhead
\endfirsthead
\multicolumn{5}{A}{\begin{small}Continued on next page\end{small}}
\endfoot
\endlastfoot
\toprule
Dist.A (cm) & Dist.B (cm) & Size/shape of beam & Average intensity across LCOS & Noise Pattern\\
\midrule
1.2 &  8.0 &  Too big - too much energy lost & Good in the middle half, faded at edges & Very poor image - strong dominant lines of about \unit[1-2]{cm} separation \\
3.5 &  5.5 &  Just the right size & Majority is good, but faded steps at top and bottom & On the order of a few pixels.  Mediocre quality. 10 big repeating structures = \unit[14.8]{cm}.  Very sensitive to movement.\\
4.0 &  
4.2 &  
Just the right size & 
Almost perfect - just slight fading at very top and bottom. & 
Structure at the pixel level.  10 big repeating structures = \unit[12.5]{cm}  \\
4.5 &  
4.0 &  
Just the right size & 
Almost perfect - just slight fading at very top and bottom.  & 
Structure at the pixel level.  10 big repeating structures = \unit[12.4]{cm}  \\ 
4.7 &  
2.1 &  
Just the right size & 
Almost perfect - just slight fading at very top and bottom.  & 
Structure at the pixel level.  10 big repeating structures = \unit[11.1]{cm}.   Best fidelity and low contrast image.\\
5.0 &  
1.5 &  
Just the right size & 
Almost perfect - just slight fading at very top and bottom.  & 
Structure at the pixel level.  10 big repeating structures = \unit[11.5]{cm}  \\ 
5.2 &  
0.0 &  
Just the right size & 
Almost perfect - just slight fading at very top and bottom.  & 
Structure at the pixel level.  10 big repeating structures = \unit[10.5]{cm}  \\
5.5 &  
-0.8 &  
Just the right size & 
Almost perfect - just slight fading at very top and bottom.  & 
Structure at the pixel level.  10 big repeating structures = \unit[10.0]{cm}. Fidelity isn't too bad. \\
5.7 &  
-1.7 &  
Just the right size & 
Almost perfect - unevenness is just starting to be noticeable by eye & 
Structure at the pixel level.  10 big repeating structures = \unit[10.0]{cm}. Fidelity isn't too bad. \\
6.0 &  
-4.0 &  
Just the right size & 
Almost perfect - just slight fading at very top and bottom. Unevenness is a small bit more noticeable at top and bottom. & 
Structure at the pixel level.  10 big repeating structures = \unit[9.0]{cm}. Fidelity isn't too bad. \\ 
6.5 &  
-6.8 &  
Slightly too small & 
Visible darkening bars at top and bottom quarters. printing.  Possibly even desirable. Software correctable. & 
Structure at the pixel level.  10 big repeating structures = \unit[7.8]{cm}. Loosing fidelity.  Possibly still okay. \\
7.0 &  
< -19 &  
Too small & 
Much more visible darkening bars. Dark lines on top and bottom around the bars. Not suitable for holography. & 
Structure at the pixel level.  10 big repeating structures = \unit[4.3]{cm}. Loosing fidelity.  Possibly still okay. \\
7.5 &  
< -19 &  
Too small & 
Not even at all. Looks like a spherical waffle. Not suitable for holography. & 
Structure at the pixel level.  30 big repeating structures = 13.5,14.9,15.2,\unit[15.1]{cm}. Terrible fidelity. \\
8.0 &  
< -19 &  
Too small & 
Not even at all. Looks like a spherical waffle. Not suitable for holography. & 
Structure at the pixel level.  10 big repeating structures = \unit[3.0]{cm}. Terrible fidelity. \\ \\
\multicolumn{5}{c}{\parbox{\textwidth}{From \unit[7.5]{cm}  it became more and more difficult to measure the repeating pattern.  The pattern no longer repeated so obvious and was too hard to count. At \unit[8.5]{cm}  the image fidelity is awful. Somewhere around \unit[9.0]{cm}  each lenslet focuses to a point on the LCOS.  This was not measured too accurately for fear of damaging the LCOS.  As the pixel becomes noticeably larger now and has a square shape, despite no aperture at all.}}\\ \\
9.5 &  
Before LCOS & 
Too small & 
Beautifully smooth spherical Gaussian shape & 
Bright points were just off of focus point of each lenslet \\
10.0 &  
Before LCOS & 
Too small & 
Beautifully smooth spherical Gaussian shape & 
Beautifully smooth illumination but only illuminating a small middle section \\ \\
\multicolumn{5}{c}{\parbox{\textwidth}{Past \unit[9.7]{cm}  the beam becomes too large for the first lens in the lens relay telescope.  This causes the energy to drop off quickly and gives a dim useless image.}}\\
\bottomrule
\caption[Raw data results for first lens array]{Raw data results for first lens array.  Each lenslet is rectangular with a size of \unit[0.40]{mm}$\times$\unit[0.46]{mm}.  The lenslets are spherical lenses with a radius of \unit[1.3]{mm} }
\label{tab:lensarray1}
\end{longtable}
\end{center}

\clearpage
\section{Second microlens array}
Each lenslet is rectangular with a size of \unit[0.20]{mm}$\times$\unit[0.23]{mm}.  The lenslets are spherical lenses with a radius of \unit[0.65]{mm}.
Note that the pattern from the lens array on the projected image was much much smaller and harder to see.
\begin{center}
\singlespacing
\begin{longtable}{R{1.3cm}R{1.2cm}L{3.5cm}L{4cm}p{10.9cm}}
\multicolumn{5}{B}{\begin{small}...raw data for second lens array continued\end{small}}
\endhead
\endfirsthead
\multicolumn{5}{B}{\begin{small}continued on next page..\end{small}}
\endfoot
\endlastfoot
\toprule
Dist.A (cm)& 
Dist.B (cm)&Size/shape of beam& 
Average intensity across LCOS & 
Noise Pattern \\
\midrule
2.2 & 
10.5 & 
Just the right size & 
Beautifully smooth. No complaints & 
7 big repeating structures = \unit[25.5]{cm}.  Worst fidelity yet. Horrible \\
3.0 &  
8.0 &  
Just the right size & 
Beautifully smooth. No complaints & 
4 big repeating structures = \unit[11.5]{cm}  but hard to measure.  Bad fidelity but getting better certainly \\
3.5 &  
8.0 &  
Just the right size & 
Beautifully smooth. No complaints & 
4 big repeating structures = \unit[10.0]{cm}  but hard to measure.  fidelity is getting a bit better but still worse then previous lensarray \\
4.0 &  
7.4 &  
Just the right size & 
Beautifully smooth. No complaints & 
5 big repeating structures = \unit[10.0]{cm}  but hard to measure.  fidelity is getting better slowly \\
4.5 &  
6.4 &  
Just the right size & 
Beautifully smooth. No complaints & 
10 big repeating structures = \unit[22.0]{cm}  but hard to measure.  fidelity is getting better slowly \\ 
5.0 &  
5.0 &  
Just the right size & 
Beautifully smooth. No complaints & 
10 big repeating structures = \unit[21.0]{cm}  but hard to measure.  fidelity is still getting better slowly \\
5.5 &  
2.0 &  
Just the right size & 
Beautifully smooth. No complaints & 
10 big repeating structures = \unit[19.0]{cm}  but hard to measure.  fidelity is still getting better slowly \\ 
6.0 &  
0.0 &  
Just the right size & 
Beautifully smooth. No complaints & 
10 big repeating structures = \unit[17.0]{cm}  but hard to measure.  fidelity is looking very nice \\ 
6.5 &  
-3.5 &  
Just the right size & 
Beautifully smooth. No complaints & 
10 big repeating structures = \unit[15.5]{cm}  but hard to measure.  fidelity is looking very nice \\
7.0 &  
-10.0 &  
Just the right size & 
Beautifully smooth. No complaints & 
10 big repeating structures = \unit[11.2]{cm}  but hard to measure.  fidelity is looking very nice \\
7.5 &  
< -17 &  
Starting to shrink & 
Beautifully smooth. No complaints & 
10 big repeating structures = \unit[9.2]{cm}  but hard to measure.  fidelity is getting worse again \\
8.0 &  
< -17 &  
A bit too small & 
Beautifully smooth but fading around the edges & 
30 big repeating structures = \unit[19.0]{cm}  but hard to measure.  fidelity is now horrible \\
8.5 &  
< -17 &  
Smaller & 
Beautifully smooth, almost Gaussian now & 
Can't measure repeating structure.  Awful fidelity! Strong circular pattern appearing \\ \\
\multicolumn{5}{c}{\parbox{\textwidth}{Now it again becomes close to where each lenslet focuses to a point on the LCOS.  Skipping this bit to protect the screen.}}\\ \\
10.1 &  
Before LCOS & 
Too small. & 
 & 
Bright points where just off of focus point of each lenslet \\
10.5 &  
Before LCOS & 
Too small. & 
Smooth spherical Gaussian shape & 
Fidelity isn't that good - repeating small structure.  only illuminating a small middle section \\
\bottomrule
\caption[Raw data results for second lens array]{Raw data results for second lens array.  Each lenslet is rectangular with a size of \unit[0.2]{mm}$\times$\unit[0.23]{mm}.  The lenslets are spherical lenses with a radius of \unit[0.65]{mm}}
\label{tab:lensarray2}
\end{longtable}
\end{center}
\clearpage
\section{Third microlens array}
Each lenslet is rectangular with a size of \unit[0.040]{mm}$\times$\unit[0.046]{mm}.  The lenslets are spherical lenses with a radius of \unit[0.13]{mm}.
\begin{center}
\singlespacing
\begin{longtable}{R{1.3cm}R{1.2cm}L{3.5cm}L{4cm}p{10.9cm}}
\multicolumn{5}{C}{\begin{small}...raw data for third lens array continued\end{small}}
\endhead
\endfirsthead
\multicolumn{5}{C}{\begin{small}continued on next page..\end{small}}
\endfoot
\endlastfoot
\toprule
Dist.A (cm)& 
Dist.B (cm)&Size/shape of beam& 
Average intensity across LCOS & 
Noise Pattern \\
\midrule
0.0 &  
$\sim$12.5 &  
A bit too big & 
Beautifully smooth & 
6 difficult to count repeating structures = \unit[21.5]{cm}.  Good fidelity. \\
2.0 &  
$\sim$10.0 &
Still a bit too big & 
Beautifully smooth & 
6 repeating structures = \unit[19.7]{cm}.  Structure is slight stronger, but still good fidelity. \\
2.5 &  
$\sim$9.5 &  
Good & 
Beautifully smooth & 
7 repeating structures = \unit[23.6]{cm}.  Good fidelity. \\
3.0 &  
$\sim$8.5 &  
Good size  & 
Beautifully smooth & 
8 repeating structures = \unit[25.2]{cm}.  Good fidelity. \\
3.5 &  
$\sim$8.0 &  
Good size & 
Beautifully smooth & 
5 repeating structures = \unit[11.4]{cm}.  Good fidelity. \\
4.0 &  
$\sim$7.5 &  
Good size & 
Beautifully smooth & 
8 repeating structures = \unit[18.7]{cm}.  Good fidelity. \\
4.5 &  
5.5 &  
Good size & 
Beautifully smooth & 
8 repeating structures = \unit[17.9]{cm}.  Good fidelity. \\
5.0 &  
2.1 &  
Good size & 
Beautifully smooth & 
10 repeating structures = \unit[22.0]{cm}.  Good fidelity. \\ 
5.5 &  
0.2 &  
A bit too small & 
Beautifully smooth & 
8 repeating structures = \unit[15.4]{cm}.  Very nice fidelity. \\
6.0 &  
-1.8 &  
3/4 Too small & 
Some gradient appearing & 
 5 repeating structures = \unit[7.8]{cm}.  Very nice fidelity. \\ 
6.5 &  
-5.0 &  
3/4 Too small & 
Some gradient appearing & 
6 repeating structures =\unit[9.3]{cm}.  Great fidelity. \\
7.0 &  
-9.8 &  
3/4 Too small & 
more gradient & 
10 repeating structures = \unit[12.3]{cm}.  Great fidelity. \\
7.5 &  
-15.0 &  
3/4 Too small & 
more gradient & 
8 repeating structures = \unit[8.7]{cm}.  Great fidelity. \\
8.0 &  
< -17   & 
1/2 Too small & 
more gradient & 
10 repeating structures = \unit[7.1]{cm}.  Great fidelity. \\ 
8.5 &  
< -17 &  
1/2 Too small & 
unacceptable gradient & 
10 repeating structures = \unit[5.0]{cm}.  Bad fidelity because we start to see curved patterns from lens array.\\
9.0 &  
< -17 &  
1/3 Too small & 
unacceptable gradient & 
6 repeating structures = \unit[17.7]{cm}  and sub repeating at 10 repeating structures = \unit[2.9]{cm}.  stronger curved patterns from lens array, but actually acceptable probably \\ \\
\multicolumn{5}{c}{\parbox{\textwidth}{Around \unit[9.5]{cm} is close to where the lenslet focuses to a point on the LCOS. Skipping this bit to protect the screen.}} \\ \\
10.0 &  
< -17 &  
1/3 Too small & 
unacceptable gradient & 
Wonderful fidelity.  No repeating structures. \\
10.5 &  
< -17 &  
1/3 Too small & 
smooth but covers a small part & 
Wonderful fidelity.  No repeating structures. \\
11.0 &  
< -17 &  
1/2 Too small & 
smooth but covers a small part - too big for first lens in lens array & 
Wonderful fidelity.  No repeating structures. \\
13.5 &  
< -17 &  
Good size & 
covers only a small part because first lens in lens array is too small & 
Okay fidelity - repeating structure again. \\
\bottomrule
\caption[Raw data results for third lens array]{Raw data results for third lens array.  Each lenslet is rectangular with a size of \unit[0.040]{mm}$\times$\unit[0.046]{mm}.  The lenslets are spherical lenses with a radius of \unit[0.13]{mm}.}
\label{tab:lensarray3}
\end{longtable}
\end{center}

\end{landscape}
\chapter{Hologram printer photographs}

A few photographs taken while working on the printer.  Note that the photographs do not necessarily reflect the final design.  In particular the \ac{LCOS} display and surrounding compound lenses and beam polariser are shown mounted individually, while the final design placed them inside a single enclosed metal tube.

\begin{figure}[p]
\centering
\includegraphics[width=\hsize]{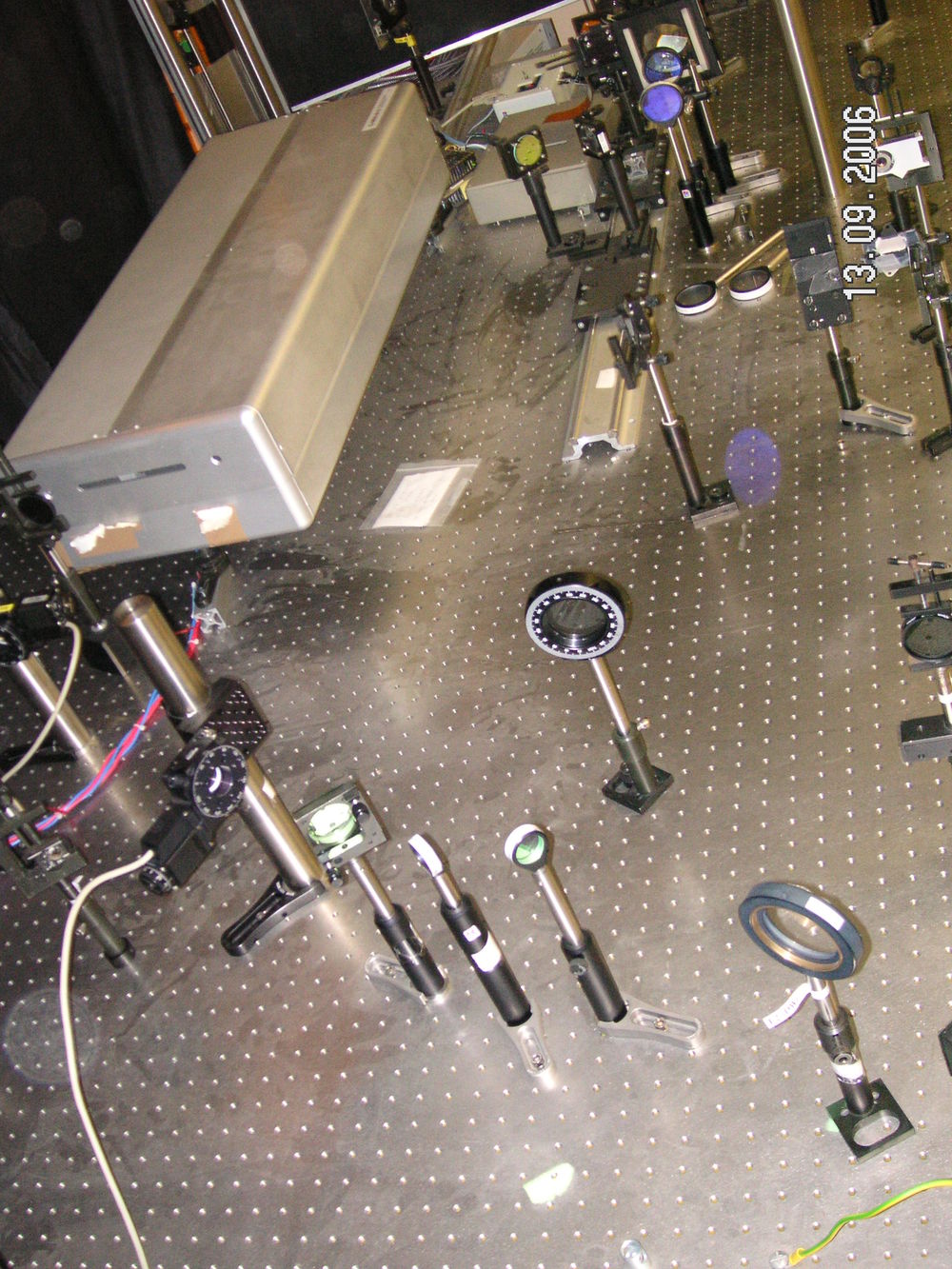}
\caption[Photograph of part of the hologram printer]{Photograph of part of the hologram printer}
\label{fig:topend_photo}
\end{figure}

\begin{landscape}

\begin{figure}[p]
\centering
\includegraphics[height=\vsize-28.9pt]{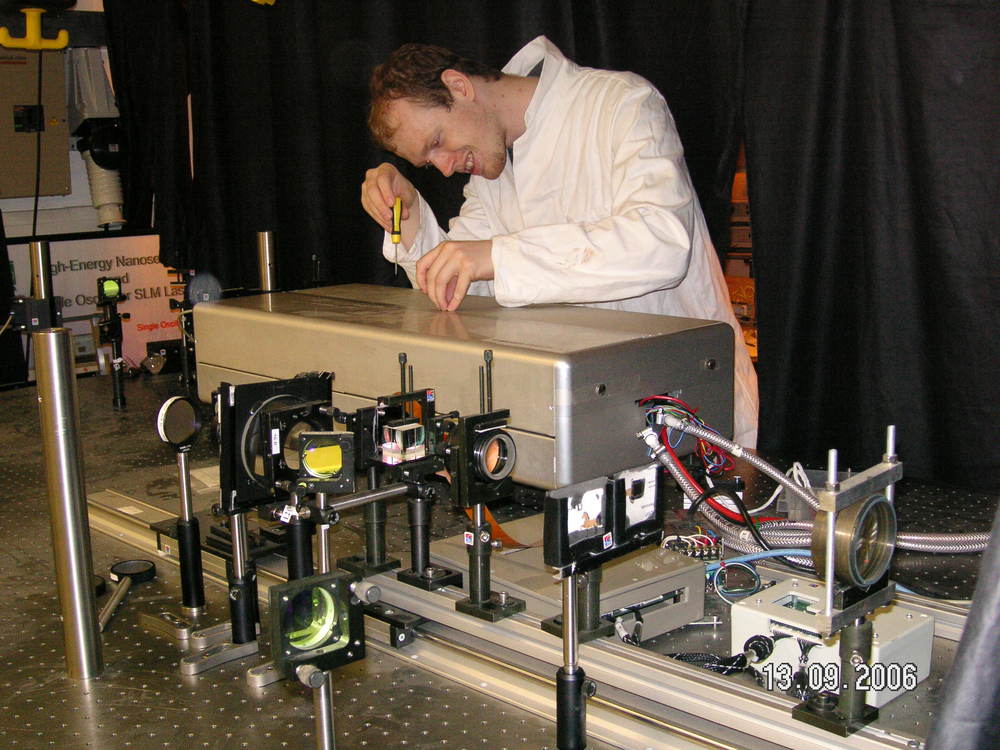}
\caption[Photograph of me working on the hologram printer]{Photograph of me working on the hologram printer}
\label{fig:me_working}
\end{figure}

\begin{figure}[p]
\centering
\includegraphics[height=\vsize-28.9pt]{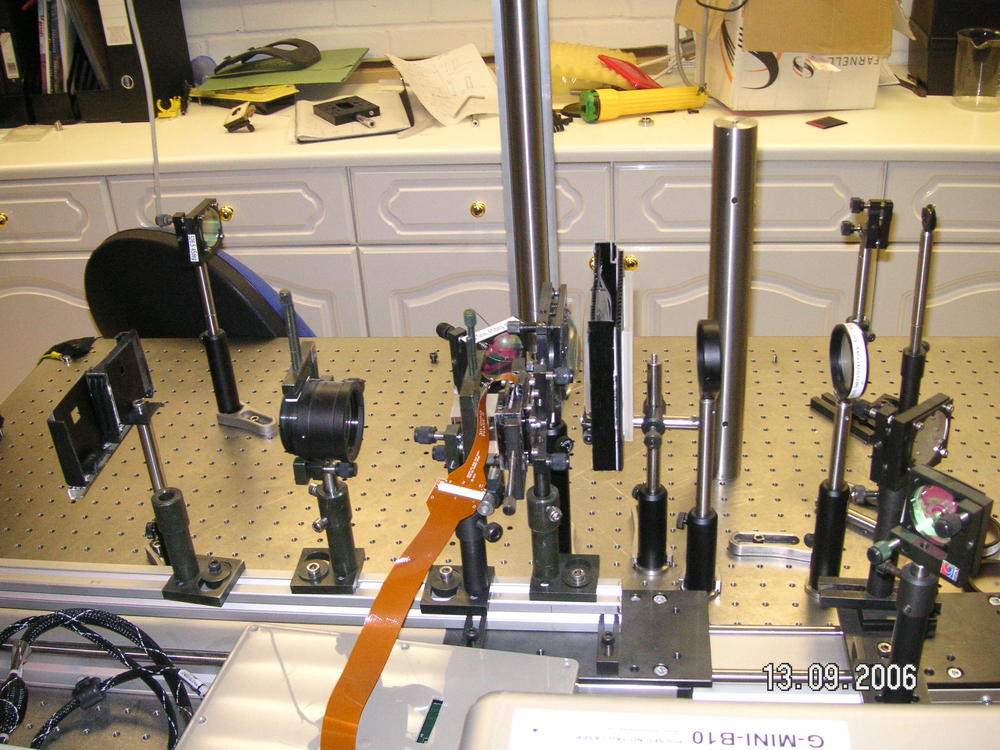}
\caption{Photograph of LCOS section of the hologram printer}
\label{fig:printer_photo}
\end{figure}

\begin{figure}[p]
\centering
\includegraphics[height=\vsize-28.9pt]{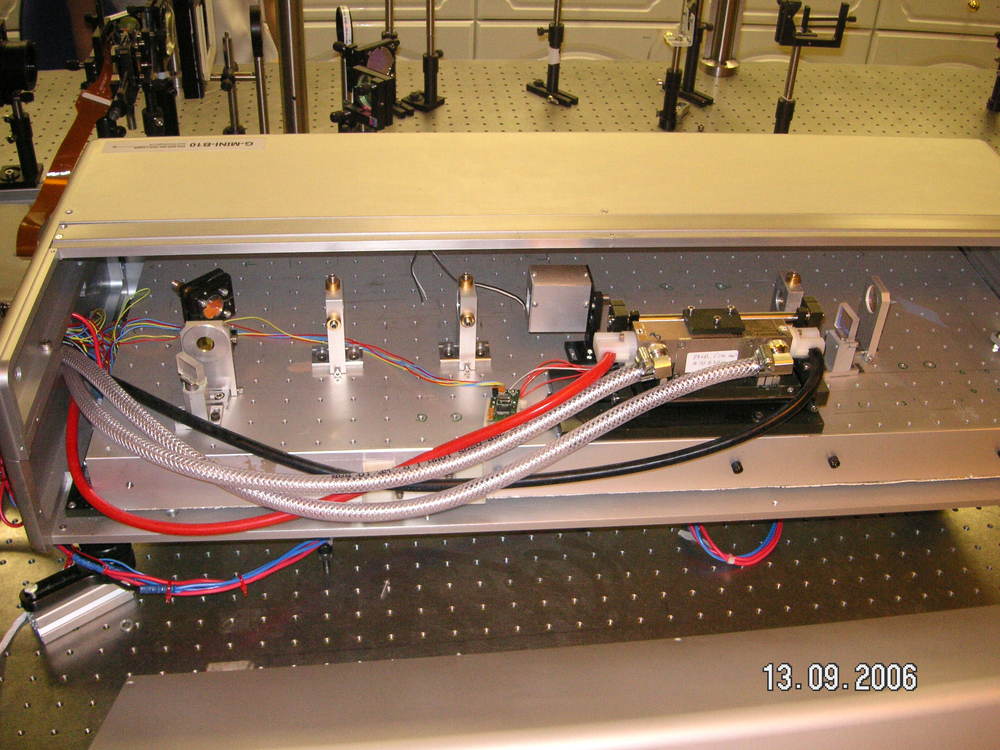}
\caption[Photograph of short cavity laser]{Photograph of frequency-doubled short cavity \ac{NDYAG} laser}
\label{fig:laserpicture}
\end{figure}

\clearpage
\end{landscape}
\end{document}